%% file: Thesis_for_arXiv.tex
\documentclass[letterpaper,12pt,titlepage,oneside,final]{book}
 
\newcommand{\href}[1]{#1} 

\newcommand{\on}[2]{\stackrel{\phantom{}_{#1}}{#2}}

\usepackage{sidecap}

\usepackage{cite}
\usepackage{ifthen}
\newboolean{PrintVersion}
\setboolean{PrintVersion}{false} 

\usepackage{amsmath,amssymb,amstext} 
\usepackage[pdftex]{graphicx} 
\usepackage{epsfig}
\usepackage{times}
\usepackage{epstopdf}
\usepackage[pdftex,pagebackref=false,pdfpagelabels=true,bookmarks=true]{hyperref} 
\hypersetup{
    plainpages=false,       
    unicode=false,          
    pdftoolbar=true,        
    pdfmenubar=true,        
    pdffitwindow=false,     
    pdfstartview={FitH},    
    pdftitle={Rigid Quasilocal Frames},    
	pdfauthor={Paul McGrath},    
	pdfsubject={Gravitational Physics},  
	pdfkeywords={Relativity} {Gravity} {Rigid}, 
    pdfnewwindow=true,      
    colorlinks=true,        
    linkcolor=blue,         
    citecolor=green,        
    filecolor=magenta,      
    urlcolor=cyan           
}
\ifthenelse{\boolean{PrintVersion}}{   
\hypersetup{	
    citecolor=black,%
    filecolor=black,%
    linkcolor=black,%
    urlcolor=black}
}{} 

\let\origdoublepage\cleardoublepage
\newcommand{\clearemptydoublepage}{%
  \clearpage{\pagestyle{empty}\origdoublepage}}
\let\cleardoublepage\clearemptydoublepage

\begin{document}


\input{FrontPages}


\chapter{Introduction}\label{chIntro}

\section{A Brief Introducton to Rigid Quasilocal Frames}\label{secIntroIntro}

For over two hundred years, Newton's laws provided the foundation for how we thought about the physical universe.  Then, with the advent of special and general relativity, it became apparent that we needed to drastically rethink some very basic notions of how physics works at a fundamental level.  We had to take a step back and decide which of these long-standing notions needed to be abandoned and which could be retained.  As we will see in this thesis, one particularly useful Newtonian concept - rigid motion - has very much been forgotten for the past one hundred years but, as it turns out, actually has a natural and useful second life in Einstein's universe.  On the other hand, we will argue that another aspect of the Newtonian mindset - locality - has inadvertently been carried forward when it should have been left behind.

Although not obvious, these two notions - rigidity and locality - are intimately related in the relativistic world.  In fact, as we will see, they are generically in conflict with one another.  Given that we are so deeply ingrained with a local view of the universe, it is easy to see how rigidity was so quickly cast aside.  If one had to make a choice between locality and rigidity, the appeal of locality would always win.  However, in our study of rigid quasilocal frames, we have uncovered enormous benefits to having a useful notion of rigidity.  At the same time, it has become apparent that if you want to take advantage of these benefits, then you have to abandon locality.  We will revisit this point in the next section but first let us look more precisely at what a rigid quasilocal frame is.  In doing so we will be able to better understand this point.

A {\it rigid quasilocal frame} (RQF) is essentially a geometrically natural way to define an extended ``system" in the context of the dynamical spacetime of general relativity.  More specifically, it is defined as a two-parameter family of timelike worldlines comprising the worldtube boundary (topologically $\mathbb{R}\times \text{S}^2$) of the history of a finite spatial volume, with the rigidity conditions that the congruence of worldlines is expansion-free (the ``size'' of the system is not changing) and shear-free (the ``shape'' of the system is not changing).  This definition of a system yields simple, exact geometrical insights into the problem of motion in general relativity.  This is because it begins by answering, in a precise way, the questions {\it what} is in motion (a rigid two-dimensional system boundary with topology $\text{S}^2$, and whatever matter and/or radiation it happens to contain at the moment), and what motions of this rigid boundary are possible.  Furthermore, the mathematics and physics describing the dynamics of the system are then simplified because we can separate important fluxes - those which are the result of some non-trivial interaction - from superficial ones due to a change in the size or shape of the boundary.  For example, in a region of space with uniform energy density, if the boundary is allowed to increase in size, it will encapsulate more energy simply because it is encapsulating more space.  Such effects camouflage the interesting fluxes at work in the problem of motion.  Analyzing the system from the frame of a rigid shell of observers eliminates this possibility.

At the turn of the last century, physicists were very aware of Newtonian rigid body formalism and its advantages.  When special relativity came along, it quickly became apparent that a finite speed of light and thus a finite speed of communication between the particles making up a solid meant that the notion of a rigid {\it body} had to be discarded.  However, working from the viewpoint of a hypothetical rigid set of observers would still seem like an obvious strategy to try.  In 1909, Born defined a notion of rigid {\it motion} for precisely this reason \cite{Born1909}.  According to Born, rigid motion in relativity means that the orthogonal spacetime distance between each pair of infinitesimally separated observers remains constant in time.  This is the same definition of rigid motion that an RQF uses.  Why then are we only exploring rigidity in a relativistic context now?  The reason is due to our desire to think about the universe in a local way.  That is, we only sought out a notion of  rigidity for a {\it volume} of observers.  And almost immediately after Born introduced his definition, Herglotz and Noether showed that a three-parameter family of timelike worldlines in Minkowski space cannot exhibit the full six motional degrees of freedom we are familiar with from Newtonian mechanics, but rather only a smaller number - essentially only three\cite{Herglotz1910,Noether1910}. The situation only gets worse when you move to curved spacetime.  Roughly speaking, we can see why this cannot work by recognizing that Born's notion of rigidity implies that a volume described by the spatial three-metric $h_{ab}$ must appear constant from the viewpoint of the congruence of rigid observers whose four-velocity we denote $u^a$.  In other words, the Lie derivative of this metric with respect to the observers' four-velocity must vanish
\begin{align}
\mathcal{L}_u h_{ab} = 0.
\end{align}
This represents six constraints when $u^a$, being timelike (i.e., $u \cdot u = -1$), only has three free parameters to work with.  Obviously this cannot be satisfied in general and this result curtailed, to a large extent, subsequent study of rigid motion in special and (later) general relativity.  However, with the arrival of RQFs, the door to rigidity has reopened.

It turns out that we {\it can}, in fact, implement Born's notion of rigid motion in, not just flat spacetime, but any arbitrary curved spacetimes using the rigid quasilocal frame described above.  The trick to circumventing the Herglotz-Noether theorem is defining the system {\it quasilocally}; that is, as the two-dimensional set of points comprising the {\it boundary} of a finite spatial volume, rather than the three-dimensional set of points within the volume.  We can see why this works using the same argument as we did for the three-dimensional case.  In particular, our rigid frame consists of timelike observers using a spatial two-metric, $\sigma_{ab}$, to describe the distances between nearest neighbours.  Thus we need only satisfy the weaker condition
\begin{align}
\sigma_a^{\phantom{a}c} \sigma_b^{\phantom{b}d}  \mathcal{L}_u \sigma_{cd} = 0.
\end{align}
This now constitutes just three constraints on three functions and, as we will argue in this thesis, can always be solved.  Furthermore, the resulting frame exhibits precisely three translational and three rotational Newtonian motional degrees of freedom (with arbitrary time dependence).  Interestingly, the fact that the RQF exhibits these six degrees of freedom is a consequence of the fact that any two-surface with $\mathbb{S}^{2}$ topology always admits precisely six conformal Killing vector fields which generate an action of the Lorentz group on the sphere.

With a useful notion of rigid motion in relativity in hand, we now have to ask the question: what can we do with it?  The first step is to test it out in familiar territory and this is the focus of this thesis. Throughout this investigation, we will see that RQFs serve as a tool for cleanly analyzing the dynamics of spatially extended, relativistic systems and also provide a natural geometric understanding of the mechanisms behind these dynamics.  We will also contrast a quasilocal approach to the more traditional local approach. In doing so, it will become apparent that the local approach has severe limitations as a means of properly understanding general relativity.  In essence, this is because a local stress-energy-momentum (SEM) tensor (1) cannot include gravitational contributions (since gravitational energy is not localizable) and (2) cannot be used to construct useful definitions of the total energy, momentum, and angular momentum of matter {\it plus} gravity without the presence of spacetime symmetries.  On the other hand, an approach using RQFs does not encounter either of these problems.  It accomplishes this by first replacing the usual local matter SEM tensor with the quasilocally defined Brown-York SEM which includes matter {\it and} gravity\cite{BY1993}.  Next, and this is where rigidity becomes crucial, we make use of the six conformal Killing vectors mentioned above to define the energy, momentum, and angular momentum of the system irrespective of the spacetime symmetries that may (or may not) be present.  All of this provides the ground work for a ``Gauss' law''-style scenario where we can study the dynamics of a system in terms of the matter and gravitational fluxes of energy, momentum, and angular momentum across the RQF boundary.

Of course, as we said above, all of this is just the first step.  In this thesis we will build a case for the generic existence of RQFs, study their usefulness for analyzing relativistic systems, and use them to develop a conceptual understanding of the underlying physics.  The next step will be to apply the RQF formalism to the problem of motion more generally and try to better understand what a quasilocal mindset can tell us about general relativity and the universe.

\section{The Bigger Picture}\label{secBiggerPicture}

In making the transition from Newtonian physics to general relativity, a deeper understanding of how the universe works was gained.  Before general relativity, motion was thought to be governed by Newton's laws - laws which really did not tell us much about the universe itself.  They explained (to a good approximation) how things moved but only in terms of an empirically motivated set of rules.  On the other hand, Einstein's theory of general relativity was based on the single, very elegant principle that particles will naturally follow straight lines in a curved spacetime.  Or, as the great John Wheeler succinctly put it \cite{Wheeler},
\begin{quotation}
{\it ``Spacetime tells matter how to move; Matter tells spacetime how to curve.''}
\end{quotation}
In fact, one can show that the Einstein field equations imply the geodesic equation so we really only need to postulate the second half of this statement.  The point is, however, that general relativity actually gives us a conceptually deeper explanation for where Newton's laws come from; from Einstein's basic idea one can, in the appropriate limits, derive all of Newton's laws.  Thus, we have moved a level deeper in our understanding of nature.  An important question remains though: what {\it else} does general relativity have left to teach us about the universe?

It might seem like we have learned all we can at a conceptual level from general relativity.  However, from our study of RQFs, we believe this is {\it not} the case.  This is where the question of locality reappears.  We have had a lot of success analyzing general relativity by perturbing, for example, around the Newtonian or Minkowskian limits, so it has become commonplace to try to interpret general relativity in the mindset of these limits.  The problem with doing this is nicely summarized by Damour, an expert of the post-Newtonian approach \cite{Damour},
\begin{quotation}
{\it ``The danger of these approaches lies in their overall conceptual framework which misses the richness and suppleness of the full Einsteinian theory and which amounts nearly to a `neo-Newtonian interpretation' of Einstein's theory.''}
\end{quotation}
In other words, if we stick to perturbing around Newtonian or special relativistic theory then we will be forced to suffer a conceptual reduction from general relativity.  We have to do better than this.  Of course, the Einstein equations have also been explored without intentionally appealing to any of these limits but we argue that because we are so comfortable with the notion of locality from Newtonian physics - the idea that we can define physical quantities like energy, momentum, and angular momentum with locally defined tensors - it has persisted in our way of thinking to this day.  As we will show, this is problematic and so it is precisely the notion of locality which we must discard.  This is a bold claim, and one that will be justified more carefully throughout this thesis, but perhaps not so difficult to believe.

It is well known from the equivalence principle that gravitational energy cannot be localized.  One can always find a frame of reference in which the Christoffel symbols $\Gamma^a_{\phantom{a}bc}$ vanish locally.  This implies no local gravitational field and, therefore, no local gravitational energy.  The only workaround to this problem is to take advantage of spacetime symmetries.  For example, the ADM mass \cite{ADM} ``captures'' the total energy by assuming we have a flat spacetime infinitely far away (and one can perform similar tricks to include momentum and angular momentum).  More generically speaking, one relies on having some set of spacetime symmetries present so as to have a Killing vector with which to define a conserved quantity.  However, spacetimes with abundant symmetries are few and far between and, worse still, as soon as one has any sort of interesting dynamics, we lose spacetime symmetries altogether.  Therefore, an approach which relies on spacetime symmetries for defining fundamental physical quantities cannot hope to succeed.

The fact that an approach using rigid quasilocal frames avoids the pitfalls of the local approach described above says something very important about the nature of our universe.  What exactly is it saying though?  It is shining light on a very basic tool we use to understand the universe: frames of reference.  It is saying that you {\it can} define extended frames of reference with all of the usual Newtonian degrees of freedom but only {\it quasilocally}; they {\it cannot} be defined locally.  That implies that we should really be thinking about all of physics from a quasilocal or {\it holographic} perspective.  This is an idea that physicists have encountered already with the holographic principle but what is particularly surprising here is that we have found evidence for this interpretation from purely classical arguments\footnote{Interestingly, one of the most important results to come from quantum mechanics is that of Bell's theorem which states that ``No physical theory of local hidden variables can ever reproduce all of the predictions of quantum mechanics'' \cite{Bell}.  The consequence of this theorem is that we must give up either realism or locality.  Most often, it is realism that is abandoned but, based on what we have seen in our study of RQFs, we would kill two birds with one stone by condemning the latter.  If we are to adhere to Occam's razor, then we should really be taking a quasilocal approach to physics in general.}.

If a quasilocal approach to gravity really {\it is} the right way to go, then further study of RQFs should also help us to better understand the problem of motion.  In fact, we will see early progress toward this goal in this thesis when we construct quasilocal conservation laws to analyze various spacetimes.  Conservation laws allow us to analyze how a system changes in terms of fluxes across the boundary (e.g., how does the momentum inside our system change due to a stress on the boundary?) so they are, in effect, tied to the equations of motion.  Therefore, conservation laws are a means for understanding how and ultimately {\it why} things move.  Why, again, do we need our quasilocal frames to be rigid to accomplish this?  The reason is that we will need the six conformal Killing vectors that come with the RQF to construct physically sensible definitions of energy, momentum, and angular momentum.  As a bonus, the idea that an RQF is a frame in which fluxes are naturally defined will make it easier to interpret the nature of the fluxes at a fundamental level.  In fact, the RQF analysis has shed light on some very basic everyday problems.  For instance, we will show that the kinetic energy that an apple gains when it is dropped is due to a flux of gravitational energy involving frame-dragging which is analogous to the electromagnetic Poynting vector flux.  While the frame-dragging may be a small effect, it ends up being multiplied by a very large number, $c^4/G$, to amplify it to a macroscopic observable.  Another example will involve energy and momentum transfer in electromagnetism.  It turns out that  half of the transfers are due to standard electromagnetic Poynting fluxes while half are due to geometrical effects arising from the electromagnetic fields warping the spacetime.  It is usually thought that you can do electromagnetism in flat spacetime and understand what's going on, but this demonstrates that you actually need general relativity to understand the whole story.  In that sense, we have already increased our fundamental understanding of nature at a really basic level.   

In terms of moving forward, what are the next steps?  So far much of what we have done has been taking known solutions, finding RQFs, and interpreting.  We have learned a lot but at this stage it has primarily been a conceptual advance.  That's not to say we haven't derived practical results already.  In this thesis alone we will see a resolution to Ehrenfest's paradox, a generalization of Archimedes' law to curved spacetime, compelling arguments for the reality of the gravitational vacuum, deeper insight into the nature of tidal interactions, and more.  We expect, though, that the bulk of the practical value will come later - that is, when we turns things around; instead of starting with a known solution to Einstein's equations and embedding an RQF, we will start with the RQF equations and construct an ``initial value'' problem on a timelike worldtube (integrating radially from the worldtube as opposed to in time from a spacelike hypersurface).  For example, we expect this to be of particular importance for the self-force problem.  In reference \cite{Poisson}, Poisson breaks down the various hurdles in calculating the self-force to second-order so as to obtain waveforms that can be used in gravitational wave astronomy.  In a nutshell, the second-order problem is significantly more difficult to solve than the first order one due to singularity issues arising when integrating down to $r=0$.  Working with an RQF, we avoid singularities altogether.  In fact, in his review, Poisson discusses a conservation equation based approach first formulated by Dirac \cite{Dirac} that has been unsuccessful for the self-force problem because of the lack of Killing vectors in a dynamical, curved spacetime.  This sounds ideally suited to an RQF approach, but - along with many other problems - will have to be left for another day.

\section{Outline}\label{secOutline}

This thesis aims to establish the existence and usefulness of rigid quasilocal frames by using them to explore familiar spacetimes to develop a geometrical understanding of general relativistic dynamics.  The chapters in this thesis are largely based on references \cite{EMM2009,EMM2011,EMM2012,EMM2013,EMM2014} and it should be pointed out that Richard Epp is responsible for coming up with the notion of an RQF and first appreciating the idea's potential.

In Chapter \ref{chRigidRevisited}, we will start by providing a rigorous definition of an RQF in the context of the dynamical spacetime of general relativity.  We will then proceed to construct two simple examples of RQFs in the relative safety of Minkowski spacetime to illustrate the two types of rigid motion allowed by the Herglotz-Noether theorem.  In particular, we will construct round sphere RQFs: first, with arbitrary time-dependent acceleration and, later, with constant rotation.  The Herglotz-Noether theorem states that these are the most general motions one can exhibit while maintaining rigidity amongst a volume of points.  Therefore, we next demonstrate that a quasilocal approach can do better than this by considering arbitrary time-dependent infinitesimal perturbations about a non-accelerating and non-rotating round sphere seed solution in flat spacetime.  The tangent space to the seed solution is shown to be spanned by precisely six arbitrary functions of time and these degrees of freedom turn out to be intimately related the action of the Lorentz group on the sphere.  We follow this up with a consideration of infinitesimal perturbations about a generic RQF in curved spacetime, which reveals a peculiar ``nonlocality" in time inherent in RQFs with finite time-dependent rotation.  This is because the presence of twist in a congruence introduces a nonlinear term in the RQF equations which changes the basic nature of the partial differential equations involved.  As we will see, this revelation also sheds light on the famous ``Ehrenfest's paradox''.  Given that time-dependent rotation is so tricky, as a proof of principle that RQFs can be found in general in flat spacetime, we conclude by iteratively solving the RQF equations for the case of highly relativistic time-dependent rotation in powers of the rotation rate and its time derivatives.

With the notion of an RQF firmly established in flat spacetime, we next extend our results to curved spacetime in Chapter \ref{chNova}.  In particular, using a Fermi normal coordinates approach, we explicitly construct, in powers of areal radius, the general solution to the RQF rigidity equations in a generic curved spacetime.  We find that the resulting RQFs possess exactly the same six motional degrees of freedom as in flat spacetime.  In this context, we then discuss how RQFs provide a natural formalism with which to understand the flow of energy, momentum and angular momentum into and out of a system.  Focusing on the case of energy, we then derive a simple, exact expression for the flux of gravitational energy (a gravitational analogue of the Poynting vector) across the boundary of an RQF in terms of operationally-defined geometrical quantities on the boundary.  Finally, we use this new gravitational (or ``geometrical'') energy flux to resolve an apparent paradox involving electromagnetism in flat spacetime.  By the end of this chapter, we will see strong evidence that RQFs which exhibit the full six Newtonian translational and rotational degrees of freedom can be found in an arbitrary curved spacetime.  Furthermore, we will begin to see how a quasilocal conservation law approach leads to a deeper understanding of the dynamics of a system in terms of fluxes across the boundary.

In Chapter \ref{chLocalQuasilocal}, we further develop the quasilocal conservation law based approach for analyzing systems in general relativity and contrast it with the traditional local approach.  We argue that conservation laws based on the {\it local} matter-only stress-energy-momentum tensor (characterized by energy and momentum per unit {\it volume}) cannot adequately explain a wide variety of even very simple physical phenomena because they fail to properly account for gravitational effects. However, we see in more detail that our general {\it quasi}local conservation law which uses the Brown and York {\it total} (matter {\it plus} gravity) stress-energy-momentum tensor (characterized by energy and momentum per unit {\it area}), does properly account for gravitational effects.  Using the energy form of our quasilocal conservation law, we then demonstrate the explanatory power of the quasilocal approach asking what happens when we accelerate toward a freely-floating massive object.  Clearly, the kinetic energy of that object increases (relative to our frame) but how, exactly, does the object acquire this increasing kinetic energy?  With the quasilocal approach we see precisely the actual mechanism by which the kinetic energy increases: It is due to a bona fide gravitational energy flux that is exactly analogous to the electromagnetic Poynting flux, and involves the general relativistic effect of frame dragging caused by the object's motion relative to us.  

Next, we turn our attention to quasilocal momentum conservation as the subject of Chapter \ref{chArch}.  Using the RQF approach, we construct in a generic spacetime completely general conservation laws for the six components of momentum (three linear and three angular) of a finite system of matter and gravitational fields.  Again, we compare in detail this quasilocal RQF approach to constructing conservation laws with the usual local one based on spacetime symmetries, and discuss the shortcomings of the latter.  On the other hand, the RQF conservation laws lead to a deeper understanding of physics in the form of simple, exact, operational definitions of gravitational energy and momentum fluxes, which in turn reveal, for the first time, the exact, detailed mechanisms of gravitational energy and momentum transfer taking place in a wide variety of physical phenomena, including a simple falling apple.  Moreover, we argue that since the RQF based conservation laws include both matter and gravitational fields and do not rely on any spacetime symmetries while the local approach fails in both of these respects, we begin to see first hand the advantage to the quasilocal approach.  Finally, we use the quasilocal approach to derive a general relativistic version of Archimedes' law and then apply it to understand electrostatic weight and buoyant force in the context of a Reissner-Nordstr\"{o}m black hole.

In Chapter \ref{chPN} we turn our focus towards the practical utility of the quasilocal approach to constructed conservation laws.  To this end, we expand these laws in a post-Newtonian approximation and find that we can characterize the flows of gravitational energy and angular momentum each in terms of one simple, physically sensible flux. We next apply the resulting post-Newtonian conservation laws to the problem of tidal interactions.   Using RQFs we show that we can obtain the Newtonian formulas for tidal heating and tidal torque without the need to introduce pseudotensors.  As a final demonstration that the quasilocal approach has practical uses, we look at two examples of tidally interacting systems within our solar system.  In particular, we compute the tidal heating of Jupiter's moon Io and the angular momentum transfer in the Earth-Moon system which is the culprit behind the Moon's gradual recession from Earth.  In both examples we find agreement with observation thereby verifying that the RQF approach is not just useful for developing a better understanding of the the universe at a fundamental level but that it also a tool for analyzing everyday problems. 

Finally, in Chapter \ref{chConclusions}, we summarize our key results and draw conclusions about what RQFs tell us about the nature of the universe.


\chapter{Introduction to Rigid Quasilocal Frames}\label{chRigidRevisited}

Rigid motion in Newtonian space-time has six degrees of freedom: three translations and three rotations.  In other words, there are six arbitrary time-dependent degrees of freedom in constructing a three-parameter congruence of ``timelike" worldlines such that the distance between each pair of infinitesimally separated worldlines remains constant.  This fact greatly simplifies the description of the motion of (rigid) extended bodies, and motivated M. Born \cite{Born1909} to propose a similar definition of rigid motion in the context of special relativity, with ``distance" now defined as the orthogonal distance between neighbouring worldlines measured with the Minkowski line element.  Soon afterwards, G. Herglotz \cite{Herglotz1910} and F. Noether \cite{Noether1910} proved that such Born-rigid motions exist, but they have essentially only three degrees of freedom.  More precisely, there are two types of Born-rigid motion in special relativity: (1) arbitrary time-dependent translations with {\it no} rotation (so-called {\it plane} motions), and (2) motions generated by a Killing field, i.e., the repeated action of one element of the Poincar\'{e} group (so-called {\it group} motions) \cite{Salzman+Taub1954,Eriksen+Mehlen1982}.  As the simplest representative example of the latter, the only possible motion of a Born-rigid body with one point fixed is an eternal unchangeable rotation \cite{Eriksen+Mehlen1982}.  Born-rigid motion in the context of general relativity, especially the rotating case, is considerably more subtle - see \cite{Mason+Pooe1987} and references therein.

A body of literature has grown out of exploring various relaxations or modifications of Born's notion of rigidity (for examples, see \cite{Llosa+Soler2004,Llosa+Soler2000,Bel+Llosa1995,Bel+Martin+Molina1994,Bona1982}), but, prior to the introduction of rigid quasilocal frames, no proposal has emerged that recovers the full set of six arbitrary time-dependent degrees of freedom in a geometrically natural (coordinate independent) way.

In this chapter we will introduce the notion of a {\it rigid quasilocal frame} (RQF), which is simply Born's notion of rigidity applied not to a {\it three}-parameter congruence (history of a spatial volume-filling set of points), but to a {\it two}-parameter congruence (history of the set of points on the surface bounding a spatial volume).  This volume-to-surface, or quasilocal, relaxation provides a simple and geometrically natural way around the restrictions found by Herglotz and Noether: Whereas Born-rigidity for a three-parameter congruence involves six differential constraints on three functions (an over determined system) \cite{Llosa+Soler2000}, we will see that an RQF involves only {\it three} differential constraints on three functions, and argue that the space of solutions is parameterized by precisely six arbitrary time-dependent degrees of freedom.  Remarkably, the existence of these degrees of freedom is intimately connected with the well known fact that a two-sphere (as opposed to a closed two-surface of any other genus), regardless of its geometry (the size and shape of the rigid ``box" bounding the volume), always admits precisely six conformal Killing vectors, which generate an action of the Lorentz group on the sphere: three rotations and three boosts \cite{Held+Newman+Posadas1970}.  A single observer undergoing arbitrary acceleration and rotation can be thought of as being acted upon by a time-dependent sequence of local Lorentz transformations.  In essence, an RQF extends this notion to a two-sphere's worth of observers being acted upon by a time-dependent sequence of ``quasilocal Lorentz transformations."

The chapter is organized as follows.  In {\S}\ref{secDefinition} we define the notion of an RQF in a general (3+1)-dimensional spacetime.  In {\S}\ref{SimpleExamples} we construct two simple, representative examples of RQFs in flat spacetime: (1) a round sphere undergoing arbitrary time-dependent translations, with no rotation, and (2) a round sphere undergoing {\it constant} rotation, with no translation. These examples illustrate the two types of rigid motion allowed by the Herglotz-Noether theorem.  In {\S}\ref{ArbitraryPerturbations} we begin to go beyond these types by considering arbitrary time-dependent infinitesimal perturbations about the simplest RQF - a non-accelerating and non-rotating round sphere in flat spacetime.  We demonstrate that the tangent space to the RQF solution space, at the point of this simplest solution, is spanned by precisely six arbitrary functions of time and, moreover, establish the connection between these degrees of freedom and the natural action of the Lorentz group on the sphere, mentioned above.  We close {\S}\ref{ArbitraryPerturbations} with a consideration of infinitesimal perturbations about a generic RQF in curved spacetime, which reveals a peculiar ``nonlocality" in time inherent in RQFs with finite time-dependent rotation.

Indeed, this is where the real difficulty lies: constructing, even in flat spacetime, RQFs with arbitrary finite time-dependent rotation.  The reason can be traced to the relativity of simultaneity, which has the most severe consequences for congruences with {\it twist}, i.e., rotating systems, for which the congruence is not hypersurface orthogonal.  As we shall see, this necessarily activates a certain nonlinear term in the rigidity equations that involves a time derivative, changing the basic nature of the partial differential equations involved.  Thus the simplest example that would nevertheless provide a strong proof of principle is a flat spacetime RQF undergoing an arbitrary finite time-dependent rotation, with no translation. (This problem is essentially the quasilocal analogue of the well known ``Ehrenfest's paradox," in which a rigid body at rest can never be brought into uniform rotation \cite{Stachel1980}).  In {\S}\ref{GeneralRotation} we construct precisely such an example, solving the rigidity equations iteratively in powers of the rotation rate and its time derivatives.  Computing the first few terms in the series we find that our approximate solution can be pushed with confidence to the rather extreme case of a round sphere RQF spinning up from rest to angular velocities for which observers on the sphere's equator are moving at 1/3 the speed of light, on a time scale less than the time it takes the sphere to rotate a small fraction of one revolution.  Finally, in {\S}\ref{Conclusions2} we summarize the results of this chapter and look ahead at how we can make use of this construction for a wider range of problems.

\section{Definition of an RQF}\label{secDefinition}

We will begin by introducing some notation.  Let $M$ be a smooth four-dimensional manifold endowed with a Lorentzian spacetime metric, $g_{ab}$, with signature $+2$.  Naturally associated with $g_{ab}$ is its torsion-free, metric-compatible covariant derivative operator, $\nabla_{a}$, and volume element $\epsilon_{abcd}$.  Let $\mathcal B$ denote a two-parameter family of timelike worldlines with topology $\mathbb{R}\times \text{S}^2$, i.e., a timelike worldtube that represents the history of a two-sphere's-worth of observers bounding a finite spatial volume.  Let $u^a$ be the future-directed unit vector field tangent to this congruence, representing the observers' four-velocity.  The spacetime metric, $g_{ab}$, induces on $\mathcal B$ a spacelike outward-directed unit normal vector field, $n^{a}$, and a Lorentzian three-metric, $\gamma_{ab}:=g_{ab}-n_{a}n_{b}$.  At each point $p\in {\mathcal B}$ we have a {\it horizontal} subspace, $H_p$, of the tangent space to $\mathcal B$ at $p$, consisting of vectors orthogonal to both $u^a$ and $n^a$.  Let $\sigma_{ab}:=\gamma_{ab}+u_a u_b$ denote the spatial two-metric induced on $H:=\bigcup_p H_p$. Finally, let $\epsilon_{ab} := \epsilon_{abcd}u^c n^d$ denote the corresponding volume element associated with $H$.  The time development of our congruence is described by the tensor field $\theta_{ab}:=\sigma_a^{\phantom{a}c}\sigma_b^{\phantom{b}d}\nabla_c u_d$.  We adopt the usual terminology: the {\it expansion} is $\theta := \sigma^{ab}\theta_{ab}$ (the trace part); the {\it shear} is $\theta_{< ab >} := \theta_{(ab)}-\frac{1}{2}\theta\sigma_{ab}$ (the symmetric trace-free part, here and elsewhere denoted by angle-brackets); and the {\it twist} is $\nu:=\frac{1}{2}\epsilon^{ab}\theta_{ab}$ (the antisymmetric part).

A {\it rigid quasilocal frame} is defined as a congruence of the type just described, with the additional conditions that the expansion and shear both vanish, i.e., the size and shape, respectively, of the boundary of the finite spatial volume - as seen by our observers, do not change with time:
\begin{equation} \label{eq:RigidityCondition}
\theta = 0 = \theta_{<ab>}\;\;\;\Longleftrightarrow\;\;\; \theta_{(ab)}=0 .
\end{equation}
These three differential constraints ensure that $\sigma_{ab}$ is a well defined two-metric on the quotient space of the congruence, $\mathcal Q\simeq \text{S}^2$, i.e., the space of the observers' worldlines.  It describes the intrinsic geometry of the rigid ``box" bounding the volume, as measured locally by our two-sphere's-worth of observers.  Notice that there is no restriction on $\nu$ - the twist of the congruence - since we want to allow for the possibility of our rigid box to rotate, in which case the subspaces comprising $H$ are not integrable, i.e., $u^a$ is not hypersurface orthogonal as a vector field in $\mathcal B$.

Both to clarify this construction, and to establish notation for the examples in subsequent sections, let us restore the speed of light, $c$, (which was hitherto set to 1) and introduce a coordinate system adapted to the congruence.  Thus, let two functions $x^i$ on ${\mathcal B}$ locally label the observers, i.e., the worldlines of the congruence. Let $t$ denote a ``time" function on ${\mathcal B}$ such that the surfaces of constant $t$ form a foliation of ${\mathcal B}$ by two-surfaces with topology $\text{S}^2$.  Collect these three functions together as a coordinate system, $x^\mu := (t,x^i)$, and set $u^\mu := N^{-1}\delta_t^\mu$, where $N$ is a lapse function ensuring that $u\cdot u=-c^2$. The general form of the induced metric $\gamma_{ab}$ then has adapted coordinate components:
\begin{equation} \label{eq:InducedMetric}
\gamma_{\mu\nu} = \left(
\begin{array}{cc}
-c^2 N^2 & N u_j \\
N u_i & \sigma_{ij}-\frac{1}{c^2}u_i u_j
\end{array}
\right).
\end{equation}
Here $\sigma_{ij}$, and the shift covector $u_i$, are the $x^i$ coordinate components of $\sigma_{ab}$ and $u_{a}$, respectively.  Note that $\sigma_{ij}\,dx^i\,dx^j$ is the radar ranging, or orthogonal distance between infinitesimally separated pairs of observers' worldlines, and it is a simple exercise to show that the RQF rigidity conditions in equation (\ref{eq:RigidityCondition}) are equivalent to the three conditions $\partial\sigma_{ij}/\partial t =0$.  The resulting time-independent $\sigma_{ij}$ is the metric induced on ${\mathcal Q}\simeq \text{S}^2$.

In other words, an RQF is a {\it rigid} frame in the sense that each observer sees himself to be permanently at rest with respect to his nearest neighbours.  The idea is that this is true even if, for example, a gravitational wave is passing through the RQF, in which case neighbouring observers must undergo different proper four-accelerations, $a^a := u^b \nabla_{b}u^a$, in order to maintain nearest-neighbour rigidity.  They will also, in general, observe different precession rates of inertial gyroscopes.  Indeed, these inertial accelerations and rotations encode information about both the motion of their rigid box and the nontrivial nature of the spacetime it is immersed in as we will explore in more detail in later chapters.

It is not obvious that the rigidity conditions (\ref{eq:RigidityCondition}) can, in general, be satisfied.  In this chapter we will explore this possibility in the simplest possible context of RQFs in flat spacetime (in Chapter \ref{chNova} we will begin to consider more general spacetimes).  However, assuming that these conditions {\it are} satisfied, we are then free to perform a time-{\it in}dependent coordinate transformation amongst the $x^i$ (a relabelling of the observers) such that $\sigma_{ij}$ takes the form $\sigma_{ij}=\Omega^2 \, \mathbb{S}_{ij}$, where $\Omega^2$ is a time-independent conformal factor encoding the size and shape of the rigid box, and $\mathbb{S}_{ij}$ is the standard metric on the unit round sphere. For example, if the observers' two-geometry is a round sphere of area $4\pi r^2$, and the observers are labelled by the standard spherical coordinates $x^i=(\theta,\phi)$, then $\mathbb{S}_{ij}=\text{diagonal}(1,\sin^2\theta)$ and $\Omega=r$.  We are also free to change the time foliation of ${\mathcal B}$ such that $N=1$, i.e., $t$ is proper time for the observers.

Thus we see that the intrinsic three-geometry of an RQF has two functional degrees of freedom that - with the choice of coordinate-fixing described above - are encoded in the two components of the shift covector field, $u_j$ (which are functions of $t$ and $x^i$), as well as the time-{\it in}dependent conformal factor, $\Omega$, encoding our choice of size and shape of the rigid box.  We may also think of the dynamical degrees of freedom, $u_j$, as being encoded in the observers' (coordinate independent) proper acceleration tangential to $\mathcal B$, $\alpha^{a} := \sigma^{a}_{\phantom{a}b}a^b$, whose covariant components are
\begin{equation}\label{eq:ObserversAcceleration}
\alpha_{j}=\frac{1}{N}\,\dot{u}_{j}+c^2\partial_{j}\ln N,
\end{equation}
in the adapted coordinate system.  (Here, and throughout this work, an over-dot denotes partial derivative with respect to $t$, and $\partial_{j}:=\partial/\partial x^{j}$.)  More precisely, in addition to $\alpha_j$ we are free to specify the twist, $\nu$, on one cross section of $\mathcal B$, where
\begin{equation}\label{eq:nu}
\nu = \frac{1}{2}\epsilon^{ij}(\partial_i u_j - \frac{1}{c^2}\alpha_i u_j ),
\end{equation}
and $\epsilon^{ij}$ are the $x^i$ coordinate components of $\epsilon^{ab}$.

A full discussion of the extrinsic geometry of RQFs - their kinematics and dynamics, respectively, will be given in Chapter \ref{chNova}.  Our goal for now is only to construct some representative examples of RQFs, and argue that RQFs have the same degrees of freedom of motion as a Newtonian rigid body.\footnote{The two dynamical functional degrees of freedom in the RQF three-geometry, $u_j$, should not be confused with the six time-dependent degrees of freedom of the rigid motion.  The former - together with extrinsic geometrical data - encode the latter, as well as information about fluxes of energy, momentum and angular momentum through the system boundary.}

\section{Simple examples of RQFs}\label{SimpleExamples}

We will construct two representative examples of RQFs in flat spacetime: (1) a round sphere undergoing arbitrary time-dependent translations, with {\it no} rotation, and (2) a round sphere rotating at a {\it constant} rate, with no translation.

\subsection{Translation Only}\label{SimpleExamples-TranslationOnly}

For this example we let $X^a =(cT,X,Y,Z)$ denote Minkowski coordinates in an inertial reference frame in flat spacetime, with metric $g_{ab}=\text{diagonal}(-1,1,1,1)$.  Let $X^a = \xi^a (t)$ define an arbitrary timelike worldline ${\mathcal C}_0$, parameterized by proper time, $t$, around which we will construct the timelike worldtube, $\mathcal B$, of our accelerating RQF.  Let $U^a =  \dot{\xi}^a$ be the four-velocity along ${\mathcal C}_0$, such that $U\!\cdot\! U = -c^2$, and define the timelike unit vector $e_{0}^{\phantom{0}a}:=U^a /c$.  At some point along ${\mathcal C }_0$, say $t=0$, choose a spatial triad $e_{I}^{\phantom{I}a}$, $I=1,2,3$, orthogonal to $U^a$.  Define $e_{I}^{\phantom{I}a}$ all along ${\mathcal C}_0$ by Fermi-Walker transport (no rotation of the spatial triad):\cite{MTW}
\begin{equation}\label{FermiWalker}
\nabla_{e_{0}}e_{A}^{\phantom{A}a}=-\Omega^{a}_{\phantom{a}b}\,e_{A}^{\phantom{A}b}.
\end{equation}
Here we have collected $e_{0}^{\phantom{0}a}$ and $e_{I}^{\phantom{I}a}$ into a tetrad, $e_{A}^{\phantom{A}a}$, defined along ${\mathcal C}_0$, and defined $\Omega^{ab}:=\frac{1}{c^3}(A^a U^b - U^a A^b )$, where $A^a:=U^b \nabla_b U^a=\dot{U}^a$ is the acceleration along ${\mathcal C}_0$ (and of course $U\!\cdot\! A=0$).  In particular, from equation (\ref{FermiWalker}) we have $\dot{e}_{I}^{\phantom{I}a}=\frac{1}{c^2}A_{I}U^a$, where $A_{I}:=e_{I}^{\phantom{I}a}A_a$ are the triad components of the proper acceleration of ${\mathcal C}_0$.

Let us now embed, in our spacetime, a two-parameter family of timelike worldlines around ${\mathcal C}_0$:
\begin{equation}\label{eq:AccelerationOnly}
X^a = \Xi^a(t,\theta,\phi):=
\xi^a(t)+r\,r^I(\theta,\phi)\,e_I^{\;\;a}(t),
\end{equation}
representing a two-sphere's worth of observers (labelled with spherical coordinates $\theta$, $\phi$) comprising the timelike worldtube, $\mathcal B$.  Here $r^I(\theta,\phi):=(\sin\theta\cos\phi, \sin\theta\sin\phi, \cos\theta)$ are the standard direction cosines of a radial unit vector in spherical coordinates in Euclidean 3-space, and $r$ is a variable parameter that will turn out to be the areal radius of our round sphere RQF.  A simple calculation reveals that, in the adapted coordinate system $x^\mu=(t,\theta,\phi)$, the components of the metric induced on $\mathcal B$ have the form of equation (\ref{eq:InducedMetric}) with $N=1+\frac{r}{c^2}\,r^I A_I$, $u_i=0$, and $\sigma_{ij} = r^2\,\mathbb{S}_{ij}$, where $\mathbb{S}_{ij}$ is the unit round sphere metric introduced in the previous section.  The components of the observers' proper acceleration tangential and normal to $\mathcal B$ are then easily found to be
\begin{align}
\alpha_j &= \frac{1}{N}\,r \,  A_{I}\,\partial_j r^{I},\\
n\cdot a &= \frac{1}{N}\,r^{I}A_{I},
\end{align}
and obviously the twist, $\nu$, vanishes.  Note that the normal acceleration, $n\cdot a$, is a component of the extrinsic curvature of $\mathcal B$.  Insofar as we are not developing the formalism to analyze extrinsic curvature in this chapter, the result is stated for completeness.

Thus we have constructed an RQF that depends on three arbitrary functions of time: the three independent components of $\xi^{a}(t)$, or, if you will, the three components of the acceleration, $A_{I}(t)$, which describes a rigid round sphere of area $4\pi r^2$ undergoing arbitrary time-dependent translations.  Despite the proper accelerations the observers experience, both tangential and normal to the spherical frame they define, they may consider themselves to be ``stationary" in the sense described earlier: each observer sees himself to be permanently at rest with respect to his nearest neighbours.  In the spirit of Einstein's principle of equivalence\cite{EinsteinStudies-EquivalencePrinciple} they can consider themselves to be at rest in a time-dependent gravitational field that varies in strength and direction from one observer to the next.

There are two points worth noting.  First, observe that our construction is valid only if $N  >  0$, i.e., $\text{max}\{|A_I(t)|\}<\frac{c^2}{r}$.  In other words, our RQF is restricted in size by a ``Rindler horizon.'' This is not surprising, and obviously must be a generic property of RQFs: for a given size of bounding box there must be a maximum value of some acceleration parameter (in this case $A_I$) in order that the proper acceleration of each observer remain finite.  To the extent that RQFs provide a general description of physical systems, we speculate that quantization may introduce a minimum size for RQFs, and hence a maximum acceleration in quantum gravity.  Second, observe that there is a ``temporal stress'' associated with the fact that different observers require different proper accelerations to ensure rigidity, and thus different observers' clocks record proper time at different relative rates.  This is analogous to the well known fact that the back of a rocket must have a greater proper acceleration than the front for the rocket to maintain constant proper length (rigidity) as measured by co-moving observers, and that these observers then necessarily experience a ``temporal stress'' \cite{Giannoni+Gron1978}.

\subsection{Constant Rotation Only}\label{SimpleExamples-ConstantRotationOnly}

\begin{SCfigure} 
\centering
\includegraphics[width=0.65\textwidth]{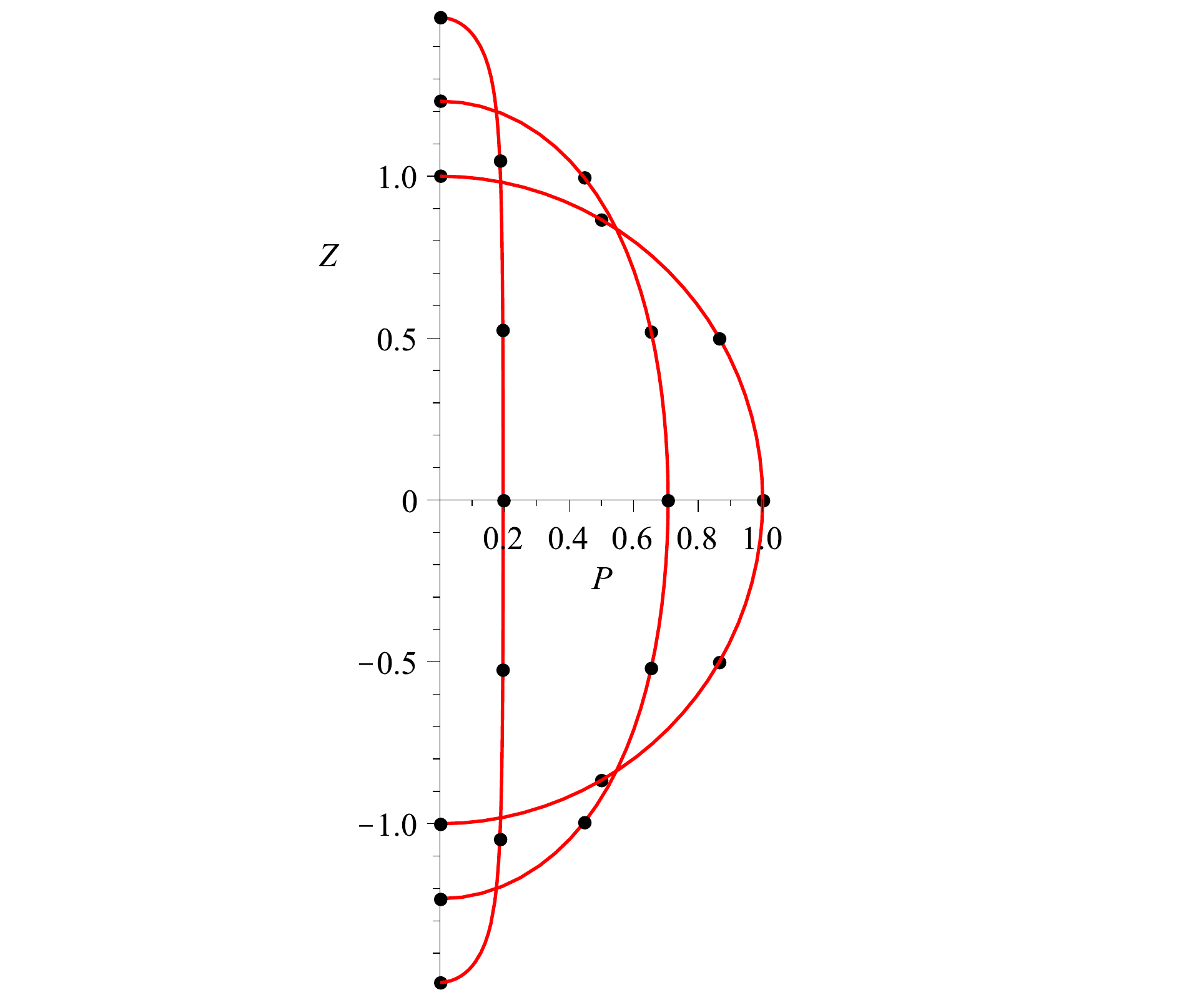}
\caption{The shape of the rotating observers' round sphere as seen by observers in the inertial reference frame: plots of the parametric curve $P=\rho(\theta)$, $Z=z(\theta)$ for $r=1$ and $\gamma =$ 0, 1 and 5.  The last case corresponds to $v_\text{equator}\approx 0.98\,c$. The dots show the locations of observers with $\theta=0,\pi/6,2\pi/6,\ldots,\pi$.  Since there is no length contraction in a $\phi=$ constant plane, the dots are equally spaced along each curve.}\label{figConstantRotation}
\end{SCfigure}

For this example we let $X^a =(cT,P,\Phi,Z)$ denote cylindrical coordinates in an inertial reference frame in flat spacetime, with metric $g_{ab}=\text{diagonal}(-1,1,P^2,1)$.  As in the previous example, we wish to embed a two-parameter family of timelike worldlines, labelled by coordinates $x^i=(\theta,\phi)$ and representing observers on a rigid round sphere of areal radius $r$, but this time with each observer rotating with constant angular velocity $\omega$ about the $Z$-axis ($\omega$ as measured by observers at rest in the inertial reference frame). Beginning with the ansatz:
\begin{equation}\label{eq:ConstantRotationAnsatz}
\begin{array}{lll}
T = & t\\
P = & \rho(\theta)\\
\Phi = & \phi + \omega t\\
Z = & z(\theta),
\end{array}
\end{equation}
a simple calculation yields an induced metric of the form given in equation (\ref{eq:InducedMetric}), with
\begin{align}
N = & \sqrt{1-\omega^2\rho^2/c^2}\\
u_i = & (\,0\,,\,\omega\rho^2/N\,)\\
\sigma_{ij} = &
\left(
\begin{array}{cc}
\rho^{\prime \, 2}+z^{\prime \, 2} & 0 \\
0 & \rho^2/N^2
\end{array}
\right),
\end{align}
where a prime denotes differentiation with respect to $\theta$.  For the rotating observers to see a round sphere of areal radius $r$ we require $\sigma_{ij}=r^2 \mathbb{S}_{ij}$ which implies:
\begin{align}
\rho(\theta) & = \frac{r\sin\theta}{\sqrt{1+\gamma^2\sin^2\theta}}\label{eq:rho}\\
z(\theta) & = -r \int_{\frac{\pi}{2}}^{\theta}\mathrm{d}\tilde{\theta}\, \sqrt{1-\frac{\cos^2\tilde{\theta}}{(1+\gamma^2\sin^2\tilde{\theta})^3}},\label{eq:z}
\end{align}
where $\gamma:=r\omega/c$ is a dimensionless measure of how relativistic the system is.  Unfortunately, the integral for $z(\theta)$ cannot be expressed in terms of elementary functions.  Figure \ref{figConstantRotation} shows the results of numerical integration for $r=1$ and three values of $\gamma$.  The sphere, which is round for our co-rotating observers, appears to inertial observers as an increasingly cigar-shaped surface as $|\gamma |$ increases.

To understand this figure, consider the tangential velocity of observers on the equator ($\theta=\pi/2$), given by
\begin{equation}\label{eq:vEquator}
\frac{v_\text{equator}}{c}=\frac{1}{c}\,\omega\,\rho(\tfrac{\pi}{2})=
\frac{\gamma}{\sqrt{1+\gamma^2}}.
\end{equation}
(Observe that $\gamma$, and hence the angular velocity, $\omega$, can range from $-\infty$ to $+\infty$.)  Recalling Einstein's famous rotating disk thought experiment \cite{Stachel1980}, in which rotating observers on the edge of the disk measure a greater circumference than inertial observers, the radius of the equator of our sphere must contract as $|\omega |$ increases in order to maintain the desired circumference of $2\pi r$.  Also note that, since there is no length contraction in a $\phi=$ constant plane, the length of each of the curves in figure \ref{figConstantRotation} is simply $\pi r$.  Thus, in the ultra relativistic limit $|\omega|\rightarrow\infty$, $z(\theta)$ ranges from $-\pi r/2$ to  $+\pi r/2$.

Substituting these results into
equations (\ref{eq:ObserversAcceleration}) and (\ref{eq:nu}) we find
\begin{align}
\alpha_\theta &=
-\frac{r^2\omega^2\sin\theta\cos\theta}{1+\gamma^2\sin^2\theta}, \quad \alpha_\phi=0\\
\nu &= \frac{\omega\cos\theta}{\sqrt{1+\gamma^2\sin^2\theta}}.
\end{align}
Notice that the magnitude of the observers' tangential proper acceleration along the lines of longitude is maximum around mid latitudes and is directed towards the poles, and that the magnitude of the twist of the congruence is maximum at the poles, as one might expect.  As in the previous example, for completeness we also provide the component of the observers' proper acceleration normal to $\mathcal B$ which turns out to be
\begin{equation}
n\cdot a=-{\frac {r {\omega}^{
2}  \sin^2 \theta   \sqrt {
 1+3\gamma^2+3\gamma^4
\sin ^{2}  \theta +\gamma^{6} \sin^4 \theta  }}{1+\gamma^2 \sin^2  \theta }}.
\end{equation}
In the slow rotation limit this reduces to $-r\omega^2\sin^2\!\theta$ as expected.  Observe that even in this very simple example of constant rotation, compared to the arbitrary time-dependent acceleration in the previous example, the expressions for acceleration (and twist) are significantly more complicated here.  In the context of relativistic rigidity, rotation is inherently much more subtle than translation.

\section{RQFs have Six Degrees of Freedom}\label{ArbitraryPerturbations}

In the previous section we wrote down the general solution for a round sphere RQF in flat spacetime undergoing arbitrary acceleration but no rotation.  In addition, we saw that even {\it constant} rotation is considerably more complicated, and anticipate that any closed form solution for time-{\it dependent} rotation, which is of the most interest to us, may well be intractable.  Thus, our strategy in this section is to consider an arbitrary {\it infinitesimal} perturbation about the trivial solution - a
non-accelerating, non-rotating round sphere RQF in flat spacetime, which is described by equation (\ref{eq:AccelerationOnly}) with $A_{I}=0$.  Perturbing about this solution we begin with the ansatz:
\begin{equation}\label{eq:GeneralPerturbation}
X^a =
\xi^a+r ( r^I +\lambda f^{I}) e_I^{\phantom{I}a}+  \lambda r f^{0}e_0^{\phantom{0}a},
\end{equation}
where $\lambda$ is an infinitesimal parameter and $f^{I}(t,\theta,\phi)$ are three arbitrary functions allowing complete freedom for the observers to ``wiggle around.''  A fourth arbitrary function, $f^{0}(t,\theta,\phi)$, is added to allow for a change in the time foliation of $\mathcal B$.  While the RQF rigidity conditions are obviously invariant under such a time reparametrization (which manifests itself in $\sigma_{ij}$ being independent of $f^0$ in equation (\ref{eq:PerturbedSigma}) below), a particular choice of $f^{0}$ will prove convenient later.  In our calculations we will retain only terms linear in $\lambda$.  The goal is to determine the dimension of the RQF solution space at the trivial solution point.  Under the fairly mild assumption that the dimension is a continuous function on the solution space, we will thus determine the dimension of the RQF solution space in general, both for flat and curved spacetimes.

Calculating the induced metric as before we now find:
\begin{align}
N & =  1+\lambda \frac{r}{c}\dot{f}^0  +O(\lambda^2)\label{eq:PerturbedN}\\
u_i & =  \lambda r^2 ( \dot{f}_{i} - \frac{c}{r}\partial_{i}f^{0}) + O(\lambda^2)\label{eq:Perturbedu_i}\\
\sigma_{ij} & = r^2\,[\,(1+2\lambda F)\,\mathbb{S}_{ij} + 2\lambda \mathbb{D}_{(i}f_{j)}]+ O(\lambda^2).\label{eq:PerturbedSigma}
\end{align}
Here we have made the decomposition
\begin{equation}\label{eq:fIdecomposition}
f^{I}(t,\theta,\phi)=F(t,\theta,\phi)\,r^{I}(\theta,\phi) + f^{i}(t,\theta,\phi)\,\mathbb{B}_{i}^{I}(\theta,\phi),
\end{equation}
where $\mathbb{B}_{i}^{I}(\theta,\phi) := \partial_i r^{I}(\theta,\phi)$ and  $f_{i}:=\mathbb{S}_{ij}f^{j}$, and we have made use of the completeness relation $\delta^{IJ}=r^{I}r^{J}+\mathbb{S}^{ij} \mathbb{B}_{i}^{I}\mathbb{B}_{j}^{J}$ (see equation (\ref{completeness})), where $\mathbb{S}^{ij}$ is the matrix inverse of $\mathbb{S}_{ij}$.  Finally, $\mathbb{D}_i$ appearing in equation (\ref{eq:PerturbedSigma}) is the covariant derivative operator associated with the unit round sphere metric, $\mathbb{S}_{ij}$.

We now demand that $\sigma_{ij}=r^2\,\mathbb{S}_{ij}$, i.e., the same rigid box we began with (a round sphere of areal radius $r$), but possibly in a state of motion different from rest.  Taking the trace and trace-free parts of equation (\ref{eq:PerturbedSigma}) yields three linear partial differential equations
\begin{align}
F & = -\frac{1}{2}\,\mathbb{D}\!\cdot\! f\label{eq:Fequation}\\
\mathbb{D}_{<i}f_{j>}&=0\label{eq:CKVequation}
\end{align}
in the three unknown functions $F$ and $f^i$.  The first equation tells us that $F$ (the radial perturbation) is determined by the vector field $f^i$ (the tangential perturbation), and the second tells us that $f^i$ must be a conformal Killing vector (CKV) field on the unit round sphere.  It is well known that any two-surface with the topology $S^2$ admits precisely six CKVs (compared with two CKVs for a torus, and zero CKVs for any closed surface of higher genus \cite{Nakahara}), and as generators of infinitesimal diffeomorphisms they form a representation of the Lorentz algebra \cite{Held+Newman+Posadas1970}.

To construct them explicitly, let $\mathbb{E}_{ij}$ denote the volume element associated with $\mathbb{S}_{ij}$.  Taking the three functions $r^I$ as a basis for the $\ell=1$ spherical harmonics we construct two sets of $\ell=1$ spherical harmonic covector fields: three {\it boost} generators, $\mathbb{B}^I_i := \mathbb{D}_i r^I$ (which were defined above), and three {\it rotation} generators, $\mathbb{R}^I_i := \mathbb{E}_{i}^{\phantom{i}j}\mathbb{B}_{j}^I$.  It is easy to verify that these are the desired CKVs: $\mathbb{D}_{<i}\mathbb{B}_{j>}^J=0=\mathbb{D}_{<i}\mathbb{R}_{j>}^J$, and that their commutators yield a representation of the Lorentz algebra (see appendix \ref{AppendixCKV} for more details).  Note also that $\mathbb{D}\cdot\mathbb{B}^J=-2r^J$ and $\mathbb{D}\cdot\mathbb{R}^J=0$, so we see from equation (\ref{eq:Fequation}) that the boost generators contribute to $F$ (the radial perturbation) whereas the rotation generators do not, as expected.

Thus, the most general infinitesimal perturbation about the trivial RQF in flat spacetime is given by
\begin{align}
f^i(t,\theta,\phi) & = \alpha^{I}(t)\,\mathbb{B}_{I}^{i}(\theta,\phi) + \beta^{I}(t)\,\mathbb{R}_{I}^{i}(\theta,\phi)\label{eq:fi}\\
F(t,\theta,\phi) & = \alpha^{I}(t)\,r_{I}(\theta,\phi),
\end{align}
where $\alpha^{I}(t)$ and $\beta^{I}(t)$ are six arbitrary functions of time.  Substituting these expressions into equation (\ref{eq:fIdecomposition}) yields
\begin{equation}
f^{I}(t,\theta,\phi)=\alpha^{I}(t)+\epsilon^{I}_{\phantom{I}JK} \beta^{J}(t) r^{K}(\theta,\phi) ,\label{eq:fI}
\end{equation}
where $\epsilon_{IJK}$ is the alternating symbol, and we made use of the identity $\mathbb{S}^{ij}\,\mathbb{B}_{i}^{I}\,\mathbb{R}_{j}^{J}
=\epsilon^{IJ}_{\phantom{IJ}K}\,r^{K}$ (see equation (\ref{completeness})).  By inspection of equation (\ref{eq:GeneralPerturbation}) it is clear that $\alpha^{I}(t)$ and $\beta^{I}(t)$ correspond to time-dependent translations and rotations, respectively.

To understand the corresponding proper acceleration and twist the observers experience we begin by substituting equation (\ref{eq:fi}) into equation (\ref{eq:Perturbedu_i}):
\begin{equation}
u_i = \lambda r^2 \,[\, \mathbb{D}_i ( \,\dot{\alpha}_{I}r^{I}-\frac{c}{r}f^{0}   ) + \mathbb{E}_{i}^{\phantom{i}j}\mathbb{D}_j ( \dot{\beta}_{I}r^{I} ) ] + O(\lambda^2).
\end{equation}
We are free to choose $f^0$, so let us choose $\frac{c}{r}f^0=\dot{\alpha}_{I}r^{I}$, which is sufficient to make the exact (gradient) part of $u_i$ vanish, leaving only a co-exact (curl) part.  This is the part responsible for the failure of the time foliation to be hypersurface orthogonal.  We then have
\begin{align}
N & = 1+\lambda \frac{r^2}{c^2} \ddot{\alpha}_{I}r^{I}  +O(\lambda^2)\\
u_i & = \lambda r^2 \dot{\beta}_{I}\mathbb{R}^{I}_{i} + O(\lambda^2).
\end{align}
Comparing with $N=1+\frac{r}{c^2} r^{I} A_{I}$ from {\S}\ref{SimpleExamples-TranslationOnly} we see that the perturbed solution effectively has an acceleration parameter $A_I=\lambda r \ddot{\alpha}_I$, proportional to the second time derivative of the translation parameter.

In physical (coordinate covariant) terms, the observers' proper acceleration and twist are found using equations (\ref{eq:ObserversAcceleration}) and (\ref{eq:nu}), and turn out to be
\begin{align}
\alpha_i &= \lambda r^2\,( \ddot{\alpha}_{I}\mathbb{B}^{I}_{i} + \ddot{\beta}_{I}\mathbb{R}^{I}_{i} )+O(\lambda^2)\\
n\cdot a &= \lambda r \,\ddot{\alpha}_{I}r^{I}+O(\lambda^2)\\
\nu &= \lambda \, \dot{\beta}_{I}r^{I}+O(\lambda^2).
\end{align}
(Again, the result for the normal acceleration, $n\cdot a$, is given for completeness.)  From the first equation we see that a nonzero second time derivative of the translation parameter $\alpha_I$ (respectively, rotation parameter $\beta_I$) is associated with a tangential proper acceleration having the pattern of a boost diffeomorphism generator, $\mathbb{B}_{I}^{i}$ (respectively, rotation diffeomorphism generator, $\mathbb{R}_{I}^{i}$).  The second and third equations tell us that translation (only) is associated with a proper acceleration normal to $\mathcal B$, and that rotation (only) is associated with a twisting congruence, at least in this lowest order (linear) approximation.

These results are intuitively sensible, and show that the most general RQF in the solution space neighbourhood of the trivial RQF in flat spacetime has precisely the same six time-dependent degrees of freedom of motion as a rigid body in Newtonian space-time.  Moreover, these degrees of freedom can be identified with time-dependent Lorentz transformations acting on our quasilocal frame of observers via the CKVs, as mentioned above.  It is plausible - by continuity - that these properties are true for {\it general} RQFs (in both flat and curved spacetimes) and we will see strong evidence for this beginning in the next chapter.  In the remainder of this section we will point out some subtleties involved in the general case.

Returning to the general notation introduced at the beginning of {\S}\ref{secDefinition} (and taking $c=1$) suppose that we have constructed an RQF in a generic spacetime, $(M,g_{ab})$.  This means we have a hypersurface ${\mathcal B}\approx \mathbb{R}\times \text{S}^2$ with spatial unit normal vector field $n^a$, and timelike unit tangent vector field $u^a$, such that the quotient space of the congruence (the space of integral curves of $u^a$) admits a well defined two-metric, $\sigma_{ab}$.  A perturbation of the RQF means a perturbation of the congruence, but this is equivalent to an infinitesimal active diffeomorphism of the spacetime metric, $g_{ab}$, in the neighbourhood of $\mathcal B$, leaving the congruence fixed.  Let $\psi^a$ denote an arbitrary infinitesimal vector field defined in the neighbourhood of $\mathcal B$ that effects this diffeomorphism: $\delta g_{ab} = 2\nabla_{(a}\psi_{b)}$, where $\nabla_a$ is the covariant derivative operator associated with the original spacetime metric.  On $\mathcal B$ we decompose $\psi^a$ as $\psi^a = \chi u^a + \Phi n^a +\phi^a$, where $\phi^a$ is tangent to the horizontal subspace, $H$, defined at the beginning of {\S}\ref{secDefinition}.  A simple exercise then shows that the corresponding change in $\sigma_{ab}$ is given by
\begin{equation}\label{eq:GeneralRQFPerturbation}
\delta\sigma_{ab}=2\left(\chi\theta_{(ab)}+\Phi k_{ab} + \tilde{\mathcal D}_{(a}\phi_{b)}\right),
\end{equation}
where $k_{ab}:=\sigma_{a}^{\phantom{a}c}\sigma_{b}^{\phantom{a}d} \nabla_{c}n_{d}$ is the $H$-projection of the (symmetric) extrinsic curvature of $\mathcal B$, and $\tilde{\mathcal D}_{a}$ is the covariant derivative operator induced on $H$, i.e., $\tilde{\mathcal D}_{a}\phi_{b}:= \sigma_{a}^{\phantom{a}c}\sigma_{b}^{\phantom{a}d} \nabla_{c}\phi_{d}$.  By assumption we have an RQF to begin with, so $\theta_{(ab)}=0$ and the first term on the right hand side of equation (\ref{eq:GeneralRQFPerturbation}) vanishes, which tells us that the observers' two-metric is independent of the time reparametrization, $\chi$, as expected.

We now demand that the perturbation leave the observers' two-metric invariant, $\delta\sigma_{ab}=0$, i.e., while the observers may be in a different state of motion, they measure the same nearest-neighbour distances as in the original RQF. Taking the trace and trace-free parts of equation (\ref{eq:GeneralRQFPerturbation}) (with respect to $\sigma_{ab}$) yields three linear partial differential equations
\begin{align}
\Phi & = -\frac{1}{k}\,\tilde{\mathcal D}\!\cdot\! \phi\label{eq:Phi_equation}\\
\tilde{\mathcal D}_{<a}\phi_{b>}&=\frac{k_{<ab>}}{k}\tilde{\mathcal D}\!\cdot\! \phi\label{eq:phi_equation},
\end{align}
in the three unknown functions $\Phi$ and $\phi^a$.  Here $k:=\sigma^{ab}k_{ab}$ and $\tilde{\mathcal D}\!\cdot\! \phi:= \sigma^{ab}\tilde{\mathcal D}_{a}\phi_{b}$.  As for the analogous equation (\ref{eq:Fequation}), the normal perturbation, $\Phi$, is determined by the $H$-perturbation, $\phi^a$.  The problem is to determine the general solution to equation (\ref{eq:phi_equation}) and show, in analogy with equation (\ref{eq:CKVequation}), that the solution space is spanned by six arbitrary functions of time.

The principle subtlety here is that when the congruence is not hypersurface orthogonal the derivative operator $\tilde{\mathcal D}_a$ necessarily involves a time derivative. To see this explicitly, let us introduce, as before, a coordinate system $x^\mu :=(t,x^i)$ adapted to the congruence.  It can be shown that the $x^i$ coordinate components of equation (\ref{eq:phi_equation}) are
\begin{align}
{\mathcal D}_{<i}\phi_{j>} & = \frac{k_{<ij>}}{k}\,\left[  {\mathcal D}\!\cdot\! \phi+U^{k}\left( 2\nu\epsilon_{k}^{\phantom{k}l}\phi_{l}+\frac{1}{N}\dot{\phi}_k \right) \right] -u_{<i}\left( 2\nu\epsilon_{j>}^{\phantom{j}k}\phi_{k} + \frac{1}{N}\dot{\phi}_{j>} \right) \label{eq:phi_equation_ij},
\end{align}
where the quantity in square brackets is $\tilde{\mathcal D}\!\cdot\! \phi$.  Here $U^{i}:=\sigma^{ij}u_{j}$ and ${\mathcal D}_i$ is the covariant derivative operator associated with the (time-independent) metric $\sigma_{ij}$, which is conformally related to the the covariant derivative operator ${\mathbb D}_i$ used earlier in this section (recall that $\sigma_{ij}=\Omega^2 {\mathbb S}_{ij}$).  In that case - perturbations about the trivial RQF in flat spacetime - $\nu$, $u_i$, and $k_{<ij>}$ are zero, and $k=2/r$.  Equations (\ref{eq:Phi_equation}) and (\ref{eq:phi_equation_ij}) then reduce to (\ref{eq:Fequation}) and (\ref{eq:CKVequation}) and we recover our previous results, with $\Phi$ and $\phi_i$ being simply related to $F$ and $f_i$ by the constant conformal factor $\Omega=r$.

To gain some insight into the structure of equation (\ref{eq:phi_equation_ij}), let us begin again with the trivial RQF in flat spacetime, except with a non-round bounding box, i.e., $\Omega \not=$ constant.  Then $u_i =0$, but $\hat{k}_{<ij>}:=k_{<ij>}/k\not=0$ and we need to solve the equation ${\mathcal D}_{<i}\phi_{j>} = \hat{k}_{<ij>}\,{\mathcal D}\!\cdot\! \phi$.  If we begin by setting $\phi_j$ equal to an arbitrary time-dependent linear combination of the six CKVs of the non-round bounding box (which are just the CKVs of the round sphere multiplied by $\Omega^2$), the left hand side is zero but the right hand side is, in general, not zero.  However, $\hat{k}_{<ij>}$, and the right hand side in general, is necessarily a linear combination of $\ell=2$ or higher (symmetric trace-free) tensor spherical harmonics.  So for the left hand side to equal the right hand side we must add to the ($\ell=1$) CKV part of $\phi_j$ vector spherical harmonics higher order in $\ell$.  But this in turn generates still higher order terms on the right hand side, and the process must be iterated, ultimately generating an infinite hierarchy of coupled linear equations for the time-dependent coefficients in a vector spherical harmonic expansion of $\phi_j$.  The key point is that this hierarchy of equations places no constraints on the starting ``seed'' - the arbitrary time-dependent linear combination of six CKVs.  Thus, if for each such seed, the infinite hierarchy of equations yields a convergent, unique solution, we recover precisely six time-dependent degrees of freedom of motion for a non-round rigid box, the same as in Newtonian space-time.

The more important case to consider is when the congruence is not hypersurface orthogonal, in which case $\nu$ and $u_i$ are necessarily non-vanishing.  Beginning with a CKV seed as before, equation (\ref{eq:phi_equation_ij}) again generates an infinite hierarchy of coupled linear equations for the time-dependent coefficients in a vector spherical harmonic expansion of $\phi_j$.  The difference is that, because of the time derivative of $\phi_j$ occurring on the right hand side of the equation, the coefficients for higher $\ell$ will depend on higher order time derivatives of the coefficients in the CKV seed - in principle continuing up to infinite order.  This introduces a ``nonlocality" in time in the sense that the solution for $\phi_j$ (and thus also $\Phi$) on any time slice depends not only on the CKV data on that slice, but also an arbitrary number of time derivatives of this data.  Roughly speaking, this suggests that while we are free to specify how the bounding box is to accelerate and tumble as time passes (encoded in the CKV data), the higher order spherical harmonics that contribute to precisely locating the observers in space at any instant of time depend, in principle, on the entire history of the ($\ell=1$ component of the) box's specified motion.  In other words, unlike in Newtonian space-time, we cannot simply specify the position and velocity of the observers at an initial instant of time, and then integrate given their acceleration.  However, a detailed analysis of the linearized RQF equations, (\ref{eq:Phi_equation}) and (\ref{eq:phi_equation}), would be needed to properly clarify these subtleties.

In the next section we will dispense with these linearized rigidity equations and solve, to the first few orders in a certain perturbation expansion, the full nonlinear equations in a highly nontrivial example involving a finite time-dependent rotation.  We will see the aforementioned nonlocality in time emerge.  This example also provides evidence that the hierarchy of equations in the linearized case should, indeed, yield a convergent, unique solution.

\section{RQF Resolution of Ehrenfest's Paradox}\label{GeneralRotation}

In this section we will provide a quasilocal resolution to the famous ``Ehrenfest's paradox," in which a rigid body at rest can never be brought into uniform rotation \cite{Stachel1980}.  In a certain perturbation expansion we will construct an RQF representing a round sphere in flat spacetime that begins at rest, is spun up to a relativistic angular velocity, and is then brought to rest again.  Roughly speaking, this means that concentric shells of a body can be spun up rigidly, but rigidity between neighbouring shells cannot be maintained.

In {\S}\ref{SimpleExamples-ConstantRotationOnly} we considered a round sphere in flat spacetime rotating at a constant rate, $\omega$.  As a first approximation to a non-constant rotation rate, $\omega(t)$, we take the ansatz in equation (\ref{eq:ConstantRotationAnsatz}), with $\rho(\theta)$ and $z(\theta)$ replaced by $\rho(t,\theta)$ and $z(t,\theta)$ given by equations (\ref{eq:rho}) and (\ref{eq:z}) with $\gamma$ replaced by $\gamma(t):=r\omega(t)/c$.  This is {\it not} an RQF, but if $\gamma(t)$ and its time derivatives are sufficiently small, it is a good approximation.  To this approximate solution we add a perturbation to the spatial embedding part (leaving the time foliation the same):
\begin{equation}\label{eq:TimeDependentRotationAnsatz}
\begin{array}{lll}
T & = t\\
P & = \rho(t,\theta)+\delta P(t,\theta)\\
\Phi & = \phi + \int_0^t\omega(\tilde{t})\,\text{d}\tilde{t}+
\delta\Phi(t,\theta)\\
Z & = z(t,\theta)+\delta Z(t,\theta),
\end{array}
\end{equation}
where $\delta P$, $\delta\Phi$, and $\delta Z$ are ``small'' arbitrary functions, and we have replaced $\omega t$ with $\int_0^t\omega(\tilde{t})\,\text{d}\tilde{t}$.  We then compute the corresponding induced metric, $\gamma_{\mu\nu}$, in particular $\sigma_{ij}$, and demand that $\sigma_{ij}=r^2\mathbb{S}_{ij}$.  This results in three algebraic or differential equations for the three unknown functions, which we solve iteratively in powers of $\gamma(t)$ and its time derivatives.  (In reference to the previous section, the unperturbed congruence - which is not an RQF - is analogous to the CKV ``seed'' specifying how the box is to rotate, and the perturbation is analogous to the higher order spherical harmonic corrections required to achieve this motion in a manner that maintains relativistic rigidity.)

For example, at the lowest order we find the three equations:
\begin{align}
0 &=: \sigma_{\theta\theta}-r^2 \label{eq:sigma22}\\ &= r^2\,\left\{ 2\cos\theta\frac{\delta P^\prime}{r} - 2\sin\theta\frac{\delta Z^\prime}{r} +\frac{\tau^2}{4}\sin^2{2\theta}  \gamma^2(t)\dot{\gamma}^2 (t)     \right\}\nonumber\\
0 &=: \sigma_{\theta\phi}=r^2\sin^2\theta\left\{  \delta\Phi^\prime-\tau\sin\theta\cos\theta\, \gamma^2(t)\dot{\gamma}(t)  \right\}\label{eq:sigma23}\\
0 &=: \sigma_{\phi\phi}-r^2\sin^2\theta \nonumber \\
&= 2r^2\sin^2\theta \left\{  \frac{\delta P}{r} + \tau\sin^3\theta\,\gamma(t)\,\delta\dot{\Phi} \right\}
\label{eq:sigma33}
\end{align}
where, as before, a prime denotes differentiation with respect to $\theta$, and $\tau := r/c$ is the characteristic time for light to cross the system.  Solving equation (\ref{eq:sigma23}) for $\delta\Phi$ with Dirichlet boundary conditions at $\theta=0$ and $\pi$ yields
\begin{equation}\label{eq:SolutionForDeltaPhi}
\delta\Phi(t,\theta)=\frac{1}{2}\tau\sin^2\!\theta\,\gamma^2(t)\dot{\gamma}(t).
\end{equation}
Substituting this result into equation (\ref{eq:sigma33}) and solving for $\delta P$ yields
\begin{equation}\label{eq:SolutionForDeltaP}
\delta P(t,\theta) = -\frac{1}{2}r\tau^2\,\sin^5\!\theta\,\gamma^2(t) \left[
2\dot{\gamma}^2(t)+\gamma(t)\ddot{\gamma}(t)           \right].
\end{equation}
Finally, substituting this result into equation (\ref{eq:sigma22}) and solving for $\delta Z$, with the condition that $\delta Z$ be an odd function about $\theta=\pi/2$, yields
\begin{align}\label{eq:SolutionForDeltaZ}
\delta Z(t,\theta) = \frac{1}{2}r\tau^2\,\cos^3\!\theta\,\gamma^2(t) \left[\,
\left(3-2\cos^2\!\theta\right)\dot{\gamma}^2(t)\right. \nonumber \\ \left.+\left( 5/3-\cos^2\!\theta \right)\gamma(t)\ddot{\gamma}(t)           \right].
\end{align}

From equation (\ref{eq:SolutionForDeltaPhi}) we see that an angular acceleration, $\dot{\gamma}(t)\not= 0$, requires the set of observers on any given meridian of the rotating round sphere to suffer an azimuthal displacement that ``leads'' the acceleration; this {\it bending} of the meridian lines in the azimuthal direction, $\delta\Phi\propto \sin^2\!\theta$, is seen by the {\it inertial} observers only, not the co-rotating observers.  For this to not distort the shape of the round sphere as seen by the co-rotating observers, the shape of the embedded sphere in the inertial frame must change.  This is accounted for by the terms proportional to $\dot{\gamma}^2(t)$ in $\delta P(t,\theta)$ and $\delta Z(t,\theta)$: relative to figure \ref{figConstantRotation} there is an additional pulling in near the equator, and pushing out near the poles.

The most intriguing aspect of this perturbation away from the constant angular velocity solution is that in general it does {\it not} vanish when $\dot{\gamma}(t)= 0$: there are terms proportional to $\gamma^3(t)\ddot{\gamma}(t)$ in $\delta P(t,\theta)$ and $\delta Z(t,\theta)$.  For instance, consider beginning with a round sphere at rest, spinning it up to some angular velocity parameter $\gamma(0)>0$ at, say, $t=0$, and then bringing it to rest again, in a time-symmetric fashion so that $\dot{\gamma}(0)=0$ but $\ddot{\gamma}(0)<0$.  Although at $t=0$ the sphere is spinning with an angular velocity that is momentarily constant, the shape of the sphere as seen by the inertial observers is {\it not} that of a sphere {\it eternally} rotating with the same angular velocity parameter ($\gamma(0)$ in both cases): relative to figure \ref{figConstantRotation} there is an additional pushing out near the equator, and pulling in near the poles.  (Note, however, that the bending of the meridian lines, $\delta\Phi$, {\it does} vanish at the time-symmetric point, as might be expected.)  The origin of this peculiar behaviour is the dependence of $\delta P$ on $\delta\dot{\Phi}$ (the time {\it derivative} of $\delta\Phi$ - see equation (\ref{eq:sigma33})), which can be traced back to the nonlinear $u_i u_j$ term in the relation $\sigma_{ij}=\gamma_{ij}+\frac{1}{c^2}u_i u_j$, and is also related to the $u_{<i}\dot{\phi}_{j>}$ term on the right hand side of equation (\ref{eq:phi_equation_ij}).

Moreover, using GRTensor II \cite{GRTensor} running under Maple it is not difficult to iterate this perturbation expansion to higher powers in $\gamma(t)$ and its derivatives.  Indeed, we have iterated ``two and a half'' more times, up to terms in $\delta\Phi(t,\theta)$ involving the seventh time derivative of $\gamma(t)$, and there appears to be no obstruction to continuing the iterations indefinitely, except that the expressions grow in size exponentially with successive iterations.  Thus it seems that the shape of the sphere at any instant of time, $t$, as seen by the inertial observers, depends not only on $\gamma(t)$ at that instant, but also on all of its time derivatives up to infinite order, i.e., it depends on the entire history of $\gamma(t)$.  This is the ``nonlocality'' in time discussed in the context of the linearized rigidity equations in the previous section.  We did not see this behaviour in the case of finite time-dependent translation (with no rotation) because in that case the congruence is hypersurface orthogonal, so at worst $u_i$ is exact (a gradient); in fact, we used a time foliation of $\mathcal B$ for which $u_i$ simply vanishes.  For the case of finite time dependent rotation the twist is nontrivial and, according to equation (\ref{eq:nu}), $u_i$ contains an irremovable co-exact (curl) part.  And then the $u_i u_j$ term in the relation $\sigma_{ij}=\gamma_{ij}+\frac{1}{c^2}u_i u_j$ - which is present because of the relativity of simultaneity - comes into play.  In the end, though, this nonlocality in time is perhaps not surprising if one recalls that rigid motion already involves nonlocality in space in the sense that the acceleration at one point on the bounding box depends on the accelerations at causally disconnected points: equation (\ref{eq:CKVequation}) is an elliptic differential equation.

As an example to test the robustness of our solution, we considered the particular choice of time dependent angular velocity given by
\begin{equation}
\omega \left( t \right) =\Omega\,{e^{-{{t}^{2}/{T}^{2}}}},
\end{equation}
that describes a time-symmetric situation in which the sphere spins up to a maximum angular velocity $\Omega$ and then back down to zero with a time scale $T$.  Considering both the absolute and relative magnitudes of the successively higher order terms in the solution, we can, with confidence, push our approximate solution to the rather extreme case of $\Omega\approx 1/3$ and $T\approx 1$ (with $r=1=c$).  This corresponds to observers on the sphere's equator spinning up to $v_\text{equator}\approx c/3$ (see equation (\ref{eq:vEquator})) on a time scale about equal to the time it takes light to cross the system, in this case a small fraction of one revolution.

\section{Discussion}\label{Conclusions2}

Rigid motion in Newtonian space-time has six time-dependent degrees of freedom: three translations and three rotations.  Rigid motions also exist in both special and general relativity, but they are severely restricted, as outlined in the first paragraph of the Introduction section.

In this chapter we have introduced the concept of a {\it rigid quasilocal frame} (RQF), which opens up the possibility of rigid motion in both special and general relativity with the full six time-dependent degrees of freedom we have in Newtonian spacetime.  The definition of an RQF, applicable in both flat and curved spacetimes, is identical to Born's definition of rigidity \cite{Born1909}, except for one key difference: to consider not a three-parameter congruence of timelike worldlines (a swarm of observers filling a three-dimensional volume of space), but a two-parameter congruence (the observers on the topologically $S^2$ boundary of the volume - a {\it quasilocal} frame).

As proofs of principle we have constructed, either exactly or approximately, several examples of RQFs in flat spacetime, including one that directly addresses the difficult problem of finite time-dependent rotations.  Two key aspects of quasilocal rigidity emerged.  The first is that the existence of six degrees of freedom in the rigid motion of our two-sphere's worth of observers is intimately connected with the fact that any two-surface with the topology $S^2$ always admits precisely six conformal Killing vectors (CKVs), which generate a representation of the Lorentz algebra.  In contrast to the usual case of the Lorentz group acting locally on a single observer (rotations and boosts of his tetrad along his worldline), here we have the Lorentz group acting {\it quasi}locally on a two-sphere's worth of observers along their worldtube.

The second is that finite time-dependent rotations are subtle precisely because they introduce a ``nonlocality" in time: unlike in Newtonian space-time it is not possible to specify a cross section of the (twisting) congruence in space, on a given time slice, without knowing the entire history of the motion.  Of course RQFs are also nonlocal in space in the sense that rigid motion requires observers at different points on the boundary to act in concert.  Thus an RQF is an inherently nonlocal construction in spacetime.  (It is perhaps worth emphasizing that in any given spacetime, an RQF is simply a congruence of worldlines with certain geometrical properties; the word ``observer'' is used in an abstract sense.  We are not requiring or implying that a two-sphere's worth of physical observers in a dynamical spacetime can actually adjust their accelerations with physical thrusters in concert to achieve an RQF.  Also, the existence of RQFs does not require or imply the existence of rigid bodies.)

We have introduced here a notion of a rigid frame which we claim exists and is useful in both flat and curved spacetimes.  However, thus far, our investigation and discussion has remained largely in the arena of flat spacetime.  The next step, then, is to consider the existence and usefulness of RQFs in curved spacetime.  It is this task which is the focus of the remainder of this thesis.  While a rigorous proof of the existence of RQFs remains elusive, we provide strong evidence for this hypothesis by considering small-sphere RQFs in Chapters \ref{chNova} - \ref{chArch} and post-Newtonian RQFs in Chapter \ref{chPN}.  The analyses which follow indicate that the transition from flat to curved spacetime introduces no obvious obstructions; our RQFs still retain the full six Newtonian time-dependent degrees of freedom.  Moreover, as alluded to in the introduction, the application of RQFs to the problem of motion yields several interesting new insights.

\chapter{Rigid Quasilocal Frames in Curved Spacetime}\label{chNova}

In the previous chapter we demonstrated that one {\it can} implement Born's notion of rigid motion in flat spacetime - with precisely the desired three translational and three rotational degrees of freedom (with arbitrary time dependence) - provided the system is defined {\it quasilocally} as the two-dimensional set of points comprising the {\it boundary} of a finite spatial volume, rather than the three-dimensional set of points within the volume.  To accomplish this we introduced the notion of a {\it rigid quasilocal frame} (RQF) as a geometrically natural way to define a system in the context of the dynamical spacetime of general relativity.  An RQF is defined as a two-parameter family of timelike worldlines comprising the worldtube boundary (topologically $\mathbb{R}\times \mathbb{S}^{2}$) of the history of a finite spatial volume, with the rigidity conditions that the congruence of worldlines is expansion-free (the ``size'' of the system is not changing) and shear-free (the ``shape'' of the system is not changing).  Furthermore, by constructed several representative examples of RQFs in flat spacetime two key results emerged: (1) RQFs exhibit precisely the six motional time-dependent degrees of freedom that we are familiar with from Newtonian physics.  Moreover, these degrees of freedom exist due to the six conformal Killing vector fields admitted on any closer two surface. (2) Of the familiar Newtonian motions, it is well known that the time-{\it dependent} rigid rotations are the most problematic in both special relativity (e.g., Ehrenfest's paradox \cite{Stachel1980}) and general relativity.  Using the RQF approach were able to address this problem and demonstrate the usefulness of RQFs in flat spacetime.

The main purpose of this chapter is twofold: (1) To show that the notion of an RQF can be easily extended from flat spacetime to a generic curved spacetime, and (2) to show that RQFs are useful for the better understanding of various fluxes, in particular gravitational energy flux.  In {\S}\ref{secExtrinsic} we analyze the extrinsic geometry of an RQF and, using Brown and York's quasilocal definition of a {\it total} (matter plus gravitational) energy-momentum tensor \cite{BY1993}, we derive an energy conservation equation that relates the change in energy of an RQF to the usual matter energy flux, plus a certain simple, operationally-defined ``geometrical'' energy flux across the boundary.  We argue that the latter is to be interpreted as a gravitational energy flux.  In {\S}\ref{secCurved} we use a Fermi normal coordinates approach to explicitly construct, to the first few orders in powers of areal radius, the general solution to the RQF rigidity equations in a generic curved spacetime. We show that the resulting RQFs possess exactly the same six motional degrees of freedom as in flat spacetime; all that changes is that the inhomogeneous terms in the rigidity equations become more complicated as they now incorporate curvature effects.  There appears to be no obstruction to iterating the solution to the RQF equations to arbitrarily high orders in powers of areal radius, which suggests that RQFs exist in a quite general context and possess precisely the same degrees of freedom of motion as rigid bodies in Newtonian space-time.

In {\S}\ref{secCurved} we also analyze the energy conservation equation introduced in {\S}\ref{secExtrinsic} in the context of this general solution, both to gain insight into the nature of this new geometrical energy flux, and to provide evidence for its interpretation as a gravitational energy flux.  Three key results emerge: (1) Even in flat spacetime there is a nonzero geometrical energy flux, which has the form of the cross product of the RQF's $\ell=1$ acceleration and rotation rate.  This flux integrates to zero over the closed surface of the RQF (as it should), and can be interpreted, via the equivalence principle, in terms of the gravitoelectromagnetic analogue of the Poynting vector familiar from linearized gravity.  It represents a form of gravitational energy flowing through the RQF system boundary analogous to the electromagnetic energy flowing through a sphere containing crossed, static electric and magnetic dipoles in electromagnetism.  (2) We construct a simple apparent paradox involving an accelerating box in flat spacetime, immersed in a uniform electric field, with co-moving observers attempting to understand the changing electromagnetic energy inside the box in terms of the Poynting vector flux through the boundary of the box.  The paradox is that considering only the Poynting flux accounts for only {\it half} of the rate of change of electromagnetic energy inside the box.  We show that an extra geometrical energy flux which naturally appears in the RQF approach precisely resolves this discrepancy.  We interpret this extra flux as a flux of gravitational energy due to the curvature of spacetime produced by the uniform electric field, which cannot be properly understood in the context of special relativity.  (3) Pushing the analysis to the next higher order in powers of areal radius (orders now involving terms quadratic in the curvature), we find that the geometrical energy flux takes the form of the cross product of the electric and magnetic parts of the Weyl tensor, which we take as additional evidence for the interpretation of the geometrical energy flux as a gravitational energy flux.  Finally, at the next higher order, the geometrical energy flux takes a form that bears some resemblance to (but is different from) the time derivative of the ``0000'' component of the Bel-Robinson tensor. So in general, the expansion of the geometrical energy flux in powers of areal radius results in an infinite series of increasingly complicated curvature tensor terms. We believe this explains why various curvature tensor expressions have, over the years, been tentatively associated with gravitational energy, but that ultimately the curvature tensor is {\it not} the correct language for gravitational energy. We contend that the correct language is the simple coupling between the intrinsic and extrinsic curvature of an RQF represented by the geometrical energy flux, which we propose is the true gravitational energy flux.

\section{RQF Extrinsic Geometry and Conservation Laws}\label{secExtrinsic}

In this section we will review how conservation laws are usually constructed in general relativity, and how the RQF approach improves on this construction (we will elaborate on this point in Chapter \ref{chLocalQuasilocal}).  To begin with, it is customary to require that the matter energy-momentum tensor, $T^{ab}_\text{mat}$, be covariantly conserved: $\nabla_a T_{\rm mat}^{ab}=0$.  As is well known \cite{Wald1984}, this condition may be interpreted as expressing local conservation of matter energy-momentum, but does not in general lead to an integrated conservation law.  An exception occurs if a Killing vector, $\Psi^a$, is present in the spacetime.  Then in the identity
\begin{equation}\label{MatterConservationEquation}
\nabla_a ( T_{\rm mat}^{ab}\Psi_b ) = ( \nabla_a T_{\rm mat}^{ab} ) \Psi_b + T_{\rm mat}^{ab} \nabla_{(a} \Psi_{b)},
\end{equation}
the two terms on the right-hand side vanish, and $J^a_\text{mat}:=-T_{\rm mat}^{ab}\Psi_b$ is a conserved current.  In other words, if $\Sigma$ is a spacelike three-surface with unit, future-directed normal vector field $u^a_{\Sigma}$ (where $u_{\Sigma} \cdot u_{\Sigma} =-c^2$), the quantity
\begin{equation}\label{ConservedCharge}
Q := -\frac{1}{c}\int_\Sigma \, d \Sigma \,\, u^a_{\Sigma} J_a^\text{mat}=\frac{1}{c}\int_\Sigma \, d \Sigma \,\, u^a_{\Sigma} \Psi^b T^{\rm mat}_{ab}
\end{equation}
is conserved, i.e., is independent of the choice of $\Sigma$ for fixed spatial boundary, $\partial\Sigma$, and evolves in the usual way according to a corresponding flux crossing $\partial\Sigma$.

If, for example, we are interested in an energy conservation law, which will be the primary focus of this section, there are three obvious shortcomings to this construction: (1) A generic spacetime does not admit a timelike Killing vector (or any Killing vectors, for that matter), and so such a construction is of limited value.  (2) Even if the spacetime {\it does} admit a timelike Killing vector, $\Psi^a$, and we are interested in, say, the matter energy contained in $\Sigma$, one might expect the integrand on the right-hand side of equation (\ref{ConservedCharge}) to be $\frac{1}{c^2} u^a_{\Sigma} u^b_{\Sigma} T^{\rm mat}_{ab}$, where we have imagined a three-parameter family of observers filling the volume, whose four-velocity is taken to be $u^a_{\Sigma}$ (i.e., hypersurface orthogonal).  Energy is, after all, an observer-dependent quantity, and $\frac{1}{c^2} u^a_{\Sigma} u^b_{\Sigma} T^{\rm mat}_{ab}$ is the local energy density of matter as measured by observers ``at rest'' with respect to $\Sigma$.  The problem is that, even in the simplest case where $\Psi^a$ is hypersurface orthogonal, and we choose $\Sigma$ to be such an orthogonal hypersurface, i.e., the observers are moving along integral curves of the Killing vector, with their four-velocity $u^a_{\Sigma}$ parallel to $\Psi^a$, $\Psi^a$ is not a unit vector in general, i.e., it differs from $u^a_{\Sigma}$ by a nonconstant scale factor, and equation (\ref{ConservedCharge}) is then not the expected expression for energy. (3) In any case, the matter energy in any finite volume cannot, in general, be separated out from the total energy (matter plus gravitational), and so any construction based on $T_{\rm mat}^{ab}$ is bound to be problematic at best.  We should really be seeking a {\it total} energy conservation law.  We will see in this section how the RQF approach resolves all three of these shortcomings.

In 1993,  Brown and  York \cite{BY1993} took an important step towards addressing the third shortcoming when they suggested, based on a careful Hamilton-Jacobi-type analysis of general relativity, that the total energy-momentum tensor (matter plus gravitational) is {\it quasilocal} in nature: it is a tensor defined in the {\it boundary} of the history of a finite spatial volume (which in our RQF approach we are denoting as $\mathcal B$), and is simply\footnote{\label{VacuumEnergyFootnote}There is the tricky question of a ``reference subtraction'' required to remove ``vacuum'' contributions to $T_{\mathcal B}^{ab}$, and also to regulate $T_{\mathcal B}^{ab}$ for infinite volumes.  However, insofar as: (1) we are dealing here with finite volumes; (2) we are primarily interested in {\it changes} in energy (and momentum and angular momentum), and so any vacuum contributions cancel out; (3) we will give some evidence that these vacuum contributions might, in fact, have some physical significance; and (4) ignoring a possible reference subtraction leads to no apparent inconsistencies or other problems, we will take the ``unreferenced'' quasilocal energy-momentum tensor in equation (\ref{BoundarySEM}) at face value and explore the consequences.}
\begin{equation} \label{BoundarySEM}
T_{\mathcal B}^{ab} := -\frac{1}{\kappa}\,\Pi^{ab},
\end{equation}
where $\Pi_{ab}:=\Theta_{ab} - \Theta \gamma_{ab}$ is the momentum canonically conjugate to the three-metric $\gamma_{ab}$ on $\mathcal B$, $\Theta_{ab}:=\gamma_a^{\phantom{a}c}\nabla_c n_b$ is the (symmetric) extrinsic curvature of $\mathcal B$ (and $\Theta$ its trace), and $\kappa = 8 \pi G / c^4$.  For example, in the context of our RQF observers, whose four-velocity is $u^a$, the quantity ${\mathcal E}:= \frac{1}{c^2} u^a u^b T^{\mathcal B}_{ab}$ is to be considered as the quasilocal total energy {\it surface} density (energy per unit area) measured by the observers; ``quasilocal'' in the sense that it has meaning only when integrated over a closed two-surface.  Roughly speaking, and stated more precisely following equation (\ref{LHS}) below, integrating this quasilocal energy density over a two-sphere slice of $\mathcal B$ yields a measure of the total energy (matter plus gravitational) inside any spatial volume spanning this slice.

How can we construct conservation laws associated with such a quasilocal energy-momentum tensor? An obvious approach to try is to simply write down the quasilocal analogue of equation (\ref{MatterConservationEquation}), wherein we replace the {\it matter} energy-momentum tensor, $T_{\rm mat}^{ab}$, defined in the four-dimensional spacetime $(M,g_{ab})$, with the {\it total} energy-momentum tensor, $T_{\mathcal B}^{ab}$, defined in the three-dimensional spacetime $({\mathcal B},\gamma_{ab})$.  It might appear that this would result in conservation laws involving charges and fluxes {\it within} $\mathcal B$, which would have to be trivial because the cross sections of $\mathcal B$ are {\it closed} surfaces (two-spheres).  But because of the quasilocal interpretation of $T_{\mathcal B}^{ab}$, and the nature of the Einstein equation and the RQF conditions, we will see that we will actually be analyzing the exchange of energy, momentum and angular momentum between the RQF system and the universe external to it, i.e., charges associated with spatial volumes spanning two-sphere slices of $\mathcal B$, and fluxes passing {\it through} $\mathcal B$.

Thus, in analogy with $\Psi^a$ in equation (\ref{MatterConservationEquation}), let $\psi^a$ be an arbitrary vector field tangent to $\mathcal B$, and consider the identity
\begin{equation}\label{Basic_Conservation_Equation}
D_a ( T_{\mathcal B}^{ab}\psi_b ) = ( D_a T_{\mathcal B}^{ab} ) \psi_b + T_{\mathcal B}^{ab} D_{(a} \psi_{b)},
\end{equation}
where $D_a$  is the covariant derivative with respect to the metric, $\gamma_{ab}$, induced in $\mathcal B$.  Integrating this identity over a portion, $\Delta {\mathcal B}$, of $\mathcal B$, between initial (${\mathcal S}_i$) and final (${\mathcal S}_f$) two-sphere slices of $\mathcal B$, we have
\begin{equation}\label{Basic_Integrated Conservation_Equation}
\frac{1}{c}\negthickspace\negthickspace\int\limits_{{\mathcal S}_f-{\mathcal S}_i}\negthickspace\negthickspace d{\mathcal S}\, u^a_{\mathcal{S}} \psi^b T^{\mathcal B}_{ab} = -\negthickspace\int\limits_{\Delta {\mathcal B}}d{\mathcal B}\,\left[( D_a T_{\mathcal B}^{ab} ) \psi_b + T_{\mathcal B}^{ab} D_{(a} \psi_{b)}\right],
\end{equation}
where $d{\mathcal S}$ and $d{\mathcal B}$ are the volume elements on ${\mathcal S}_{i,f}$ and ${\mathcal B}$, respectively, and $u^a_{\mathcal{S}}$ is the unit, future-directed vector field normal to ${\mathcal S}_{i,f}$, analogous to $u^a_{\Sigma}$ in equation (\ref{ConservedCharge}).
As will be discussed in detail below, the left-hand side of equation (\ref{Basic_Integrated Conservation_Equation}) represents the change in some physical quantity of the RQF (energy, momentum or angular momentum) between ${\mathcal S}_{i}$ and ${\mathcal S}_{f}$.  The right-hand side represents two types of flux crossing the timelike boundary, $\Delta {\mathcal B}$, spanning ${\mathcal S}_{i}$ and ${\mathcal S}_{f}$, that account for this change: a matter flux, represented by the term $( D_a T_{\mathcal B}^{ab} ) \psi_b$, and a ``geometrical,'' or gravitational, flux, represented by the term $T_{\mathcal B}^{ab} D_{(a} \psi_{b)}$.  The choice of which physical quantity we are concerned with (energy, momentum or angular momentum) depends on our choice of $\psi^a$.  (Note that $( D_a T_{\mathcal B}^{ab} ) \psi_b$ and $T_{\mathcal B}^{ab} D_{(a} \psi_{b)}$ are outward-directed fluxes that cause a decrease in the corresponding physical quantity, which explains the presence of the negative sign in equation (\ref{Basic_Integrated Conservation_Equation}))

To begin with, let us establish some notation.  We follow Brown and York \cite{BY1993} in splitting the total energy-momentum tensor into physically distinct parts:
\begin{equation}\label{SurfaceSEMcomponents}
\begin{split}
{\mathcal E} & := \frac{1}{c^2} u^a u^b T^{\mathcal B}_{ab}\\
c{\mathcal P}_a & := - \frac{1}{c} \sigma_{a}^{\phantom{a}b}u^{c}T^{\mathcal B}_{bc}\\
{\mathcal S}_{ab} & := - \sigma_{a}^{\phantom{a}c}\sigma_{b}^{\phantom{b}d}T^{\mathcal B}_{cd},
\end{split}
\end{equation}
which denote, respectively, the energy surface density (energy per unit area), the momentum surface density (momentum per unit area), and the spatial stress\footnote{Note that the sign convention here for the quasilocal stress is opposite to that of Brown and York \cite{BY1993}. Our sign convention aligns with that of electromagnetism, and makes various arguments which we will present in Chapters \ref{chLocalQuasilocal} and \ref{chArch} more physically sensible.} (force per unit length) as seen by our RQF observers, whose four-velocity is $u^a$.

Next, let us choose $\psi^a=u^a /c$, which will give us an energy conservation law associated with the quasilocal energy-momentum current $J^a_{\mathcal B}:=-T_{\mathcal B}^{ab}\psi_b = {\mathcal E}u^a /c + c{\mathcal P}^a$.  In this case the integrand on the left-hand side of equation (\ref{Basic_Integrated Conservation_Equation}) becomes $\frac{1}{c^2} u^a_{\mathcal{S}} u^b T^{\mathcal B}_{ab}$.  When the observers are ``at rest'' with respect to ${\mathcal S}_{i,f}$ (i.e., $u^a=u^a_{\mathcal{S}}$), this integrand reduces to $\frac{1}{c^2} u^a_{\mathcal{S}} u^b_{\mathcal{S}}  T^{\mathcal B}_{ab}=\frac{1}{c^2} u_{a} u_b T_{\mathcal B}^{ab}={\mathcal E}$, which is the expression for (now {\it total}) energy density one might expect - recall the discussion of ``shortcomings'' in the paragraph following equation (\ref{ConservedCharge}).  In general, however, the observers are {\it not} ``at rest'' with respect to ${\mathcal S}_{i,f}$, and $u^a$ and $u^a_{\mathcal{S}}$ are related by a boost transformation: $u^a_{\mathcal{S}}=\gamma (u^a + v^a )$.  Here $v^a\in H$ is a ``shift'' vector that can be interpreted as the tangential, spatial two-velocity of the RQF observers ``gliding over the sphere,'' and $\gamma$ is an inverse ``lapse'' function that corresponds to the usual ``$\gamma$-factor'' of the associated Lorentz transformation. In this, the general case, the $-u^a_{\mathcal{S}} /c$ projection of the energy-momentum current $J^a_{\mathcal B}$ suffers a Lorentz transformation, and the left-hand side (LHS) of equation (\ref{Basic_Integrated Conservation_Equation}) becomes
\begin{equation}\label{LHS}
\text{ LHS of equation (\ref{Basic_Integrated Conservation_Equation})} =\negthickspace\negthickspace\negthickspace\int\limits_{{\mathcal S}_f-{\mathcal S}_i}\negthickspace\negthickspace\negthickspace d{\mathcal S}\,\gamma \left({\mathcal E} - v_a {\mathcal P}^{a}\right).
\end{equation}
It is worth noting that $v^a$ contains the same information as the dynamical degrees of freedom of the RQF intrinsic geometry, $u_j$, discussed in the previous section.  To see this, we can choose our adapted coordinate system such that ${\mathcal S}_{i,f}$ are surfaces of constant $t$, and then clearly we require $u^{\mathcal{S}}_j=0$, i.e., $v_j = - u_j$ (and note that $v_t =0$ by construction).

There are at least two general arguments suggesting the plausibility of ${\mathcal E}:=\frac{1}{c^2}u_{a}u_b T_{\mathcal B}^{ab}$ as a total energy surface density, and thus equation (\ref{LHS}) as a total system energy (matter plus gravitational).  First, note that
\begin{equation}\label{SpatialExtrinsicCurvature}
{\mathcal E}:=\frac{1}{c^2}u_{a}u_b T_{\mathcal B}^{ab}=-\frac{1}{\kappa c^2}u^a u^b (\Theta_{ab}-\Theta\gamma_{ab})=-\frac{1}{\kappa}\sigma^{ab}\Theta_{ab}=-\frac{c^4}{8\pi G}k,
\end{equation}
where $k:=\sigma^{ab}\Theta_{ab}$.  Consider the simplest case, in which $u^a$ is hypersurface orthogonal (as a vector field in $\mathcal B$), and let $\mathcal S$ be a two-sphere slice of $\mathcal B$ to which $u^a$ is everywhere orthogonal (i.e., the observers are ``at rest'' with respect to $\mathcal S$).  Then $k$ is the trace of the extrinsic curvature of $\mathcal S$ in the $n^a$ direction normal to $\mathcal B$.  In other words, $k$ measures the fractional rate at which the surface area elements of $\mathcal S$ increase as we move a unit proper distance radially outwards.  For example, if $\mathcal S$ is a round sphere of areal radius $r$ in Euclidean three-space, then $k=2/r$.  If we now put some mass-energy inside the sphere, it will ``warp'' the three-space such that, for a sphere of the same areal radius, its surface area will increase {\it less rapidly} than we expect based on Euclidean intuition.  (Think of the standard funnel-shaped embedding diagram for the warped three-space surrounding a compact object - the more mass-energy inside, the steeper the slope of the funnel, and the less rapidly the area of a spatial two-sphere slice will increase as we move a unit proper distance radially outwards.) Thus, the presence of mass-energy inside the sphere {\it decreases} $k$ (increases $-k$), and in this way $\mathcal E$ is a quasilocal measure of the mass-energy inside the sphere.  This measure naturally includes a gravitational energy contribution.  See reference \cite{BY1993} for a detailed discussion and examples (in particular, see the discussion surrounding their equation (6.15)).

As a second general plausibility argument for the interpretation of ${\mathcal E}:=\frac{1}{c^2}u_{a}u_b T_{\mathcal B}^{ab}$ as a total energy surface density, note that $\Theta = (\sigma^{ab}-\frac{1}{c^2}u^a u^b )\nabla_a n_b = k+\frac{1}{c^2}n\cdot a$.  Defining the pressure as one-half the trace of the spatial stress, $\mathbb{P}:= \frac{1}{2}\sigma^{ab}{\mathcal S}_{ab}$, and using equation (\ref{SpatialExtrinsicCurvature}), we have
\begin{equation}\label{EPnDOTaRelation}
{\mathcal E} = \frac{c^2}{4\pi G} n\cdot a + 2 \mathbb{P}.
\end{equation}
The first term on the right-hand side is intuitively satisfying in that the normal component of acceleration, $n\cdot a$, required for observers to hover a fixed distance from a static, compact object is clearly a measure of the mass-energy of the object.  For example, in the Newtonian limit of a two-sphere's worth of observers hovering a fixed radial distance $r$ from a point mass $M$, the normal component of their acceleration is $n\cdot a = GM/r^2$, and integrating the first term in equation (\ref{EPnDOTaRelation}) over the two-sphere yields a contribution of $Mc^2$ to the total energy.  More generally, for any spherically symmetric mass distribution, any mass outside the radius of the RQF sphere has no effect on $n\cdot a$. In other words this method of measuring mass is {\it inherently quasilocal} in nature, not referring to anything outside of the system.  And of course it is closely related to Komar's definition of mass (see, e.g., section 11.2 of reference \cite{Wald1984} for a discussion of the Komar mass), except that here we are not restricted to stationary (and asymptotically flat) spacetimes.  So one might wonder why this is not the end of the story - why there is a $2 \mathbb{P}$ term added to the $n\cdot a$ term.  It has to do with the distinction between (inertial) mass and energy.  To see this, it helps to rewrite the equation as $\frac{c^2}{4\pi G} n\cdot a = {\mathcal E} - 2 \mathbb{P}$.  It is well known that, in relativity, pressure contributes to the inertia of a system, and it is plausible that this equation is the quasilocal, {\it total} energy (matter plus gravitational) version of this phenomenon.

Let us now turn our attention to the right-hand side of equation (\ref{Basic_Integrated Conservation_Equation}), beginning with the first term - the matter flux term.  Using the identity $D_a \Pi^{ab}=\gamma^{bc}n^{d}G_{cd}$, where $G_{cd}$ is the Einstein tensor, and the definition of $T_{\mathcal B}^{ab}$ in equation (\ref{BoundarySEM}), this term is:
\begin{equation} \label{eq:MatterFlux}
( D_a T_{\mathcal B}^{ab} ) \psi_b=-{\frac{1}{\kappa}}n^{a} \psi^{b} G_{ab}= -n^{a} \psi^{b} T^{\rm mat}_{ab},
\end{equation}
where for the last equality we used the Einstein equation.  Unlike in equation (\ref{MatterConservationEquation}), where $\nabla_a T_{\rm mat}^{ab}=0$, in equation (\ref{Basic_Conservation_Equation}) the divergence of the quasilocal total energy-momentum tensor is not zero.  If $\nabla_a T_{\rm mat}^{ab}$ was not zero it would signal the presence of ``external sources,'' i.e., matter fields not accounted for in $T_{\rm mat}^{ab}$ that are nevertheless interacting with the system represented by $T_{\rm mat}^{ab}$.  In our case, the nonvanishing of $D_a T_{\mathcal B}^{ab}$ likewise corresponds to external sources, now in the form of various matter fluxes passing {\it through} $\mathcal B$ that interact with the RQF system represented by $T_{\mathcal B}^{ab}$.  These fluxes of energy, momentum, and angular momentum (depending on our choice of $\psi^a$) are intimately connected with the {\it motion} of the RQF system, by which we mean inertial accelerations and precession rates of inertial gyroscopes.

For example, let us again take $\psi^a= u^a /c$ (so we are dealing with energy conservation), and consider an electromagnetic (em) field present at $\Delta {\mathcal B}$.  Then, setting $T^{\rm mat}_{ab}$ to $T^{\rm em}_{ab}$, we have
\begin{equation}\label{Electromagnetic Flux}
-n^{a} u^{b} T^{\rm em}_{ab} =  \frac{c}{4\pi} \epsilon_{ab}e^{a}b^{b} ,
\end{equation}
where $e^a$ and $b^a$ denote the electric and magnetic fields experienced by observers with four-velocity $u^a$ on $\Delta \mathcal B$.  As expected, this expression is the $n^a$ (outward-directed) component of the Poynting vector, i.e., if electromagnetic energy is leaving the system, the energy of the RQF will decrease - the energy change in equation (\ref{LHS}) will be negative.  If, on the other hand, $\psi^a$ is taken to be in a spatial direction (tangential to $\mathcal B$ and orthogonal to $u^a$), equation (\ref{eq:MatterFlux}) represents forces and torques (e.g., components of the Maxwell stress tensor in the case of electromagnetism) that cause changes in the momentum and angular momentum of the RQF.\footnote{This is, of course, not surprising.  In the same way that $\nabla_a T_{\rm mat}^{ab}=0$ is intimately connected with the equations of motion of matter, equation (\ref{eq:MatterFlux}) is intimately connected with the equations of motion of an RQF system.}  This will be discussed in more detail in Chapters \ref{chArch} and \ref{chPN}.

Finally, let us turn our attention to the most interesting, and last, term on the right-hand side of equation (\ref{Basic_Integrated Conservation_Equation}).  To begin with, recall that in equation (\ref{MatterConservationEquation}) we required both terms on the right-hand side to vanish in order for $J^a_\text{mat}=-T_{\rm mat}^{ab}\Psi_b$ to be a conserved current.  We have just seen that in the quasilocal analogue of this equation, equation (\ref{Basic_Conservation_Equation}), the first term on the right-hand side need {\it not} vanish in order for $J^a_{\mathcal B}=-T_{\mathcal B}^{ab}\psi_b$ to be a conserved current.  Indeed, its nonvanishing simply represents the matter flux naturally associated with the time evolution of the corresponding charge.  Together with the fact that, in the quasilocal case, we are dealing with the {\it total} energy-momentum tensor (matter plus gravitational), it is thus reasonable to guess that the second term on the right-hand side of equation (\ref{Basic_Conservation_Equation}) need not vanish in order for $J^a_{\mathcal B}$ to be a conserved current, and that, in fact, it will represent a gravitational contribution to the associated flux.

So while it needn't, and in general doesn't, vanish, there is a good reason - that will become evident shortly - to see how far we can go to making the second term on the right-hand side of equation (\ref{Basic_Conservation_Equation}) vanish.  In analogy with $\Psi^a$ in equation (\ref{MatterConservationEquation}), it would obviously vanish if $\psi^a$ was a Killing vector of the boundary three-geometry.  But the boundary three-geometry will not, in general, admit any Killing vectors - and certainly not in the most interesting case when gravitational radiation is crossing the boundary.  So we will do the next best thing and ask, ``To what extent - in the {\it general} case - can $\psi^a$ have Killing vector-{\it like} properties?''

For example, while the boundary three-geometry will not, in general, admit a timelike Killing vector - a time symmetry of the {\it full} three-geometry - can the observers' congruence of worldlines always be chosen such that it admits a time symmetry of at least the spatial two-geometry of the bounding box of the RQF?  Taking $\psi^a \propto u^a$, this is equivalent to asking if the {\it spatial projection}, $\sigma_{a}^{\phantom{a}c}\sigma_{b}^{\phantom{b}d}D_{(c}u_{d)}=0$, of the boundary Killing equation can be satisfied {\it in general}.  Indeed, these are precisely the RQF rigidity conditions in equation (\ref{eq:RigidityCondition}).  We will argue below, and in the next section, that the failure of $u^a$ to satisfy the {\it full} Killing equation in the boundary is associated in a simple and precise way with gravitational energy passing through that boundary.  Also, it is worth noting that this spatially projected Killing equation is invariant under a rescaling (``conformal'' transformation) of $u^a$ by an arbitrary function.  This is the underlying reason explaining how the RQF approach resolves the second of the three ``shortcomings'' discussed in the paragraph following equation (\ref{ConservedCharge}).

Similarly, the boundary three-geometry will not, in general, admit a spacelike Killing vector. However, as already emphasized, the spatial two-geometry of the bounding box of an RQF always admits precisely six {\it conformal} Killing vectors (CKVs): three boost-like and three rotation-like.  Taking $\psi^a$ to be such a boost- or rotation-like CKV results in simple, exact expressions for gravitational momentum or angular momentum, respectively, passing through the bounding box (and corresponding expressions for total - matter plus gravitational - momentum or angular momentum charges).  This case will be discussed in detail in Chapters \ref{chArch} and \ref{chPN}.  For now we consider only the case $\psi^a \propto u^a$.

In particular, taking $\psi^a = u^a /c$, as before, it is easy to see from the definitions in equation (\ref{SurfaceSEMcomponents}) that, when the RQF rigidity conditions in equation (\ref{eq:RigidityCondition}) are satisfied,
\begin{equation}\label{GravitationalFluxesRQF}
T_{\mathcal B}^{ab} D_{(a} u_{b)} = \alpha_a {\mathcal P}^a .
\end{equation}
We claim that the quantity $\alpha_a {\mathcal P}^a$ is a simple, exact expression for the outward-directed ``geometrical'' flux of gravitational energy across the boundary of an RQF.  We will provide detailed evidence for this interpretation in the next section, but to immediately see that this is plausible, observe that, starting from the definition in equation (\ref{SurfaceSEMcomponents}),
\begin{equation}\label{InterpretationOfScriptP}
{\mathcal P}^a :=-\frac{1}{c^2}\sigma^{ab}u^c T^{\mathcal B}_{bc}=\frac{1}{\kappa c^2}\sigma^{ab}u^c (\Theta_{bc}-\Theta\gamma_{bc})=\frac{c^2}{8\pi G}\sigma^a_{\phantom{a}b}u^c\nabla_c n^b=:\frac{c^2}{8\pi G}\epsilon^a_{\phantom{a}b}\omega^b ,
\end{equation}
where $\omega^a$ is the precession rate of inertial gyroscopes as measured by the RQF observers (i.e., the $\epsilon^a_{\phantom{a}b}$ projection of the rotation rate of $n^a$ under parallel transport along $u^a$).  So ${\mathcal P}^a$ can be interpreted as either a tangential (to $\mathcal B$) momentum surface density, or a gyroscope precession rate (projected tangentially to $\mathcal B$ and rotated 90 degrees by an $\epsilon^a_{\phantom{a}b}$ tensor).  Hence, our geometrical, or gravitational energy flux can be expressed in the more suggestive form:
\begin{equation}\label{ACrossOmegaFormOfGeometricalFlux}
\alpha_a{\mathcal P}^a = \frac{c^2}{8\pi G}\epsilon_{ab}a^a \omega^b ,
\end{equation}
(where we have replaced, with impunity, the tangential acceleration, $\alpha^a$, with the full four-acceleration, $a^a$).  The analogy with electromagnetic energy flux is striking - recall equation (\ref{Electromagnetic Flux}), with the identifications $e^a\leftrightarrow a^a$ and $b^a\leftrightarrow \omega^a$.  It is well known that, in the gravitoelectromagnetic interpretation of linearized general relativity, the gravitoelectric field is associated with acceleration, and the gravitomagnetic field with rotation \cite{Mashhoon}.  So if one had to guess an expression for gravitational energy flux, analogous to that in electromagnetism, one might well try an expression like $\epsilon_{ab}a^a \omega^b$, multiplied by $c^2/G$ to get the units right. The question would then be, ``What do we put for $a^a$ and $\omega^a$?''  The RQF approach provides a natural and precise answer: $a^a$ is the concerted two-parameter family of four-accelerations that observers must undergo in order to maintain constant radar ranging distances to all nearest neighbour observers, i.e., to ``actively'' (e.g., with rockets) compensate for the geodesic deviations they would otherwise experience in freefall in a dynamical spacetime.  And $\omega^a$ is derived uniquely from the resulting $a^a$.  It is remarkable that this simple expression for gravitational energy flux is actually {\it exact}, in the full nonlinear theory.  Moreover, it is {\it operational}: RQF observers can directly measure $a^a$ and $\omega^a$ using accelerometers and gyroscopes, and thus determine the quasilocal gravitational energy flux at their respective positions.

In summary, substituting $\psi^a = u^a /c$ into the general conservation equation (\ref{Basic_Integrated Conservation_Equation}) gives us an energy conservation law that relates the change in {\it total} energy of the RQF system (matter plus gravitational), between initial and final time slices ${\mathcal S}_{i}$ and ${\mathcal S}_{f}$ of $\mathcal B$, to two types of energy flux, matter and gravitational, through the timelike three-surface, $\Delta {\mathcal B}$, spanning ${\mathcal S}_{i}$ and ${\mathcal S}_{f}$:
\begin{equation}\label{eq:SimpleConservationEq}
\int\limits_{{\mathcal S}_f-{\mathcal S}_i} \negthickspace\negthickspace d{\mathcal S}\,\gamma \left({\mathcal E} - v_{a}{\mathcal P}^{a}\right)=
\frac{1}{c}
\int\limits_{\Delta {\mathcal B}} d{\mathcal B} \, \left(  n^{a} u^{b} T^{\rm mat}_{ab} - \alpha_a {\mathcal P}^a \right).
\end{equation}
On the left-hand side of this equation, $\gamma$ and $-v_a$ represent a Lorentz boost from the observers' four-velocity, $u^a$, to the four-velocity $u^a_{\mathcal{S}}$ of observers momentarily ``at rest'' with respect to ${\mathcal S}_{i}$ and ${\mathcal S}_{f}$.  In the next section we will examine this energy conservation law in detail, highlighting the interpretation and significance of the geometrical, or gravitational energy flux term, $\alpha_a {\mathcal P}^a$.

\section{RQFs in Curved Spacetime}\label{secCurved}

In Chapter \ref{chRigidRevisited} we addressed the existence and utility of RQFs in flat spacetime.  Based on the results of the previous section, we will now extend this work to a generic curved spacetime, proceeding perturbatively in powers of the areal radius of a round sphere RQF.  Let us begin by writing down the components of the metric, $g_{ab}$, in Fermi normal coordinates\footnote{Fermi normal coordinates are the natural choice for this purpose because they describe the {\it proper} reference frame of a non-inertial observer in an arbitrary curved spacetime near the observers worldline.}, $X^{a} := (cT,X^{I})$, $I=1,\,2,\,3$, in the neighbourhood of a timelike worldline, $\mathcal C$, (with arbitrary acceleration and rotation) in a generic spacetime:\cite{standardFNCreference}
\begin{align}
g_{00} &= - \left(1 + \frac{1}{c^2} A_{K} X^{K}\right)^2 + \frac{1}{c^2} R^2 W_{K} W_{L} P^{KL} - \overset{\mathtt{o}}{R}_{0K0L} X^{K} X^{L} + \mathcal{O}(R^3),\label{eq:FNCmetric00}\\
g_{0J} &=  \frac{1}{c} \epsilon_{JKL}  W^{K} X^{L} - \frac{2}{3} \overset{\mathtt{o}}{R}_{0KJL} X^{K} X^{L} + \mathcal{O}(R^3),\label{eq:FNCmetric0J}\\
g_{IJ} &= \delta_{IJ} - \frac{1}{3} \overset{\mathtt{o}}{R}_{IKJL} X^{K} X^{L} + \mathcal{O}(R^3),\label{eq:FNCmetricIJ}
\end{align}
where $ R^2 := \delta_{IJ} X^I X^J$, $A_{K}(T)$ is the proper acceleration along $\mathcal C$, $W_{K}(T)$ is the proper rate of rotation of the spatial axes (triad) along $\mathcal C$, $P^{KL} := \delta^{KL} - X^{K} X^{L} / R^2$ projects vectors perpendicular to the radial direction, and $ \overset{\mathtt{o}}{R}_{abcd}(T)$ are the Fermi normal coordinate components of the Riemann curvature tensor evaluated on $\mathcal C$.  Note that $T$ is the proper time along $\mathcal C$ ($R=0$), and an overset circle indicates a quantity evaluated on $\mathcal C$ (except for $A_{K}$ and $W_{K}$, which obviously refer to $\mathcal C$).

Let us now embed, into this coordinate system, a two-parameter family of worldlines in the neighbourhood of $\mathcal C$ that will represent a two-sphere's worth of observers, i.e., a fibrated timelike worldtube, $\mathcal B$, surrounding $\mathcal C$.  To do this, we introduce a second set of coordinates, $x^\alpha := (t,\,r,\,x^i )$, which are the coordinates $x^\mu := (t,\,x^i )$ on $\mathcal B$ introduced in {\S}\ref{secDefinition}, augmented by a radial coordinate, $r$.  Then, taking $x^i =(\theta,\phi)$, we introduce the coordinate transformation:
\begin{align}
T(t,\,r,\,\theta,\phi) &= t,\label{T=t}\\
X^I(t,\,r,\,\theta,\phi) &= r r^I(\theta,\phi) + r^3 f^I(t,\theta,\phi) + \mathcal{O}(r^4),\label{X^I=...}
\end{align}
where $r^{I}(\theta,\phi) := (\sin{\theta}\cos{\phi},\,\sin{\theta}\sin{\phi},\,\cos{\theta})$ are the standard direction cosines of a radial unit vector in spherical coordinates in Euclidean three-space.  The idea is that the three arbitrary functions $f^I(t,\theta,\phi)$ (and their counterparts at higher order in $r$) allow us the full freedom to ``wiggle'' the observers' worldlines (defined by $r,\,\theta,\,\phi=$ constant) arbitrarily in the three spatial directions, and are to be chosen such that the three RQF rigidity conditions in equation (\ref{eq:RigidityCondition}) are satisfied.  Specifically, we will demand that the observers' radar ranging two-metric, $\sigma_{ij}$, induced by the embedding, be equal to $r^2 \mathbb{S}_{ij}$, so that the observers find themselves on a round sphere of areal radius $r$.  (Recall {\S}\ref{secDefinition} for a reminder of our notation.)

There are three points worth noting: (1) The RQF conditions, which are equivalent to $\partial_t \sigma_{ij} = 0$ in our adapted coordinate system, are clearly invariant under a time reparametrization, and so to simplify the analysis as much as possible we have chosen surfaces of constant $t$ to coincide with surfaces of constant $T$, i.e., $T=t$ in equation (\ref{T=t}); (2) the RQF conditions are obviously trivially satisfied at lowest order ($X^I = r r^I$ in equation (\ref{X^I=...})), and we find that the first nontrivial order is two orders of $r$ higher, which explains the absence of an $\mathcal{O}(r^2)$ term in equation (\ref{X^I=...}); and (3) for technical reasons it proves useful to decompose $f^I$ as follows:
\begin{align}\label{f^Idecomposition}
f^I (t,\theta,\phi) = F(t,\theta,\phi) r^I (\theta,\phi) + f^{i} (t,\theta,\phi) \mathbb{B}^{I}_{i}(\theta,\phi).
\end{align}
Here $F$ encodes a radial, or normal perturbation of the observers' worldlines, and $f^i$ encodes an angular, or tangential perturbation, together comprising three functional degrees of freedom.  Recall, $\mathbb{B}^{I}_{i} := \partial_{i} r^I $ are the {\it boost generators} (see appendix \ref{AppendixCKV} for more details).

With this construction, we find that the induced radar ranging two-metric seen by the RQF observers is:
\begin{align}\label{DiffEqRQF}
\sigma_{ij} =& r^2 \mathbb{S}_{ij} + r^4 \left(2 \mathbb{D}_{(i} f_{j)} + 2 F \mathbb{S}_{ij} - \frac{1}{3} \overset{\mathtt{o}}{R}_{IKJL} \mathbb{B}^I_i \mathbb{B}^J_j r^K r^L  + \frac{1}{c^2} W_I W_J \mathbb{R}^I_i \mathbb{R}^J_j\right)
+ \mathcal{O}(r^5) ,
\end{align}
where here, and in what follows, the quantities $A_K$, $W_K$ and $\overset{\mathtt{o}}{R}_{abcd}$, which in equations (\ref{eq:FNCmetric00}) to (\ref{eq:FNCmetricIJ}) are functions of $T$, are now functions of $t$, according to equation (\ref{T=t}).
We have defined $f_i :=\mathbb{S}_{ij}f^j$, and $\mathbb{D}_i$ is the covariant derivative operator associated with the unit round sphere metric, $\mathbb{S}_{ij}$.  Letting $\mathbb{E}_{ij}$ denote the volume form associated with $\mathbb{S}_{ij}$, we have also have $\mathbb{R}^I_i := \mathbb{E}_i^{\phantom{i}j} \mathbb{B}^I_j$ as the {\it rotation generator} counterparts to $\mathbb{B}^{I}_{i}$.

Inspection of equation (\ref{DiffEqRQF}) reveals that to satisfy the RQF rigidity conditions ($\sigma_{ij}=r^2 \mathbb{S}_{ij}$) to lowest nontrivial order in $r$ requires that $F$ and $f_i$ satisfy the three differential equations
\begin{align}\label{StartDE}
\mathbb{D}_{(i} f_{j)} +  F \mathbb{S}_{ij} = I_{ij},
\end{align}
where we have set
\begin{align}\label{Iij}
I_{ij} := \frac{1}{6} \overset{\mathtt{o}}{R}_{IKJL} \mathbb{B}^I_i \mathbb{B}^J_j r^K r^L  - \frac{1}{2c^2} W_I W_J \mathbb{R}^I_i \mathbb{R}^J_j .
\end{align}
Taking the trace and trace-free parts of these equations yields three equivalent equations:
\begin{align}
F &= - \frac{1}{2}\mathbb{D} \cdot f+\frac{1}{2} I , \label{eq:traceDE} \\
\mathbb{D}_{<i} f_{j>} &=  I_{<ij>}, \label{eq:tracefreeDE}
\end{align}
where $I := \mathbb{S}^{ij} I_{ij}$ is the trace, and $I_{<ij>} := I_{ij} - \frac{1}{2} \mathbb{S}_{ij} I$ the (symmetric) trace-free part of $I_{ij}$.  With the inhomogeneous ``source'' term $I_{ij}$ specified, equation (\ref{eq:traceDE}) tells us that $F$ (the radial perturbation) is determined uniquely once $f_i$ (the angular perturbation) is known.  Thus, our focus will be on solving equation (\ref{eq:tracefreeDE}) for $f_i$.  To do so, we expand $f_i$ as a sum of independent vector spherical harmonics with arbitrary coefficients, calculate $\mathbb{D}_{<i} f_{j>}$, decompose both $\mathbb{D}_{<i} f_{j>}$ and  $I_{<ij>}$ into independent tensor spherical harmonics, and then read off the required coefficients.  The result here is:
\begin{align}
F =\,& \alpha_{I}(t) r^{I} +\frac{\kappa}{18} \overset{\mathtt{o}}{T}_{00} -\frac{1}{6c^2}  W^2 +\mathbb{F}, \label{eq:F} \\
f_i =\,&  \alpha_{I}(t) \mathbb{B}^{I}_{i} + \beta_{I}(t) \mathbb{R}^{I}_{i} +\frac{1}{4} \partial_{i} \mathbb{F} , \label{eq:fi2}
\end{align}
where
\begin{equation}\label{script F}
\mathbb{F}:= Q^{IJ}\left( \frac{1}{c^2} W_I W_J + \frac{1}{3} \overset{\mathtt{o}}{\mathcal{E}}_{IJ} + \frac{\kappa}{6} \overset{\mathtt{o}}{T}_{IJ}\right)
\end{equation}
is a pure $\ell=2$ spherical harmonic.
Here $\alpha_I(t)$ and $\beta_I(t)$ are six arbitrary, time-dependent functions; $Q^{IJ} := r^I r^J - \frac{1}{3} \delta^{IJ}$ is trace-free and represents the five independent pure $l=2$ spherical harmonics; $W^2 := \delta^{IJ} W_I W_J$; and we have decomposed the Riemann tensor into the electric part of the Weyl tensor, $\mathcal{E}_{IJ} := C_{0I0J}$, the magnetic part of the Weyl tensor, $\mathcal{B}_{IJ} := \frac{1}{2} \epsilon_{I}^{\phantom{I}KL} C_{0JKL}$ (which we will need later), and the Ricci tensor, $\overset{\mathtt{o}}{R}_{ab}=\kappa(\overset{\mathtt{o}}{T}_{ab}-\frac{1}{2}\overset{\mathtt{o}}{T}\overset{\mathtt{o}}g_{ab})$, where we have dropped the superscript ``mat'' on the matter energy-momentum tensor, $\overset{\mathtt{o}}{T}_{ab}$.

It is instructive to take a moment to analyze this solution.  We begin with the homogeneous part, i.e., the solution to (\ref{StartDE}) when $I_{ij}=0$.  It is given by the first term on the right-hand side of equation (\ref{eq:F}) and the first two terms on the right-hand side of equation (\ref{eq:fi2}), i.e., $F =\alpha_{I}(t) r^{I}$ and $f_i =  \alpha_{I}(t) \mathbb{B}^{I}_{i} + \beta_{I}(t) \mathbb{R}^{I}_{i}$.  As we saw in {\S} \ref{ArbitraryPerturbations}, $\alpha^I(t)$ and $\beta^I(t)$ correspond to time-dependent translations and rotations of the RQF, respectively, and impart the RQF with the six degrees of freedom of rigid body motion we are familiar with in Newtonian space-time.

It is important to note that: (1) While we are working at order $r^3$ here - see equation (\ref{X^I=...}), i.e., the lowest order with a nontrivial particular solution, we find a homogeneous solution of the form discussed above for perturbations at both lower orders ($r$ and $r^2$) and higher orders, with no obvious reason this would change at arbitrarily high orders.  Thus, the general solution to the RQF rigidity equations has six arbitrary functions of time, $\alpha^I(t)$ and $\beta^I(t)$, at every order in $r$, or equivalently, six arbitrary functions of $t$ and $r$.  In other words, if we have ``nested'' RQFs, we are free to specify the ``Newtonian, $\ell=1$ vector spherical harmonic motion'' of each one {\it independently}.  (2) When we worked out various geometrical quantities (e.g., $\alpha_i$ and ${\mathcal P}_i$) at lowest order in the perturbation (order $r$), we noticed that $\alpha_I$ and $\beta_I$ were always paired with the Fermi frame proper acceleration, $A_I$, and proper rotation rate, $W_I$, in the combinations:
\begin{equation}\label{alpha-A and beta-W}
(A_I+r\ddot{\alpha}_I)\hspace{.5in}\text{and}\hspace{.5in} (W_I+\dot{\beta}_I).
\end{equation}
So at the lowest order, at least, for an RQF of given areal radius $r$, the perturbation generated by $\alpha_I$ (respectively, $\beta_I$) is equivalent to the corresponding acceleration (respectively, rotation rate) of the Fermi frame that the RQF is tied to.  Although we have not checked it at higher order, this is a natural result, and for simplicity's sake we will henceforth set the homogeneous solution at {\it all} orders in $r$ to zero,\footnote{It should be pointed out, however, that setting the homogeneous solution to the order $r^3$ perturbation to zero has no effect on any of our results.  One can show that the $\alpha_I$ and $\beta_I$ arising at this order of the perturbation do not appear in the results quoted below to the orders in $r$ to which they are displayed.} and take $A_I (t)$ and $W_I (t)$ as the six arbitrary, time-dependent degrees of freedom of the RQF.

Moving on to the particular solution, there are three points worth making in order to appreciate the physical significance of the various terms in equations (\ref{eq:F}) and (\ref{eq:fi2}): (1) Positive mass-energy matter inside an RQF will ``warp'' the spatial slice spanning the round two-sphere boundary of the RQF (of areal radius $r$) in such a way that the proper radial distance to the ``centre'' of the RQF will be {\it larger} than $r$ (think of a standard funnel-shaped embedding diagram).  This explains the presence of the matter mass-energy term proportional to $\overset{\mathtt{o}}{T}_{00}$ in equation (\ref{eq:F}).  (2) In Chapter \ref{chRigidRevisited} we considered a round sphere RQF of areal radius $r$ spinning with constant angular velocity, $\omega$, in flat spacetime.  We found that inertial observers outside the system would see a rotating, ``cigar''-shaped sphere with radial perturbation (at order $r^3$) given by $F = \frac{\omega^2}{c^2} (\cos^2 \theta - \frac{1}{2})$, where $\theta=0$ defines the rotation axis.  This $F$ corresponds to a radial contraction near the equator (to compensate for a circumferential Lorentz contraction) and a radial expansion near the poles (to maintain a pole-to-pole distance of $\pi r$ in spite of the radial contraction near the equator).  It is a simple exercise to check that equation (\ref{eq:F}) (including the $W_I W_J$ term in $\mathbb{F}$) reduces to this expression for $F$ in this case.  This explains the presence of the rotation terms in equation (\ref{eq:F}).  (3) The pure $\ell =2$ spherical harmonic term, $\mathbb{F}$, in equations (\ref{eq:F}) and (\ref{eq:fi2}), also includes contributions from the electric part of the Weyl tensor ($\overset{\mathtt{o}}{\mathcal{E}}_{IJ}$), i.e., tidal forces, and spatial matter stresses ($\overset{\mathtt{o}}{T}_{IJ}$).  Both of these spatial curvature effects clearly need to be present in the coordinate perturbation required to achieve a round sphere RQF.

Having found the general solution to the RQF rigidity equations, we can now compute the intrinsic geometry of a generic RQF.  Recall from {\S}\ref{secDefinition} that the two intrinsic geometrical degrees of freedom of an RQF can be encoded, in a coordinate independent manner, in the observers' proper acceleration tangential to ${\mathcal B}$. Computing the lapse and shift functions, $N$ and $u_i$, in the induced three-metric, equation (\ref{eq:InducedMetric}), and substituting these into equation (\ref{eq:ObserversAcceleration}), we find:
\begin{align}\label{eq:TangentialAcceleration}
&\alpha_i = r A_I \mathbb{B}^{I}_{i}  + r^2 \dot{W}_I \mathbb{R}^{I}_{i}
+ r^2 \left[ - \frac{1}{c^2} A_I A_J + W_I W_J + c^2 \overset{\mathtt{o}}{\mathcal{E}}_{IJ} -\frac{c^2\kappa}{2} \overset{\mathtt{o}}{T}_{IJ} \right] \mathbb{B}^{I}_{i} r^J + \mathcal{O}(r^3).
\end{align}
If we let $\mathbf{A}$ and $\mathbf{W}$ denote the vectors $A^I\partial_I$ and $W^I\partial_I$ in the Fermi spatial coordinate system $X^I$ (with $\partial_I:=\partial/\partial X^I$), then the contribution of the first term on the right-hand side to $\alpha^i\partial_i$ is the projection of $\mathbf{A}$ tangential to the RQF sphere, and the contribution of the second is $\mathbf{R}\times\dot{\mathbf{W}}$, where $\mathbf{R}$ is the radial vector from the origin of the coordinate system to observers on the RQF sphere.  Thus, these parts of $\alpha^i\partial_i$ are the direct result of the acceleration and rotation rate of the Fermi frame, to which the RQF is tied.  Of the other terms in equation (\ref{eq:TangentialAcceleration}), the one involving the electric part of the Weyl tensor is interesting: it represents tidal forces, that is, tangential accelerations that the RQF observers must undergo in order to maintain rigidity, i.e., to compensate for the geodesic deviations they would otherwise experience in freefall. In the framework of gravitoelectromagnetism (GEM), we may follow Mashhoon in reference \cite{Mashhoon} and define (at this order in $r$) the GEM electric and magnetic fields in the neighbourhood of $\mathcal C$ as:
\begin{align}
E_I^\texttt{GEM}:= \,& c^2 \overset{\mathtt{o}}{\mathcal{E}}_{IJ}X^J, \label{GEM E Field} \\
B_I^\texttt{GEM}:= - \,& c^2 \overset{\mathtt{o}}{\mathcal{B}}_{IJ}X^J . \label{GEM B Field}
\end{align}
Then the part of $\alpha^i\partial_i$ that arises from the electric part of the Weyl tensor term in equation (\ref{eq:TangentialAcceleration}) is easily seen to be given by the projection $P^{IJ}E_I^\texttt{GEM}\partial_J$, i.e., the component of the GEM electric field (which is essentially acceleration in the GEM framework) tangential to the RQF sphere.

For completeness we also give the twist of the RQF congruence, computed using equation (\ref{eq:nu}):
\begin{equation} \label{Twist}
\nu = W_I r^I  + r \bigg[ - \left( c \overset{\mathtt{o}}{\mathcal{B}}_{IJ} + \frac{2}{c^2} W_I A_J \right) r^I r^J + \frac{1}{c^2} \delta^{IJ} A_I W_J \bigg] + \mathcal{O}(r^2).
\end{equation}
The twist measures the proper rotation rate of observers' spatial dyads relative to inertial gyroscopes.  As one would expect, at lowest order the twist is the (negative) of the radial component of the rotation rate of the Fermi frame.  At next order in $r$, the most interesting term is the one involving the magnetic part of the Weyl tensor; it is interesting because rotation is believed to be one of the sources of ${\cal B}_{IJ}$ \cite{Bonnor1995}. In terms of Mashhoon's definition of the related GEM magnetic field in equation (\ref{GEM B Field}), the part of $\nu$ in question is easily seen to be $r^I B_I^\texttt{GEM}/c$, which, in the GEM framework, is the radial component of the rotation vector, $ B_I^\texttt{GEM}/c$ \cite{Mashhoon}.

We mentioned in {\S}\ref{secDefinition} that the RQF intrinsic geometrical degrees of freedom are essentially encoded in $\alpha_i$, discussed above, but that we are also free to specify the twist on one cross section of $\mathcal B$. To see this at lowest order, notice that if we specify the $\ell=1$ component of $\alpha_i$, i.e., $\alpha_i = r A_I \mathbb{B}^{I}_{i}  + r^2 \dot{W}_I \mathbb{R}^{I}_{i}$, then this determines $A_I(t)$ and $\dot{W}_I(t)$.  To know $W_I(t)$, i.e., the full six degrees of freedom, we must also specify $W_I(0)$ at some initial time $t=0$, which we do when we specify the $\ell=1$ component of $\nu$ on an initial slice of $\mathcal B$.

Let us now turn our attention to the extrinsic geometrical quantities associated with an RQF, beginning with the momentum surface density, ${\mathcal P}_a$, appearing in the energy conservation law in equation (\ref{eq:SimpleConservationEq}). Starting at equation (\ref{SurfaceSEMcomponents}), we find:
\begin{equation}\label{eq:SurfaceMomentumAW}
{\mathcal P}_i = \frac{1}{c\kappa} r W_I \mathbb{R}^{I}_{i} + r^2 \bigg[ - \frac{1}{c \kappa}\left( c \overset{\mathtt{o}}{\mathcal{B}}_{IJ} + \frac{2}{c^2} W_I A_J \right) \mathbb{R}^{I}_{i} r^J + \frac{1}{2} \overset{\mathtt{o}}{T}_{0I} \mathbb{B}^{I}_{i}\bigg] + \mathcal{O}(r^3).
\end{equation}
The last term on the right-hand side is clearly associated with the matter momentum density projected tangentially to the RQF sphere, and so makes sense intuitively.  However, from equation (\ref{SurfaceSEMcomponents}) we recall that ${\mathcal P}^a$ can also be interpreted in terms of the precession rate of inertial gyroscopes (projected tangentially to $\mathcal B$ and rotated 90 degrees by an $\epsilon_i^{\phantom{i}j}$ tensor).  Apart from a common factor of $-r/c\kappa$, the first three terms in equation (\ref{Twist}) are identical to the first three terms in equation (\ref{eq:SurfaceMomentumAW}), except in the former case we have the contraction of a rotation vector with $r^I$ (projection normal to the RQF sphere), and in the latter case we have the contraction of the {\it same} rotation vector with $\mathbb{R}^I_i$ (projection tangential to the RQF sphere and rotated by 90 degrees).

We now proceed to calculate both the matter and geometrical energy fluxes appearing on the right-hand side of our energy conservation law in equation (\ref{eq:SimpleConservationEq}).  Note that $d{\mathcal B}/c=r^2 \, d\mathbb{S}\,dt\,N$, where $d\mathbb{S}$ is the area element on a unit round sphere, and $N$ is the lapse function associated with our choice of time foliation of $\mathcal B$.  So we are really interested in the matter and geometrical energy fluxes {\it times} the lapse function.  A straightforward but tedious calculation reveals (recall that we have dropped the superscript ``mat'' on the matter energy-momentum tensor, $T_{ab}$):
\begin{align}
N \left( n^a u^b T_{ab}\right) &=  -r^I \overset{\mathtt{o}}{S}_I + r \left( \frac{1}{3} \frac{\partial \overset{\mathtt{o}}{\rho}}{\partial t} + \frac{1}{3 c^2} \overset{\mathtt{o}}{S}_I A^I + \Psi_{\texttt{mat}} \right) + \mathcal{O}(r^2)  \label{ndotS} , \\
N\left(-\alpha \cdot \mathcal{P} \right)&=   \frac{c^2}{8\pi G} \epsilon_{IJK} r^I A^J W^K
  + r \left( -\frac{1}{3 c^2} \overset{\mathtt{o}}{S}_I A^I  -\frac{1}{3}\frac{c^2}{8\pi G}  \frac{\partial W^2}{\partial t} + \Psi_{\texttt{geo}} \right) + \mathcal{O}(r^2)   \label{adotP},
\end{align}
where $\overset{\mathtt{o}}{\rho} := \overset{\mathtt{o}}{T}_{00} $ is the matter energy density (energy per unit volume) evaluated on $\mathcal C$, i.e., at the ``centre'' of the sphere; $\overset{\mathtt{o}}{S}_I := -c \overset{\mathtt{o}}{T}_{0I}$ is the matter energy flux (power per unit area) in the $X^I$ direction, evaluated at the ``centre'' of the sphere; and
\begin{align}
& \Psi_{\texttt{mat}} := Q^{IJ} \left( -\overset{\mathtt{o}}{S}_{I;J} - \frac{1}{c^2} \overset{\mathtt{o}}{S}_I A_J + \overset{\mathtt{o}}{T}_{IK} \epsilon^{K}_{\phantom{K}JL} W^L \right), \label{psi_mat} \\
& \Psi_{\texttt{geo}} := Q^{IJ} \left( \frac{1}{2c^2} \overset{\mathtt{o}}{S}_I A_J + \frac{1}{2} \overset{\mathtt{o}}{T}_{IK} \epsilon^{K}_{\phantom{K}JL} W^L + \frac{c^2}{16\pi G} \frac{\partial}{\partial t} (W_I W_J)\right.  \nonumber \\
& \hspace{30mm} \left. - \frac{1}{\kappa} \epsilon_{J}^{\phantom{J}KL} \left[\frac{2}{c^4} A_K W_L A_I - \frac{1}{c} A_K \overset{\mathtt{o}}{\mathcal{B}}_{LI} + W_K \overset{\mathtt{o}}{\mathcal{E}}_{LI} \right] \right) \label{psi_geo}.
\end{align}
Equations (\ref{ndotS}) and (\ref{adotP}) represent the {\it negative} of outgoing fluxes, and according to equation (\ref{eq:SimpleConservationEq}), if we multiply these by $r^2 \, d\mathbb{S}\,dt$, add them, and integrate over the angles of the sphere, and time, we will get the change in the total energy of the RQF (matter plus gravitational) between initial and final time slices (${\mathcal S}_i$ and ${\mathcal S}_f$) of $\mathcal B$.

Let us try to understand the physical significance of the various individual flux terms in these equations.  First, recall that $Q^{IJ} := r^I r^J - \frac{1}{3} \delta^{IJ}$ is trace-free and represents the five independent pure $\ell=2$ spherical harmonics, so $\Psi_{\texttt{mat}}$ and $\Psi_{\texttt{geo}}$ both vanish when integrated over the angles.  These are similar in character to the near-field energy fluxes in an electromagnetically radiating system, in that there is energy flowing inwards and outwards, with no net flux.  For example, in $\Psi_{\texttt{geo}}$, there are cross products of acceleration with the magnetic part of the Weyl tensor, and rotation with the electric part of the Weyl tensor.  Considering the close relationship between acceleration and electric-like effects of gravity, and rotation and magnetic-like effects of gravity, which we will see more of below, these terms are similar in spirit to a gravitational analogue of the electromagnetic Poynting vector.  However, as interesting as they may be, they do not contribute to the integrated flux so we will not study them in detail here.

The first term on the right-hand side of equation (\ref{ndotS}) is the inward radial projection of the matter energy flux evaluated on the RQF sphere, to lowest order in $r$; the latter is constant, and equal to its value at the centre of the sphere, i.e., $\overset{\mathtt{o}}{S}_I$.  The result is obviously a pure $\ell=1$ spherical harmonic that integrates to zero over the angles.  For example, if $\overset{\mathtt{o}}{S}_I$ is in the $z$-direction, then $-r^I \overset{\mathtt{o}}{S}_I$ will be proportional to $-\cos\theta$, and the fact that it integrates to zero just says that, to lowest order in $r$, whatever matter flux enters through the bottom half of the sphere must leave the top half of the sphere.

The corresponding lowest order term in the geometrical energy flux - the first term on the right-hand side of equation (\ref{adotP}) - is similarly a pure $\ell =1$ spherical harmonic that integrates to zero over the angles.  However, its interpretation is worth discussing.  In particular, comparing the lowest order terms between the matter and geometrical energy fluxes we find the correspondence: $\overset{\mathtt{o}}{S}_I$ (matter flux) $\leftrightarrow \frac{c^2}{8\pi G} \epsilon_{IJK} A^J W^K$ (geometrical flux). Thus, at lowest order, the geometrical energy flux is proportional to the cross product of the Fermi frame acceleration and rotation rate.  Interestingly, this flux exists even in flat spacetime, and can be motivated from the equivalence principle as follows.

We imagine an RQF in flat spacetime undergoing arbitrary, but slow motion, time-dependent acceleration and rotation.  Retaining terms only linear in the acceleration and rotation, and setting curvature and matter terms to zero, the observers' tangential acceleration can be read off from equation (\ref{eq:TangentialAcceleration}): $\alpha_i = rA_I(t)\mathbb{B}_i^I+r^2\dot{W}_I(t)\mathbb{R}_i^I$.  We now consider a spacetime in general, linearized gravity, with line element:\cite{Mashhoon}
\begin{equation}\label{GEM Line Element}
ds^2 = -c^2\left(1+2\frac{\Phi}{c^2}\right)\,dT^2+\frac{4}{c}{\mathcal A}_I\,dX^I\,dT+ \left(1-2\frac{\Phi}{c^2}\right)\delta_{IJ}\,dX^I\,dX^J,
\end{equation}
where, in the Newtonian limit, $\Phi$ reduces to the Newtonian gravitational potential, and ${\mathcal A}_I$ is a vector potential associated with rotation of the spacetime.  Comparing with equations (\ref{T=t}) to (\ref{f^Idecomposition}), we now embed RQF observers who are `at rest' in this spacetime via the coordinate transformation:
\begin{equation}\label{GEM Embedding}
T=t\hspace{.5 in}\text{and}\hspace{.5 in}X^I=r\left(1+F\right)r^I.
\end{equation}
\label{GEMDiscussion}A quick calculation shows that the RQF rigidity equations are satisfied when we choose $F=\Phi/c^2$.  Computing $N$ and $u_i$, and substituting these into equation (\ref{eq:ObserversAcceleration}), we find that observers `at rest' in this linearized gravitational field experience a tangential gravitational force per unit mass given by $-\alpha_i=-\partial_i\Phi-2r\dot{\mathcal A}_I\mathbb{B}_i^I/c$.  In the spirit of the equivalence principle, we now ask, ``Can we find gravitational potentials $\Phi$ and ${\mathcal A}_I$ such that RQF observers `at rest' in this gravitational field experience the {\it same} tangential gravitational force per unit mass as they do while accelerating and tumbling in flat spacetime, and so cannot distinguish between these two situations?''  In other words, we wish to equate $-\alpha_i=-rA_I(t)\mathbb{B}_i^I-r^2\dot{W}_I(t)\mathbb{R}_i^I$ (inertial gravitational field in flat spacetime) with $-\alpha_i=-\partial_i\Phi-2r\dot{\mathcal A}_I\mathbb{B}_i^I/c$ (gravitational force per unit mass associated with remaining `at rest' in a linearized gravitational field).  Equating (the negative of) these two accelerations results in the required gravitational potentials: $\Phi=A_I(t)x^I$ and ${\mathcal A}_I=c\,\epsilon_{IJK}x^J W^K (t)/2$, where $x^I := rr^I$. Now we ask, ``Is there a gravitational energy flux associated with these gravitational potentials?''  According to the gravitoelectromagnetic (GEM) interpretation of linearized gravity, these gravitational potentials are associated with gravitoelectric and gravitomagnetic vector fields.  Using the formulas in reference \cite{Mashhoon} we find (in obvious boldface vector notation): ${\mathbf E}^\texttt{GEM}={\mathbf A}(t)+\frac{1}{4}{\mathbf r}\times\dot{{\mathbf W}}(t)$ and ${\mathbf B}^\texttt{GEM}=c{\mathbf W}(t)$.  Within this same interpretation, there `ought' to be an associated GEM Poynting vector, ${\mathbf S}^\texttt{GEM}$, proportional to ${\mathbf E}^\texttt{GEM}\times {\mathbf B}^\texttt{GEM}$ \cite{Mashhoon}.  We can determine this proportionality constant by comparing our expression for ${\mathbf E}^\texttt{GEM}\times {\mathbf B}^\texttt{GEM}$ with the lowest order result in equation (\ref{adotP}); we find:\footnote{Notice that the proportionality constant here (determined using our coordinate invariant RQF approach) differs from that obtained in reference \cite{Mashhoon} (determined using a coordinate-dependent pseudotensor approach).}
\begin{equation}\label{GEMPoynting}
{\mathbf S}^\texttt{GEM}=\frac{c}{8\pi G} {\mathbf E}^\texttt{GEM}\times {\mathbf B}^\texttt{GEM} = \frac{c^2}{8\pi G}\left[{\mathbf A}\times{\mathbf W}+\frac{1}{4}\left({\mathbf r}\times\dot{{\mathbf W}}\right)\times{\mathbf W}\right].
\end{equation}
Thus we see some justification in the above argument, based on the equivalence principle, for both the existence of the geometrical energy flux even in flat spacetime, and its interpretation as a gravitational energy flux.

Related to the previous discussion, the geometrical energy flux in equation (\ref{adotP}) contains a term proportional to the time derivative of $W^2$, which does not vanish upon integration over the angles.  We will see below, when we evaluate the left-hand side of equation (\ref{eq:SimpleConservationEq}), that there is a correctly matching $W^2$ term contributing to the energy of the RQF - see equation (\ref{LHSintergral}).  Two comments on this rotational contribution to the RQF energy are worth making. (1) It can be accounted for using our GEM Poynting vector in equation (\ref{GEMPoynting}), but not perfectly.  When we compute the radial component of the term proportional to $({\mathbf r}\times\dot{{\mathbf W}})\times{\mathbf W}$, we find a part that integrates to zero over the angles, and a part that does not.  The latter is proportional to the time derivative of $W^2$, but the numerical factor in the proportionality constant does not match what we have in equation (\ref{adotP}).  So it agrees in spirit, but not in detail. However, this is not unexpected, since the GEM calculation is in the context of linearized gravity, and nonlinear effects could very well contribute a term of this form.  We emphasize that equations (\ref{adotP}) (and \ref{ndotS}) are {\it exact} (to the displayed order in $r$), accounting fully for the nonlinearity of general relativity.  The GEM calculation, on the other hand, is approximate, and used here for motivational purpose only.  (2) \label{VacuumW2Argument}Inspection of the sign in equation (\ref{adotP}), or (\ref{LHSintergral}) below, reveals that the rotational contribution to the RQF energy is {\it negative}: if $W^2$ increases, the RQF energy decreases.  One possibly plausible explanation for the sign (and a second argument for the very existence of this rotational energy) is that the (``unreferenced'') quasilocal energy density, $\mathcal E$, contains a {\it negative} vacuum contribution.  Looking ahead to equation (\ref{eq:SurfaceEnergyAW}), the vacuum energy surface density is the first term on the right-hand side, ${\mathcal E}^\texttt{vac}:=-2/\kappa r$, which integrates to $E^\texttt{vac}=-c^4 r/G$ over the surface of the RQF sphere. As mentioned in footnote (\ref{VacuumEnergyFootnote}) in {\S}\ref{secExtrinsic}, this vacuum energy is irrelevant when computing {\it changes} in energy, but it may, after all, be {\it indirectly} relevant.  If this energy is actually present ``inside'' an RQF, even when the RQF is in flat spacetime and not rotating, then after spinning up the RQF observers, perhaps the RQF observers are rotating relative to this vacuum energy, and as such `ought' to perceive this as a kind of rotational kinetic energy.  Since the vacuum energy is negative, presumably any moment of inertia that might be associated with $E^\texttt{vac}$ would also be negative, and hence the negative rotational kinetic energy.  Turning the argument around, we might say that the existence of a negative rotational kinetic energy indirectly implies the existence of a negative vacuum energy.

We now turn to what might be considered the main, and perhaps most interesting, flux terms in equations (\ref{ndotS}) and (\ref{adotP}).  These are the matter energy flux terms: $r \left[ \frac{1}{3} \frac{\partial \overset{\mathtt{o}}{\rho}}{\partial t} + \frac{1}{3 c^2} \overset{\mathtt{o}}{S}_I A^I \right]$, and the geometrical energy flux term: $r \left[ -\frac{1}{3 c^2} \overset{\mathtt{o}}{S}_I A^I  \right]$.  Integrating the first matter energy flux term over the RQF sphere, i.e., multiplying by $4\pi r^2$, gives $V\partial \overset{\mathtt{o}}{\rho}/\partial t$ (where $V=4\pi r^3/3$ is the proper volume of the RQF sphere), i.e., the proper time rate of change of the matter energy inside the RQF, to lowest order in $r$.  This is an expected result, correctly matched by the corresponding matter energy term on the right-hand side of equation (\ref{LHSintergral}) below.  However, there is a second matter energy flux term, which couples the `standard' matter energy flux ($\overset{\mathtt{o}}{S}_I$) with the acceleration of the (rigid quasilocal) frame.  This term does not, in general, integrate to zero, and so if we used only $N\left( n^a u^b T_{ab} \right)$ to evaluate the change in matter energy of an accelerating system, we would get the wrong answer.  However, being of the opposite sign, the geometrical energy flux term is exactly what is required to cancel this extra acceleration-induced flux term, resulting in the correct answer.  To see the physical significance of this cancellation process, and the necessity of the geometrical energy flux term, we will now construct an apparent paradox in special relativity and resolve it using these RQF results (a more thorough analysis of this apparent paradox is given in the next chapter).

Consider a right cylinder of length $L$ and cross sectional area $A$, whose axis is parallel to the $z$-axis of an inertial reference frame in flat spacetime, with Minkowski coordinates $(t,x,y,z)$.  The cylinder sits in a constant, uniform electric field with magnitude $E$ in the positive $x$-direction, and thus contains electromagnetic energy $E^2/8\pi$ times the volume of the cylinder, $AL$.  We now subject the cylinder to a constant proper acceleration in the positive $z$ direction in such a way that it is an RQF, and ask how the electromagnetic energy in the cylinder changes with time. We will compute this change using two methods: (1) the change in the volume energy density (times the volume), and (2) the net Poynting flux integrated over the surface.  The paradox is that these two methods will give different answers.  The resolution of this apparent paradox will involve the RQF geometrical energy flux, which we will interpret as a bona fide gravitational effect. While it can be understood superficially in the context of special relativity, its deeper explanation lies in general relativity.

First, we need to accelerate the cylinder in such a way that it is an RQF.  Let the bottom of the cylinder have constant proper acceleration, $a$.  It is well known that, in order for the length of the cylinder to remain constant for co-moving (RQF) observers (the requirement for an RQF), the top of the cylinder must experience {\it less} proper acceleration, namely, $a^\prime=a/(1+aL/c^2)$.  This simple fact is usually called Bell's spaceship paradox \cite{Bell} (and is not the paradox we are concerned with here). Since the dimensions of the cross sections of the cylinder are not affected by this acceleration, we thus have an RQF. There are two important facts to note about this RQF. (1) Proper time moves at different relative rates for observers at the bottom and top of the cylinder. If, between two simultaneities for the RQF observers, a proper time $\Delta\tau$ elapses for observers at the bottom, a {\it greater} proper time, $\Delta\tau^\prime =(1+aL/c^2)\Delta\tau$ elapses for observers at the top. (2) While the relative velocity, $v$, between RQF observers and the inertial reference frame is of course changing (increasing), on any given RQF simultaneity all RQF observers see the {\it same} instantaneous relative velocity.  So we can use the relative velocity $v$ to label the RQF simultaneities.

Next, let us consider the electromagnetic field the RQF observers see.  Since they are moving perpendicular to an electric field, they will see, in addition to a stronger electric field, also a magnetic field: $\vec{E} = \hat{x} \gamma E $ and $\vec{B} = - \hat{y}\beta \gamma E $, where $\beta=v/c$ and $\gamma=1/\sqrt{1-\beta^2}$.  Note that since these fields depend only on $v$, all RQF observers on any given RQF simultaneity will instantaneously see the {\it same} electric and magnetic fields.  They will thus see the same Poynting vector, $\vec{S} = \frac{c}{4\pi}\vec{E} \times \vec{B} = - \hat{z}\frac{c}{4 \pi} \beta \gamma^2 E^2 $, and the same volume energy density, $u = \frac{1}{8 \pi} (E^2 + B^2) = \frac{1}{8 \pi} (1 + \beta^2) \gamma^2 E^2$.  Note that, according to this expression for $u$, the total electromagnetic energy inside the cylinder is clearly increasing with time.  The question is, ``What is the mechanism responsible for this increase?''

Now we will calculate the change in the electromagnetic energy in the cylinder as seen by the RQF observers.  The natural way to parameterize this change is to consider the change in the electromagnetic energy between a pair of infinitesimally separated RQF simultaneities, labelled by, say, proper time $\tau$ and $\tau+\Delta\tau$ as experienced by observers at the bottom of the cylinder.  (A bit of thought shows that it does not matter whose proper time we use to parametrize the simultaneities.)  As mentioned above, we can then calculate this change using two different methods: (1) The volume energy density method, and (2) the Poynting flux method.  For method (1), note that the proper volume of the RQF is constant (by the nature of it being an RQF) and equal to $AL$, which we will denote as $V$.  Thus, $\Delta E_{\rm{Total}} = V(du /d\tau)\Delta\tau$.  Now $u = \frac{1}{8 \pi} (1 + \beta^2) \gamma^2 E^2$ depends only on $v$, so to calculate $du /d\tau$ we need to know $dv /d\tau$, which is equal to $a/\gamma^2$ for an observer experiencing constant proper acceleration, $a$.  A simple calculation then yields:
\begin{equation}\label{E(1)}
\Delta E_{\rm{Total}} = \frac{1}{2\pi}\beta\gamma^2 E^2 V\frac{a}{c}\Delta\tau.
\end{equation}
This result is analogous to using the first of the two matter energy flux terms discussed above: $r \left[ \frac{1}{3} \frac{\partial \overset{\mathtt{o}}{\rho}}{\partial t} \right]$, and is the correct answer.

For method (2) - the Poynting flux method, recall that all observers on a given RQF simultaneity (in particular, those at the bottom and top of the cylinder) see the {\it same} Poynting flux, $\vec{S} = - \hat{z}\frac{c}{4 \pi} \beta \gamma^2 E^2 $.  On first thought this may seem to be a problem, since wouldn't an equal flux flowing in through the top and out through the bottom mean no net change in the electromagnetic energy?  What saves us is the fact that, due to the differing proper accelerations, proper time flows more quickly at the top of the cylinder relative to the bottom.  As noted above, if - between two RQF simultaneities - a proper time $\Delta\tau$ elapses at the bottom, a greater proper time, $\Delta\tau^\prime =(1+aL/c^2)\Delta\tau$, elapses at the top.  Thus, between two RQF simultaneities, more proper time elapses at the top, allowing more energy to enter through that surface than exits through the bottom.  This is apparently the mechanism explaining {\it how} the electromagnetic energy inside the cylinder increases with time.  We say ``apparently'' because it doesn't quite give the right answer.  With the magnitude of the Poynting vector given by $\frac{c}{4 \pi} \beta \gamma^2 E^2$, we have $\Delta E_{\rm{Poynting}} = A \frac{c}{4 \pi} \beta \gamma^2 E^2 \left(\Delta\tau^\prime - \Delta\tau\right)$, and so:
\begin{equation}\label{E(2)}
\Delta E_{\rm{Poynting}}=\frac{1}{4\pi}\beta\gamma^2 E^2 V\frac{a}{c}\Delta\tau.
\end{equation}
Clearly, this accounts for only half of the correct answer: $\Delta E_{\rm{Poynting}}=\Delta E_{\rm{Total}} / 2$.  This approach is analogous to using {\it both} of the matter energy flux terms: $r \left[ \frac{1}{3} \frac{\partial \overset{\mathtt{o}}{\rho}}{\partial t} + \frac{1}{3 c^2} \overset{\mathtt{o}}{S}_I A^I \right]$.  In fact, if we replace the cylinder in this example with a round sphere RQF, one can show that, numerically, $\frac{1}{3 c^2} \overset{\mathtt{o}}{S}_I A^I = - \frac{1}{2} (\frac{1}{3} \frac{\partial \overset{\mathtt{o}}{\rho}}{\partial t})$ so that $\frac{1}{3} \frac{\partial \overset{\mathtt{o}}{\rho}}{\partial t} + \frac{1}{3 c^2} \overset{\mathtt{o}}{S}_I A^I = \frac{1}{2} (\frac{1}{3} \frac{\partial \overset{\mathtt{o}}{\rho}}{\partial t})$, which explains the factor of one-half.
Furthermore, using these results, and the fact that $N=1+\frac{1}{c^2}rr^I A_I + \mathcal{O}(r^2)$, it is easy to show that
$n^a u^b T_{ab}$ (the flux without the lapse function in front) is equal to
$$
-r^I \overset{\mathtt{o}}{S}_I + r \left[ \frac{1}{3} \frac{\partial \overset{\mathtt{o}}{\rho}}{\partial t} + \frac{2}{3 c^2} \overset{\mathtt{o}}{S}_I A^I  \right]+ \mathcal{O}(r^2) = -r^I \overset{\mathtt{o}}{S}_I + \mathcal{O}(r^2)
$$
(ignoring $\Psi_{\texttt{mat}}$).  In other words, $-n^a u^b T_{ab}$ is analogous to $\vec{S}^\prime$ in the cylinder example.  Multiplying $n^a u^b T_{ab}$ by the lapse function, $N$, is equivalent to taking into account the difference in proper times, $\Delta\tau^\prime =(1+aL/c^2)\Delta\tau$, in the cylinder example.  In short, the Poynting flux method is analogous to using $N\left( n^a u^b T_{ab}\right)$ to compute the change in electromagnetic energy, which one might {\it think} is the correct thing to do, but it is not.  This is the paradox.

This paradox is resolved by including the geometrical energy flux term, $r \left[ -\frac{1}{3 c^2} \overset{\mathtt{o}}{S}_I A^I  \right]$, coming from $\alpha\cdot {\mathcal P}$.  There are two senses in which this geometrical energy flux can be thought of as a bona fide gravitational energy flux. (1) The mechanism behind the Poynting flux method here relies entirely on the fact that in an accelerating frame, proper time flows more quickly at the top relative to the bottom.  According to the equivalence principle, this situation is in essence the same as gravity.  So in our special relativity calculation above, we are really encroaching on the domain of gravity.  But to do it properly, we must use general relativity, not accelerating frames in flat spacetime.  The difference amounts to adding the geometrical flux term, which is thus seen to be a bona fide gravitational effect; so being in the context of an energy flux, it must be a {\it gravitational} energy flux.  It is amusing to compare this situation with the gravitational deflection of light.  It is well known that using the principle of equivalence to calculate the deflection of light gives exactly one-half of the correct result calculated using general relativity \cite{Comer1978}. (2) In the ``real world'' we have $G_{ab}=\kappa T_{ab}$.  In special relativity with an electromagnetic field, on the other hand, we have $T_{ab}\not= 0$, but $G_{ab}=0$. In going from special to general relativity we allow the electromagnetic field to curve the geometry.  It is not unreasonable to imagine that an electromagnetically curved geometry gives rise to gravitational (curvature) effects that account for at least some of the effects of the electromagnetic field. In fact, in {\S} \ref{secQuasiParadox} we will see a more detailed analysis starting with the metric for a homogeneous electromagnetic field in general relativity \cite{Stephani} which reveals that this is exactly what is happening - the geometrical energy flux {\it is} a gravitational energy flux. Precisely half of the energy entering the cylinder is due to a traditional matter energy flux (Poynting vector), and the other half is due to a novel gravitational energy flux associated with the spacetime curvature created by the electromagnetic field. In general relativity all forms of energy (e.g., electromagnetic and gravitational) are equivalent, and the sum yields the correct total energy.  This will be explored in more detail in the next chapter.

Having discussed the matter and gravitational flux terms appearing on the right-hand side of our energy conservation law in equation (\ref{eq:SimpleConservationEq}), we now turn our attention to the left-hand side of this equation, both for its own sake, and to provide a useful check of the (integrated) flux expressions in equations (\ref{ndotS}) and (\ref{adotP}). A short calculation reveals that the quasilocal energy surface density is given by
\begin{align}\label{eq:SurfaceEnergyAW}
&\mathcal{E} = - \frac{2}{\kappa r} - \frac{r}{\kappa} \left[ \left( \frac{3}{c^2} W_I W_J + \delta^{KL} \overset{\mathtt{o}}{R}_{IKJL} \right) r^I r^J -\frac{1}{2} \delta^{IJ} \delta^{KL} \overset{\mathtt{o}}{R}_{IKJL} \right].
\end{align}
The first term is a negative vacuum energy, discussed earlier. Using the fact that $\gamma\,d{\mathcal S}=r^2 \, d\mathbb{S}$, the result $- v_i = u_i = r^2 W_I \mathbb{R}^{I}_{i} + \mathcal{O}(r^3)$,
and the earlier result for ${\mathcal P}_i$, a straightforward calculation yields
\begin{align}\label{LHSintergral}
\int\limits_{{\mathcal S}_f-{\mathcal S}_i} d{\mathcal S}\, \gamma \left({\mathcal E} - v \cdot {\mathcal P} \right) =
\left[ \frac{4 \pi r^3}{3} \overset{\mathtt{o}}{\rho} - r^3 \frac{c^2}{6G} W^2 \right]_{t_i}^{t_f},
\end{align}
consistent with the integral of the matter and gravitational fluxes discussed earlier.  It is worth noting that both $\mathcal E$ and  $- v \cdot {\mathcal P}$ contribute to give the correct numerical factor for the $W^2$ rotational kinetic energy term on the right-hand side.

To further strengthen the evidence that RQFs can be constructed in generic spacetimes, and to further explore the interpretation of the geometrical energy flux as a gravitational energy flux, we have carried out calculations to two higher orders in powers of $r$.  To make the calculations tractable we have have gone to a nonrotating Fermi normal coordinate system centered on a geodesic, i.e., $A_I = 0 = W_I$.  We have also turned off the matter sources, $R_{ab} = 0$, leaving only the electric and magnetic parts of the Weyl tensor.  With the matter sources off, the matter energy flux will vanish ($n^a u^b T_{ab}=0$), leaving only the geometrical energy flux.  At order $r^2$ (order $r^4$ when integrated over the RQF sphere), i.e., one order higher than in equation (\ref{adotP}), the (outward) geometrical energy flux is found to be
\begin{align}
N\left(\alpha \cdot \mathcal{P}\right) =   \frac{c}{8\pi G}   \,\epsilon^{IJK}\,r_I\, E_J^\texttt{GEM} B_K^\texttt{GEM}
\end{align}
where $E_I^\texttt{GEM}$ and $B_I^\texttt{GEM}$ are the Weyl tensor-type GEM fields defined in equations (\ref{GEM E Field}) and (\ref{GEM B Field}).  At this order, this result is in agreement with Mashhoon's definition of a GEM Poynting vector  \cite{Mashhoon}, which again adds more weight to the interpretation of the geometrical energy flux as a gravitational energy flux.  Note that the flux at this order is composed only of pure $\ell=3$ and $\ell=1$ spherical harmonics, and thus integrates to zero over the RQF sphere.  It represents a ``near field-like'' energy flux, flowing into and out of the RQF sphere with no net energy flow.

At the next order in $r$, the flux is composed of $\ell=4$, $\ell=2$ and $\ell=0$ parts.  For simplicity, we give only the integrated (outward) flux (i.e., the $\ell=0$ part):
\begin{align}\label{GravEnergyAtFifthOrder}
&\frac{1}{c}\int_{ \Delta {\mathcal B}} d{\mathcal B} \, \alpha \cdot \mathcal{P} = \left[ \frac{1}{60}\frac{c^4}{G} r^5  \left(\overset{\mathtt{o}}{\mathcal{E}} {}^2 - 2 \overset{\mathtt{o}}{\mathcal{B}} {}^2\right) \right]_{t_i}^{t_f},
\end{align}
where $\overset{\mathtt{o}}{\mathcal{E}} {}^2 = \overset{\mathtt{o}}{\mathcal{E}} {}_{IJ} \overset{\mathtt{o}}{\mathcal{E}} {}^{IJ}$ and $\overset{\mathtt{o}}{\mathcal{B}} {}^2 = \overset{\mathtt{o}}{\mathcal{B}} {}_{IJ} \overset{\mathtt{o}}{\mathcal{B}} {}^{IJ}$.

\label{GEMFailureDiscussion}At first sight, the relative factor of $-2$ between $\overset{\mathtt{o}}{\mathcal{E}} {}^2$ and $\overset{\mathtt{o}}{\mathcal{B}} {}^2$ may seem troubling, both in magnitude and in sign.  For example, based on both the energy density in electromagnetism, and the ``0000'' component of the Bel-Robinson tensor, one might have expected an expression proportional to $(\overset{\mathtt{o}}{\mathcal{E}} {}^2 + \overset{\mathtt{o}}{\mathcal{B}} {}^2)$.  However, it is actually not clear {\it what} to expect.  For example, in Chapter \ref{chArch} we discuss how the left-hand side of equation (\ref{eq:SimpleConservationEq}) is exactly analogous to the {\it covariant} definition of electromagnetic energy given in equation (16.44) of Jackson \cite{Jackson3rdEdition}. In the case of a purely electrostatic system, viewed by a moving observer, Jackson shows that the correct integrand for the electromagnetic energy is proportional not to $(\mathbf{E}^2+\mathbf{B}^2 )$, but rather $(\mathbf{E}^2-\mathbf{B}^2 )$ (see his equation (16.46)), and in the case of a non-purely electrostatic system the integrand is more complicated. So the fact that our expression for gravitational energy at order $r^5$ in equation (\ref{GravEnergyAtFifthOrder}) is not proportional to $(\overset{\mathtt{o}}{\mathcal{E}} {}^2 + \overset{\mathtt{o}}{\mathcal{B}} {}^2)$ is perhaps not troublesome at all. Understanding this result more fully is an open problem we hope to pursue in the future. In any case, the main point we would like to make is that the RQF approach gives an extremely simple, operational definition for gravitational energy flux: $\alpha\cdot {\mathcal P}$.  When we expand it in powers of $r$ we get curvature tensor expressions that strongly suggest we are dealing with a bona fide gravitational energy flux, but the terms in the series will clearly get increasingly more complicated at higher orders in $r$.  Perhaps this is simply because ``curvature tensor expressions'' is not the correct language for gravitational energy. The RQF approach suggests that the correct language is a coupling between intrinsic and extrinsic curvature of the system boundary in the form $\alpha\cdot {\mathcal P}$.  This is a simple, exact, operational definition that is physically well-motivated.

\section{Discussion}

In this chapter we have provided strong evidence that the notion of a {\it rigid quasilocal frame} can be extended from flat to curved spacetime.  We have presented a completely general solution of the RQF rigidity equations in an expansion in areal radius, based on Fermi normal coordinates, up to third order.  In the case of vanishing acceleration, rotation, and sources we were able to push this solution up to fifth order.  While the amount of algebra involved in such calculations grows very quickly, there do not appear to be any technical obstructions to extending these solutions to any order.  In other words, for all practical purposes it seems that the RQF equations can be satisfied in an arbitrary curved spacetime, at least out to the radius at which acceleration horizons form.

One of the motivations for introducing RQFs is to provide a new approach to the problem of motion, in particular, to allow the motion of a system to be analyzed in terms of natural, well-defined fluxes passing through the system boundary. Here we have seen that, within the context of both flat and curved spacetimes, the notion of an RQF allows for the construction of simple conservation laws and, in the case of energy conservation, a natural definition for the flux of gravitational energy, namely, $\alpha \cdot \mathcal{P}$.  We provided several arguments, some in the context of gravitoelectromagnetism (GEM), that this energy flux is, indeed, gravitational in nature. Moreover, this definition is simple, exact, and operational in nature - it can be measured by RQF observers using accelerometers and gyroscopes.

Finally, to demonstrate the importance of this new gravitational energy flux, we considered an apparent paradox that arises in a simple electromagnetism problem in special relativity. The paradox is that the increase in electromagnetic energy inside a rigid, accelerating box cannot be accounted for by the Poynting flux alone. We need to add another flux - the gravitational energy flux, $\alpha \cdot \mathcal{P}$, to get the correct answer.  The latter flux cannot be properly understood in the context of special relativity; it involves geometrical effects at the boundary of the system, namely, a coupling between the intrinsic ($\alpha$) and extrinsic ($\mathcal{P}$) geometry of the boundary, which is properly in the domain of general relativity.

In the next two chapters we will explore more deeply the construction of conservation laws in general relativity.  In fact, in the energy case (Chapter  \ref{chLocalQuasilocal}), we will see that the apparent paradox above {\it can} alternatively be resolved using a more familiar local approach.  However, in doing so we will uncover significant drawbacks to the local approach.  When we move to the case of momentum (Chapter \ref{chArch}) the applicability of the local approach only gets worse.  Fortunately, a quasilocal approach using RQFs does {\it not} suffer from any of the same pitfalls.  In effect, we will see in the next two chapters that a quasilocal approach to conservation laws in general relativity is vastly superior to the traditional local approach.

\chapter{Local vs. Quasilocal Conservation Laws} \label{chLocalQuasilocal}

A conservation law ought to explain the change in some physical quantity contained inside a volume of space (e.g., total energy) in terms of related fluxes passing through the bounding surface of that volume. The standard approach to constructing conservation laws is based on the identity: $\nabla_a ( T^{ab}\Psi_b ) = ( {\nabla_a T^{ab}} ) \Psi_b + T^{ab} \nabla_{(a} \Psi_{b)}$, where $T^{ab}$ is the matter stress-energy-momentum tensor, and $\Psi^a$ is a vector that determines the type of conservation law, viz., energy, momentum, or angular momentum. In the context of general relativity, matter energy-momentum is locally covariantly conserved, i.e., $\nabla_a T^{ab}=0$, and the identity reduces to $\nabla_a ( T^{ab}\Psi_b ) = T^{ab} \nabla_{(a} \Psi_{b)}$. As is well known, the problem with this local conservation law is that the right hand side is, in general, not zero (or even a covariant divergence), resulting in a bulk term in the integrated conservation law that spoils what a conservation law ought to be.

This problem is essentially gravitational in nature. There are several ways to see this: (1) The offending bulk term disappears when the spacetime has a suitable symmetry, i.e., admits a Killing vector, $\Psi^a$, but dynamically interesting spacetimes (e.g., ones containing gravitational effects due to objects in motion) generically do not. The idea of relying on a spacetime symmetry to construct a conservation law is a throwback to pre-general relativity days, e.g., special relativity, where spacetime is maximally symmetric. The same goes for relying on asymptotic spacetime symmetries, where we are still in essentially a special relativistic mindset. Given that gravity {\it is} nontrivial spacetime geometry, an approach relying on spacetime symmetries cannot hope to properly incorporate gravitational effects in general. (2) The local conservation law above is {\it homogeneous} in $T^{ab}$. In any matter-free region it is vacuous, even if that region contains interesting gravitational physics, e.g., gravitational waves; it is essentially blind to gravitational physics. We need a conservation law that is nontrivial even when $T^{ab}=0$. (3) Because of the equivalence principle, gravitational effects, e.g., gravitational energy, are not localizable, so we have no hope of capturing gravitational physics with a conservation law based on a local stress-energy-momentum tensor. For example, there is no such thing as a local gravitational energy density (energy per unit volume), that when integrated over a volume gives the total gravitational energy in that volume (see, e.g., {\S}20.4 of reference \cite{MTW}). (4) A bulk term in a local conservation law can be a symptom of the presence of fields that are not being accounted for in the stress-energy-momentum tensor. For example, in the standard Poynting theorem, the $\vec{j}\cdot\vec{E}$ bulk term is present because $T^{ab}$ excludes the charged matter field that is the source of the electromagnetic field, and represents an energy transfer mechanism between the electromagnetic field and the charged matter field. We contend that the bulk term in the local conservation law is, similarly, a result of $T^{ab}$ not properly accounting for the physics of the gravitational field.

A solution to this problem is to move from local to {\it quasi}local conservation laws, which {\it can} properly account for the gravitational physics. In this chapter we construct a general quasilocal conservation law based not on the local {\it matter} stress-energy-momentum tensor, but on the Brown and York quasilocal {\it total} stress-energy-momentum tensor (matter {\it plus} gravity) \cite{BY1993}. Here, ``quasilocal'' means that the differential conservation law is integrated not over the history of a volume of space, but over the history of the {\it boundary} of that volume. We focus on the case of energy conservation, and show that in the quasilocal approach, the quasilocal analogue of the offending $T^{ab} \nabla_{(a} \Psi_{b)}$ bulk term becomes a {\it surface flux} term, which immediately solves the main problem mentioned in the opening paragraph above. Moreover, this surface flux term has two components: (I) The first component is a ``stress times strain'' term that can always be made to vanish by a suitable choice of frame.  In particular, this is accomplished by moving to a {\it rigid quasilocal} frame which, regarding point (1) in the previous paragraph, satisfies a certain ``quasilocally projected'' form of the timelike Killing vector condition for stationary spacetimes that allows us to move just far enough away from the spacetime symmetry mindset to include generic (i.e., non-stationary) spacetimes in conservation laws for energy, momentum, and angular momentum. (II) The second component is an ``acceleration times momentum" term. This term, first encountered in the previous chapter, is familiar from classical mechanics, and represents the rate at which the kinetic energy of an object increases due to one's acceleration toward it. Motivated by a simple equivalence principle argument, we show that this second term is actually a {\it gravitational} energy flux involving the general relativistic effect of frame dragging. We thus show precisely how quasilocal conservation laws resolve the bulk term problem in local conservation laws by properly accounting for the physics of the gravitational field.

This chapter is organized as follows. In {\S}\ref{secParadox} we revisit the apparent paradox from the previous chapter as a very simple example of energy conservation in the context of special relativity, for the purpose of having a concrete example with which to illustrate the development of the general ideas.  Recall, this involved considering a variant of Bell's spaceship paradox in which a box accelerates rigidly in a transverse, uniform electric field. Obviously, the electromagnetic energy inside the box increases, but how would co-moving observers explain this increase? Paradoxically, only {\it half} of the increasing energy came from a net Poynting flux, and, as we will see, according to the local energy conservation law, the other half comes from a bulk ``acceleration times momentum'' term integrated over the volume of the box. In {\S}\ref{secLocal} we examine local conservation laws in general, with a particular focus on the role played by the $T^{ab} \nabla_{(a} \Psi_{b)}$ bulk term in our paradox example. This provides a point of comparison for {\S}\ref{secQuasilocal}, in which we construct a general quasilocal conservation law and argue how it properly accounts for gravitational physics. We also apply the energy form of this quasilocal conservation law to the general relativistic version of our paradox example to concretely illustrate how, what we would normally think of as a bulk ``acceleration times momentum'' term, is actually a gravitational energy flux entering through the boundary of the box. In {\S}\ref{Conclusions3} we present a complementary summary, and argue that quasilocal conservation laws are necessary to understand more deeply a wide variety of phenomena, including the simple example of dropping an apple.

\section{An Apparent Paradox and its Resolution}\label{secParadox}

\subsection{Paradox Outline}

\begin{figure}
\begin{center}
\includegraphics[scale=1.5]{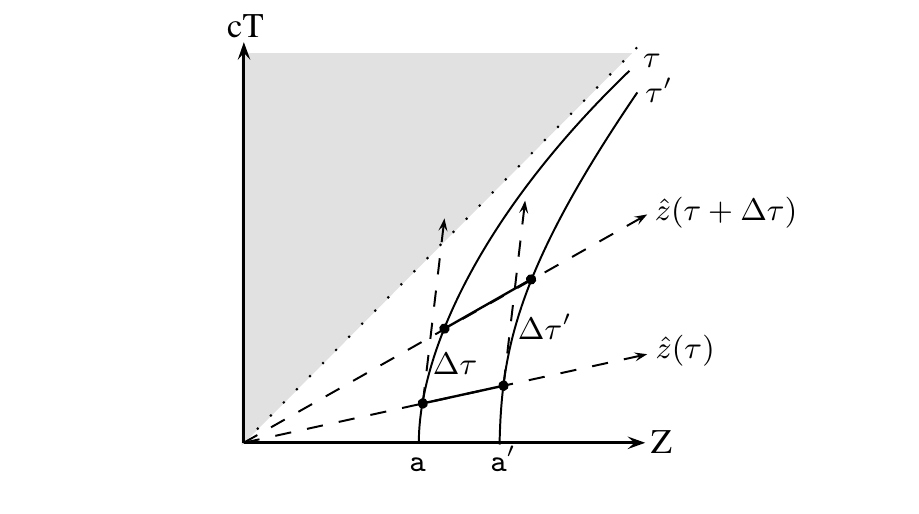}
\caption{The trajectory of a rigidly accelerating cylinder}\label{trajectory}
\end{center}
\end{figure}

Consider the right cylinder from {\S}\ref{secCurved} of length $L$ and cross-sectional area $A$, whose axis is parallel to the $Z$-axis of an inertial reference frame in flat spacetime with Minkowski coordinates $(cT,X,Y,Z)$.  The cylinder is immersed in a constant, uniform electric field of magnitude $E$ in the positive $X$-direction, and thus contains an electromagnetic energy density equal to $E^2/8\pi$. We gave this cylinder constant proper acceleration in the positive $Z$-direction such that its length (and volume) remain fixed for co-moving observers. As we know from Bell's spaceship paradox, such Born rigidity requires the top of the cylinder (represented by the hyperbola on the right in figure \ref{trajectory}) to experience {\it less} proper acceleration compared to the bottom (the hyperbola on the left) \cite{Bell}.  More precisely, $\mathtt{a}^\prime=\mathtt{a}/(1+\mathtt{a}L/c^2)$, where $\mathtt{a}^\prime$ and $\mathtt{a}$ denote the proper accelerations at the top and bottom of the cylinder, respectively.\footnote{\label{footnoteParadox} This can be shown via the usual transformation from Minkowski to Rindler (accelerated) coordinates \cite{Rindler}: $cT = z\, \sinh (ct/L)$, $X=x$, $Y=y$, $Z=z\, \cosh (ct/L)$, from which it follows that $ds^2 = - N^2(z) c^2 dt^2 + dx^2 + dy^2 + dz^2$, where the lapse function $N(z)=z/L$. The proper acceleration at co-moving position $z$ along the cylinder is given by $a(z) = c^2 \partial \log N/\partial z = c^2/z$. Taking the bottom of the cylinder to be located at $z=c^2/\mathtt{a}$ [note that $a(c^2/\mathtt{a})=\mathtt{a}$], the top of the cylinder is then located at $z=c^2/\mathtt{a}+L$. The proper acceleration at the top of the cylinder is thus $\mathtt{a}^\prime = a(c^2/\mathtt{a}+L)= \mathtt{a}/(1 + \mathtt{a}L/c^2)$. Later we also use the fact that the differential proper time at co-moving position $z$ along the cylinder is given by $d\tau=N(z)\,dt$, so when a proper time $\Delta\tau$ elapses at the bottom of the cylinder, a proper time $\Delta\tau^\prime = \left[ N(c^2/\mathtt{a}+L) / N(c^2/\mathtt{a})\right]\,\Delta\tau=(1 + \mathtt{a}L/c^2)\,\Delta\tau$ elapses at the top.} The following two facts, also illustrated in the diagram, will be important to us:  (1) The straight lines passing through the origin represent a natural choice for the surfaces of simultaneity for the co-moving observers: on any such simultaneity, all of the accelerating co-moving observers see the {\it same} instantaneous velocity, $v$, relative to observers at rest in the inertial frame. In other words, this is a ``constant $v$" time foliation, with $v$ monotonically increasing with time. (2) Co-moving observers at different positions along the length of the cylinder will see proper time flowing at different rates relative to one another. That is, between two co-moving simultaneities, if a proper time $\Delta\tau$ elapses for an observer at the bottom of the cylinder, a {\it greater} proper time, $\Delta\tau^\prime =(1+\mathtt{a}L/c^2)\Delta\tau$, elapses for an observer at the top.$\phantom{}^\text{\ref{footnoteParadox}}$

With these facts in mind, let us consider the proper electric and magnetic fields seen by the co-moving observers. Let $\tau$ (respectively, $\tau^\prime$) denote the proper time of observers at the bottom (respectively, top) of the cylinder, and $(\hat{x},\hat{y},\hat{z})$ denote the natural choice of co-moving spatial Cartesian unit vectors. As we argued in {\S}\ref{secCurved}, since the observers are moving in a direction perpendicular to the electric field, they will see, in addition to a stronger electric field, a magnetic field; specifically, $\vec{E} = \hat{x} \gamma E $ and $\vec{B} = - \hat{y}\beta \gamma E $, where $\beta=v/c$ and $\gamma=1/\sqrt{1-\beta^2}$. It is important to remember that since these fields depend only on $v$, all observers on any given co-moving simultaneity will see the {\it same} instantaneous electric and magnetic fields. They will thus see the {\it same} proper Poynting vector, $\vec{S} = \frac{c}{4\pi}\vec{E} \times \vec{B} = - \hat{z}\frac{c}{4 \pi} \beta \gamma^2 E^2$, and the {\it same} proper electromagnetic energy density, $u = \frac{1}{8 \pi} (|\vec{E}|^{2} + |\vec{B}|^{2}) = \frac{1}{8 \pi} (1 + \beta^2) \gamma^2 E^2$. Notice that, according to this expression for $u$, the total electromagnetic energy inside the cylinder is clearly increasing with time. We ask: How would the co-moving observers explain this increasing electromagnetic energy? One might expect that the change in energy between two co-moving simultaneities is just equal to the net Poynting flux into the cylinder over that time interval; this, it turns out, is only half correct.

Consider a pair of infinitesimally separated co-moving simultaneities labelled by proper time $\tau$ and $\tau+\Delta\tau$ as experienced by observers at the bottom of the cylinder.  Between these two simultaneities the total electromagnetic energy inside the cylinder changes by an amount
\begin{align} \label{E(1)2}
\Delta E_{\rm Total}= V \left( \frac{du}{d\tau} \right) \Delta \tau = 4 \beta\gamma^2 \left( \frac{E^2}{8\pi } V \right)   \frac{\mathtt{a}\Delta\tau}{c},
\end{align}
where $V=AL$ is the volume of the cylinder (which is constant for the co-moving observers). In this calculation we have made use of the relation $dv /d\tau = \mathtt{a}/\gamma^2$ for an observer at the bottom of the cylinder, who is experiencing constant proper acceleration $\mathtt{a}$.

Naively, we ought to be able to arrive at the same result by considering just the net Poynting flux crossing the boundary of the cylinder. Before we calculate this, however, it is interesting to understand the mechanism by which the Poynting vector carries energy into the cylinder. As noted earlier, on any given co-moving simultaneity the relative velocity, $v$, is constant along the cylinder and, thus, the proper Poynting vector is the same at the top and bottom of the cylinder. Since no flux leaves or enters the sides of the cylinder, this suggests that the net Poynting flux is zero. Recall, however, that because of the acceleration, proper time advances more quickly at the top of the cylinder relative to the bottom. This results in a greater proper time-integrated flux entering the top of the cylinder compared to that leaving the bottom of the cylinder - in other words, a net accumulation of electromagnetic energy. Net electromagnetic energy enters the cylinder because of the time dilation effect associated with the acceleration. With the magnitude of the Poynting vector given above, we find
\begin{align} \label{E(2)2}
\Delta E_{\rm Poynting} = |\vec{S}| A \left( \Delta \tau^\prime - \Delta \tau \right)= 2 \beta\gamma^2 \left( \frac{E^2}{8\pi } V\right) \frac{\mathtt{a}\Delta\tau}{c}.
\end{align}
Observe that the net Poynting flux, equation (\ref{E(2)2}), accounts for only {\it half} of the change in electromagnetic energy inside the cylinder, equation (\ref{E(1)2}). This is the apparent paradox introduced in section {\S}\ref{secCurved}.

\subsection{Paradox Resolution}\label{secParadoxResolution}

To understand the missing piece of this puzzle, let us first consider a simple problem in classical mechanics.  Imagine an object with rest mass $m$ from the point of view of a reference frame moving towards it with instantaneous speed $v$. In this reference frame, we would consider the object to have total energy $E =\sqrt{ m^2 c^4 + c^2 p^2}$, where $p = \gamma m v$ is the object's instantaneous momentum relative to the frame. Now, if our frame is accelerating at a proper rate $\mathtt{a}$, it is easy to show that, in a proper time $\Delta \tau$, the object's total energy changes by an amount $\Delta E = \mathtt{a} p \Delta \tau$. Thus, the total energy of the object increases solely as a result of the acceleration of our frame relative to the object.

The resolution to the paradox follows the same line of reasoning: as the cylinder accelerates relative to the existing momentum inside, there is an increase in energy proportional to this momentum times the acceleration. The momentum in this case belongs to the electromagnetic field contained in the cylinder, and is proportional to the Poynting vector:
\begin{align}
p =  \frac{1}{c^2} |\vec{S}| V = \beta \gamma^2 \left(\frac{E^2}{4\pi c } V\right).
\end{align}
(As a consistency check, observe that as the electromagnetic energy, $e = uV$, inside the cylinder increases, so does the electromagnetic momentum, $p$, such that $c^2$ times the invariant mass,
\begin{align}
m c^2 = \sqrt{e^2 - c^2 p^2} = \frac{E^2}{8\pi} V,
\end{align}
remains constant, as expected, and equal to the electromagnetic energy inside the cylinder when $v=0$.)

Now, for the problem at hand, the contribution to the energy change due to this effect is just
\begin{align}  \label{E(3)}
\Delta E_{\rm Bulk}  = \mathtt{a} p\, \Delta \tau = 2 \beta\gamma^2 \left(\frac{E^2}{8\pi } V\right)    \frac{\mathtt{a}\Delta\tau}{c}.
\end{align}
Notice that this energy change combined with the accumulation of Poynting flux due to the acceleration-induced time dilation along the length of the cylinder, $\Delta E_{\rm Poynting}$, now precisely matches that calculated from the energy density, i.e., $\Delta E_{\rm Poynting} + \Delta E_{\rm Bulk} = \Delta E_{\rm Total}$. Thus we have found the missing piece of the puzzle in the apparent paradox. It is not sufficient to consider the change in energy inside the cylinder based solely on the accumulation of Poynting flux across the surface; one must also take into account a ``momentum times acceleration'' term due to the frame accelerating relative to an existing momentum. The inclusion of this {\it bulk} term then resolves the apparent paradox but comes with a cost - it does away with the usual {\it ``change in energy equals net energy flux through the boundary''} picture of a conservation law.

\section{Local Conservation Laws}\label{secLocal}

\begin{figure}
\begin{center}
\includegraphics[scale=0.9]{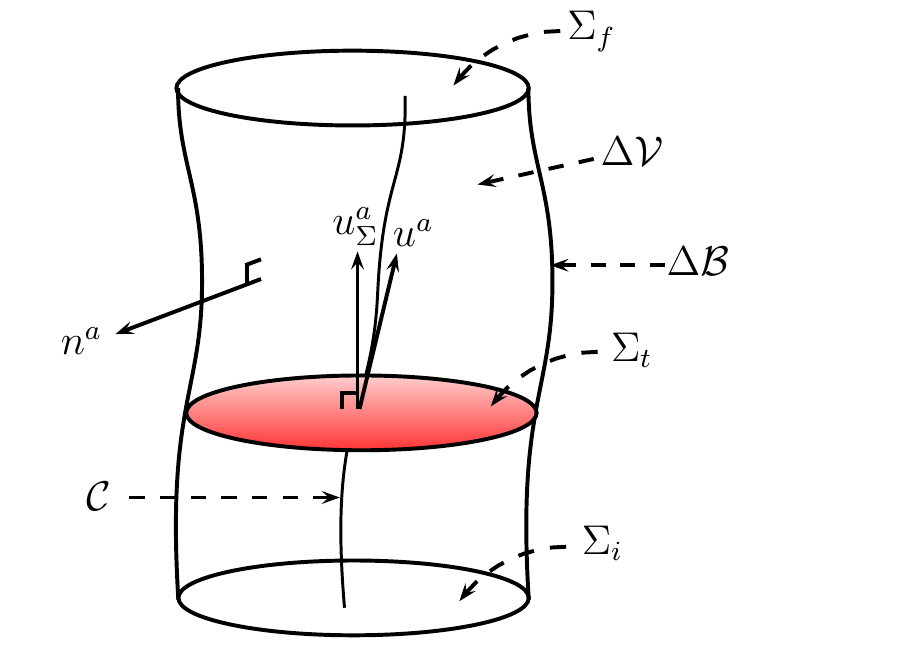}
\caption{An observer in the three-dimensional spatial volume $\Sigma_t$ follows a timelike worldline $\cal C$ with tangent four-velocity vector $u^a$, which is not necessarily parallel to the timelike vector field $u^a_{\Sigma}$ orthogonal to $\Sigma_t$.  The one-parameter family of spatial volumes, $\Sigma_t$, foliate the four-dimensional spacetime region, $\Delta \mathcal{V}$, whose timelike worldtube boundary, $\Delta\mathcal{B}$, has spacelike unit normal vector field $n^a$.}\label{LocalWorldtube}
\end{center}
\end{figure}

Let us now look more generally at the notion of conservation laws in the context of both special and general relativity.  Consider a smooth four-dimensional manifold, $\mathcal{M}$, endowed with a Lorentzian spacetime metric, $g_{ab}$, and associated covariant derivative operator, $\nabla_a$. In the presence of a non-zero matter stress-energy-momentum tensor, $T^{ab}$, we can construct the identity
\begin{equation}\label{BulkDifferential}
\nabla_a ( T^{ab}\Psi_b ) = ( {\nabla_a T^{ab}} ) \Psi_b + T^{ab} \nabla_{(a} \Psi_{b)}.
\end{equation}
This identity gives a differential conservation law for the current $Q^a = - T^{ab} \Psi_b$, whose physical interpretation depends on the choice of the weighting vector, $\Psi^a$. Integrating both sides of this identity over a finite four-volume, $\mathcal{V}$, (see figure \ref{LocalWorldtube}) we have
\begin{align} \label{BulkIntegrated}
\frac{1}{c} \int\limits_{\Sigma_f - \Sigma_i}  d{\Sigma}\,  T^{ab} u^\Sigma_a \Psi_b = \int\limits_{\Delta \mathcal{B}}  d \mathcal{B} \, T^{ab} n_a \Psi_b - \int\limits_{\Delta {\mathcal V}}  d{\mathcal V}\, \left[ \left( \nabla_a T^{ab} \right) \Psi_b + T^{ab} \left( \nabla_{(a} \Psi_{b)} \right)  \right].
\end{align}
On the left hand side, $\Sigma_i$ and $\Sigma_f$ are the initial and final three-dimensional spatial volume ``end caps" of $\mathcal{V}$, with timelike future-directed unit normal vector $\frac{1}{c}u^a_\Sigma$. On the right hand side, $\mathcal B$ is a three-dimensional timelike worldtube spanning the boundaries of the end caps, $\partial \Sigma_i$ and $\partial \Sigma_f$, with spacelike outward-directed unit normal vector, $n^{a}$, and induced Lorentzian three-metric $\gamma_{ab}=g_{ab}-n_{a}n_{b}$.

In general, the left hand side of equation (\ref{BulkIntegrated}) gives the change in a quantity contained inside a spatial volume (e.g., electromagnetic energy) over some time interval, while the first term on the right hand side corresponds to matter fluxes across the boundary of that spatial volume during that time interval (e.g., electromagnetic Poynting flux). This is the desired form of a conservation law. However, there are two additional terms on the right hand side that are four-dimensional bulk integrals. The first of these involves $\nabla_a T^{ab}$, which is zero in general relativity; however in special relativity it will in general not vanish when $T^{ab}$ does not include all of the matter fields (e.g., when it does not include the electric four-current source of an electromagnetic field). The second term is present when $\Psi^a$ is not a Killing vector and, as it turns out, is crucial in resolving the apparent paradox above. To gain a better understanding of the terms in equation (\ref{BulkIntegrated}) let us look at an example.

\subsection{Example: Electromagnetism}

Consider a general electromagnetic field, which can be decomposed as
\begin{align}
F^{ab} = \frac{2}{c} u^{[a} E^{b]} + \epsilon^{ab}_{\phantom{ab}c} B^c,
\end{align}
where $u^a$ is the four-velocity of a volume-filling, three-parameter family of observers who see proper electric and magnetic fields $E^a$ and $B^a$, respectively, and $\epsilon_{bcd}=\frac{1}{c}u^{a}\epsilon_{abcd}$ is the spatial three-volume element orthogonal to the observers' worldlines. Denoting the spatial three-metric orthogonal to the observers' worldlines as $h_{ab} = g_{ab} + \frac{1}{c^2} u_a u_b$, the electromagnetic stress-energy-momentum tensor can be decomposed as
\begin{align}\label{EMStressTensor}
T^{ab} = \frac{1}{4\pi} \left\{ \frac{1}{2 c^2} u^a u^b \left( E^2 + B^2 \right) + \frac{2}{c} u^{(a} \epsilon^{b)}_{\phantom{b)}cd} E^c B^d - \left[ E^a E^b +B^a B^b -\frac{1}{2} h^{ab} \left( E^2 + B^2 \right) \right] \right\}
\end{align}
Also, it follows from Maxwell's equations that
\begin{align}
\nabla_a T^{ab} = j_a F^{ab},
\end{align}
where $j^a$ is the electric four-current source of the electromagnetic field.

For simplicity we will work with a congruence that has zero twist, i.e., one for which the four-velocity $u^a$ is hypersurface orthogonal, and assume that $u^a = u_\Sigma^a$ on $\Sigma_i$ and $\Sigma_f$. This reduction in generality does not affect our results - it just makes the formulas simpler and more transparent. Choosing as our weighting vector $\Psi^a = \frac{1}{c} u^a$, equation (\ref{BulkIntegrated}) becomes an energy conservation equation, the individual terms of which are
\begin{align}
\frac{1}{c} T^{ab} u^\Sigma_a \Psi_b &= u, \label{Energy}\\
T^{ab} n_a \Psi_b &= - \frac{1}{c} \, n_a S^a, \\
\left( \nabla_a T^{ab} \right) \Psi_b &=  j_a E^a, \\
T^{ab} \left( \nabla_{(a} \Psi_{b)} \right) &= \frac{1}{c} \,  a_a \mathbb{P}^a - \mathbb{T}^{ab} K_{ab}.  \label{BulkTransfer2}
\end{align}
Here $u = \frac{1}{8\pi} \left( E^2 + B^2 \right)$ and $S^a = \frac{c}{4\pi} \epsilon^a_{\phantom{a}bc} E^b B^c$ are the proper electromagnetic energy density and Poynting vector, respectively, and $j^a$ is the electric four-current vector, as before. $\mathbb{P}^a = \frac{1}{c^2}  S^a$ is the proper electromagnetic momentum density, and $\mathbb{T}^{ab} = - h^a_{\phantom{a}c} h^b_{\phantom{b}d} T^{cd}$ is the spatial, three-dimensional Maxwell stress tensor.  . Recall, $a^a = u^b \nabla_b u^a$ is the observers' four-acceleration.  Additionally, $K_{ab} = \frac{1}{c} h_{(a}^{\phantom{(a}c} h_{b)}^{\phantom{b)}d} \nabla_c u_d$ is the observers' spatial, three-dimensional strain rate tensor, measuring the expansion and shear of their congruence. Inserting equations (\ref{Energy}-\ref{BulkTransfer2}) into equation (\ref{BulkIntegrated}) gives a general relativistic version of Poynting's theorem for a hypersurface orthogonal congruence of observers:
\begin{align} \label{EMBulkIntegrated}
\int\limits_{\Sigma_f - \Sigma_i}  d{\Sigma}\,  u = - \frac{1}{c} \int\limits_{\Delta \mathcal{B}}  d \mathcal{B} \,   n_a S^a -  \int\limits_{\Delta {\mathcal V}}  d{\mathcal V}\, \left[  j_a E^a +  \frac{1}{c} \,  a_a \mathbb{P}^a - \mathbb{T}^{ab} K_{ab}    \right].
\end{align}
Notice that the standard form of Poynting's theorem is recovered when $u^a$ is a timelike Killing vector, in which case the last two bulk terms on the right hand side, coming from equation (\ref{BulkTransfer2}), vanish. (The bulk term involving $j_a E^a$ remains, however, and represents an energy transfer between the electromagnetic field and its electric four-current source.)  When $u^a$ is {\it not} a Killing vector, however, these two bulk terms provide additional mechanisms for energy transfer. The $a_a \mathbb{P}^a$ term represents change in energy due to the frame accelerating relative to an existing electromagnetic momentum in the system (as was seen in resolving the apparent paradox in \S\ref{secParadoxResolution}). The $\mathbb{T}^{ab} K_{ab}$ term represents the Maxwell stress, $\mathbb{T}^{ab}$, doing work against the strain, $K_{ab}$, of the three-parameter congruence. Since we would like to generalize beyond the case where $u^a$ is a Killing vector, both of these energy transfer mechanisms will be important.

\subsection{Apparent Paradox Revisited}

We will now illustrate the use of the local conservation law, equation (\ref{BulkIntegrated}), to resolve the apparent paradox discussed in \S\ref{secParadoxResolution}, generalized slightly to allow for a time-dependent acceleration along the $Z$-axis.\footnote{The analysis can be generalized to arbitrary acceleration; the limit to acceleration along the $Z$-axis is chosen purely for notational simplicity.} Moreover, to facilitate comparison with the general relativistic calculation in the next section (on {\it quasi}local conservation laws), we will switch from a right circular cylinder of length $L$ and cross-sectional area $A$ to a round sphere of areal radius $r$. The reason for this switch is that the general relativistic calculation involves the extrinsic curvature of the boundary of our spatial volume, and the sharp corners of a cylinder introduce an unnecessary technical complication.

Let us denote the Minkowski coordinates of an inertial reference frame in flat spacetime by $X^a = (X^0, X^I)=(cT, X,Y,Z)$, $I=1,2,3$. Let the embedding $X^a=\xi^a(\tau)$ define an accelerated, timelike worldline, ${\cal C}_0$, where $\tau$ is proper time, and let $e_0^{\phantom{0}a}(\tau)=\frac{1}{c}d\xi^a(\tau)/d\tau$ denote the unit vector tangent to ${\cal C}_0$. We will construct a three-parameter family of accelerated observers in the neighbourhood of this worldline using the coordinate transformation
\begin{align} \label{Worldline}
X^a (x) = \xi^a (\tau) + r r^I (\theta, \phi) e_I^{\phantom{I}a}(\tau).
\end{align}
Here $x^\alpha = (x^0,x^1,x^i)=(\tau, r, \theta, \phi)$ are coordinates adapted to the congruence of observers: the spherical coordinates $(r, \theta, \phi)$ are parameters that label the observers' worldlines, and $\tau$ is a time parameter along the worldlines which in general is {\it not} proper time, except for the observer at $r=0$ (i.e., there is a nontrivial lapse function, $N$, which will be given shortly). Also, $r^I (\theta, \phi) = (\sin{\theta} \cos{\phi}, \sin{\theta} \sin{\phi}, \cos{\theta})$ are the usual direction cosines in a spherical coordinate system, and $e_I^{\phantom{I}a}(\tau)$ is a Fermi-Walker transported spatial triad defined on ${\cal C}_0$ that is orthogonal to $e_0^{\phantom{0}a}(\tau)$.

Specializing to the case relevant to our apparent paradox, we let ${\cal C}_0$ (i.e., the observer at $r=0$) undergo proper acceleration $\mathtt{a}(\tau)$ along the $X^3$ (i.e., $Z$) axis, in which case our tetrad has the form:
\begin{align}
e_0^{\phantom{0}a}(\tau) &= \cosh{\alpha (\tau)} \, \delta_0^a + \sinh{\alpha (\tau)} \, \delta_3^a,\nonumber\\
e_1^{\phantom{1}a}(\tau) &= \delta_1^a, \qquad e_2^{\phantom{2}a}(\tau) = \delta_2^a, \label{tetrad}\\
e_3^{\phantom{3}a}(\tau) &= \sinh{\alpha (\tau)} \, \delta_0^a + \cosh{\alpha (\tau)} \, \delta_3^a, \nonumber
\end{align}
where $\alpha (\tau) = \frac{1}{c}\int_0^\tau \, \mathtt{a}(t) \,dt$. Explicitly, the coordinate transformation in equation (\ref{Worldline}) then reads
\begin{align}
cT  &= c \int_0^\tau \cosh\alpha(t) \, dt  + r \cos\theta \sinh\alpha(\tau),  \nonumber \\
X & = r \sin\theta \cos\phi,   \qquad Y =  r \sin\theta \sin\phi, \label{Rindtmf} \\
Z &=  c \int_0^\tau \sinh\alpha(t) \, dt  + r \cos\theta \cosh\alpha(\tau), \nonumber
\end{align}
which in turn yields the metric
\begin{equation}\label{AdaptedCoordMetric}
g_{\alpha\beta}=
\left(
\begin{array}{ccc}
-c^2 N^2 & 0 & 0 \\
0 & 1 & 0 \\
0 & 0 & r^2\mathbb{S}_{ij}
\end{array}
\right),
\end{equation}
in the observer-adapted coordinate system,
where $\mathbb{S}_{ij}={\rm diagonal}(1,\sin^2\theta)$ is the metric on the unit round sphere, and $N(x) = 1 + \frac{1}{c^2}\mathtt{a}(\tau) r \cos \theta$ is the lapse function. At a spacetime point $x^\alpha$,
the observers' four-velocity is defined as $u^a =N^{-1}(x)(\partial/\partial \tau)^a=c e_0^{\phantom{0}a}(\tau)$, with resulting four-acceleration $a^a(x)=N^{-1}(x)\mathtt{a}(\tau)e_3^{\phantom{3}a}(\tau)$. Comparing with the discussion of the paradox given in {\S}\ref{secParadoxResolution}, in the context of an accelerating cylinder, the nontrivial lapse function, $N(x)$, is the analogue of our earlier relation $\Delta\tau^\prime =(1+\mathtt{a}L/c^2)\Delta\tau$, i.e., when $\mathtt{a}(\tau)>0$, proper time passes more quickly for observers above the plane $\theta=\pi/2$ relative to those below it. Also, the magnitude of the observers' proper acceleration, $N^{-1}(x)\mathtt{a}(\tau)$, is the analogue of our earlier relation $\mathtt{a}^\prime=\mathtt{a}/(1+\mathtt{a}L/c^2)$, i.e., when $\mathtt{a}(\tau)>0$, observers above the plane $\theta=\pi/2$ experience less proper acceleration compared to those below it. Finally, by inspection of the metric in equation (\ref{AdaptedCoordMetric}), it is clear that the observers are accelerating {\it rigidly}, in the sense that the orthogonal distance between all nearest neighbour pairs of observers remains constant in time (despite the time-dependent acceleration), in accordance with Born rigidity. Thus, $K_{ab}$ in equation (\ref{BulkTransfer2}) is zero. (This is a very special case. In general, it is not possible to find a volume-filling congruence of observers for which $K_{ab}=0$, a point we will return to in the next section.)

As in {\S}\ref{secParadoxResolution}, we now introduce a constant, uniform electric field of magnitude $E$ in the positive $X$-direction, perpendicular to the observers' motion: $E^a = E \delta_1^a$ (and $B^a$ and $j^a$ vanish, at least in the region of the observers' congruence). It follows from equation (\ref{EMStressTensor}) that the electromagnetic stress-energy-momentum tensor is $T^{ab}=(E^2/8\pi)\times{\rm diagonal}(1,-1,1,1)$. Choosing $\Psi^a = \frac{1}{c} u^a$ in equation (\ref{BulkIntegrated}) makes this an energy conservation equation. With $u_{\Sigma}^a=u^a$ and $n^a=r^I e_I^{\phantom{I}a}$ we find that the individual terms in equation (\ref{BulkIntegrated}) are
\begin{align}
\frac{1}{c} T^{ab} u^\Sigma_a \Psi_b &= \left( 1 + \beta^2 \right) \gamma^2 \frac{E^2}{8\pi}, \\
T^{ab} n_a \Psi_b &= \beta \gamma^2 \frac{E^2}{4\pi} \cos \theta, \\
\left( \nabla_a T^{ab} \right) \Psi_b &= 0, \\
T^{ab} \left( \nabla_{(a} \Psi_{b)} \right) &= -\beta \gamma^2 \frac{E^2}{4\pi} \frac{\mathtt{a}(\tau)}{c^2}\frac{1}{N}  , \label{BulkEM2}
\end{align}
where we have introduced the usual Lorentz transformation parameters: $\gamma(\tau) = \cosh \alpha(\tau)$ and $\beta(\tau) = \tanh \alpha(\tau)$ such that $\gamma = 1/\sqrt{1 - \beta^2}$. Observe that the right hand side of equation (\ref{BulkEM2}) is due entirely to the $a_a \mathbb{P}^a$ term in equation (\ref{BulkTransfer2}) since, as noted above, the collective rigidity of the observers' motion means that  $K_{ab}$ vanishes in this case. In other words, this last contribution is entirely the result of a change in energy due to the frame accelerating with respect to the existing electromagnetic momentum in the system. Putting these results together, the conservation law reads
\begin{align} \label{ParadoxConservationLaw}
\int\limits_{\Sigma_f - \Sigma_i}  dr \, r^2 \, d{\mathbb{S}}\,   \left( 1 + \beta^2 \right) \gamma^2 \frac{E^2}{8\pi} & = \int\limits_{\Delta \mathcal{B}}  c\,d\tau\,r^2\,d\mathbb{S} \,  \beta \gamma^2 \frac{E^2}{4\pi} \cos \theta \left(1 + \frac{1}{c^2}\mathtt{a}(\tau) r \cos \theta\right)  \nonumber \\
&\quad\quad  + \int\limits_{\Delta {\mathcal V}}  c\,d\tau\,dr\,r^2\,d\mathbb{S}\, \beta \gamma^2 \frac{E^2}{4\pi} \frac{\mathtt{a}(\tau)}{c^2},
\end{align}
where the lapse function in the bulk term, $T^{ab} \left( \nabla_{(a} \Psi_{b)} \right)$ - see equation (\ref{BulkEM2}), has been cancelled by that in the volume element, $d\mathcal{V} = c\,d\tau\,dr\,r^2\,d\mathbb{S}\,N$.  Note also that we have used $d\Sigma =  dr \, r^2 \,d\mathbb{S}$  and $ d\mathcal{B} = c\,d\tau\,r^2\,d\mathbb{S} \,N $, where $d\mathbb{S} = \sin \theta \,d\theta\, d\phi$ is the area element on the unit round sphere.

In order to compare directly with the results in {\S}\ref{secParadoxResolution}, let the time interval be infinitesimal: $\tau_f-\tau_i=\Delta\tau$. Using the relation $\frac{d}{d\tau} \left( (1 + \beta^2) \gamma^2  \right) = 4 \beta \gamma^2\frac{\mathtt{a}}{c} $, the left hand side of equation (\ref{ParadoxConservationLaw}) then integrates to precisely the same change in electromagnetic energy, $\Delta E_{\rm Total}$, given in equation (\ref{E(1)2}), except with the cylinder volume, $V=AL$, replaced with the sphere volume, $V=\frac{4}{3} \pi r^3$. Also, the $\Delta\tau$ in the equation now refers to the proper time elapsed for the observer at the center of the sphere, instead of an observer at the bottom of the cylinder. The Poynting flux integral over $\mathcal{B}$ on the right hand side of equation (\ref{ParadoxConservationLaw}) has two terms. The first term (proportional to $\cos\theta$) integrates to zero over the angles, which is analogous to the proper Poynting vector in {\S}\ref{secParadoxResolution} being the same at the top and bottom of the cylinder such that, at lowest order, flux in equals flux out. At the next order in $r$, however, the nontrivial lapse function is responsible for a $\cos^2\theta$ term that does {\it not} integrate to zero. As in {\S}\ref{secParadoxResolution}, it is this non-isotropic time dilation that allows for a nonzero accumulation of Poynting flux. Evaluating this integral we find precisely the same change in electromagnetic energy, $\Delta E_{\rm Poynting}$, given in equation (\ref{E(2)2}), except with $V$ and $\Delta\tau$ reinterpreted as discussed above. Finally, the bulk integral over $\mathcal{V}$ on the right hand side of equation (\ref{ParadoxConservationLaw}), which measures the change in energy due to the acceleration of the frame relative to the existing electromagnetic momentum in the system, evaluates to precisely the same change in electromagnetic energy, $\Delta E_{\rm Bulk}$, given in equation (\ref{E(3)}), except with $V$ and $\Delta\tau$ again reinterpreted as discussed above.

This example illustrates that any local conservation law constructed from the matter stress-energy-momentum tensor, $T^{ab}$, as in equations (\ref{BulkDifferential}) and (\ref{BulkIntegrated}), in general contains {\it two} terms responsible for the change in a physical quantity inside a volume of space: the first is a three-dimensional surface flux integral over the worldtube $\cal B$, as one might expect, and the second is a four-dimensional bulk integral over $\cal V$. It is the addition of this bulk integral, which naively we did not expect, that resolves the apparent paradox introduced in \S\ref{secParadoxResolution}. But is such a bulk integral really necessary? In the next section we will see how to naturally convert this bulk integral into a surface flux integral, and at the same time generalize the conservation law in equations (\ref{BulkDifferential}) and (\ref{BulkIntegrated}) to include gravitational effects.

\section{Quasilocal Conservation Laws}\label{secQuasilocal}

We will now argue that the local conservation law given in equations (\ref{BulkDifferential}) and (\ref{BulkIntegrated}) is defective in two respects. First, even when matter is locally covariantly conserved, i.e., $\nabla_a T^{ab}=0$ (which is always true in general relativity), the local conservation law contains a nontrivial bulk term when $T^{ab}$ is present and $\Psi^a$ is not a Killing vector. Insofar as a generic spacetime does not admit any Killing vectors, when $T^{ab}$ is present this bulk term is generically present. But this bulk term violates what we usually think of as a conservation law, i.e., that the change in some physical quantity over a period of time is equal to some related surface flux integral over that period of time. Is there a natural and general way to convert this bulk term into a surface flux term?

To help motivate this question, let us return for a moment to the ``momentum times acceleration" bulk term that resolves the apparent paradox introduced in \S\ref{secParadoxResolution}. We imagine being inside an accelerating box in flat spacetime that contains a freely-floating, massive object that appears to be accelerating toward us; the object's kinetic energy (relative to us) increases due to the acceleration of our frame. We ask: Where does the increasing kinetic energy come from? This might sound like a silly question - after all, energy is frame-dependent, and we are just changing the frame! However, the question is perhaps not so silly when we ask it in the context of the equivalence principle. Instead of being inside an accelerating box, we could imagine that the box is at rest in a uniform gravitational field, and that the object is experiencing an acceleration toward us due to the ``force" of gravity. This ``force" acting through a distance represents an energy transfer mechanism from the gravitational field energy to the kinetic energy of the object. So it might be possible to convert the ``momentum times acceleration" bulk term into some kind of surface flux term representing {\it gravitational} energy entering the box from the outside. In the context of general relativity, i.e., when we properly account for the frame dragging produced by the object in motion, we will see that this is exactly what happens.

The second defect, which is related to the first, is that the local conservation law given in equations (\ref{BulkDifferential}) and (\ref{BulkIntegrated}) cannot properly account for gravitational effects since it is {\it homogeneous} in $T^{ab}$. In any vacuum spacetime region where $T^{ab}=0$, the local conservation law has nothing to say. For example, we can imagine a vacuum region of space containing gravitational energy that is flowing in or out, to which the local conservation law is completely blind. This is, of course, to be expected, given that gravitational energy is not localizable. It seems that we need a term {\it like} $\nabla_{(a}\Psi_{b)}$, which can be thought of as a measure of the presence of interesting gravitational physics, e.g., a non-stationary spacetime, and that this term should be coupled not to $T^{ab}$, but rather some kind of {\it quasi}local stress-energy-momentum tensor that represents both matter {\it and} gravity, so that it can be nonzero even when $T^{ab}=0$.  Building upon the scheme introduced in section {\S}\ref{secExtrinsic}, we will presently construct a general quasilocal conservation law with precisely these properties, that will remove this second defect and, by its very construction, will also automatically remove the first defect, in an interesting and subtle way, exactly as anticipated in the previous paragraph.

\begin{figure}
\begin{center}
\includegraphics[scale=0.9]{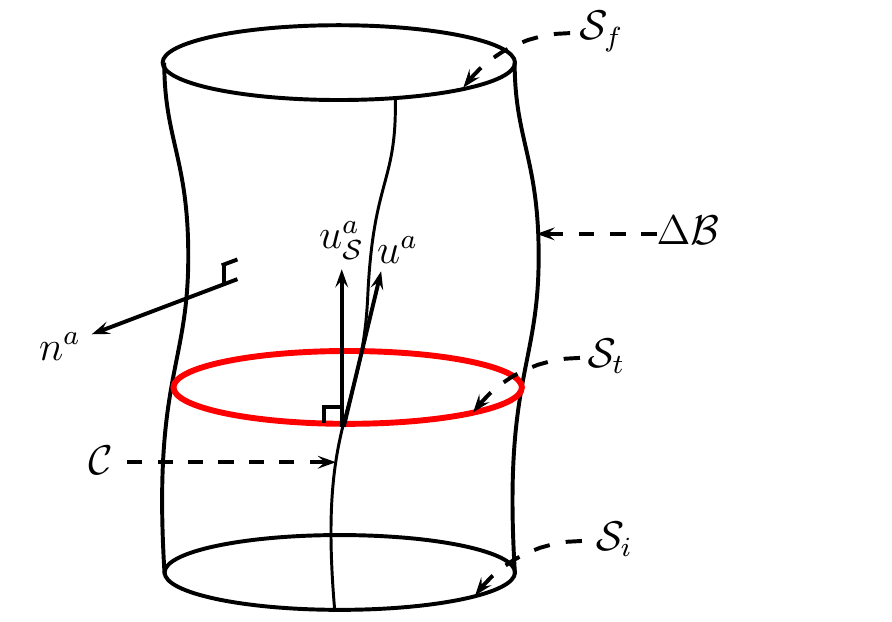}
\caption{An observer in the two-dimensional spatial surface $\mathcal{S}_t$ follows a timelike worldline $\cal C$ that lies in $\Delta\mathcal{B}$, and whose tangent four-velocity vector $u^a$ is not necessarily parallel to the timelike vector field $u^a_{\mathcal{S}}$ orthogonal to $\mathcal{S}_t$ and tangent to $\Delta\mathcal{B}$. The one-parameter family of spatial surfaces, $\mathcal{S}_t$, foliate the three-dimensional spacetime region, $\Delta \mathcal{B}$, whose boundaries are $\mathcal{S}_i$ and $\mathcal{S}_f$.}\label{QuasilocalWorldtube}
\end{center}
\end{figure}

Following a line of reasoning similar to that of reference \cite{BY1993}, let us consider an identity exactly analogous to equation (\ref{BulkDifferential}), except defined in the {\it three}-dimensional spacetime of the timelike worldtube, $\mathcal B$, i.e., the history of the boundary of a three-dimensional system. As we saw in the last chapter, for an arbitrary vector field, $\psi^a$, tangent to $\mathcal B$, we then have the identity
\begin{equation}\label{QuasilocalDifferential}
D_a ( T_{\mathcal B}^{ab}\psi_b ) = ( D_a T_{\mathcal B}^{ab} ) \psi_b + T_{\mathcal B}^{ab} D_{(a} \psi_{b)},
\end{equation}
where $D_a$  is the covariant derivative with respect to the three-metric, $\gamma_{ab} = g_{ab} - n_a n_b $, induced in $\mathcal B$. In place of the four-dimensional {\it matter} stress-energy-momentum tensor, $T^{ab}$, used in equation (\ref{BulkDifferential}), we instead insert the three-dimensional {\it total} (matter plus gravitational) stress-energy-momentum tensor, $T_{\mathcal B}^{ab}$, defined by Brown and York \cite{BY1993}.  This {\it quasi}local stress-energy-momentum tensor is defined as $T_{\mathcal B}^{ab} = -\frac{1}{\kappa}\,\Pi^{ab}$, where $\Pi^{ab}$ is the gravitational momentum canonically conjugate to the three-metric $\gamma_{ab}$ on $\mathcal B$, and $\kappa = 8 \pi G / c^4$. $\Pi^{ab}$, in turn, is equal to $\Theta_{ab} - \Theta \gamma_{ab}$, where $\Theta_{ab}=\gamma_a^{\phantom{a}c}\nabla_c n_b$ is the extrinsic curvature of $\mathcal B$. We remind the reader that the Brown-York quasilocal stress-energy-momentum tensor can conveniently be decomposed into energy, momentum, and stress surface densities, respectively (see equation (\ref{SurfaceSEMcomponents})), as:
\begin{align}\label{SurfaceSEMcomponents3}
{\mathcal E} & = \frac{1}{c^2} u^a u^b T^{\mathcal B}_{ab} & {\rm [Energy/Area]},\\
{\mathcal P}_a & = - \frac{1}{c^2} \sigma_{a}^{\phantom{a}b}u^{c}T^{\mathcal B}_{bc} & {\rm [Momentum/Area]},\\
{\mathcal S}_{ab} & = - \sigma_{a}^{\phantom{a}c}\sigma_{b}^{\phantom{b}d}T^{\mathcal B}_{cd} & {\rm [Force/Length]},
\end{align}
where $u^a$ is the four-velocity of a two-parameter family of observers residing at the boundary of a spatial volume; i.e., the worldtube boundary, $\mathcal B$, is the congruence of the integral curves of $u^a$. The spatial two-metric, $\sigma_{ab} = g_{ab} - n_a n_b + \frac{1}{c^2} u_a u_b$, projects tensors into the space orthogonal to both $u^a$ and $n^a$, i.e., tangent to the spatial two-surface the observers reside on. Integrating equation (\ref{QuasilocalDifferential}) over a section of the worldtube, $\mathcal{B}$, (see figure \ref{QuasilocalWorldtube}) bounded by initial and final spacelike slices, $\mathcal{S}_i$ and $\mathcal{S}_f$, we have
\begin{equation} \label{QuasilocalIntegrated}
\frac{1}{c} \int\limits_{\mathcal{S}_f - \mathcal{S}_i}  d{\mathcal{S}}\,  T_{\mathcal B}^{ab} u^{\mathcal{S}}_a \psi_b = \int\limits_{\Delta \mathcal{B}}  d \mathcal{B} \, \left[  T^{ab} n_a \psi_b - T_{\mathcal B}^{ab} \left( D_{(a} \psi_{b)} \right)  \right].
\end{equation}
On the left hand side, $d{\mathcal{S}}$ is the surface area element on $\mathcal{S}_i$ and $\mathcal{S}_f$, and $\frac{1}{c}u_{\mathcal{S}}^a$ is the timelike future-directed unit vector normal to $\mathcal{S}_i$ and $\mathcal{S}_f$ (and tangent to $\mathcal B$). This is the same as equation (5.2) in reference \cite{BY1993}, except that Brown and York do not include the last term on the right hand side, since they assume that $\psi^a$ is a Killing field of $\mathcal B$, or at least an approximate one. As we we will see below, this additional term turns out to be crucial.

It is interesting to compare this quasilocal conservation law to the local conservation law in equation (\ref{BulkIntegrated}). The left hand side of both equations gives the change in a physical quantity (e.g., energy) contained in a spatial three-volume. The first key difference is that the left hand side of equation (\ref{QuasilocalIntegrated}) is {\it quasilocal}: it is an integral of a {\it surface} density (e.g., energy per unit area) over the two-surface boundary of the three-volume. Unlike the local volume density, the quasilocal surface density has no meaning by itself; only the quasilocal surface density integrated over a closed two-surface is physically meaningful. The second key difference is that the integrated surface density includes contributions from both matter {\it and} gravity (e.g., gravitational energy) \cite{BY1993}.

The first term on the right hand side of equation (\ref{QuasilocalIntegrated}) - the matter flux term - is identical to that in equation (\ref{BulkIntegrated}) (when $\Psi^a=\psi^a$ is tangent to $\mathcal B$), except its origin is very different. Unlike equation (\ref{BulkDifferential}), equation (\ref{QuasilocalDifferential}) is purely geometrical, involving only the intrinsic and extrinsic geometry of $\mathcal B$, with no reference to matter. The matter stress-energy-momentum tensor, $T^{ab}$, enters equation (\ref{QuasilocalIntegrated}) by first applying the Gauss-Codazzi identity ($D_a T_{\mathcal B}^{ab} = -\frac{1}{\kappa} D_a \Pi^{ab} = -\frac{1}{\kappa} n_a G^{ab}$, where $G_{ab}$ is the Einstein tensor), and then applying the Einstein equation ($G_{ab} = \kappa T_{ab}$). Thus, equation (\ref{QuasilocalIntegrated}) forbids us from ignoring the fact that matter distorts the spacetime geometry; for example, it is meaningless to analyze an electromagnetic field in {\it flat} spacetime, as we did in {\S}\ref{secLocal}. Notice also that, because the boundary of a boundary is zero, there is actually {\it no} term in the quasilocal conservation law that is analogous to the matter flux term in equation (\ref{BulkIntegrated}). The matter flux term in equation (\ref{QuasilocalIntegrated}) arises through an entirely different mechanism: it is actually a surface flux analogue of the $j_a E^a$ bulk term in Poynting's theorem. The latter term arises when $\nabla_a T^{ab}\not=0$, which is analogous to $D_a T_{\mathcal B}^{ab} \not=0$ in the quasilocal conservation law. Specifically, $\nabla_a T^{ab}=j_a F^{ab}$ represents local energy-momentum transfer from the electromagnetic field to its sources, whereas $D_a T_{\mathcal B}^{ab} = - n_a T^{ab}$ represents quasilocal energy-momentum transfer from the matter fields to the ``system" contained in $\mathcal B$.

The last term on the right hand side of the quasilocal conservation law in equation (\ref{QuasilocalIntegrated}) is analogous to the last term in equation (\ref{BulkIntegrated}). To make the comparison more detailed, let us look at energy conservation by taking $\psi^a = \frac{1}{c} u^a$. Using the decomposition in equation (\ref{SurfaceSEMcomponents3}) we find
\begin{equation}\label{QuasilocalTransfer}
\frac{1}{c}\,T_{\mathcal B}^{ab} \left( D_{(a} u_{b)} \right) = \frac{1}{c}\,\alpha_a \mathcal{P}^a - \mathcal{S}^{ab} \theta_{ab},
\end{equation}
where $\alpha^a = \sigma^a_{\phantom{a}b} a^b$ is the projection of the observers' four-acceleration tangent to $\mathcal B$, and $\theta_{ab} = \frac{1}{c} \sigma_{(a}^{\phantom{(a}c} \sigma_{b)}^{\phantom{b)}d} \nabla_{c} u_{d}$ is the strain rate tensor describing the time development (i.e., expansion and shear) of their congruence. The structure of equation (\ref{QuasilocalTransfer}) is identical to that of equation (\ref{BulkTransfer2}), except it is one dimension lower: surface, rather than volume, i.e., quasilocal, rather than local. By the very construction of the quasilocal conservation law, the volume densities $a_a \mathbb{P}^a$ and $- \mathbb{T}^{ab} K_{ab}$ in equation (\ref{BulkTransfer2}) automatically become surface densities, $\alpha_a \mathcal{P}^a$ and $- \mathcal{S}^{ab} \theta_{ab}$, respectively, i.e., surface fluxes. This removes the first defect of local conservation laws discussed at the beginning of this section.

Let us first focus on the quasilocal ``momentum times acceleration" term, $\alpha_a \mathcal{P}^a$. Unlike the momentum volume density, $\mathbb{P}^a$, the momentum surface density, $\mathcal{P}^a$, is purely geometrical: it is a component of the extrinsic curvature of $\mathcal B$ closely related to a vector, $\omega^a$, that measures the precession rate of the observers' local ``radial" vector, $n^a$, relative to inertial gyroscopes. As noted previously in equation (\ref{InterpretationOfScriptP}), the precise relation is: $\mathcal{P}^a = \frac{c^2}{8\pi G}\,\epsilon^{a}_{\phantom{a}b}\omega^{b}$, where $\epsilon_{ab} = \frac{1}{c}\epsilon_{abcd}u^c n^d$ is the two-dimensional spatial volume form orthogonal to both $u^a$ and $n^a$. This allows us to write:
\begin{equation}\label{GravitationalPoynting}
\alpha_a \mathcal{P}^a = \frac{c^2}{8\pi G}\epsilon_{ab}\alpha^a \omega^b,
\end{equation}
which is reminiscent of the normal component of the electromagnetic Poynting vector, $\hat{n}\cdot \frac{c}{4\pi}\vec{E}\times\vec{B}$. As argued in the previous chapter, we, in fact, believe that $\alpha_a \mathcal{P}^a$ is the exact, purely geometrical, operational expression for gravitational energy flux. So the ``momentum times acceleration'' bulk term, $a_a \mathbb{P}^a$, has really become a gravitational energy surface flux term, exactly as anticipated in the equivalence principle argument given near the beginning of this section. In the context of general relativity, the presence of momentum (matter or gravitational) flowing through the system causes the observers' local ``radial'' vector to precess relative to inertial gyroscopes (i.e., a frame dragging effect), and the vector cross product between this precession rate (the gravitational analogue of a magnetic field) and the observers' acceleration (the gravitational analogue of an electric field) corresponds to a flow of gravitational energy into the system.

Of course this frame dragging mechanism of energy transfer would be difficult to confirm experimentally because the frame dragging precession rate is smaller than the momentum density by the factor $\frac{c^2}{8\pi G}$. Nevertheless, the existence of this mechanism is not surprising given that it bears a close resemblance to a similar mechanism in electrodynamics. In the context of special relativity, consider a charged particle accelerating in a uniform electric field (for simplicity we will ignore radiation reaction, which doesn't affect the point of our discussion). We ask: Where does the particle's increasing kinetic energy come from? The answer is obviously ``the field." But what is the mechanism, exactly? Consider that the velocity field part of the magnetic field due to the particle's motion is localized near the particle, and circulates around its axis of motion. A moment's thought shows that the vector cross product between the external electric field (causing the particle's acceleration) and this magnetic velocity field (moving with the particle) points toward the axis of motion, and is maximum in the plane containing the particle. In other words, there is a Poynting vector representing a flow of energy from the field to the particle. Intuitively, one might guess that this answers our question. Surprisingly, it seems that a detailed analysis of this basic energy transfer mechanism in classical electrodynamics has been done only very recently \cite{Gaidukov}. The authors of this reference verify that our intuition is correct, at least in the ultra-relativistic limit.\footnote{In the slow-motion limit, only $2/3$ of the particle's kinetic energy is supplied by the Poynting flux; $1/3$ is supplied by a peculiar ``interference" effect between the external electric field and the velocity field part of the particle's electric field \cite{Gaidukov}. This subtlety is interesting, but it does not affect the spirit of our discussion.}

To the extent that the linearized approximation to general relativity is very similar to electrodynamics, one might expect essentially the same energy transfer mechanism to occur in gravitational physics, when, e.g., we replace the massive, charged particle accelerating in a uniform electric field with a massive (charged or uncharged) particle accelerating in a uniform gravitational field (or, via the equivalence principle, a frame accelerating toward a freely-floating particle). In the context of the linearized, slow motion approximation to general relativity, sometimes called {\it gravitoelectromagnetism} (for a review, see \cite{Mashhoon}), precisely such a mechanism has been confirmed; see, e.g., \cite{Krumm,Matos}. Here we have gone beyond these approximations to confirm that this frame dragging energy transfer mechanism holds in the full, nonlinear general theory of relativity with arbitrary matter and with no approximations; the gravitational energy flux, $\alpha_a \mathcal{P}^a$, is an exact, purely geometrical contribution to the energy flux that must be considered to properly explain a wide variety of phenomena, including phenomena as basic as a falling apple.  While this mechanism was first briefly encountered in the previous chapter, we now have an equivalence principle argument for {\em why} this mechanism should exist.  We will compute an explicit example of this energy transfer mechanism in the next section.

Let us now focus on the quasilocal ``stress times strain'' term, $- \mathcal{S}^{ab} \theta_{ab}$, in equation (\ref{QuasilocalTransfer}). The boundary observers' $\theta_{ab}$ is the two-dimensional analogue of the three-dimensional $K_{ab}$ in equation (\ref{BulkTransfer2}), for a volume-filling set of observers. In the latter case, it is obvious that such a {\it three}-parameter family of observers cannot, in general, move {\it rigidly}, i.e., move in such a way as to maintain constant radar-ranging distances between all nearest neighbour pairs of observers. This is because the condition $K_{ab} = 0$ represents six differential constraints on three functions (the three independent components of the observers' four-velocity, $u^a$). In other words, the local conservation law generically contains a nontrivial bulk ``stress times strain" term, $- \mathbb{T}^{ab} K_{ab}$.

The same is {\it not} true for a {\it two}-parameter family of observers, because there the condition $\theta_{ab} = 0$ represents only {\it three} differential constraints on the same three functions. In the previous chapter, we argue that, in a generic spacetime, it is always possible to find a two-parameter congruence of integral curves of $u^a$, comprising $\mathcal B$, that is expansion- and shear-free. Moreover, the degrees of freedom of motion left to such a congruence of boundary observers are precisely the same as those for rigid body motion in Newtonian space-time, viz., three translations and three rotations, each with arbitrary time dependence. In other words, in the quasilocal conservation law (but {\it not} in the local conservation law), it is always possible to choose a family of observers (i.e., RQF observers) such that the ``stress times strain'' term vanishes.  For our present purposes it is important to recall that, since a congruence of RQF observers is expansion- and shear-free, there are no extraneous fluxes entering or leaving the system due merely to a change in the size or shape of the system boundary. So all of the interesting physics lies in the $\alpha_a \mathcal{P}^a$ surface flux term discussed above. This is a key advantage to working with a quasilocal conservation law in general, and RQFs in particular. We will compute an explicit example of an RQF in the next section.

In summary, the quasilocal conservation law introduced in this section removes two defects of the familiar local conservation law: (1) The undesirable bulk terms in the local conservation law automatically appear as surface flux terms in the quasilocal conservation law, and (2) unlike the local conservation law, the quasilocal conservation law accounts for gravitational effects. For example, the quasilocal energy conservation law contains an energy transfer mechanism related to frame dragging that is needed to understand in detail what is actually happening with regards to energy flow when, say, an apple falls. In the next section we will apply the quasilocal energy conservation law to a general relativistic analogue of the apparent paradox introduced in {\S}\ref{secParadoxResolution}.

\subsection{Quasilocal Approach to the Apparent Paradox}\label{secQuasiParadox}

To illustrate the explanatory power of the quasilocal approach let us now look at a general relativistic analogue of the apparent paradox at the start of this chapter. We will work with a spacetime whose metric is $g_{ab}$ in coordinates $X^a = (X^0, X^I)=(cT, X,Y,Z)$, $I=1,2,3$, and which contains a covariantly constant (i.e., uniform) electric field of magnitude $E$ in the positive $X$-direction \cite{Bertotti,Stephani}:
\begin{equation}\label{ElectricSpacetime}
ds^2 = -\left(1 +  X^2/L^2 \right) c^2 dT^2 + \frac{dX^2}{\left( 1 +  X^2/L^2 \right)}+ \frac{dY^2}{\left( 1 - Y^2/L^2 \right)} + \left(1 -  Y^2/L^2 \right) dZ^2,
\end{equation}
where $L = c^2/\sqrt{G} E$ is the length scale of the geometry. It is important to note that the presence of the electric field is intimately connected to the spacetime curvature - we are no longer working in flat spacetime. This is readily verified by checking that there exist non-zero curvature invariants such as $R_{a b c d} R^{a b c d} = 8 /L^4$. Alternatively, it is easy to see that this spacetime is a product of two surfaces of constant curvature \cite{Bertotti,Stephani}.

Observe that $g_{ab}$ reduces to the Minkowski metric in the $X=0,Y=0$ plane. In this plane let us construct a timelike worldline, ${\mathcal C}_0$, parameterized by proper time, $\tau$, and defined by the embedding $X^a = \xi^a (\tau)$, that represents a fiducial observer undergoing proper acceleration $\mathtt{a}(\tau)$ along the $Z$-axis. As in \S\ref{secLocal}, along ${\mathcal C}_0$ we define a tetrad, $e_0^{\phantom{0}a}(\tau)$ and $e_I^{\phantom{I}a}(\tau)$, where $e_0^{\phantom{0}a}(\tau)=\frac{1}{c}\,d\xi^a(\tau)/\,d\tau$ denotes the unit vector tangent to ${\mathcal C}_0$, and the spatial triad, $e_I^{\phantom{I}a}(\tau)$, is Fermi-Walker transported along ${\mathcal C}_0$. Because $g_{ab}$ equals the Minkowski metric on $\mathcal{C}_0$, we can use the same coordinate components for this tetrad as given in equation (\ref{tetrad}). Around $\mathcal{C}_0$ we will construct a two-parameter family (two-sphere's worth) of accelerated observers, whose congruence of worldlines comprises a timelike worldtube, $\mathcal{B}$. Our notation will follow that used in \S\ref{secLocal}: The spacetime metric, $g_{ab}$, induces on $\mathcal{B}$ the spacelike outward-directed unit normal vector field, $n^a$, the Lorentzian three-metric, $\gamma_{ab} = g_{ab} - n_a n_b$, and the covariant derivative operator, $D_a$. Also, we let $u^a$ denote the observers' four-velocity, which is tangent to $\mathcal{B}$, and $\sigma_{ab} = g_{ab} - n_a n_b + \frac{1}{c^2} u_a u_b$ denote the observers' spatial two-metric.

In \S\ref{secLocal} we constructed a three-parameter (volume-filling) family of accelerated observers in flat spacetime in the neighbourhood of ${\mathcal C}_0$ using the coordinate transformation given in equation (\ref{Worldline}). For arbitrary fixed $r$, this coordinate transformation actually defines an exact RQF, i.e., a two-parameter family (two-sphere's worth) of observers who, despite their time-dependent acceleration, maintain constant radar-ranging distance from their nearest neighbours - they are moving {\it rigidly}. Indeed, inspection of equation (\ref{AdaptedCoordMetric}) shows that $n_i$ and $u_i$ both vanish, so the observers reckon they are on a surface with constant spatial two-metric $\sigma_{ij} = g_{ij} = r^2\mathbb{S}_{ij}$, which is a round sphere of areal radius $r$. Now because spacetime is locally flat, at lowest order in $r$ we can start with the same coordinate transformation given in equation (\ref{Worldline}), but because of the spacetime curvature, we must add corrections at higher order in $r$ to achieve the same RQF, i.e., to achieve $\sigma_{ij} = r^2\mathbb{S}_{ij}$. Thus we begin with the ansatz:
\begin{equation} \label{RQFWorldTube}
X^a (x) = \xi^a (\tau) + r r^I (\theta, \phi) e_I^{\phantom{I}a}(\tau) + \frac{r^3}{L^2}\left[
F^I(\tau,\theta,\phi) e_I^{\phantom{I}a}(\tau) + F^0(\tau,\theta,\phi) e_0^{\phantom{I}a}(\tau)\right] +\mathcal{O}\left(\frac{r^4}{L^3}\right),
\end{equation}
where, as before, $x^\alpha = (x^0,x^1,x^i) = (\tau,r,\theta,\phi)$ are coordinates  adapted to the observers: the angular coordinates $x^i=(\theta,\phi)$ are parameters that label the observers' worldlines; the radial coordinate, $r$, parameterizes the size of the RQF; and the time coordinate, $\tau$, represents our choice of simultaneities on $\mathcal B$.  The first two terms on the right hand side of
equation (\ref{RQFWorldTube}) are given in equation (\ref{Rindtmf}).
The three arbitrary functions $F^I(\tau,\theta,\phi)$ allow us to perturb the observers' worldlines to satisfy the three conditions $\sigma_{ij} = r^2\mathbb{S}_{ij}$ needed to achieve a round sphere RQF of areal radius $r$. (An RQF of generic shape and size need only satisfy the three weaker conditions, $\partial \sigma_{ij}/\partial\tau = 0$.) The arbitrary function $F^0(\tau,\theta,\phi)$ allows us to perturb the choice of simultaneities on $\mathcal B$ to achieve other natural geometrical conditions, which will be discussed later. For the time being, we will keep the four functions $F^I$ and $F^0$ {\it arbitrary}.

Choosing $\psi^a=\frac{1}{c}u^a$ in equation (\ref{QuasilocalIntegrated}) makes this a quasilocal energy conservation equation. Using GRTensorII \cite{GRTensor} running under Maple, we compute the individual terms in equation (\ref{QuasilocalIntegrated}) and find:
\begin{align}	
\frac{1}{c}\, d\mathcal{S}\,  T_{\mathcal B}^{ab} u^{\mathcal{S}}_a \psi_b &=
r^2\,d\mathbb{S}\,\left\{ -\frac{c^4}{4\pi G}\frac{1}{r} + \gamma^2\frac{E^2}{16\pi}\sin^2\theta (5\cos^2\theta+1)\,r \right. \nonumber \\
& \qquad\qquad\qquad\qquad \left. +  (C + \Phi_F + \Psi_F)\, r + \frac{c^4}{GL} \times \mathcal{O}\left(\frac{r^2}{L^2}\right)  \right\},\label{RQFEnergyDensity}\\
d \mathcal{B} \,  T^{ab} n_a \psi_b &= c\,d\tau\,r^2\,d\mathbb{S}\, \beta\gamma^2\frac{E^2}{4\pi}\cos\theta\, \left\{ 1 + \frac{1}{c^2}\mathtt{a}(\tau)r\cos\theta + \mathcal{O}\left(\frac{r^2}{L^2}\right)  \right\},\label{RQFPoyntingDensity}\\
d \mathcal{B} \,  T_{\mathcal B}^{ab} \left( D_{(a} \psi_{b)} \right) & = d \mathcal{B}\, \left( \frac{1}{c}\,\alpha_a \mathcal{P}^a +  \mathcal{S}^{ab} \theta_{ab} \right) \nonumber \\
& = -c\,d\tau\,r^2\,d\mathbb{S}\, \beta\gamma^2\frac{E^2}{8\pi}\sin^2 \theta\frac{\mathtt{a}(\tau)r}{c^2}\, \left\{ 1 + \mathcal{O}\left(\frac{r}{L}\right)  \right\}
+ d \mathcal{B}\,\mathcal{S}^{ab} \theta_{ab},\label{RQFGravityDensity}
\end{align}
where $d\mathbb{S} = \sin\theta \,d\theta \,d\phi$ is the area element on the unit round sphere, as before, and $\gamma(\tau)$ and $\beta(\tau)$ are the same Lorentz transformation parameters we defined after equation (\ref{BulkEM2}). We have also inserted the relation $L = c^2/\sqrt{G} E$ where appropriate. Let us analyze each of these terms separately.

Equation (\ref{RQFEnergyDensity}) represents the quasilocal energy density. The first term on the right hand side, at order $1/r$ inside the braces, is a vacuum energy. Insofar as it is constant in time, this term does not contribute to changes in the energy inside the sphere, and so can be ignored. The quantity $C$ is a function of the angles $(\theta,\phi)$, but not $\tau$. So for the same reason just cited, it can also be ignored. The quantity $\Phi_F$ is a homogeneous function of the angular derivatives of the $F^I$, which is a total derivative. It integrates to zero for {\it any} choice of the arbitrary functions $F^I$, and so can also be ignored. The quantity $\Psi_F$ is an inhomogeneous function of the angular derivatives of the $F^I$, which is a linear combination of the three round sphere RQF conditions, $\sigma_{ij} - r^2\mathbb{S}_{ij}=0$. It is zero when we choose a set of functions $F^I$ that satisfy these conditions (which we will do later), and so can also be ignored. The only nontrivial term is the second one. Insofar as only the integrated quasilocal energy density is physically meaningful, we can subtract off the spherical harmonics that integrate to zero, leaving the {\it effective} quasilocal energy density:
\begin{equation}
\left\{\frac{1}{c}\, d\mathcal{S}\,  T_{\mathcal B}^{ab} u^{\mathcal{S}}_a \psi_b \right\}_{\rm effective}=
r^2\,d\mathbb{S}\,\left\{  \gamma^2\frac{E^2}{12\pi}\,r + \frac{c^4}{GL} \times \mathcal{O}\left(\frac{r^2}{L^2}\right)  \right\}.\label{EffectiveRQFEnergyDensity}
\end{equation}

Equation (\ref{RQFPoyntingDensity}) represents the quasilocal Poynting flux. The quantity in braces is (at least up to the order indicated) equal to the lapse function, $N(x)=1 + \frac{1}{c^2}\mathtt{a}(\tau)r\cos\theta + \mathcal{O}\left(\frac{r^2}{L^2}\right)$, which is contained in $d{\mathcal B}$. Up to the lowest nontrivial order in $r$, this lapse function agrees with that found in \S\ref{secLocal}; as before, it is this non-isotropic time dilation that allows for a nonzero accumulation of Poynting flux in the sphere. Note that, up to the order indicated, the Poynting flux is independent of our choice of arbitrary functions $F^I$ and $F^0$. Equation (\ref{RQFGravityDensity}) represents the sum of the quasilocal gravitational flux and the quasilocal ``stress times strain" term - recall equation (\ref{QuasilocalTransfer}). When we impose the RQF rigidity conditions on the functions $F^I$, the latter term vanishes (i.e.,  $\theta_{ab}$ vanishes). This leaves only the quasilocal gravitational flux term, $\alpha_a \mathcal{P}^a$, to consider. Note that, up to the order indicated, the gravitational flux is also independent of our choice of arbitrary functions $F^I$ and $F^0$.

Putting these results together, with the assumption that the RQF rigidity conditions can be satisfied (we'll demonstrate this below), the quasilocal energy conservation law in equation (\ref{QuasilocalIntegrated}) reads, at order $r^3$,
\begin{align} \label{QuasilocalIntegratedParadox}
\int\limits_{\mathcal{S}_f - \mathcal{S}_i}  r^2\,d\mathbb{S}\,\gamma^2\frac{E^2}{12\pi}\,r & = \int\limits_{\Delta \mathcal{B}}  c\,d\tau\,r^2\,d\mathbb{S}\,\left\{ \beta\gamma^2\frac{E^2}{4\pi}\cos\theta\, \left( 1 + \frac{1}{c^2}\mathtt{a}(\tau)r\cos\theta \right) \right. \nonumber \\
& \qquad \qquad \qquad  \qquad \qquad \left. + \beta\gamma^2\frac{E^2}{8\pi}\sin^2 \theta\frac{\mathtt{a}(\tau)r}{c^2} \right\}.
\end{align}
This quasilocal conservation law is to be compared with the local conservation law in equation (\ref{ParadoxConservationLaw}). As we did for the latter equation, we will take the time interval to be infinitesimal: $\tau_f -\tau_i = \Delta\tau$. Using the relation $\frac{d}{d\tau}  \gamma^2   = 2 \beta \gamma^2\frac{\mathtt{a}}{c} $ [compare this with $\frac{d}{d\tau} \left( (1 + \beta^2) \gamma^2  \right) = 4 \beta \gamma^2\frac{\mathtt{a}}{c} $, which differs by a factor of 2], the left hand side of equation (\ref{QuasilocalIntegratedParadox}) then integrates to precisely the same change in electromagnetic energy, $\Delta E_{\rm Total}$, given by the local conservation law, and also equation (\ref{E(1)2}) (with the cylinder volume, $V=AL$, replaced with the sphere volume, $V=\frac{4}{3} \pi r^3$). The quasilocal Poynting flux integral over $\mathcal{B}$ on the right hand side of equation (\ref{QuasilocalIntegratedParadox}) is, as mentioned earlier, identical to the local Poynting flux integral in equation (\ref{ParadoxConservationLaw}), and so yields the same change in electromagnetic energy, $\Delta E_{\rm Poynting}$, as found in that case, and also given in equation (\ref{E(2)2}) (with, again, the cylinder volume replaced with the sphere volume). We emphasize once more the role of the non-isotropic time dilation (the lapse function in parentheses) that allows for a nonzero accumulation of Poynting flux.

The quasilocal gravitational flux integral over $\mathcal{B}$ on the far right hand side of equation (\ref{QuasilocalIntegratedParadox}) is, as discussed earlier, the surface flux version of the bulk ``acceleration times momentum" energy transfer term in the local conservation law. The difference is that the local momentum volume density, $\mathbb{P}^a$, has been replaced with the quasilocal momentum surface density, ${\mathcal P}^a$, and the integration is not over a volume, but the boundary of that volume. We stress again that the actual mechanism of the energy transfer is a general relativistic effect, viz., the vector cross product between the boundary observers' acceleration and the precession rate of their gyroscopes due to the frame dragging caused by the electromagnetic momentum inside the system. All together, then, half of the energy flux is due to the electromagnetic field (with the flux entering mainly near the poles of the sphere, i.e., $\cos^2\theta$), and the other half is due to the gravitational field (with the flux entering mainly near the equator of the sphere, i.e., $\sin^2\theta$). Both contribute to a changing electromagnetic energy on the left hand side of the quasilocal conservation law.

One might wonder how a flux of {\it gravitational} energy through the boundary of the system becomes {\it electromagnetic} energy inside the system. It seems that in general relativity there is no distinction between the two forms of energy. All forms of energy are equivalent. This is not surprising when we look at the metric in equation (\ref{ElectricSpacetime}). The geometry is nontrivial and so presumably contains gravitational energy (in some nonlocal, or quasilocal sense), and the nontrivial geometry in turn is intimately connected to the electric field. They cannot be disentangled.

All that remains in this section is to verify that we actually {\it can} solve the RQF rigidity conditions, viz., the three differential constraints $\sigma_{ij} = r^2\mathbb{S}_{ij}$ on the three functions $F^I(\tau,\theta,\phi)$ which ensure that the observers' frame is a rigid round sphere of areal radius $r$. As argued in Chapter \ref{chNova}, such RQF solutions always exist in a generic spacetime; moreover, they are unique up to motions of the RQF equivalent to those of rigid motion in Newtonian space-time, viz., six arbitrary time-dependent degrees of freedom: three translations and three rotations. Aiming for the simplest solution, we first set $F^1(\tau,\theta,\phi)$ and $F^2(\tau,\theta,\phi)$ to zero, which eliminates translations of the RQF away from ${\mathcal C}_0$ in the $X$ and $Y$ directions. We next demand that the RQF is not rotating, i.e., that the twist of the observers' congruence is zero, i.e., that the observers' four-velocity, $u^a$, is hypersurface orthogonal (as a vector field in $\mathcal B$). This means that, by adjusting $F^0(\tau,\theta,\phi)$ appropriately, we can always find a time foliation of $\mathcal B$ (i.e., choose the observers' surfaces of simultaneity) such that $u^a$ is orthogonal to them, which is equivalent to the two conditions $u_i=0$. Using GRTensorII \cite{GRTensor} we can readily find a solution that satisfies all of these conditions, given by:
\begin{align}
F^3(\tau,\theta,\phi) &= \frac{1}{6}\,\gamma^2\,\cos\theta(3-\cos^2\theta)
-\frac{1}{6}\cos\theta(\cos^2\theta+3\sin^2\theta\cos^2\phi),\label{F3}\\
F^0(\tau,\theta,\phi) &= -\frac{1}{3}\,\beta\gamma^2\,\cos\theta(3-\cos^2\theta).\label{F0}
\end{align}
Note that the $\phi$ dependence in $F^3(\tau,\theta,\phi)$ is not surprising since the electric field, being parallel to the $\phi=0$ plane, breaks the azimuthal symmetry - the electric field affects the geometry, and hence the coordinate embedding of the RQF. One can show that these results are in agreement with the more generally derived formulas in {\S}\ref{secCurved} of the previous chapter.

\section{Summary}\label{Conclusions3}

Using the standard, local way of constructing conservation laws - see equations (\ref{BulkDifferential}) and (\ref{BulkIntegrated}) - we analyzed conservation of energy in the context of a simple example in special relativity, viz., a box, rigidly accelerating along the $Z$-axis, that is immersed in a transverse, uniform electric field. According to the local energy conservation law, the electromagnetic energy inside the box increases due to two separate mechanisms: (1) Half of the increasing energy is due to energy flowing in from outside the box via a Poynting flux. Interestingly, even though the instantaneous proper Poynting vector is uniform throughout the box (suggesting no net flux), there is a net proper time-integrated flux due to the acceleration-induced relative time dilation between observers at the top and bottom of the box. (2) The other half of the increasing energy is {\it not} due to energy flowing in from the outside; rather, it is a bulk effect due to the co-moving observers accelerating relative to the existing electromagnetic momentum inside the box, i.e., an ``acceleration times momentum" energy transfer mechanism familiar from classical mechanics, integrated over the volume of the box.

One might wonder if these two energy transfer mechanisms adequately explain what's happening. The answer is: No. First, in special relativity we assume that spacetime is flat, even though there is an electromagnetic field present. This precludes any possible general relativistic effects such as a flux of gravitational energy associated with the curvature of the spacetime caused by the electromagnetic field. Secondly, even in the context of general relativity, a {\it local} conservation law cannot properly capture all of the gravitational physics; for example, gravitational energy is not localizable - there is no such thing as a gravitational energy per unit volume, so a local conservation law cannot tell the whole story. Another way to see this problem is to notice that the standard local conservation law is {\it homogeneous} in the matter stress-energy-momentum tensor, and so has nothing to say in a matter-free region of space, even though that region may contain dynamical curvature, e.g., gravitational waves.

To address these shortcomings in general, and in particular see what's really happening with regards to the increasing energy inside the box, we constructed a {\it quasi}local conservation law based on the Brown \& York  quasilocal {\it total} (matter plus gravitational) stress-energy-momentum tensor defined on the history of the boundary of a spatial volume - see equations (\ref{QuasilocalDifferential}) and (\ref{QuasilocalIntegrated}). Although the computational approach is very similar to that of 
reference \cite{BY1993}, we assume no symmetries (i.e., $\psi^a$ is not a Killing vector) and therefore retain the final term in equation (\ref{QuasilocalDifferential}), which results in the more general quasilocal conservation law given in equation (\ref{QuasilocalIntegrated}). Using the energy form of this quasilocal conservation law, we analyzed the general relativistic analogue of the simple example described above. We found, again, that the electromagnetic energy inside the box increases due to two separate mechanisms. The first mechanism is identical in form to the Poynting flux mechanism described above [number (1)], but conceptually it has a completely different origin: it is actually analogous to a surface flux version of the $\vec{j}\cdot\vec{E}$ bulk term in the standard Poynting theorem, except instead of energy being transferred locally from the electromagnetic field to the four-current source of the field, it represents energy being transferred quasilocally from the matter fields to the ``system" contained inside the box. The second mechanism is, at first sight, conceptually identical to the ``acceleration times momentum" energy transfer mechanism described above [number (2)], except the volume integral has been converted to a {\it surface flux} integral over the boundary of the box. Going further, we argued that this surface flux is actually a {\it gravitational} energy flux exactly analogous to the electromagnetic Poynting flux, $\vec{E}\times\vec{B}$, with $\vec{E}$ replaced by the acceleration of the co-moving observers on the boundary of the box, and $\vec{B}$ replaced by the precession rate of their gyroscopes due to the frame dragging caused by the electromagnetic momentum (in general, matter and gravitational momentum) flowing through the box. In {\it both} cases, now, energy is entering from outside the box via surface fluxes: half is an electromagnetic energy flux (entering the box through its top and bottom) and the other half is a gravitational energy flux (entering the box through its {\it sides}). Because an electromagnetic field is intimately intertwined with the spacetime curvature it produces, general relativity does not distinguish between electromagnetic and gravitational energy entering the box - both contribute on the same footing to the increasing electromagnetic energy inside the box.

We can understand the second, gravitational energy flux mechanism intuitively as follows. Imagine being inside an accelerating box in empty space, which contains a freely-floating massive body that appears to be accelerating toward you. From your perspective, its kinetic energy is increasing. Where does the increasing kinetic energy come from? We could ``explain" it as simply the ``acceleration times momentum" energy transfer mechanism familiar from classical mechanics. Alternatively, we could invoke the equivalence principle and say that our box is at rest in a uniform gravitational field, and that the kinetic energy is coming from the gravitational field outside the box via some kind of gravitational energy flux passing through the boundary of the box. An analogous mechanism exists in the context of electrodynamics that explains, in detail, how an accelerating charge acquires kinetic energy from the external electric field causing the particle's acceleration \cite{Gaidukov}; moreover, the gravitational version of this mechanism has been shown to exist in the context of the linearized, weak field approximation to general relativity \cite{Krumm,Matos}. In this chapter we have established the existence of this very basic, but subtle mechanism in the full, nonlinear general theory of relativity, with arbitrary matter and no approximations.  It is also worth pointing out that our analysis made crucial use of the concept of RQFs, discussed more fully in Chapters \ref{chRigidRevisited} and \ref{chNova}, in order to properly isolate the relevant energy fluxes passing through the boundary of the box.

In summary, we have explained the bulk ``acceleration times momentum" energy transfer mechanism familiar in classical mechanics {\it exactly} in terms of a simple, operationally defined, purely geometrical, general relativistic gravitational energy flux passing through the {\it boundary} of the volume in question. Naively, one might argue that since there is no $G$ or $c$ in ``acceleration times momentum'', this cannot be a general relativistic effect. But it {\it is}. It is based on frame dragging (the gravitational analogue of ``$\vec{B}$'' in ``$\vec{E}\times\vec{B}$''), which is a general relativistic effect. We don't notice this mechanism in our day-to-day experiences because the typical gyroscopic precession rate vector due to a nearby object in motion is very tiny; but it is precisely this vector, multiplied by the huge number $\frac{c^2}{8\pi G}$, that we identify as the ``momentum'' of the object [more precisely, the quasilocal momentum surface density, rotated by $90$ degrees - see equation (\ref{GravitationalPoynting})]. This general relativistic gravitational energy flux mechanism is what's {\it really} happening in the bulk ``acceleration times momentum" energy transfer mechanism in classical mechanics. This deeper understanding would not be possible in the context of local conservation laws, and is a nice example of why we need {\it quasi}local conservation laws and a notion of rigidity in general relativity.

\chapter{Archimedes' Law for General Relativity} \label{chArch}

As we have now seen, with the advent of general relativity, the notions of the energy and momentum (linear and angular) of a system became elusive.  In the previous chapter we touched upon the two reasons for this. Firstly, when the spacetime geometry is treated as a dynamical field, energy and momentum are no longer local concepts, i.e., there is no such thing as an energy or momentum per unit volume, which when integrated over a finite volume yields the total energy or momentum inside that volume. We will explore this point in more detail below and see that this non-localizable nature applies to both gravitational {\it and} matter fields. In other words, the local stress-energy-momentum tensor of matter, $T^{ab}$, is not, in any fundamental way, related to the matter energy or momentum of a finite system. Secondly, energy and momentum are frame-dependent constructs. In Newtonian space-time we have the concept of an inertial reference frame that allows us to define the energy and momentum of a point particle or an extended system {\it relative} to such a frame. We can even use a frame that is accelerating or rotating, provided we properly account for non-inertial effects. Thus, the key concept is not that of an {\it inertial} frame, but that of a {\it rigid} frame, that is, one in which the distances between all nearest-neighbouring pairs of observers comprising the frame are constant in time. In Newtonian space-time, such rigid frames have precisely six arbitrary time-dependent degrees of freedom: three for linear velocity and three for angular velocity. However, in general relativity, such frames do not exist in general.  This presents a serious obstacle to constructing physically sensible and useful definitions of energy and momentum in the context of general relativity, and their related conservation laws.\footnote{In general relativity there exist notions of energy and momentum for an isolated system in a spacetime that admits asymptotic symmetries at infinity, but these are a throwback to the pre-general relativistic practice of relying on spacetime symmetries to construct conservation laws. This approach represents a break from that tradition.}

As we have seen already, the solution to the first problem (failure of the notion of local energy or momentum) involves shifting from a local to a {\it quasilocal} way of thinking. To see this more clearly, let $I_{\rm mat}[g,\varphi]$ denote an action functional for a set of dynamical matter fields, $\varphi$, in a spacetime $\mathcal{M}$ with non-dynamical (fixed) background metric $g$. Since the metric is not treated as a dynamical field, $I_{\rm mat}[g,\varphi]$ is the {\it total} action functional. If, as is usually done, we define the total local stress-energy-momentum tensor of the system as the functional derivative of the total action functional with respect to the metric, we have
\begin{equation}\label{T_matter}
2\delta_g \, I_{\rm mat}[g,\varphi]=\int_\mathcal{M}\epsilon_\mathcal{M} \,T^{ab}\,\delta g_{ab}
\end{equation}
(where $\epsilon_\mathcal{M}$ is the spacetime volume form), and so the total local stress-energy-momentum tensor of the system is just the matter stress-energy-momentum tensor, $T^{ab}$. Insofar as a conservation law constructed from $T^{ab}$ will be homogeneous in $T^{ab}$, it will be essentially blind to any interesting gravitational physics. However, in general relativity the metric is treated as a dynamical field, and we must add to $I_{\rm mat}[g,\varphi]$ the action functional of the gravitational field. Using the usual first order action functional for gravity we then find
\begin{equation}\label{T_matter_and_gravity}
2\delta_g \, I_{\rm mat+grav}[g,\varphi]=\int_\mathcal{M}\epsilon_\mathcal{M} \, \left(T^{ab}-\frac{1}{\kappa}G^{ab}\right)\,\delta g_{ab}+\int_{\cal B}\epsilon_{\cal B} \, \left(-\frac{1}{\kappa}\Pi^{ab}\right)\,\delta \gamma_{ab}
\end{equation}
(where $\kappa=8\pi G/c^4$ and $G^{ab}$ is the Einstein tensor), which tells us that the total local stress-energy-momentum tensor of the system is the sum, $T^{ab}-\frac{1}{\kappa}G^{ab}$, which is just zero by the Einstein equation, i.e., there is no nontrivial local notion of total stress-energy-momentum in general relativity. Note that this statement applies to both gravitational {\it and} matter fields, not just gravitational. This argument is not new. The idea that $-\frac{1}{\kappa}G^{ab}$ is the local stress-energy-momentum tensor of the gravitational field was independently put forward by both Lorentz and Levi-Civita, but was rejected by Einstein on various physical grounds, e.g., gravitational waves in vacuum could then not transport energy, and Einstein and others continued to use a pseudotensor to represent local gravitational stress-energy-momentum \cite{Cattani_and_De_Maria_1993}. We know today that what saves us is the boundary term. On the right-hand side of equation (\ref{T_matter_and_gravity}), $\Pi^{ab}$ is the gravitational momentum conjugate to the three-metric $\gamma_{ab}$ induced on the boundary, $\cal B$. In the spirit of identifying the stress-energy-momentum tensor as the functional derivative of the action with respect to the metric, one identifies $-\frac{1}{\kappa}\Pi^{ab}$ as the {\it quasi}local total stress-energy-momentum tensor in general relativity. It is defined only on the boundary (energy and momentum per unit {\it area}), and includes contributions from both the matter and gravitational fields. $T^{ab}$ is no longer involved. This is the essence of what Brown and York did in 1993 \cite{BY1993}.

The solution to the second problem (failure of the general existence of rigid frames) also involves shifting from a local to a quasilocal way of thinking.  In particular, it requires introducing the notion of a rigid quasilocal frame.  It is worth noting that this also allows us to cleanly identify the most relevant energy and momentum fluxes crossing the system boundary (i.e., eliminate those fluxes due merely to changes in the size or shape of the boundary) and, moreover, to obtain simple, exact definitions for the elusive {\it gravitational} versions of those fluxes in terms of operationally-defined geometrical quantities on the boundary.

This chapter is a fusion of these two solutions: Brown and York's quasilocal stress-energy-momentum tensor and our notion of an RQF. As we shall see, the real significance of the RQF approach is that it allows us to construct conservation laws for energy and momentum {\it without relying on any spacetime symmetries}. In the previous chapter we constructed a completely general matter plus gravity RQF energy conservation law for spatially finite systems that does not rely on the existence of a timelike Killing vector field. Here we do the same, but for linear and angular momentum, without relying on the existence of a spacelike Killing vector field.

In {\S}\ref{secLocal2} we begin with a local momentum conservation law based on the identity
\begin{equation}\label{differential_local_conservation_law}
\nabla_a (T^{ab}\Psi_b )=(\nabla_a T^{ab})\Psi_b + T^{ab}\nabla_{(a}\Psi_{b)},
\end{equation}
where $\Psi^a$ is a spatial vector field determining the particular component of linear or angular momentum we are interested in. In general relativity, $\nabla_a T^{ab}=0$, and the local conservation law reduces to $\nabla_a ( T^{ab}\Psi_b ) = T^{ab} \nabla_{(a} \Psi_{b)}$. We argue that this differential conservation law integrated over a four-dimensional worldtube volume makes no physical sense unless two conditions are satisfied: (1) the frame (three-parameter bundle of worldlines) must be rigid in the sense discussed above, and (2) $\Psi^a$ must be a Killing vector field of the spatial three-metric on the quotient space of the geometrically rigid bundle of worldlines. Neither of these conditions is satisfied in general, and so such an integrated local momentum conservation law is not general. We argue that, ultimately, this failure results because the local approach, based on only the matter stress-energy-momentum tensor, $T^{ab}$, does not (and cannot) properly account for gravitational effects. As an example to close the section, we specialize this law to a stationary context and construct a general relativistic version of Archimedes' law. While Archimedes' law \cite{Arch} forms the foundation of hydrostatics and has broad applications in a number of disciplines, its application in a general relativistic context has remained almost completely unexplored. We compare our general relativistic version of Archimedes' law to a similar law constructed by Eriksen and Gr{\o}n \cite{Eriksen2006} in the context of accelerated observers in Rindler space (or equivalently, a uniform gravitational field).

In {\S}\ref{secQuasilocal2} we properly account for gravitational effects by replacing $T^{ab}$ with Brown and York's quasilocal matter plus gravity stress-energy-momentum tensor, and the identity (\ref{differential_local_conservation_law}) with an analogous identity defined in the boundary spacetime, $\mathcal{B}$ [see equation (\ref{differential_quasilocal_conservation_law})]. The vector field $\Psi^a$ becomes a vector field $\psi^a$ tangent to $\mathcal{B}$. We argue that the integrated form of this differential conservation law always makes physical sense because: (1) the rigid quasilocal frames (RQFs) discussed above always exist, and (2) the six conformal Killing vector fields discussed above also always exist, and $\psi^a$ can always be taken to be one them (three boosts, corresponding to the three components of linear momentum, and three rotations, corresponding to the three components of angular momentum). Moreover, the resulting completely general matter plus gravity RQF momentum conservation law for spatially finite systems tells us something new about the physics of momentum conservation. Firstly, it reveals a simple, exact operational definition for gravitational momentum flux (mentioned above) that allows us to understand more deeply a wide variety of physical phenomena, including the simple example of a falling apple. Secondly, while both the local and quasilocal laws handle tangential (shear) stresses similarly, the quasilocal law treats stresses normal to the spatial boundary on a different footing, in a novel way that involves the quasilocal {\it pressure}, which can have both matter and gravitational (``geometrical") sources. We show that the quasilocal law reduces to the local law in the limit of a small-sphere RQF, but in general it involves completely new gravitational effects that are not accounted for in the local law. We close the section by deriving a quasilocal version of Archimedes' law which, unlike the local version we constructed at the end of {\S}\ref{secLocal2}, is as general as such a law can be. We apply this law to an example of electrostatic weight and buoyant force in the context of the Reissner-Nordstr\"{o}m black hole. We present a brief summary of the chapter in {\S}\ref{Summary}.

\section{Local Momentum Conservation Law in General Relativity}\label{secLocal2}

In this section we follow the standard {\it local} conservation law approach to construct an integrated momentum conservation law for matter fields in a finite volume of space in the context of general relativity. We argue that for this integrated law to make sense, physically, we must impose two conditions (a rigidity condition in time and a Killing vector condition in space) that {\it cannot} always be satisfied. When they {\it can} be satisfied, and when we can further specialize to a certain {\it stationary} context, we can construct a general relativistic version of Archimedes' law for matter fields (e.g., electromagnetism), which illustrates how Maxwell stress-like buoyant forces support the matter weight contained in a non-inertial reference frame. We compare this law with that constructed by Eriksen and Gr{\o}n \cite{Eriksen2006} for electromagnetism in the context of uniformly accelerating (Rindler) observers in flat spacetime.

\subsection{General Analysis}\label{Local-1}

Given a smooth, four-dimensional Lorentzian manifold with metric $g_{ab}$ and associated derivative operator $\nabla_a$, recall that the identity (equation (\ref{differential_local_conservation_law})) provides the differential form of a local conservation law for the current $T^{ab}\Psi_b$.  In the context of general relativity, where the matter stress-energy-momentum tensor, $T^{ab}$, includes all sources one has $\nabla_a T^{ab}=0$.  However, note that we will not impose this identity so as to be able to apply our analysis to electromagnetism in special relativity.  Now, to integrate this conservation law, we consider a spatially finite three-parameter family of observers with four-velocity vector field $u^a$ tangent to their congruence of worldlines. We use the same notation as in section \ref{secLocal}.  In particular, let $\mathcal{B}$ denote the three-dimensional Lorentzian manifold boundary of this bundle of worldlines. Let $\Sigma_{i}$ and $\Sigma_{f}$ denote two finite spacelike three-surfaces slicing through the bundle (respectively the initial and final volumes of space), and $\Delta{\cal B}$ denote the section of $\mathcal{B}$ lying between $\Sigma_{i}$ and $\Sigma_{f}$. Finally, let $\Delta {\cal V}$ denote the finite spacetime four-volume contained within these boundaries (refer to figure \ref{LocalWorldtube}). Integrating equation (\ref{differential_local_conservation_law}) over $\Delta {\cal V}$ yields the integrated form of this differential conservation law:
\begin{equation} \label{integrated_local_conservation_law}
\frac{1}{c} \int\limits_{\Sigma_f - \Sigma_i}  d{\Sigma}\,  T^{ab} u^\Sigma_a \Psi_b = \int\limits_{\Delta\mathcal{B}}  d \mathcal{B} \, T^{ab} n_a \Psi_b - \int\limits_{\Delta{\mathcal V}}  d{\mathcal V}\, T^{ab} \nabla_{(a} \Psi_{b)}.
\end{equation}
Here $\frac{1}{c}u^a_\Sigma$ denotes the timelike future-directed unit vector field orthogonal to $\Sigma_{i}$ and $\Sigma_{f}$, and $n^{a}$ denotes the spacelike outward-directed unit vector field orthogonal to $\mathcal{B}$.

Roughly speaking, for a suitably-chosen spacelike vector field $\Psi^a$, this is a momentum conservation law that says that the difference in the matter three-momentum contained in the finite volumes $\Sigma_{i}$ and $\Sigma_{f}$ equals the flux of three-momentum that entered through the system boundary $\Delta\mathcal{B}$, plus a bulk $\Delta\mathcal{V}$ contribution arising when $\Psi^a$ is not a Killing vector field. We will be more precise later in this section.

To make the physical content of this law more transparent we decompose $T^{ab}$ as:
\begin{equation}\label{MatterSEM}
T^{ab} = \frac{1}{c^2}u^a u^b\mathbb{E}+2u^{(a}\mathbb{P}^{b)}-\mathbb{S}^{ab},
\end{equation}
where
\begin{align}
\mathbb{E} &= \frac{1}{c^2}u_a u_b T^{ab}=\frac{\rm Energy}{\rm Volume}=\frac{1}{8\pi}\left(E^2+B^2\right)\;\;{\rm (e.g.,\,electromagnetic\,energy\,density)}\nonumber\\
\mathbb{P}^a &= -\frac{1}{c^2}h^{a}_{\phantom{a}b}u_c T^{bc} = \frac{\rm Momentum}{\rm Volume}=\frac{1}{4\pi c}\epsilon^{a}_{\phantom{a}bc}E^b B^c\;\;{\rm (e.g.,\,Poynting\,vector\,over\,}c^2{\rm )}\label{Maxwell_stress}\\
\mathbb{S}^{ab} &= -h^{a}_{\phantom{a}c}h^{b}_{\phantom{b}d}T^{cd}=\frac{\rm Force}{\rm Area}=\frac{1}{4\pi}\left[E^a E^b+B^a B^b-\frac{1}{2}h^{ab}\left(E^2+B^2\right)\right]\,{\rm (e.g.,\, Maxwell\,stress)}, \nonumber
\end{align}
where $h^a_{\phantom{a}b}=g^a_{\phantom{a}b}+\frac{1}{c^2}u^a u_b$ is the projection operator into the vector space orthogonal to the worldlines of the congruence, and $\epsilon_{abc}=\frac{1}{c}u^d\epsilon_{dabc}$ is the corresponding volume form in this space. The last equality corresponds to the example of electromagnetism, which will be used later when we compare our results with that in reference \cite{Eriksen2006}. In this case, $E^a =  \frac{1}{c}F^{a b} u_b$ and $B^a = \frac{1}{2}\epsilon^{abc} F_{bc} $ are the proper electric and magnetic fields seen by the observers with four-velocity $u^a$ \cite{Wald1984}.

To get a momentum conservation law we set $\Psi^a = -\frac{1}{c}\Phi^a$, where $\Phi^a$ is orthogonal to (the worldlines of) the congruence. We then arbitrarily choose a time function on $\Delta{\mathcal V}$, i.e., a foliation of $\Delta{\mathcal V}$ by spacelike three-surfaces, $\Sigma_t$, of constant time parameter, $t$ (that coincide with $\Sigma_{i}$ and $\Sigma_{f}$ at times $t_i$ and $t_f$), and set $u^a=N^{-1}(\partial/\partial t)^a$, where $N$ is the lapse function. We naturally extend the definition of $u^a_\Sigma$ to all $\Sigma_t$ surfaces (as opposed to on just $\Sigma_{i}$ and $\Sigma_{f}$) as
\begin{equation}\label{definition_of_HSO_u}
u^a_\Sigma=\Gamma (u^a+V^a),
\end{equation}
where $V^a$ is orthogonal to the congruence, and $\Gamma=(1-V^2/c^2)^{-1/2}$ is a Lorentz factor. $V^a$ represents the three-velocity of fiducial observers who are `at rest' with respect to $\Sigma_t$ (whose hypersurface-orthogonal four-velocity is $u^a_\Sigma$) as measured by our congruence of observers (whose four-velocity is $u^a$). We will refer to these as the $u^a_\Sigma$- and $u^a$-observers, respectively. Note that while $u^a_\Sigma$ is hypersurface orthogonal, $u^a$ need not be, i.e., we are allowing for a twisting congruence. With these definitions, the conservation law in equation (\ref{integrated_local_conservation_law}) reads:
\begin{align}\label{explicit_integrated_local_conservation_law}
\int\limits_{\Sigma_f - \Sigma_i}  d\hat{\Sigma} \, & \left( \mathbb{P}^a + \frac{1}{c^2}\mathbb{S}^{ab}V_b\right)\Phi_a =
\int\limits_{\Delta\mathcal{B}} N \, dt \, d\hat{\mathcal{S}}\; \mathbb{S}^{ab} n_a \Phi_b \nonumber\\
& \quad \quad - \int\limits_{\Delta{\mathcal V}}  N \, dt \, d\hat{\Sigma}\,  \left[\frac{1}{c^2}\mathbb{E}\,a^a\Phi_a +\mathbb{P}^a(\Theta_a^{\phantom{a}b}+\nu^c\epsilon_{ca}^{\phantom{ca}b} )\Phi_b-\mathbb{P}^a u^b \nabla_b \Phi_a +\mathbb{S}^{ab}\hat{\nabla}_{(a}\Phi_{b)} \right].
\end{align}
Here $d\hat{\Sigma}=\Gamma\,d\Sigma$ is the proper three-volume element seen by the $u^a$-observers. It is constructed from $h_{ab}$, the `radar ranging' metric that measures the orthogonal distance between neighbouring worldlines of the congruence. Similarly, $d\hat{\mathcal{S}}$ is the proper two-surface element constructed from $\sigma_{ab}=h_{ab}-n_a n_b$, the `radar ranging' metric between neighbouring worldlines of the congruence restricted to $\mathcal{B}$. In expanding the term $T^{ab} \nabla_{(a} \Psi_{b)}$ in equation (\ref{integrated_local_conservation_law}) we made use of the following definitions associated with properties of the $u^a$-congruence: the observers' four-acceleration is defined as $a^a=u^b \nabla_b u^a$; the strain rate tensor (i.e., expansion and shear) of the congruence is defined as $\Theta_{ab}=h_{(a}^{\phantom{(a}c}h_{b)}^{\phantom{b)}d}\nabla_{c}u_{d}$, and the twist as $\nu_c=\frac{1}{2}\epsilon_{c}^{\phantom{c}ab}\nabla_{a}u_{b}$; and the derivative operator induced in the vector space orthogonal to the congruence is defined as $\hat{\nabla}_{a}\Phi_{b}=h_{a}^{\phantom{a}c}h_{b}^{\phantom{b}d}\nabla_{c}\Phi_{d}$.

The left-hand side of equation (\ref{explicit_integrated_local_conservation_law}) is the change (between $\Sigma_i$ and $\Sigma_f$) in the $\Phi^a$-component of the matter (e.g., electromagnetic) momentum in the $\Sigma_t$ system, as measured from the $u^a$-observers' frame. One might wonder why there is a Maxwell stress-like term in the integrand. Recall that the integrand started as $-\frac{1}{c^2}\,d{\Sigma}\,  T^{ab} u^\Sigma_a \Phi_b$, where $-\frac{1}{c^2} T^{ab} u^\Sigma_a$ is the matter four-momentum per unit volume as measured by the $u_\Sigma^a$-observers, who are co-moving with the $\Sigma_t$ system (i.e., their four-velocity is orthogonal to $\Sigma_t$). Multiplying by $d\Sigma$ gives the amount of matter four-momentum (again, as measured by the $u_\Sigma^a$-observers) contained in their proper volume element $d\Sigma$ of $\Sigma_t$. Contracting the resulting four-vector with $\Phi_b$ yields the $\Phi^a$-component of three-momentum as seen by the $u^a$-observers along {\it their} space axes. Finally, integrating over $\Sigma_t$ yields the total $\Phi^a$-momentum in the $\Sigma_t$ system, as measured by the $u^a$-observers at time $t$, who see the $\Sigma_t$ system as being in motion. So we are calculating the right thing. The $V^a$ in $u_\Sigma^a$ then results in the extra Maxwell stress-like term in the integrand.

Similarly, it is easy to see that instead of just $\mathbb{E}$ we will have $(\mathbb{E} - \mathbb{P}^a V_a)$ in the integrand when we choose $\Psi^a=\frac{1}{c}u^a$, and are dealing with an energy (instead of a three-momentum) conservation law (see Chapter \ref{chLocalQuasilocal}).  This {\it covariant} definition of matter four-momentum is the same as Rohrlich's 1960 definition of electromagnetic four-momentum \cite{Rohrlich1960}, which solved the infamous ``$4/3$ problem"\footnote{The ``4/3'' problem is a seemingly paradoxical result regarding the Abraham-Lorentz classical model of the electron - a shell of radius $a$ with charge $e$ uniformly spread out over the surface.  Calculating the rest mass from the self-energy of the Coulomb field, $\vec{E} = \frac{e}{r^2} \hat{r}$, yields $m := \frac{1}{8\pi c^2} \int^{\infty}_{a} d^3 x |E|^2 = \frac{e^2}{2 a c^2}$.  Meanwhile, if the electron has velocity $v$, then its momentum should be given by the integral of its Poynting field.  This gives $\vec{p} = \frac{1}{4\pi c^2} \int^{\infty}_a \vec{E} \times \vec{B} = \frac{2 e^2}{3 a c^2} \vec{v} = \frac{4}{3} m \vec{v}$ instead of what one should expect, $\vec{p} = m \vec{v}$.  This contradiction is the result of ignoring the change in the shape of the shell due to Lorentz contraction from the rest frame where the integration is performed.} by properly defining - with respect to relativistic effects - the momentum of the classical electron model (which was solved in essentially the same way by Fermi in 1922 \cite{Fermi1922}, but apparently forgotten). Compare Rohrlich's equation (17) with our equation (\ref{explicit_integrated_local_conservation_law}).\footnote{Alternatively, see section 16.5 in Jackson \cite{Jackson3rdEdition}, in particular his equation (16.44).} We are just seeing Rohrlich's electromagnetic four-momentum in special relativity generalized to arbitrary matter four-momentum in general relativity. To help clarify the parallel between our work here and Rohrlich's work on the electron, our $u^a$-observers (who will be moving {\it rigidly} - see next subsection) correspond to Rohrlich's {\it at rest} observers; they are observing a {\it moving} $\Sigma_t$ system that corresponds to Rohrlich's moving electron. Our $u_\Sigma^a$-observers correspond to the observers co-moving with Rohrlich's electron.

To understand the extra Maxwell stress-like term physically, consider for example a set of $u^a$-observers who see the $\Sigma_t$ system moving with velocity $V^a$ in the azimuthal direction (i.e., rotating relative to them), and choose $\Phi^a$ also in the azimuthal direction. Then the additional stress term is of the form $\frac{1}{c^2}\times {\rm Pressure}\times {\rm Velocity}$. In relativity theory, pressure makes a relativistic contribution to inertia; indeed, $\frac{1}{c^2}\times {\rm Pressure}$ has the dimensions of mass per unit volume. So matter (e.g., electromagnetic) pressure in a rotating system is equivalent to a rotating mass, which must contribute to the momentum ({\it angular} momentum in this example). Hence the $\frac{1}{c^2}\mathbb{S}^{ab}V_a\Phi_b$ term in equation (\ref{explicit_integrated_local_conservation_law}).

So far, our analysis has been completely general. Before discussing the terms on the right-hand side of equation (\ref{explicit_integrated_local_conservation_law}) it will be helpful to first simplify the equation by specializing it to the case of {\it rigid motion}, which, as we shall see, is ultimately necessary to achieve a physically sensible matter momentum conservation law for a finite-sized system.

\subsection{Specialization to Local Rigid Motion}\label{LocalRigid}

For reasons that will be made clear shortly, suppose that the $u^a$-observers are moving {\it rigidly}, i.e., that the orthogonal distance between the worldlines of all nearest neighbouring pairs of $u^a$-observers is constant in time. This is equivalent to the condition $\Theta_{ab}=0$, i.e., we have a congruence with zero expansion and shear. Since this represents six differential constraints on three functions (the three independent components of $u^a$), this cannot always be realized. (However, as we will discuss in {\S}\ref{QuasilocalRigidMotion}, the quasilocal analogue of this condition {\it can} always be realized.) In what follows, however, we will simply assume we are in a context in which this rigidity condition {\it is} satisfied. As a first consequence, we obviously have that the spatial integration measures $d\hat{\Sigma}$ and $d\hat{\mathcal{S}}$ in equation (\ref{explicit_integrated_local_conservation_law}) are time-independent.

Now let $\Upsilon^a$ denote any vector field orthogonal to the congruence, which the $u^a$-observers would consider to be purely spatial, i.e., to lie along their space axes. For this vector field to appear {\it stationary} to the rigidly-moving $u^a$-observers (i.e., not change with time), $\Upsilon^a$ must be Lie-dragged along the fibres, i.e., $\mathcal{L}_u \Upsilon^a \propto u^a$. Contracting both sides with $u_a$ reveals the proportionality factor, and we find that we require:
\begin{equation}\label{stationary_vectors}
\mathcal{L}_u \Upsilon^a = \frac{1}{c^2} (\Upsilon^b a_b) u^a.
\end{equation}
We will call spatial vector fields satisfying this condition {\it stationary}. In a context in which the rigidity condition holds, such stationary vector fields can be uniquely constructed throughout $\Delta\mathcal{V}$ given their specification on any one spatial three-surface, e.g., $\Sigma_i$.

Our conservation law is for the $\Phi^a$-component of the matter momentum. Obviously, we would certainly want $\Phi^a$ to be stationary, and will assume that such a choice has been made. Using equation (\ref{stationary_vectors}) with  $\Upsilon^a =\Phi^a$ we find that $-\mathbb{P}^a u^b \nabla_b \Phi_a=\epsilon_{abc}\nu^a\mathbb{P}^b\Phi^c$, and so two of the terms in the $\Delta\mathcal{V}$ integrand in equation (\ref{explicit_integrated_local_conservation_law}) can be combined into one:
\begin{equation}\label{Coriolis}
-\left[ \mathbb{P}^a\nu^c\epsilon_{ca}^{\phantom{ca}b}\Phi_b-\mathbb{P}^a u^b \nabla_b \Phi_a \right] = -2\epsilon_{abc}\nu^a\mathbb{P}^b\Phi^c.
\end{equation}
Recalling the usual vector formula for a Coriolis force, ${\bf F}_{\rm Cor}=-2\,{\bf \Omega}\times (m{\bf v})$, equation (\ref{Coriolis}) is clearly the $\Phi^a$-component of a matter Coriolis force density, which is associated with a change in the $\Phi^a$-component of the matter momentum as seen by our $u^a$-observers in the case that they are rotating (twisting congruence). Notice that it includes the correct factor of $-2$: half of the effect arises from the position-dependence of the relative velocity of points in the rotating frame (the first term on the left-hand side), and the other half arises from the rate at which the rotating frame coordinate axes change direction (the second term on the left-hand side). Relatedly, the $-\left[\frac{1}{c^2}\mathbb{E}\,a^a\Phi_a\right]$ term in equation (\ref{explicit_integrated_local_conservation_law}) includes the $\Phi^a$-component of the sum of the matter Euler (``$-m\,{\bf \dot{\Omega}}\times {\bf r}$") and centrifugal (``$-m\,{\bf \Omega}\times {\bf \Omega}\times {\bf r}$") force densities, written in a covariant form that does not involve a radial vector (``${\bf r}")$.

Finally, we consider the term $\mathbb{S}^{ab}\hat{\nabla}_{(a}\Phi_{b)}$ in equation (\ref{explicit_integrated_local_conservation_law}). If we introduce coordinates $x^I$, $I=1,\,2,\,3$, that label the worldlines of the congruence, then it is obvious that the rigidity condition $\Theta_{ab}=0$ is equivalent to $\dot{h}_{IJ}=0$, where $h_{IJ}$ are the spatial coordinate components of $h_{ab}$, and an over-dot denotes differentiation with respect to the parameter time, $t$. Assuming rigidity of the congruence, and stationarity of $\Phi^a$, a simple calculation shows that
\begin{equation}\label{KV_condition}
\hat{\nabla}_{(I}\Phi_{J)}=\frac{1}{2}\left(\Phi^K\partial_K h_{IJ}+2\,h_{K(I}\partial_{J)}\Phi^K\right),
\end{equation}
which are the only coordinate components of $\hat{\nabla}_{(a}\Phi_{b)}$ that do not identically vanish. Here $\partial_I$ denotes partial differentiation with respect to $x^I$. Recognizing the structure of a Lie derivative on the right-hand side of equation (\ref{KV_condition}), requiring $\hat{\nabla}_{(a}\Phi_{b)}=0$ is thus equivalent to $\Phi^a$ being a Killing vector field with respect to $h_{ab}$. This is a natural condition to impose on $\Phi^a$ and $h_{ab}$; if it is not satisfied, it is not clear how meaningful it is to say we are dealing with the ``$\Phi^a$-component of the matter momentum" (more on this below). Of course this condition is not realizable in general. (However, we will see in {\S}\ref{QuasilocalRigidMotion} that the quasilocal analogue of this condition - a certain {\it conformal} Killing vector condition, {\it can} always be realized.) Here we will simply assume we are in a context in which this Killing vector condition {\it is} satisfied.

To summarize this subsection, assuming {\it rigid} motion, and a stationary {\it Killing} vector field $\Phi^a$ (Killing vector with respect to $h_{ab}$, not $g_{ab}$), equation (\ref{explicit_integrated_local_conservation_law}) reduces to
\begin{align}\label{rigid_KV_explicit_integrated_local_conservation_law}
\int\limits_{\Sigma_f - \Sigma_i}  d\hat{\Sigma} \, \left( \mathbb{P}^a+\frac{1}{c^2}\mathbb{S}^{ab}V_b\right)\Phi_a & = 
\int\limits_{\Delta\mathcal{B}} N \, dt \, d\hat{\mathcal{S}}\; \mathbb{S}^{ab} n_a \Phi_b \nonumber \\
& - \int\limits_{\Delta{\mathcal V}}  N \, dt \, d\hat{\Sigma}\,  \left[\frac{1}{c^2}\mathbb{E}\,a^a\Phi_a + 2\epsilon_{abc}\nu^a\mathbb{P}^b\Phi^c\right],
\end{align}
where $d\hat{\Sigma}$ and $d\hat{\mathcal{S}}$ are time-independent. Within these assumptions, equation (\ref{rigid_KV_explicit_integrated_local_conservation_law}) is a completely general matter momentum conservation law in the context of general relativity. It says that the change (between $\Sigma_i$ and $\Sigma_f$) in the $\Phi^a$-component of the matter momentum contained in the system, as measured by the $u^a$-observers, is equal to the $\Phi^a$-component of the impulse imparted to the system by matter stresses acting on the system boundary over that time interval (the $\Delta\mathcal{B}$ integral), plus a correction due to the non-inertial motion (acceleration and rotation) of the rigid $u^a$-frame (the $\Delta\mathcal{V}$ integral).

Before we move on, there is an important subtlety in equation (\ref{rigid_KV_explicit_integrated_local_conservation_law}) worth pointing out. In the context of special relativity (and also Newtonian space-time), we sometimes come across the integral of a vector field - e.g., imagine equation (\ref{rigid_KV_explicit_integrated_local_conservation_law}) without the contraction of the integrands with $\Phi_a$. This happens, for instance, in special relativity when we calculate the total electromagnetic force acting on the electromagnetic sources and fields in a given volume of space by integrating the Maxwell stress tensor (contracted with $n_a$) over the surface of the volume, i.e., an integral of the form $\int d\hat{\mathcal{S}}\; \mathbb{S}^{ab} n_a$. Of course such an integral makes no sense in the context of general relativity, where we cannot add vectors with different base points. But even when we contract the integrand with $\Phi_a$, so it makes mathematical sense, it won't make any physical sense unless $\Phi^a$ has certain special properties. In the context of equation (\ref{rigid_KV_explicit_integrated_local_conservation_law}), $\Phi^a$ must somehow be uniquely determined throughout $\Delta\mathcal{V}$ by its value (and possibly a finite number of its derivatives) at a {\it single point} of $\Delta\mathcal{V}$, i.e., its degrees of freedom must be {\it discrete}, or `global', and in one-to-one correspondence with the degrees of freedom of the rigid frame itself. This, of course, is precisely the nature of a Killing vector field. In the present context, it is not difficult to show that the commutator of $h_{ab}$-compatible derivative operators acting on any {\it stationary} $\Phi^a$ depends on $\Phi^a$ at only a single point, i.e.,
\begin{equation}\label{Nature_of_Local_KV}
\hat{\nabla}_a\hat{\nabla}_b\Phi_c - \hat{\nabla}_b\hat{\nabla}_a\Phi_c = \hat{R}_{abc}^{\phantom{abc}d}\Phi_d,
\end{equation}
where $\hat{R}_{abc}^{\phantom{abc}d}$ is the Riemann tensor of $h_{ab}$, i.e., the curvature of the quotient space of our geometrically rigid bundle of worldlines. If we further demand that $\Phi^a$ be a Killing vector field with respect to $h_{ab}$ and define the antisymmetric object $A_{ab}=\hat{\nabla}_{[a}\Phi_{b]}$ it is not difficult to show that, for any vector field $X^a$ orthogonal to the congruence,
\begin{align}
X^a\hat{\nabla}_a\Phi_b &= X^a A_{ab} \label{IntegratingKV1}\\
X^a\hat{\nabla}_a A_{bc} &= -\hat{R}_{bca}^{\phantom{bca}d} X^a \Phi_d \label{IntegratingKV2}
\end{align}
in analogy with equations (C.3.7-8) in reference \cite{Wald1984}. Together with the stationarity condition on $\Phi^a$, this means $\Phi^a$ is uniquely determined throughout $\Delta\mathcal{V}$ by its value, and its antisymmetrized derivative, at a single point of $\Delta\mathcal{V}$.\footnote{In the case that $u^a$ is not hypersurface orthogonal one might worry about the integrability of equations (\ref{IntegratingKV1}) and (\ref{IntegratingKV2}), i.e., their compatibility with the stationarity condition. However, a short calculation reveals that they are, in fact, compatible. If we start at a given point and integrate $\Phi^a$ to two different points on the same neighbouring fibre (by following two different paths), the resulting $\Phi^a$ will satisfy equation (\ref{stationary_vectors}).} In the case of our three-parameter family of worldlines, this integration data corresponds to six discrete degrees of freedom, resulting in six linearly independent Killing vector fields (three translational and three rotational) corresponding to six components of momentum (three linear and three angular) that can be analyzed in equation (\ref{rigid_KV_explicit_integrated_local_conservation_law}). So requiring that $\Phi^a$ be a Killing vector field with respect to $h_{ab}$ is not only ``natural" (word used earlier), it is {\it crucial} for equation (\ref{rigid_KV_explicit_integrated_local_conservation_law}) to make any physical sense (which, in turn, is predicated on the rigidity condition being satisfied). Indeed, $h_{ab}$ must admit the {\it maximal} number of Killing vector fields. The fact that this condition generically {\it cannot} be satisfied is a serious problem, not to mention the fact that the rigidity condition, too, generically {\it cannot} be satisfied. We emphasize this point because (as mentioned earlier), both the Killing (actually, {\it conformal} Killing) vector and rigidity conditions can always be realized in the quasilocal context, as we will discuss in {\S}\ref{QuasilocalRigidMotion}.

\subsection{Specialization to Local Archimedes' Law}\label{LocalArchimedes}

We now turn to formulating an Archimedes' law in general relativity.  Typically, Archimedes' law deals with hydrostatics and states that at equilibrium the weight of an object is supported by the net buoyant force acting on that object.  In the context of general relativity, we thus think of Archimedes' law as the statement: the ``weight'' of a region containing some field is supported by the stresses acting on the boundary of that region.  Since Archimedes' law is usually envisioned in a static, or at most a stationary context, which is certainly the simplest case, that is the one we will explore here. But at this point, the rigid frame defined by the $u^a$-observers may still be undergoing time-dependent acceleration and/or rotation. So to formulate an Archimedes' law, it is natural to further assume that we are in a context in which $a^a$ and $\nu^a$ (which are both orthogonal to the congruence) are stationary. It is easy to show that the spatial coordinate components of $a_a$ and $\nu_a$ are:
\begin{align}
a_I &=\frac{1}{N}\dot{u}_I+c^2\partial_I\ln N\label{local_acceleration}\\
\nu_I &=\frac{1}{2}\epsilon_{I}^{\phantom{I}JK}\left(\partial_J u_K-\frac{1}{c^2}a_J u_K\right),\label{local_twist}
\end{align}
and the time coordinate components identically vanish. Assuming rigidity of the $u^a$-frame, stationarity of $a^a$ and $\nu^a$ is equivalent to  $\dot{a}_I=0$ and $\dot{\nu}_I=0$. Naturally, we assume that $V^a$ - the three-velocity of the $u^a_\Sigma$-observers relative to the $u^a$-frame - is also stationary, i.e., $\dot{V}_I=0$. This is equivalent to $\dot{u}_I=0$ since equation (\ref{definition_of_HSO_u}) implies $u_I=-V_I$. With $\dot{u}_I=0$, the acceleration $a_I =c^2\partial_I\ln N$ is a pure gradient. Demanding stationary acceleration then implies both a time-independent lapse function (so the {\it full} integration measures in equation (\ref{rigid_KV_explicit_integrated_local_conservation_law}) are time-independent) and a stationary twist.

Having imposed conditions on the congruence of observers (rigidity, plus stationary acceleration and twist), and specified the $\Phi^a$ the observers will use (a stationary $h_{ab}$ Killing vector field), we have established a suitable, stationary framework in which the measurements of the matter (e.g., electromagnetic) field will be made. The final step is to assume that the matter field itself is stationary, which means $\mathbb{E}$, $\mathbb{P}^a$, and $\mathbb{S}^{ab}$ in equation (\ref{Maxwell_stress}) are stationary. The left-hand side of equation (\ref{rigid_KV_explicit_integrated_local_conservation_law}) then vanishes, and we are left with
\begin{equation}\label{local_Archimedes}
\int\limits_{\Delta\mathcal{B}} N \, dt \, d\hat{\mathcal{S}}\; \mathbb{S}^{ab} n_a \Phi_b=
\int\limits_{\Delta{\mathcal V}}  N \, dt \, d\hat{\Sigma}\,  \left[\frac{1}{c^2}\mathbb{E}\,a^a\Phi_a + 2\epsilon_{abc}\nu^a\mathbb{P}^b\Phi^c \right].
\end{equation}
This is an Archimedes' law for a general matter field in the context of general relativity, derived following the standard {\it local} conservation law approach. It basically says that the weight of the matter field in a non-inertial reference frame (right-hand side) is supported by a Maxwell stress-like buoyant force acting on the field inside through the boundary of the system (left-hand side).

In the case that the matter field is electromagnetism, this result is essentially identical to the main result, equation (6.15), in Eriksen and Gr{\o}n's beautiful paper \cite{Eriksen2006}, except for three key differences: (1) We have generalized the result from {\it static} to {\it stationary} contexts by allowing for stationary rotation (the addition of the Coriolis term). (2) We do not incorporate the lapse function, $N$, into the definition of the (electromagnetic) stress, energy, and momentum densities. For example, Eriksen and Gr{\o}n's definition of the Maxwell stress ($t_{ij}$ in their equation (6.12)) is multiplied by $N$ ($g_0 x$ in their notation), compared to the standard definition in our equation (\ref{Maxwell_stress}). We prefer to show the lapse function explicitly to highlight an important aspect of the actual {\it mechanism} of the buoyancy. In {\S}\ref{Example} we will exhibit a simple example in which $\int d\hat{\mathcal{S}}\; \mathbb{S}^{ab} n_a \Phi_b=0$, i.e., the Maxwell stress on one side of the surface balances that on other side, and so one might expect zero net buoyancy. That the net effect is {\it not} zero follows from the fact that the acceleration induces an inhomogeneous time dilation (encoded in $N$), so observers on one side of the surface experience the {\it same} (magnitude of) proper force, but for a {\it longer} proper time, than on the other side, resulting in a net buoyant impulse acting on the system. Similar consequences of an inhomogeneous time dilation were noticed in the previous chapter. (3) Eriksen and Gr{\o}n work in the context of Rindler observers in special relativity. Here we work more generally in the context of accelerating and/or rotating rigid frames with a local approach to general relativity. This is not to say, however, that equation (\ref{local_Archimedes}) (or, at a more general level, (\ref{rigid_KV_explicit_integrated_local_conservation_law})) is describing what is {\it really} happening in the general relativistic context. We believe it is not, because it does not properly account for gravitational effects. To do so requires not a local, but a quasilocal approach, which we turn to presently.

\section{Quasilocal Momentum Conservation Law in General Relativity}\label{secQuasilocal2}

In this section we will use a quasilocal approach to construct an integrated momentum conservation law for both matter and gravitational fields contained in a finite volume of space.   Recall that the integrated local matter momentum conservation law in equation (\ref{rigid_KV_explicit_integrated_local_conservation_law}) required two key conditions to make it physically sensible: a rigid motion condition and a Killing vector condition, neither of which can be satisfied in general. Here we show how these two serious obstacles, inherent in the local approach to constructing a momentum conservation law, are overcome in the quasilocal approach using RQFs, yielding a conservation law of general validity.  The essence of the solution is the proper inclusion of gravitational effects.

\subsection{General Analysis}\label{QuasilocalMomentumConservationLaw}

As was seen in Chapters \ref{chNova} and \ref{chLocalQuasilocal}, we begin with an identity analogous to equation (\ref{differential_local_conservation_law}), except constructed in the three-dimensional Lorentzian manifold, $\mathcal{B}$, the boundary of the bundle of worldlines defined in {\S}\ref{Local-1}, whose three-metric is $\gamma_{ab}=g_{ab}-n_a n_b$, with associated derivative operator $D_a$:
\begin{equation}\label{differential_quasilocal_conservation_law}
D_a (T_\mathcal{B}^{ab}\psi_b )=(D_a T_\mathcal{B}^{ab})\psi_b + T_\mathcal{B}^{ab}D_{(a}\psi_{b)}.
\end{equation}
Recall that $\psi^a$ is an arbitrary vector field tangent to $\mathcal{B}$, and the local matter stress-energy-momentum tensor, $T^{ab}$, has been replaced with the quasilocal total (matter plus gravity) stress-energy-momentum tensor defined by Brown and York \cite{BY1993} as $T_\mathcal{B}^{ab}=-\frac{1}{\kappa}\Pi^{ab}$. The quasilocal analogue of integrating equation (\ref{differential_local_conservation_law}) over $\Delta\mathcal{V}$ is integrating equation (\ref{differential_quasilocal_conservation_law}) over $\Delta\mathcal{B}$. Denoting the boundaries of $\Delta\mathcal{B}$ as $\mathcal{S}_i$ and $\mathcal{S}_f$, the initial and final spacelike two-surfaces where $\Sigma_i$ and $\Sigma_f$ intersect $\mathcal{B}$, this integration yields (refer to figure \ref{QuasilocalWorldtube}):
\begin{equation} \label{integrated_quasilocal_conservation_law}
\frac{1}{c} \int\limits_{\mathcal{S}_f - \mathcal{S}_i}  d{\mathcal{S}}\,  T_{\mathcal B}^{ab} u^{\mathcal{S}}_a \psi_b = \int\limits_{\Delta\mathcal{B}}  d \mathcal{B} \, \left[ \frac{1}{\kappa} G^{ab} n_a \psi_b - T_{\mathcal B}^{ab} D_{(a} \psi_{b)} \right].
\end{equation}
Here $\frac{1}{c}u_{\mathcal{S}}^a$ denotes the timelike future-directed unit vector field tangent to $\mathcal{B}$ and orthogonal to $\mathcal{S}_i$ and $\mathcal{S}_f$, and we used the Gauss-Codazzi identity, $D_a \Pi^{ab}=n_a G^{ab}$, where $G^{ab}$ is the Einstein tensor associated with $g_{ab}$.

Note that this conservation law is a purely geometrical identity relating the intrinsic and extrinsic geometry of $\mathcal{B}$ (through $D_a$ and $T_\mathcal{B}^{ab}$, respectively) to the geometry of the embedding space, $\mathcal{M}$ (through $G^{ab}$). The matter stress-energy-momentum tensor, initially absent in equation (\ref{integrated_quasilocal_conservation_law}), will enter once Einstein's equation is invoked, in which case the first term on the right-hand side becomes $\frac{1}{\kappa} G^{ab} n_a \psi_b=T^{ab} n_a \psi_b$. This term  looks very much like $T^{ab} n_a \Psi_b$, the matter stress term in the local conservation law in equation (\ref{integrated_local_conservation_law}). However, there are two important differences. First, the origins of these two terms are completely different. Unlike $T^{ab} n_a \Psi_b$, $T^{ab} n_a \psi_b$ does {\it not} come from integrating a divergence. It comes from the fact that $D_a T_\mathcal{B}^{ab} \neq 0$ (in contrast to $\nabla_a T^{ab}=0$), and so is analogous to the Lorentz force density in electrodynamics (except that, unlike in electrodynamics, where the Lorentz force density acts only on the sources of the electromagnetic field, $D_a T_\mathcal{B}^{ab}$ acts on {\it all} of the fields in the system, matter and gravitational). The second important difference is that $\psi^a$ in $T^{ab} n_a \psi_b$ is {\it tangent} to $\mathcal{B}$, whereas $\Psi^a$ in $T^{ab} n_a \Psi_b$ need not be - it can (and often does) have a component in the normal direction, $n^a$. To see the significance of this, imagine a system with a two-sphere boundary, and we are interested in the `vertical', or `$Z^a$' component of the external matter force acting on the system (we will be more precise later). While there is no problem setting $\Psi^a=Z^a$ in the local conservation law, in the quasilocal law $\psi^a$ can accommodate only the {\it tangential} component of $Z^a$, not the normal component. So the quasilocal law seems to be missing the normal (pressure) contribution to the external matter force, $T^{ab} n_a n_b$. However, through an application of the `radial' Hamiltonian constraint of general relativity, we will show that this ``missing'' normal matter force is found in the $T_{\mathcal B}^{ab} D_{(a} \psi_{b)}$ term in equation (\ref{integrated_quasilocal_conservation_law}).

Before proceeding further with the analysis of equation (\ref{integrated_quasilocal_conservation_law}), let us first recall from equation ((\ref{SurfaceSEMcomponents}) a more transparent notation for $T_{\mathcal B}^{ab}$ which follows that introduced by Brown and York \cite{BY1993}.  In particular, we resolve the quasilocal stress-energy-momentum tensor into components adapted to the $u^a$-observers:
\begin{equation}
T_\mathcal{B}^{ab} = \frac{1}{c^2}u^a u^b\mathcal{E}+2u^{(a}\mathcal{P}^{b)}-\mathcal{S}^{ab},
\end{equation}
where $\mathcal{E}$, $\mathcal{P}^a$, and $\mathcal{S}^{ab}$ are the quasilocal energy, momentum, and stress with units of energy per unit area, momentum per unit area, and force per unit length.  These equations are exactly analogous to equations (\ref{MatterSEM}) and (\ref{Maxwell_stress}), except $\mathcal{E}$, $\mathcal{P}^a$, and $\mathcal{S}^{ab}$ refer to matter {\it and} gravity, whereas $\mathbb{E}$, $\mathbb{P}^a$, and $\mathbb{S}^{ab}$ refer to matter only (e.g., electromagnetism).\footnote{We remind the reader that our sign convention for the quasilocal stress is opposite to that of Brown and York \cite{BY1993} so as to align ourselves instead with the sign conventions in electromagnetism.}

Proceeding in analogy with the local case, to get a quasilocal momentum conservation law we set $\psi^a = -\frac{1}{c}\phi^a$, where $\phi^a$ is tangent to $\mathcal{B}$ and orthogonal to $u^a$. Letting $\mathcal{S}_t$ denote the intersection of $\Sigma_t$ with $\Delta{\mathcal B}$, we inherit from our local analysis an arbitrary time function on $\Delta{\mathcal B}$ (i.e., a foliation of $\Delta{\mathcal B}$ by spacelike two-surfaces, $\mathcal{S}_t$, which we assume are topologically two-spheres), and set $u^a=N^{-1}(\partial/\partial t)^a$ as before, where $N$ is the lapse function. In analogy to equation (\ref{definition_of_HSO_u}), we extend the definition of $u^a_\mathcal{S}$ to all $\mathcal{S}_t$ surfaces as
\begin{equation}\label{definition_of_Boundary_HSO_u}
u^a_\mathcal{S}=\gamma (u^a+v^a),
\end{equation}
where $v^a$ is tangent to $\mathcal{B}$ and orthogonal to $u^a$, and $\gamma=(1-v^2/c^2)^{-1/2}$ is a Lorentz factor. Here $v^a$ represents the spatial two-velocity of fiducial observers who are `at rest' with respect to $\mathcal{S}_t$ (whose hypersurface-orthogonal four-velocity is $u^a_\mathcal{S}$) as measured by our congruence of observers (whose four-velocity is $u^a$). Note that, while $u^a_\mathcal{S}$ is hypersurface orthogonal, $u^a$ need not be. (Also note that, unlike $u_{\Sigma}^a$ appearing in equation (\ref{definition_of_HSO_u}), $u_{\mathcal{S}}^a$ is independent of the choice of $\Sigma_t$ in the interior. In the quasilocal approach we are completely decoupled from the interior.) With these definitions, the conservation law in equation (\ref{integrated_quasilocal_conservation_law}) becomes:
\begin{align}\label{explicit_integrated_quasilocal_conservation_law}
\int\limits_{\mathcal{S}_f - \mathcal{S}_i}  d\hat{\mathcal{S}} \, & \left( \mathcal{P}^a+\frac{1}{c^2}\mathcal{S}^{ab}v_b\right)\phi_a  = -\int\limits_{\Delta\mathcal{B}} N \, dt \, d\hat{\mathcal{S}}\; T^{ab} n_a \phi_b \nonumber\\
& \qquad - \int\limits_{\Delta{\mathcal B}}  N \, dt \, d\hat{\mathcal{S}}\,  \left[\frac{1}{c^2}\mathcal{E}\,\alpha^a\phi_a +\mathcal{P}^a(\theta_a^{\phantom{a}b}+\nu\epsilon_{a}^{\phantom{a}b} )\phi_b-\mathcal{P}^a u^b D_b \phi_a +\mathcal{S}^{ab}\hat{D}_{(a}\phi_{b)} \right].
\end{align}
Analogous to $d\hat{\Sigma}=\Gamma\,d\Sigma$ in the local case, we have $d\hat{\mathcal{S}}=\gamma\,d\mathcal{S}$, the proper two-surface element seen by the $u^a$-observers on $\mathcal{B}$. In expanding the term $T_{\mathcal B}^{ab} D_{(a} \psi_{b)}$ in equation (\ref{integrated_quasilocal_conservation_law}) we made use of the following definitions associated with properties of the $u^a$-congruence: the component of the observers' four-acceleration tangent to $\mathcal{B}$ is defined as $\alpha^a=\sigma^a_{\phantom{a}b} a^b$; the strain rate tensor (i.e., expansion and shear) of the congruence is defined as $\theta_{ab}=\sigma_{(a}^{\phantom{(a}c}\sigma_{b)}^{\phantom{b)}d} D_{c}u_{d}$, and the twist as $\nu=\frac{1}{2}\epsilon^{ab}D_{a}u_{b}$; and the derivative operator induced in the vector space tangent to $\mathcal{B}$ and orthogonal to the congruence is defined as $\hat{D}_{a}\phi_{b}=\sigma_{a}^{\phantom{a}c}\sigma_{b}^{\phantom{b}d}D_{c}\phi_{d}$.

Comparing this quasilocal momentum conservation law  with the local one in equation (\ref{explicit_integrated_local_conservation_law}), we see that the left-hand side of both equations gives the change, between times $t_i$ and $t_f$, of the momentum contained in the system as measured by the $u^a$-observers, including the stress term required for relativistic covariance as discussed at the end of {\S}\ref{Local-1}. The two key differences are: (1) the momentum density in equation (\ref{explicit_integrated_quasilocal_conservation_law}) is {\it quasilocal} - a momentum per unit area, versus per unit volume, and is meaningless unless it is integrated over the entire closed two-surface bounding a given volume; and (2) the quasilocal momentum density includes {\it all} contributions to the momentum in the system - matter plus gravity, versus matter only. Note that $\phi^a$ in $\mathcal{P}^a\phi_a$ is {\it tangent} to $\mathcal{B}$, whereas $\Phi^a$ in $\mathbb{P}^a\Phi_a$ need not be, so one might wonder if the quasilocal momentum density is missing a `normal' contribution to the momentum, similar to the missing normal component in $T^{ab}n_a\phi_b$ noted above. The answer is no. Unlike $T^{ab}$, $\mathcal{P}^a$ is {\it inherently quasilocal} in the sense that, to use the example introduced earlier, $\mathbb{P}^a Z_a$ corresponds to $\mathcal{P}^a\phi_a$ when $\phi^a$ is a conformal Killing vector on the two-sphere boundary representing a boost in the $Z^a$ direction. (We will be more precise below.)

As in the local approach, we will now simplify the right-hand side of equation (\ref{explicit_integrated_quasilocal_conservation_law}) by specializing to the case of a reference frame in rigid motion, which admits a maximal set of spatial {\it conformal} Killing vector fields, $\phi^a$. Unlike in the local approach, this is {\it always} possible in the quasilocal approach.

\subsection{Specialization to Quasilocal Rigid Motion}\label{QuasilocalRigidMotion}

As we argued in Chapter \ref{chNova}, in a generic spacetime it is always possible to construct a {\it rigid quasilocal frame}.  Constructing such a frame is equivalent to the condition $\theta_{ab}=0$, i.e., a congruence with zero expansion and shear. Unlike the analogous condition in the local approach, this condition represents only {\it three} differential constraints on three functions (the three independent components of $u^a$), and can always be realized. In fact, as we first saw in {\S} \ref{ArbitraryPerturbations}, the degrees of freedom remaining are precisely those of rigid frames in Newtonian space-time: three linear and three angular velocities, each with arbitrary time dependence. Let us assume that our $u^a$-observers comprise such an RQF. As a first consequence, $d\hat{\mathcal{S}}$ in equation (\ref{explicit_integrated_quasilocal_conservation_law}) is time-independent. Secondly, we are always able to construct spatial vector fields, $\upsilon^a$, in $\mathcal{B}$ (vector fields tangent to $\mathcal{B}$ and orthogonal to $u^a$) that are {\it stationary}: in analogy to equation (\ref{stationary_vectors}), they satisfy $\mathcal{L}_u \upsilon^a = \frac{1}{c^2} (\upsilon^b \alpha_b) u^a$. Naturally, we will choose $\phi^a$ to be such a stationary spatial vector field, which is uniquely determined everywhere on $\mathcal{B}$ given its specification on any one two-surface, e.g., $\mathcal{S}_i$. As with equation (\ref{Coriolis}), two of the terms in equation (\ref{explicit_integrated_quasilocal_conservation_law}) then combine into one term:
\begin{equation}\label{Quasilocal_Coriolis}
-\left[ \mathcal{P}^a\nu\epsilon_{a}^{\phantom{a}b}\phi_b-\mathcal{P}^a u^b D_b \phi_a \right] = -2\nu \epsilon_{ab}\mathcal{P}^a\phi^b,
\end{equation}
which is the $\phi^a$-component of the {\it quasilocal} Coriolis force density of both the matter and gravitational fields in the system. Relatedly, the $-\left[\frac{1}{c^2}\mathcal{E}\alpha^a\phi_a\right]$ term in equation (\ref{explicit_integrated_quasilocal_conservation_law}) includes the quasilocal versions of Euler and centrifugal force densities, again for both the matter and gravitational fields in the system.

Next, we consider the term $\mathcal{S}^{ab}\hat{D}_{(a}\phi_{b)}$ in equation (\ref{explicit_integrated_quasilocal_conservation_law}).  Introducing coordinates $x^i$, $i=1,\,2$, that label the worldlines of the congruence, the rigidity condition $\theta_{ab}=0$ is equivalent to $\dot{\sigma}_{ij}=0$, where $\sigma_{ij}$ are the spatial coordinate components of $\sigma_{ab}$. Assuming rigidity of the congruence, and stationarity of $\phi^a$, we have, in analogy to equation (\ref{KV_condition}),
\begin{equation}\label{quasilocal_KV_condition}
\hat{D}_{(i}\phi_{j)}=\frac{1}{2}\left(\phi^k\partial_k \sigma_{ij}+2\,\sigma_{k(i}\partial_{j)}\phi^k\right),
\end{equation}
which are the only coordinate components of $\hat{D}_{(a}\phi_{b)}$ that do not identically vanish. Here $\partial_i$ denotes partial differentiation with respect to $x^i$. Clearly, requiring $\hat{D}_{(a}\phi_{b)}=0$ is equivalent to $\phi^a$ being a Killing vector field of the `radar ranging' metric $\sigma_{ab}$. As discussed at the end of {\S}\ref{LocalRigid}, in the local approach we demanded that $\Phi^a$ be a Killing vector of the `radar ranging' metric $h_{ab}$ (i.e., $\hat{\nabla}_{(a}\Phi_{b)}=0$). There, this was the right thing to do, since a three-dimensional space admits at most six linearly independent Killing vectors, corresponding to three linear and three angular components of momentum. The problem was that such Killing vectors do not always exist. In the quasilocal approach, on the other hand, this would {\it not} be the right thing to do, since a closed two-dimensional space admits at most three Killing vectors, corresponding essentially to three components of angular momentum - we would be missing the the three components of linear momentum. We need a weaker condition.

The natural and appropriate condition turns out to be the {\it conformal} Killing vector (CKV) condition:
\begin{equation}\label{quasilocal_CKV_condition}
\hat{D}_{(a}\phi_{b)}=\frac{1}{2}\sigma_{ab}\hat{D}_c\phi^c.
\end{equation}
This condition represents two differential constraints on two functions of two variables, $\phi^i (x^j )$. Assuming the topology of $\mathcal{B}$ is $\mathbb{R}\times\mathbb{S}^2$ (as indicated earlier), then it is well-known that equation (\ref{quasilocal_CKV_condition}) admits precisely six linearly independent CKV solutions, representing the action of the Lorentz group on the two-sphere (three boosts and three rotations). This is true regardless of $\sigma_{ab}$ - the geometry of the quotient space of the rigid bundle, i.e., the size and shape of the topological two-sphere boundary of the system. It is this fact that gives RQFs the same six degrees of freedom as a rigid frame in Newtonian space-time. If we choose $\phi^a$ to be a boost (respectively, rotation) generator we are dealing with a linear (respectively, angular) momentum conservation equation.

Summarizing, in the quasilocal approach it is always possible to choose the $u^a$-observers to be in {\it rigid} motion (i.e., to comprise an RQF), and to choose a spatial vector field, $\phi^a$, that is a stationary conformal Killing vector field (CKV with respect to $\sigma_{ab}$, not $\gamma_{ab}$). Thus, equation (\ref{explicit_integrated_quasilocal_conservation_law}) can always be reduced to the form:
\begin{equation}\label{reduced_integrated_quasilocal_conservation_law}
\int\limits_{\mathcal{S}_f - \mathcal{S}_i} \hspace{-5pt} d\hat{\mathcal{S}} \, \left( \mathcal{P}^a+\frac{1}{c^2}\mathcal{S}^{ab}v_b\right)\phi_a =
\int\limits_{\Delta\mathcal{B}} \hspace{-5pt} N \, dt \, d\hat{\mathcal{S}} \left\{ \mathbb{S}^{ab} n_a \phi_b
- \left[\frac{1}{c^2}\mathcal{E}\,\alpha^a\phi_a +2\nu\epsilon_{ab}\mathcal{P}^a\phi^b + {\rm P}\,\hat{D}_a\phi^a \right]\right\},
\end{equation}
where ${\rm P}=\frac{1}{2}\sigma_{ab}\mathcal{S}^{ab}$ is the quasilocal {\it pressure} (force per unit {\it length}) between the worldlines of $\mathcal{B}$. Its physical interpretation will be discussed in the next subsection. Note that since $\phi^a$ is orthogonal to $u^a$ we may replace $-T^{ab} n_a \phi_b$ in equation (\ref{explicit_integrated_quasilocal_conservation_law}) with $+\mathbb{S}^{ab} n_a \phi_b$, which emphasizes that this is a purely spatial stress term. This completely general RQF momentum conservation law for matter {\it and} gravitational fields is to be compared with the less general (valid only in special spacetimes that admit both a rigid three-parameter congruence and a spatial Killing vector field orthogonal to the congruence) and incomplete (includes matter fields only, not gravitational) local momentum conservation law in equation (\ref{rigid_KV_explicit_integrated_local_conservation_law}).

There are two key differences between equations (\ref{reduced_integrated_quasilocal_conservation_law}) and (\ref{rigid_KV_explicit_integrated_local_conservation_law}). The first is the obvious shift from integrations over volume densities to integrations over surface densities.  To understand the physical significance of this shift, let us return to the example at the end of {\S} \ref{Conclusions3} where we imagined being inside an accelerating box in flat spacetime that contains a freely-floating, massive object.  The object appears to accelerate toward us but only because the object's momentum (relative to us) changes due to the acceleration of our frame.  A quick calculation reveals that if the instantaneous proper acceleration of our frame is $a$, the change in momentum during an infinitesimal proper time interval $\Delta\tau$ is given by $\Delta p=-\frac{1}{c^2}E a \,\Delta\tau$, where $E$ is the instantaneous relativistic energy of the object (and the minus sign reflects the fact that $\Delta p$ and $a$ are in opposite directions). This explains the bulk term, $-\frac{1}{c^2}(\mathbb{E}\,d\hat{\Sigma})\,(a^a\Phi_a)\,(N\,dt)$, in the local law, equation (\ref{rigid_KV_explicit_integrated_local_conservation_law}). But wait, momentum is a conserved quantity and, by the equivalence principle, we have no way of telling if the box is accelerating versus the whole system being immersed in a uniform gravitational field.  In the latter case, the ``force'' of gravity acts over time and this impulse is what gives the object momentum.  This represents a transfer of the momentum of {\it something} to the momentum of the object. If it was an electromagnetic force we would say that the ``something" is the electromagnetic field (and ultimately, the source of that field). Since the ``force" is gravitational, the ``something" must be the gravitational field (and ultimately, the source of that field). There must be some kind of surface flux representing {\it gravitational} momentum entering the box from the outside. According to the general relativistic equation (\ref{reduced_integrated_quasilocal_conservation_law}), that gravitational momentum flux is $-\frac{1}{c^2}\mathcal{E}\,\alpha^a\phi_a$, which has dimensions of momentum per unit area per unit time. So the ``mass times acceleration" bulk term, $-\frac{1}{c^2}\mathbb{E}\,a^a\Phi_a$, in the local conservation law has become a bona fide gravitational momentum flux term, $-\frac{1}{c^2}\mathcal{E}\,\alpha^a\phi_a$, in the quasilocal conservation law, exactly as anticipated in our equivalence principle argument.

In the context of general relativity, the presence of mass-energy (matter or gravitational) inside a system causes a change in the spatial trace of the extrinsic curvature, $k=\sigma^{ab}K_{ab}$, of the two-sphere the RQF observers reside on. Since $\mathcal{E}=-k/\kappa$, measuring this change in extrinsic curvature (using a ruler) is operationally how the RQF observers measure the mass-energy in the system. For an everyday mass, $m$, and areal radius of the RQF, $r$, the magnitude of this change is exceedingly tiny, of order $Gm/c^2 r^2$ (as one can see by dimensional analysis). Nevertheless, multiplying this usually tiny general relativistic effect by the large number $c^2/G$ (and then integrating over the sphere) converts it into what we understand in classical mechanics as the mass of the object inside our frame of reference. This everyday mass, times an everyday acceleration, integrated over an everyday time interval, then gives an everyday change in momentum. But it is important to appreciate that the mass of an object times the acceleration of the frame in classical mechanics really represents a general relativistic transfer of gravitational momentum through the boundary of our reference frame, effected by a coupling between extrinsic curvature and acceleration. In other words, we claim that the simple expression, $-\frac{1}{c^2}\mathcal{E}\,\alpha^a\phi_a$, is actually the exact operational definition of gravitational momentum flux in general relativity (and similar comments apply to the $-2\nu\epsilon_{ab}\mathcal{P}^a\phi^b$ expression). So the RQF momentum conservation law in equation (\ref{reduced_integrated_quasilocal_conservation_law}) leads us to a deeper understanding of physics: it explains in detail what is actually happening with regards to momentum transfer when, say, an apple falls. We discovered an exactly analogous result for energy transfer in Chapter \ref{chLocalQuasilocal}, except in place of a tiny change in extrinsic curvature we had a tiny frame dragging effect. But both are {\it general relativistic} effects.

The second key difference between the local and quasilocal momentum conservation laws is the extra ${\rm P}\,\mathcal{D}_a\phi^a$ term on the right-hand side of equation (\ref{reduced_integrated_quasilocal_conservation_law}). In the next subsection we will show that it is this term that contains, in a subtle and interesting way, the missing normal ($-T^{ab} n_a n_b=+\mathbb{S}^{ab} n_a n_b$ pressure) contribution to the external matter force acting on the system, as alluded to earlier. (Remember that in equation (\ref{reduced_integrated_quasilocal_conservation_law}), $-T^{ab}n_a\phi_b$ was replaced with $+\mathbb{S}^{ab}n_a\phi_b$, and so we are comparing $\mathbb{S}^{ab}n_a\Psi_b$ in the local law with $\mathbb{S}^{ab}n_a\psi_b$ in the quasilocal law. The {\it difference}, we claim, is contained in the extra ${\rm P}\,\mathcal{D}_a\phi^a$ term.)

\subsection{Role and Interpretation of the Quasilocal Pressure} \label{Interpretation}

General relativity imposes four constraint equations that intertwine the intrinsic and extrinsic geometry of $\mathcal{B}$ with matter. The RQF momentum conservation law in equation (\ref{reduced_integrated_quasilocal_conservation_law}) represents two of these constraint equations. The RQF energy conservation law (see equation (\ref{eq:SimpleConservationEq})) represents the third. The fourth constraint equation - the `radial' Hamiltonian constraint, originates in the geometrical identity:
\begin{equation}\label{RadialHamiltonianIdentity}
-2\,G_{ab}n^a n^b={}^\mathcal{B}\!R +\Pi_{ab}\Pi^{ab}-\frac{1}{2}\Pi^2,
\end{equation}
where ${}^\mathcal{B}\!R$ is the Ricci scalar of the intrinsic geometry of $\mathcal{B}$, $\Pi_{ab}=K_{ab}-K\gamma_{ab}$ (defined earlier) represents the extrinsic geometry of $\mathcal{B}$, and $\Pi=\gamma^{ab}\Pi_{ab}$. Substituting the Einstein equation ($G_{ab}=\kappa T_{ab}$) and the definition of the quasilocal stress-energy-momentum tensor ($\Pi_{ab}=-\kappa T^\mathcal{B}_{ab}$), we find that the radial Hamiltonian constraint is {\it linear} in the quasilocal pressure:
\begin{equation}\label{RadialHamiltonianPhysical}
2\kappa\, \mathbb{S}^{ab}n_a n_b={}^\mathcal{B}\!R +\kappa^2\left(\frac{1}{2}\mathcal{E}^2-2c^2\mathcal{P}^2+\tilde{\mathcal{S}}^2-2\mathcal{E}{\rm P}\right),
\end{equation}
where again we used the fact that $-T^{ab}n_a n_b=+\mathbb{S}^{ab}n_a n_b$. Here $\mathcal{P}^2=\mathcal{P}_a \mathcal{P}^a$ and $\tilde{\mathcal{S}}^2=\tilde{\mathcal{S}}_{ab} \tilde{\mathcal{S}}^{ab}$, where $\tilde{\mathcal{S}}_{ab}=\mathcal{S}_{ab}-{\rm P}\sigma_{ab}$ is the trace-free part of the quasilocal stress tensor. It is clear from this equation that there is a close relationship between the quasilocal pressure, $\rm P$, and the ``missing" external normal matter pressure, $\mathbb{S}^{ab}n_a n_b$, acting on the system. Solving for $\rm P$ we have:
\begin{equation}
{\rm P} = {\rm P}_{\rm mat} + {\rm P}_{\rm geom},\label{P_split}
\end{equation}
where we have split $\rm P$ into separate matter and geometry terms:
\begin{align}
{\rm P}_{\rm mat} &= -\frac{1}{\kappa\mathcal{E}}\, \mathbb{S}^{ab}n_a n_b,\label{MatterPressure}\\
{\rm P}_{\rm geom} &= \frac{1}{2\mathcal{E}}\left[ \frac{1}{\kappa^2} {}^\mathcal{B}\!R+\left( \frac{1}{2}\mathcal{E}^2 -2c^2\mathcal{P}^2+\tilde{\mathcal{S}}^2 \right) \right].\label{GeometricalPressure}
\end{align}

Before we can continue, it is important to appreciate that the quasilocal momentum conservation law in equation (\ref{reduced_integrated_quasilocal_conservation_law}) includes gravitational effects, whereas the local momentum conservation law in equation (\ref{rigid_KV_explicit_integrated_local_conservation_law}) does not. This means we can hope to meaningfully compare the two laws only in the limit where gravitational effects do not play a role, i.e., the limit of a small-sphere RQF, so that spacetime is nearly flat in the neighbourhood of the RQF. For simplicity, we will take the RQF to be a round sphere of areal radius $r$, and construct series expansions of ${\rm P}_{\rm mat}$ and ${\rm P}_{\rm geom}$ in the first few leading powers of $r$. While this will get a bit messy, the messiness is only a result of unnaturally trying to cast a quasilocal law in the form of a local law; the quasilocal law itself is very simple and elegant.

For a small round-sphere RQF containing a smooth (non-singular) matter distribution, the quasilocal energy density has the general expansion:
\begin{equation}\label{QuasilocalEnergyExpansion}
\mathcal{E}=-\frac{2}{\kappa r} + \mathcal{E}_1\,r+\mathcal{O}(r^2).
\end{equation}
The dominant term for small $r$ comes from the fact, noted earlier, that $\mathcal{E}=-k/\kappa$, where $k=\sigma^{ab}K_{ab}$ is the spatial trace of the extrinsic curvature. For a round-sphere RQF of areal radius $r$ in flat spacetime, $k=2/r$. This leading term is often called the {\it vacuum energy density}; its role will be discussed in more detail at the end of this subsection. The next term in equation (\ref{QuasilocalEnergyExpansion}), at order $r$, represents the lowest order at which matter can make a contribution to the energy of the system: when $\mathcal{E}$ is integrated over the two-sphere, this term becomes of order $r^3$, i.e., proportional to the spatial volume of the system. The lowest order at which gravity can make a contribution to the energy is order $r^3$ (i.e., a term of order $r^5$ when integrated over the two-sphere). We will verify both of these statements in the explicit example in {\S}\ref{Example}.

Substituting equation (\ref{QuasilocalEnergyExpansion}) into equation (\ref{MatterPressure}), and using the fact that for a non-singular matter distribution we must have $\mathbb{S}^{ab}n_a n_b =\mathcal{O}(1)$, we get
\begin{equation}\label{MatterPressureSeries}
{\rm P}_{\rm mat}=\frac{r}{2}\,\mathbb{S}^{ab}n_a n_b +\mathcal{O}(r^3 ).
\end{equation}
The factor of $r$ is obviously needed on dimensional grounds to convert a local pressure (force per unit area) into a quasilocal pressure (force per unit length). That the factor is precisely $r/2$ follows from a simple physical argument. Imagine that the round-sphere RQF is immersed in a matter field that is exerting a normal-normal stress $\mathbb{S}^{ab}n_a n_b$ that is {\it negative}, and for simplicity is uniform over the sphere. A negative $\mathbb{S}^{ab}n_a n_b$ corresponds to a local pressure pushing radially {\it inwards} on the surface of the sphere. The work done {\it by} the system (thought of in the local approach as the contents of the volume inside the sphere) in expanding the areal radius of the sphere from $r$ to $r+dr$ is then {\it positive}, and equal to $-4\pi r^2 \, \mathbb{S}^{ab}n_a n_b \, dr$. In the quasilocal approach, which doesn't ``know" anything about the contents of the volume of the sphere, the system is the surface of the sphere. For $\mathbb{S}^{ab}n_a n_b$ negative, we can imagine this surface to be like the elastic surface of a balloon, with an effective pressure, ${\rm P}_{\rm eff}$ (force per unit length), that is {\it negative}, i.e., the surface is under tension. Then the work done {\it by} the system against this tension in expanding the areal radius of the two-sphere from $r$ to $r+dr$ will be {\it positive}, and equal to $-{\rm P}_{\rm eff}\,d(4\pi r^2)=-8\pi r \, {\rm P}_{\rm eff}\,dr$. Equating the local and quasilocal expressions for the work done by the system, we have ${\rm P}_{\rm eff}=(r/2)\,\mathbb{S}^{ab}n_a n_b$, which explains the leading term in equation (\ref{MatterPressureSeries}).

Returning to equation (\ref{reduced_integrated_quasilocal_conservation_law}), we wish to show that the term $-{\rm P}_{\rm mat}\,\hat{D}_a\phi^a$ adds the correct normal matter pressure term to $\mathbb{S}^{ab} n_a \phi_b$. Using equation (\ref{MatterPressureSeries}) we have
\begin{equation} \label{NormalMatterCorrection}
\mathbb{S}^{ab} n_a \phi_b -{\rm P}_{\rm mat}\,\hat{D}_a\phi^a = \mathbb{S}^{ab} n_a \Phi_b + \mathcal{O}(r^2),\;\;{\rm where}\;\; \Phi_b =\phi_b-\frac{r}{2}(\hat{D}_a\phi^a )n_b.
\end{equation}
The notation ``$\Phi_b$" is suggestive of the fact that $\Phi_b$  here can, indeed, be identified with the $\Phi_b$ appearing in the local momentum conservation law, equation (\ref{rigid_KV_explicit_integrated_local_conservation_law}). For example, if we choose $\phi^a$ to be a CKV that generates a boost in the $Z$-direction, we can make $\Phi^a$ here equal the $Z^a$ spatial unit vector discussed in the second paragraph of {\S}\ref{QuasilocalMomentumConservationLaw}, with the usual components tangential and normal to the sphere. Introducing a spherical coordinate system adapted to the RQF, $x^a = (t,r,\theta,\phi)$, it is not difficult to show that we must have $\phi_a = (0,\mathcal{O}(r^2),-r\sin\theta,0)$, with $\hat{D}_a\phi^a =-(2/r)\cos\theta$ - an exact result, and $n_a = (0,1+\mathcal{O}(r^2),0,0)$, which results in $Z_a = (0,\cos\theta+\mathcal{O}(r^2),-r\sin\theta,0)$, as required. Thus, we can write the right-hand side of our exact RQF momentum conservation law in equation (\ref{reduced_integrated_quasilocal_conservation_law}) in the approximate form
\begin{equation}\label{reduced_integrated_quasilocal_conservation_law_series}
{\rm R.H.S.} =
\int\limits_{\Delta\mathcal{B}} N \, dt \, d\hat{\mathcal{S}} \left\{ \left[ \mathbb{S}^{ab} n_a \Phi_b +\mathcal{O}(r^2) \right]
- \left[\frac{1}{c^2}\mathcal{E}\,\alpha^a\phi_a +2\nu\epsilon_{ab}\mathcal{P}^a\phi^b +{\rm P}_{\rm geom}\,\hat{D}_a\phi^a \right]\right\}.
\end{equation}
Comparing with the right-hand side of the local momentum conservation law in equation (\ref{rigid_KV_explicit_integrated_local_conservation_law}) we see that the matter stress terms are now {\it identical}, at least to the two leading orders in $r$.

But what of the ${\rm P}_{\rm geom}\,\hat{D}_a\phi^a$ term on the right-hand side? There is no such analogous term in the local momentum conservation law. What role does it play? In part, it serves to provide a `geometrical buoyant force' that supports (cancels) the dominant-in-$r$ vacuum weight coming from the vacuum energy density in the term $\frac{1}{c^2}\mathcal{E}\,\alpha^a\phi_a$. (Note that, since the vacuum energy density is {\it negative}, this weight is `up' and the buoyant force is `down'.) Of course such a cancellation {\it must} happen because there are no other terms in the conservation law at this order in $r$, and the conservation law is an identity. Nevertheless, it is instructive to see how this cancellation works in detail.

First, note that to lowest order in $r$, the lapse function must have the form
\begin{equation}\label{LapseSeries}
N=1+\frac{1}{c^2}\,{\bf A}\cdot {\bf r} +\mathcal{O}(r^2)=1-\frac{r^2}{2c^2}\hat{D}_a \alpha^a+\mathcal{O}(r^2),
\end{equation}
where ${\bf A}$ is the acceleration of a fiducial point at the center of the small round-sphere RQF in usual boldface vector notation. This will induce a tangential acceleration, $\alpha^a$, experienced by the RQF observers located on the surface of the sphere such that $\hat{D}_a \alpha^a=-(2/r^2)\,{\bf A}\cdot {\bf r}$ (as a simple calculation reveals); hence the alternative form of $N$, which will be more useful to us below. Thus we have, for the vacuum weight density:
\begin{equation}\label{VacuumWeight}
-N\,\frac{1}{c^2}\mathcal{E}\,\alpha^a\phi_a=+\frac{2}{\kappa c^2 r}\,\alpha^a\phi_a +\mathcal{O}(1),
\end{equation}
where $\alpha^a\phi_a=\mathcal{O}(1)$. What we need to show is that
\begin{equation}\label{VacuumBuoyancy}
-N\,{\rm P}_{\rm geom}\,\hat{D}_a\phi^a = -\frac{2}{\kappa c^2 r}\,\alpha^a\phi_a + \Omega +\mathcal{O}(1),
\end{equation}
where $\Omega$ represents a possible term of order $1/r$ or lower that integrates to zero over the sphere (i.e., is the divergence of a vector field). Note that to lowest order in $r$, the lapse function plays no role in equation (\ref{VacuumWeight}), but it {\it will} play a critical role in equation (\ref{VacuumBuoyancy}).

To prove equation (\ref{VacuumBuoyancy}) we must examine the dominant terms in equation (\ref{GeometricalPressure}). A bit of thought (and experience with RQFs) shows that the $\mathcal{P}^2$ and $\tilde{\mathcal{S}}^2$ terms will not play a dominant role, but the other two terms will. A short calculation reveals that, in the context of an RQF, the Ricci scalar associated with $\gamma_{ab}$ (the intrinsic geometry of $\mathcal{B}$) can be written as
\begin{equation}\label{RicciScalar}
{}^\mathcal{B}\!R=\hat{R}-\frac{2}{c^2}\,\left( \hat{D}_a\alpha^a + \frac{1}{c^2} \alpha_a\alpha^a -\nu^2 \right),
\end{equation}
where $\hat{R}$ is the Ricci scalar associated with $\sigma_{ab}$ (the geometry of the two-sphere quotient space), which, for the case of a round-sphere RQF of areal radius $r$, is $2/r^2$ (exact). The next-to-leading order term is $\hat{D}_a\alpha^a=\mathcal{O}(1/r)$, and the rest are higher order. Combining these results with the series expansion for $\mathcal{E}$ given in equation (\ref{QuasilocalEnergyExpansion}) we find:
\begin{equation}\label{GeometricalPressureSeries}
{\rm P}_{\rm geom}=-\frac{1}{\kappa r}+\frac{r}{2\kappa c^2}\,\hat{D}_a\alpha^a +\mathcal{O}(r).
\end{equation}
The dominant term for small $r$ is a negative {\it vacuum pressure}, whose existence is intimately connected with the existence of the vacuum energy density through the geometrical identity $\mathcal{E}-2{\rm P}=(2/c^2\kappa)\,n_a a^a$ discussed more fully in {\S} \ref{secExtrinsic}.  If we think of the negative vacuum pressure as a positive surface tension in the balloon analogy used earlier, then a similar calculation shows that the system must do an amount of positive work equal to $(8\pi/\kappa)\,dr$ to expand the areal radius by an amount $dr$, which is exactly the amount by which the (negative) vacuum energy of the system is reduced during this expansion. So the two vacuum entities are logically self-consistent.

Given that the vacuum pressure is {\it uniform} over the surface of the sphere (no gradient) one might assume that it does not create a buoyant force. However, when multiplied by the lapse function in equation (\ref{LapseSeries}) it picks up a dipole cross term that accounts for precisely half of the proper time-integrated buoyant force (buoyant impulse) necessary to support the vacuum weight. (This is due to the acceleration-induced inhomogeneous time dilation mechanism mentioned in the last paragraph of {\S}\ref{LocalArchimedes}.) The next-to-leading order term in ${\rm P}_{\rm geom}$ is a dipole term that accounts for the other half of the buoyant force supporting the vacuum weight. Putting all of these results together we have:
\begin{equation}\label{VacuumBuoyancyActual}
-N\,{\rm P}_{\rm geom}\,\hat{D}_a\phi^a=\left[ \frac{1}{\kappa r}-\frac{r}{\kappa c^2} \, \hat{D}_a\alpha^a +\mathcal{O}(r) \right]\,\hat{D}_b\phi^b.
\end{equation}
Integrating $(\hat{D}_a\alpha^a)\,(\hat{D}_b\phi^b)$ by parts results in equation (\ref{VacuumBuoyancy}) with
\begin{equation}
\Omega = \hat{D}_a \left[  (1/\kappa r)   \, \phi^a - ( r/  \kappa c^2 ) \, \alpha^a \hat{D}_b \phi^b \right].
\end{equation}
Note that this part of the analysis has been purely geometrical. Matter contributions begin to appear in the quasilocal pressure only at order $r$ - compare equations (\ref{MatterPressureSeries}) and (\ref{GeometricalPressureSeries}).

Presumably we can continue the expansions started in equations (\ref{VacuumWeight}) and (\ref{VacuumBuoyancy}) to the order in $r$ at which matter begins to contribute to the quasilocal weight density and check that, when the sum is integrated over the two-sphere, it gives the same result as the local matter weight density $-N\frac{1}{c^2}\mathbb{E}\,a^a\Phi_a$ in equation (\ref{rigid_KV_explicit_integrated_local_conservation_law}) integrated over the three-volume inside the sphere. However, this would be prohibitively tedious (and besides, in the next subsection we will verify this explicitly in a nontrivial example). In any case, at this point it is clear that the quasilocal momentum conservation law in equation (\ref{reduced_integrated_quasilocal_conservation_law}) is saying essentially the same thing as the local momentum conservation law in equation (\ref{rigid_KV_explicit_integrated_local_conservation_law}), but in a novel way that has two key (and intimately related) advantages: (1) unlike in the local law, the rigidity and Killing vector-cum conformal Killing vector conditions can always be realized, and (2) unlike in the local law, the quasilocal law properly includes the effects of gravity. For a small-sphere RQF, the quasilocal law reduces to the local one; but as the system gets larger, the quasilocal law clearly departs from the local law in what it is saying.

Thus, the local law {\it does not explain what is really happening}. The local law treats the tangential (shear) components of force and weight on the same footing as the normal components. The quasilocal law does not. The normal components are accounted for in an entirely different way - through the quasilocal pressure term, $N\, {\rm P}\,\hat{D}_a\phi^a$, which has no analogue in the local law. When $\phi^a$ is a rotational CKV (so we are dealing with angular momentum), shear effects alone are sufficient, which is consistent with the fact that $\hat{D}_a\phi^a=0$ for a rotational CKV. But when $\phi^a$ is a boost CKV (so we are dealing with linear momentum), shear effects alone are {\it not} sufficient, which is consistent with the fact that $\hat{D}_a\phi^a$ is {\it not} zero in this case. The two-sphere integral of ${\rm P}\,\hat{D}_a\phi^a$ is then a measure of the corresponding $\ell=1$ spherical harmonic component of ${\rm P}$, i.e., the dipole component. A nontrivial dipole component represents a {\it gradient} in pressure, which, according to the quasilocal law, is how general relativity accounts for normal forces or weights. We examined only the dominant terms in equation (\ref{GeometricalPressure}). There are clearly higher order nonlinear geometrical (gravitational) corrections that take us completely outside of the physics described by the local law.

As a final note, it is sometimes thought that the vacuum energy density in equation (\ref{QuasilocalEnergyExpansion}) should be removed using the freedom in the definition of $T_\mathcal{B}^{ab}$ (interpreted as a freedom to choose the zero of energy for the system) to subtract a suitable reference stress-energy-momentum tensor (see, e.g., reference \cite{BY1993}). However, we have seen here that the vacuum energy density and related vacuum pressure not only work together in a physically sensible way, but are actually {\it necessary} for the quasilocal momentum conservation law to have the correct small-sphere limit. The fact that $\mathcal{E}\rightarrow -2/\kappa r$ as $r\rightarrow 0$ is what allows the factor $-1/\kappa\mathcal{E}$ in equation (\ref{MatterPressure}) to approach $r/2$ as $r\rightarrow 0$, which in turn leads to equation (\ref{MatterPressureSeries}). This then allows $-{\rm P}_{\rm mat}\,\hat{D}_a\phi^a$ to add the correct normal matter pressure term to $\mathbb{S}^{ab} n_a \phi_b$ in equation (\ref{NormalMatterCorrection}). Without this mechanism, we would have only a shear matter stress, and the quasilocal law would not reduce to the local law in the small-sphere limit. But this leads to a potential problem: a seemingly nonphysical dominant-in-$r$ vacuum weight coming from the term $\frac{1}{c^2}\mathcal{E}\,\alpha^a\phi_a$ in equation (\ref{reduced_integrated_quasilocal_conservation_law}). As we have just seen, what comes to the rescue is the geometrical pressure in equation (\ref{GeometricalPressure}), in particular the vacuum pressure plus the first subleading term, i.e., the two leading terms in $\rm P$ before matter begins to contribute. This provides further evidence\footnote{Recall, on page \pageref{VacuumW2Argument} we argued that the vacuum energy might be of real physical importance based on its apparent (negative) contribution to the moment of inertia of a small-sphere RQF.} that perhaps the vacuum energy (and attendant vacuum pressure) should not be be freely removed, and might actually be physically real in some sense.

\subsection{Specialization to Quasilocal Archimedes' Law with Example}\label{Example}

We begin with the completely general RQF momentum conservation law for matter and gravitational fields in a finite volume given in equation (\ref{reduced_integrated_quasilocal_conservation_law}), and specialize to a stationary context suitable for Archimedes' law. As in {\S}\ref{LocalArchimedes}, the first step is to assume certain properties of the frame of reference, namely, that the RQF observers are experiencing at most {\it stationary} acceleration and twist. In the RQF-adapted coordinates $x^a=(t,r,x^i )$ introduced earlier (where $x^i$ label the two-parameter family of worldlines), the spatial coordinate components of acceleration and twist are
\begin{align}
\alpha_i &=\frac{1}{N}\dot{u}_i+c^2\partial_i\ln N\label{quasilocal_acceleration}\\
\nu &=\frac{1}{2}\epsilon^{ij}\left(\partial_i u_j-\frac{1}{c^2}\alpha_i u_j\right);\label{quasilocal_twist}
\end{align}
compare with equations (\ref{local_acceleration}) and (\ref{local_twist}). In further analogy to {\S}\ref{LocalArchimedes}, assuming also that $v^a$ (the two-velocity of the fiducial $u^a_\mathcal{S}$-observers relative to the $u^a$-frame) is also stationary implies that $\dot{u}_i=0$, in which case demanding stationary acceleration ($\dot{\alpha}_i=0$) implies that the lapse function (and thus the full integration measure) is time-independent and, in turn, the twist is stationary ($\dot{\nu}=0$). Recall also that the component of momentum the observers are measuring is specified by a {\it stationary} CKV, $\phi^a$. The second step is to assume properties of the matter and gravitational fields themselves, namely, that $\mathcal{E}$, $\mathcal{P}^a$ and $\mathcal{S}^{ab}$ are all stationary. The left-hand side of equation (\ref{reduced_integrated_quasilocal_conservation_law}) then vanishes (the momentum in the system, if any, does not change with time) and we are left with
\begin{equation}
\int\limits_{\Delta\mathcal{B}} N \, dt \, d\hat{\mathcal{S}} \left[ \mathbb{S}^{ab} n_a \phi_b - {\rm P}_{\rm mat}\,\hat{D}_a\phi^a\right] =
\int\limits_{\Delta\mathcal{B}} N \, dt \, d\hat{\mathcal{S}} \left[\frac{1}{c^2}\mathcal{E}\,\alpha^a\phi_a +2\nu\epsilon_{ab}\mathcal{P}^a\phi^b +{\rm P}_{\rm geom}\,\hat{D}_a\phi^a \right],\label{quasilocal_Archimedes}
\end{equation}
where we used equation (\ref{P_split}) and rearranged some terms.

We contend that equation (\ref{quasilocal_Archimedes}) is the fully general relativistic analogue of Archimedes' law for matter and gravitational fields contained in a finite volume.  It says essentially the same thing as the local Archimedes' law in equation (\ref{local_Archimedes}), but does not rely on any spacetime symmetries.  To achieve this generality, the bulk integral on the right-hand side is replaced with a surface integral, and there are additional quasilocal pressure terms that account for normal forces and weights in a general relativistically correct way. Roughly speaking, it says that the weight of the matter and gravitational mass-energy in a non-inertial reference frame (right-hand side) is supported by a Maxwell stress-like buoyant force acting on the fields inside the system through the boundary of the system (left-hand side).

\begin{figure}
\begin{center}
\includegraphics[scale=0.8]{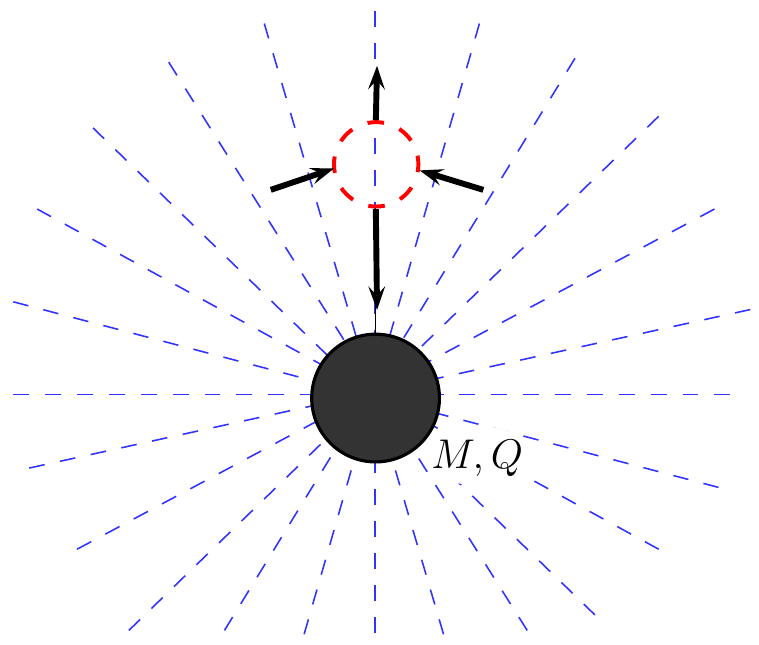}
\caption{The weight of the electrostatic field in a volume of space a fixed distance from a Reissner-Nordstr\"{o}m black hole is supported by the tensions along the electric field lines and pressures orthogonal to them at the boundary of the volume. For this to be true one might expect that the electric field lines exert a net upward force. Actually, they do not. Instead, there is a net buoyant {\it impulse} arising from an acceleration-induced inhomogeneous time dilation that enhances the upward impulses relative to the downward ones.}\label{RQFinRN}
\end{center}
\end{figure}

As a concrete example of this Archimedes' law, and one that is a close general relativistic analogue of the special relativistic work done in reference \cite{Eriksen2006}, we consider a small round-sphere RQF hovering a fixed distance above a Reissner-Nordstr\"{o}m black hole. See figure \ref{RQFinRN}. First, let us give a brief description of the physical mechanism behind Archimedes' law in this example and later the mathematical details of the analysis. We assume that the round-sphere RQF has an areal radius $r$, and conduct the analysis in powers of $r$ far enough to include the first two terms in the energy density expansion given in equation (\ref{QuasilocalEnergyExpansion}), i.e., enough to be able to calculate the gravitational vacuum energy and lowest order electrostatic energy contained inside the sphere. Operationally, the RQF observers measure this energy by measuring (with a ruler) its local effect on the extrinsic geometry of the sphere, and then integrate this local effect over the entire sphere. Moreover, by observing a constant precession rate of their local gyroscopes, they conclude that the momentum in the system is not changing.\footnote{In the previous chapter we discussed how mass-energy in motion produces a relatively tiny frame dragging effect that causes the RQF observers' local gyroscopes to precess. Roughly speaking, rotating this relatively tiny precession rate vector through 90 degrees and multiplying by the large number $c^2/8\pi G$ yielded the quasilocal momentum density, $\mathcal{P}^a$, which when integrated over the sphere gives what we understand in classical mechanics as the momentum of the mass-energy in motion.} On the other hand, using local accelerometers, the RQF observers measure the proper acceleration required for them to hover a fixed distance above the black hole. Thus seeing that their (rigid quasilocal) frame of reference is accelerating, but that the momentum of the mass-energy inside is not changing, the RQF observers conclude that there must be a Maxwell stress at the surface of the sphere acting on the electrostatic field inside, causing it to accelerate along with their frame. By measuring the local electrostatic field they are at first surprised to see that this Maxwell stress actually averages to {\it zero} over the surface of the sphere. It produces no net buoyant force. But then they realize that, because of their acceleration (or equivalently, the spacetime curvature), proper time elapses at a greater rate at the top of the sphere (the point furthest from the black hole) than at the bottom, and this results in a nonzero net buoyant {\it impulse}. This important effect of an acceleration-induced inhomogeneous proper time was mentioned earlier, and will be discussed in more detail below.\footnote{Incidentally, the quasilocal pressure in equation (\ref{reduced_integrated_quasilocal_conservation_law}) can be measured indirectly using the identity $2{\rm P}=\mathcal{E} - (2/c^2\kappa)n_a a^a$ (discussed in Chapter \ref{chNova}), i.e., operationally, {\rm P} is essentially the difference between the quasilocal energy density (measured with a ruler) and the normal component of proper acceleration (measured with an accelerometer). The split between ${\rm P}_{\rm mat}$ and ${\rm P}_{\rm geom}$ in equation (\ref{quasilocal_Archimedes}) is then determined by measuring ${\rm P}_{\rm mat}$ through its relation to the normal matter pressure - see equation (\ref{MatterPressure}). Alternatively, and probably closer to the spirit of the quasilocal approach, we would move the ${\rm P}_{\rm mat}$ term in equation (\ref{quasilocal_Archimedes}) to the right-hand side and deal with the full quasilocal pressure, $\rm P$.}

We begin with the line element for the Reissner-Nordstr\"{o}m (RN) black hole spacetime given in quasi-Minkowskian coordinates $X^a = (X^0, X^I) = (cT, X,Y,Z),\, I=1,2,3$:
\begin{align}\label{RNCartesian}
ds^2 = - F^2 (R) c^2 dT^2 + \delta_{IJ} dX^I dX^J + \frac{1}{R^2} \left( \frac{1}{F^2 (R)} - 1 \right) \delta_{IK} \delta_{JL} X^I X^J dX^K dX^L ,
\end{align}
where $\delta_{IJ}$ is the Kronecker delta function, $R^2 = X^2 + Y^2 + Z^2$, $F(R) = \sqrt{1 - \frac{2GM}{c^2R} + \frac{G Q^2}{c^4 R^2}}$, and $M$ and $Q$ are the mass and charge of the black hole, respectively. We then construct a small spherical RQF about the point $Z=R_0$ on the $Z$-axis via the coordinate transformation
\begin{align}\label{RQF_coordinate_embedding}
T = \frac{1}{F(R_0)}\, t, \qquad X = r \sin \theta \cos \phi, \qquad Y = r \sin \theta \sin \phi, \qquad Z = R_0 + \sum^{\infty}_{n=1} f_{n} (\theta) \, r^n,
\end{align}
where $(t,r,x^i )$ are the RQF-adapted coordinates. The angular coordinates $x^i=(\theta,\phi)$ label the two-parameter family of RQF observers on the sphere (and their worldlines in the congruence, $\mathcal{B}$). The radial coordinate, $r\ll R_0$, is a small parameter related to the size of the sphere. The time coordinate, $t$, labels a natural choice of two-spheres, $\mathcal{S}_t$, foliating $\mathcal{B}$. In the case of flat spacetime, i.e., $F(R)=1$, the choice of functions $f_1(\theta)=\cos\theta$ and $f_{n>1}(\theta)=0$ represents a set of RQF observers sitting on a round sphere of areal radius $r$, with standard spherical coordinates $(\theta,\phi)$, centered on a point along the $Z$-axis a distance $R_0$ from the origin. This constitutes an RQF with two-surface metric $\sigma_{ij} = r^2 \mathbb{S}_{ij}$, where $\mathbb{S}_{ij}$ is the standard spherical coordinate metric on a unit two-sphere. As we turn on the mass and charge of the RN black hole, i.e., $F(R)\neq 1$, we would like the RQF observers to maintain the {\it same} round sphere two-surface metric ($\sigma_{ij} = r^2 \mathbb{S}_{ij}$, for which $d\hat{\mathcal{S}}=r^2 d \mathbb{S} = r^2\sin\theta\,d\theta\,d\phi$), but the presence of spacetime curvature means that in the RQF coordinate embedding in equation (\ref{RQF_coordinate_embedding}) we will have to rescale $f_1(\theta)$, and will need an infinite number of higher order corrections, $f_{n>1}(\theta)\neq 0$.\footnote{Note that the functions $f_n(\theta)$ do not depend on $\phi$ owing to the symmetry of the RN spacetime under rotations about the $Z$-axis.}
Using GRTensorII \cite{GRTensor} running under Maple, we have solved for these corrections up to order $r^4$. At the two lowest orders we find:\footnote{Note that at each order the functions $f_n(\theta)$ are determined only up to a constant of integration by the round-sphere RQF conditions (being differential in nature). We fix this residual gauge freedom so as to recover the flat spacetime result as $r\rightarrow 0$. For example, one actually finds $f_1 (\theta) = F(R_0) \cos \theta + C_1$, but we take $C_1 = 0$ to make $g_{rr} \rightarrow 1$ as $r\rightarrow 0$. Also note that we rescale $t$ by $1/F(R_0)$ in equation (\ref{RQF_coordinate_embedding}) so that $g_{tt}\rightarrow -c^2$ as $r\rightarrow 0$.}
\begin{align}
f_1 (\theta) &= F(R_0) \cos \theta, \label{f1} \\
f_2(\theta) &= \frac{1}{4}\left[ F(R_0)F^\prime (R_0) + \frac{1-F^2(R_0)}{R_0} \right](2\cos^2\theta-1)+\frac{1}{4}\left[ F(R_0)F^\prime (R_0) - \frac{1-F^2(R_0)}{R_0} \right],\label{f2}
\end{align}
where $F^\prime(R_0)=dF(R_0)/dR_0$. The rescaling of $f_1 (\theta)$ by $F(R_0)$ represents a uniform contraction of the embedded $r={\rm constant}$ coordinate sphere along the $Z$-axis (but not the $X$- or $Y$-axes) to account for the fact that the spatial line element at the center of the sphere is $ds^2=dX^2+dY^2+dZ^2 / F^2 (R_0)$, i.e., proper distance along the $Z$-axis at $R=R_0$ is given by $ds = dZ / F(R_0)$; the coordinate contraction thus maintains a geometrically round sphere of areal radius $r$. The correction at next order in $r$, $f_2(\theta)$, which goes to zero in the flat spacetime limit $F(R) = 1$, includes monopole and quadrupole terms, and in general there are an infinite number of higher order multipole corrections at higher order in $r$.

Our Maple calculation reveals that the quasilocal energy density is given by
\begin{equation}\label{script_E_RN}
\mathcal{E} = - \frac{2}{\kappa r} + \left[ \frac{1}{3}\mathbb{E}_\mathtt{EM} + \left( \frac{M c^2}{8\pi R_0^3} - \frac{2}{3}\mathbb{E}_\mathtt{EM} \right)\left( 3 \cos^2 \theta - 1 \right) \right] r
 + \mathcal{O}(r^2)
\end{equation}
where $\mathbb{E}_\mathtt{EM}=|\vec{E}_0|^2/8\pi$ is the energy density (energy per unit {\it volume}) of the electrostatic field, evaluated at the center of the sphere, and $|\vec{E}_0|=Q/R_0^2$ is the magnitude of the electrostatic field, also evaluated at the center. Observe that the quasilocal energy density has the general form proposed in equation (\ref{QuasilocalEnergyExpansion}). Multiplying $\mathbb{E}_\mathtt{EM}$ by $r/3$ converts the electrostatic volume energy density to an effective surface energy density (energy per unit area). Upon integrating $\mathcal{E}$ over the surface of the sphere, the $\ell=2$ spherical harmonic term makes no net contribution and we find that the total energy inside the RQF is
\begin{equation}\label{TotalEnergy}
E = - \frac{8 \pi}{\kappa} r + \mathbb{E}_\mathtt{EM}\left(\frac{4\pi}{3} r^3\right)+ \mathcal{O}(r^5).
\end{equation}
The first term on the right-hand side is the (negative) vacuum energy, and the second is the electrostatic field energy (at lowest order in $r$). Note: the Maple calculation shows that the order $r^2$ term in equation (\ref{script_E_RN}) is actually a sum of $\ell=1$ and $\ell=3$ spherical harmonics that integrate to zero, so there is no order $r^4$ contribution to equation (\ref{TotalEnergy}). General relativistic curvature effects begin to appear only at order $r^5$, as mentioned earlier.

Next, we compute the radial electrostatic pressure exerted on the field inside the RQF (i.e, the normal-normal component of the Maxwell stress tensor) and find
\begin{equation} \label{TnnRN}
\mathbb{S}^{ab} n_{a} n_{b} = \mathbb{E}_\mathtt{EM}\, (2\cos^2\theta-1)-4\,\mathbb{E}_\mathtt{EM}\,\frac{F(R_0)}{ R_0}\, (3\cos^3\theta-2\cos\theta) \,r+ \mathcal{O}(r^2)
\end{equation}
Observe that at zeroth order in $r$ this pressure is directed radially outwards near the poles of the sphere (where the electric field lines are mainly orthogonal to the surface, and under tension) and inwards near the equator (where the electric field lines are mainly parallel to the surface, and repel each other) - see figure \ref{RQFinRN}. The correction at order $r$ accounts for the non-uniform (radial) nature of the electrostatic field around the black hole, and reduces to the special relativistic result when $F(R)=1$. Observe also that $\mathbb{S}^{ab}n_a n_b =\mathcal{O}(1)$, which is required for equations (\ref{MatterPressureSeries}) and (\ref{NormalMatterCorrection}) to be valid. We have similarly calculated $\mathbb{S}^{ab} n_{a} \phi_{b}$, the electrostatic shear stress exerted on the field inside the RQF. Substituting equation (\ref{TnnRN}) into equation (\ref{MatterPressureSeries}), and recalling that $\hat{D}_a\phi^a =-(2/r)\cos\theta$ (an exact result), we find that the integrand on the left-hand side of equation (\ref{quasilocal_Archimedes}) is given by
\begin{equation}\label{TnPhiRN}
\mathbb{S}^{ab} n_a \phi_b - {\rm P}_{\rm mat}\,\hat{D}_a\phi^a = \mathbb{S}^{ab} n_a \Phi_b + \mathcal{O}(r^2) = \mathbb{E}_\mathtt{EM}\cos\theta -2\,\mathbb{E}_\mathtt{EM}\,\frac{F(R_0)}{ R_0}\, (3\cos^2\theta-1) \,r+ \mathcal{O}(r^2).
\end{equation}
As argued in the discussion surrounding equation (\ref{NormalMatterCorrection}), this is the net vertical stress (force per unit area) exerted on the field inside the RQF, and thus represents the buoyant forces due to matter. Surprisingly, at least to the order calculated, it is a sum of spherical harmonics that integrates to {\it zero} over the surface of the sphere. Setting $F(R)=1$ in equation (\ref{TnPhiRN}) yields the corresponding result for a small round-sphere RQF sitting a fixed distance $R_0$ from a point charge in flat spacetime: no net buoyant force.

In going from flat spacetime to a RN black hole the net buoyant force remains zero, but there is an interesting temporal effect arising from the lapse function in the integrand on the left-hand side of equation (\ref{quasilocal_Archimedes}). We find, in accordance with equation (\ref{LapseSeries}), that
\begin{equation} \label{LapseRN}
N = 1 + \frac{A_0\cos\theta}{c^2} r  + \mathcal{O} (r^2),\;\;{\rm where}\;\;A_0=c^2F^\prime (R_0)=\frac{1}{F(R_0)}\left(\frac{GM}{R_0^2}-\frac{GQ^2}{c^2 R_0^3}\right).
\end{equation}
Here, $A_0$ is the acceleration of a fiducial point at the center of the sphere. The first term in parentheses is the corresponding Newtonian acceleration, and the second is a post-Newtonian correction that represents the well-known repulsive effect of the electric charge of a RN black hole. This lapse function says that, due to the acceleration of the RQF (or equivalently, the spacetime curvature), more proper time elapses at points in the top half of the sphere (furthest from the black hole), in a given parameter time interval, $dt$, than corresponding points in the bottom half. So although the magnitude of the upward buoyant forces in the top half is equal to the magnitude of the corresponding downward buoyant forces in the bottom half (at lowest order in $r$), there is a net upward buoyant {\it impulse}. Mathematically, the $\ell=1$ part of $N$ combines with the $\ell=1$ part of equation (\ref{TnPhiRN}) to yield an $\ell=0$ term that does {\it not} vanish upon integration. We find
\begin{equation}
\int\limits_{\mathcal{S}_t} N \, d\hat{\mathcal{S}} \left[ \mathbb{S}^{ab} n_a \phi_b - {\rm P}_{\rm mat}\,\hat{D}_a\phi^a\right] = \int\limits_{\mathcal{S}_t} N \, d\hat{\mathcal{S}} \left[ \mathbb{S}^{ab} n_a \Phi_b + \mathcal{O}(r^2) \right]= \frac{\mathbb{E}_\mathtt{EM}}{c^2}\left(\frac{4\pi}{3} r^3\right) A_0+ \mathcal{O} (r^4),\label{LHS_quasilocal_Archimedes}
\end{equation}
which is the relativistic mass of the electrostatic field inside the RQF (to order $r^3$) times the mean acceleration of the frame, i.e., just equal in magnitude to the {\it weight} of the field inside, as we might expect. Integrating this {\it effective} net vertical buoyant force over parameter time $t$, as directed in equation (\ref{quasilocal_Archimedes}), results in a nonzero net vertical buoyant {\it impulse} acting on the electrostatic field inside the RQF and supporting its weight.

The weight is calculated using the right-hand side of equation (\ref{quasilocal_Archimedes}). In our example the twist is obviously zero ($\nu=0$), so we are left with two terms: $\frac{1}{c^2}\mathcal{E}\,\alpha^a\phi_a$ and ${\rm P}_{\rm geom}\,\hat{D}_a\phi^a$. The former is similar to the weight density in the local approach, $\frac{1}{c^2}\mathbb{E}\, a^a \Phi_a$, except that it contains vacuum contributions. We argued in the previous subsection that, in part, the geometrical pressure term provides a `geometrical buoyant force' that supports (cancels) these vacuum weight contributions. Let us quickly verify, in the present example, the facts on which this argument hinged. Beginning with equation (\ref{GeometricalPressure}), a Maple calculation reveals that $\mathcal{P}^2 = 0$ (of course) and $\tilde{\mathcal{S}}^2 = \mathcal{O}(r^2)$, so these terms can be neglected relative to $^\mathcal{B}\mathcal{R}$ and $\mathcal{E}^2$, which both start at order $1/r^2$. Similarly, regarding the boundary Ricci scalar in equation (\ref{RicciScalar}), we verify that we can neglect the last two terms on the right-hand side since $\alpha_a \alpha^a = \mathcal{O}(1)$ (and $\nu = 0$), while $\hat{R} = 2/r^2$ (exactly) and $\hat{D}_a  \alpha^a$ starts at order $1/r$. With these details confirmed, we arrive at equation (\ref{VacuumBuoyancyActual}), which verifies that the sum of the left-hand sides of equations (\ref{VacuumWeight}) and (\ref{VacuumBuoyancy}) vanishes at orders $1/r^2$ and $1/r$, modulo spherical harmonic terms that integrate to zero. Using Maple we take the analysis two orders in $r$ higher, up to and including the order at which matter begins to contribute. The expressions contain a large number of spherical harmonic terms, but upon integration we arrive at a simple result for the (effective\footnote{The term ``effective" reminds us that the lapse function plays an important role here, i.e., properly speaking we are dealing with an impulse rather than a force.}) weight of the electrostatic field inside the RQF:
\begin{equation}\label{RHS_quasilocal_Archimedes}
\int\limits_{\mathcal{S}_t} N \, d\hat{\mathcal{S}} \left[\frac{1}{c^2}\mathcal{E}\,\alpha^a\phi_a  +{\rm P}_{\rm geom}\,\hat{D}_a\phi^a \right]= \frac{\mathbb{E}_\mathtt{EM}}{c^2}\left(\frac{4\pi}{3} r^3\right) A_0+ \mathcal{O} (r^4),
\end{equation}
in agreement with the effective buoyant force in equation (\ref{LHS_quasilocal_Archimedes}).

\section{Summary and Conclusions}\label{Summary}

The local approach to constructing integrated energy and momentum conservation laws begins with the differential identity given in equation (\ref{differential_local_conservation_law}), which reduces to $\nabla_a ( T^{ab}\Psi_b ) = T^{ab} \nabla_{(a} \Psi_{b)}$ in general relativity. The problems with this approach are essentially two-fold: (1) being local, this approach cannot account for gravitational effects, which are nonlocal (this is reflected in the fact that the identity is empty when $T^{ab}=0$), and (2) we get the intuitive form of a conservation law (namely, that the change in a physical quantity over time equals the net corresponding flux through the system boundary during that time, with no bulk term integrals) only when $\Psi^a$ is a Killing vector field, i.e., the approach relies on the existence of spacetime symmetries, which rules out spacetimes with interesting matter or gravitational dynamics.

As we have seen, a natural solution to the first problem is to replace the local matter stress-energy-momentum tensor, $T^{ab}$, with the Brown and York quasilocal stress-energy-momentum tensor, $T_\mathcal{B}^{ab}$, which represents both matter {\it and} gravitational fields, and employ the differential identity $D_a (T_\mathcal{B}^{ab}\psi_b ) = (D_a T_\mathcal{B}^{ab})\psi_b + T_\mathcal{B}^{ab}D_{(a}\psi_{b)}$ [equation (\ref{differential_quasilocal_conservation_law})] in the boundary, $\mathcal{B}$ \cite{BY1993}. The solution to the second problem is to replace the notion of a rigid local frame (which, we argued, is required to make sense of an integrated conservation law, but does not always exist) with a rigid quasilocal frame (which {\it always} exists).  If $\psi^a$ is chosen to be $\frac{1}{c}u^a$ (where $u^a$ is the RQF observers' four-velocity) we get a completely general RQF {\it energy} conservation law for matter and gravitational fields in a finite system in general relativity (discussed in the previous chapter).  As shown in the present chapter, if $\psi^a$ is chosen to be $-\frac{1}{c}\phi^a$, where $\phi^a$ is a stationary conformal Killing vector (CKV) field orthogonal to $u^a$ (the full set of six of which always exists) we get the completely general RQF {\it momentum} conservation law given in equation (\ref{reduced_integrated_quasilocal_conservation_law}):
\begin{equation}
\int\limits_{\mathcal{S}_f - \mathcal{S}_i}  \hspace{-8pt} d\hat{\mathcal{S}} \, \left( \mathcal{P}^a+\frac{1}{c^2}\mathcal{S}^{ab}v_b\right)\phi_a =
\int\limits_{\Delta\mathcal{B}} \hspace{-6pt} N \, dt \, d\hat{\mathcal{S}} \left\{ \mathbb{S}^{ab} n_a \phi_b
- \left[\frac{1}{c^2}\mathcal{E}\,\alpha^a\phi_a +2\nu\epsilon_{ab}\mathcal{P}^a\phi^b + {\rm P}\,\hat{D}_a\phi^a \right]\right\},
\end{equation}
which is a conservation law for  linear or angular momentum according to whether $\phi^a$ is one of the three boost or three rotation CKV fields. The left-hand side of this equation gives the change in the momentum contained in the system as measured by the  RQF observers between the initial and final times.  The right-hand side is the flux of momentum  (including the gravitational momentum flux,  $-\frac{1}{c^2}\mathcal{E}\,\alpha^a\phi_a$) entering the system from the outside during this time. In short, in the last two chapters, we have constructed completely general energy and momentum conservation laws that {\it do not rely on any spacetime symmetries}.  The notion of an RQF plays a crucial role in the construction of these laws.

These new energy and momentum conservation laws teach us some new physics. They allow us to identify simple, exact, operational definitions for fluxes of gravitational energy [equation (\ref{ACrossOmegaFormOfGeometricalFlux})] and momentum (linear and angular), and these fluxes in turn provide a deeper insight into what's really happening in a wide variety of physical phenomena. For instance, some simple, everyday effects in Newtonian mechanics are actually very tiny, subtle general relativistic effects multiplied by a large number (like $c^2/G$ or $c^4/G$) to become effects we see in the everyday world. When we drop an apple, for example, the apple gains energy and momentum relative to our accelerated frame of reference at rest on the Earth. The gain in energy is due to a gravitational energy flux involving the general relativistic effect of frame dragging; the gain in momentum is due to a gravitational momentum flux involving the general relativistic effect that mass-energy has on the extrinsic curvature of the boundary of the reference frame.

Insofar as energy and momentum have traditionally been useful concepts in physics, these new completely general RQF energy and momentum conservation laws will further deepen our understanding of energy and momentum, and find useful applications.  Towards this end, we derived a general relativistic version of Archimedes' law, which we applied to understand how the weight of a small volume of electrostatic field, located a fixed distance from a Reissner-Nordstr\"{o}m black hole, is supported by the Maxwell stress buoyant forces of the surrounding electrostatic field. To our surprise, we found that the net buoyant force is actually zero, but the net buoyant {\it impulse} is not. The nonzero buoyant impulse is due to an acceleration-induced inhomogeneous time dilation at the boundary of the reference frame, as revealed by the RQF momentum conservation law.

\chapter{Post-Newtonian RQFs}\label{chPN}

Up until now we have been primarily concerned with developing the RQF formalism.  In doing so we have made many conceptual advances with regards to understanding the importance of the RQF structure and what it is telling us about the nature of the universe.  In this chapter we will shift our focus more towards the {\it practical} utility of RQFs.  For this purpose we will take the quasilocal conservation laws that we have derived in the previous chapters and expand them in a post-Newtonian approximation.  We provide general expressions for the various fluxes we have encountered in this approximation and employ them to understand measurable gravitational effects.  More specifically, in {\S}\ref{secPNRQFLaws} we start with a standard post-Newtonian metric and embed in it an RQF to derive a new metric describing the post-Newtonian spacetime as seen by RQF observers.  It is in this metric that we expand the general conservation laws of the previous section and ultimately derive expressions for the time rate of change of the energy and angular momentum inside the RQF in terms of fluxes through the RQF boundary in a post-Newtonian context.  We also show that, at this post-Newtonian order, the change in the linear momentum inside the RQF is trivial - that is, there is no measurable net flux of linear momentum through the RQF boundary.

In {\S}\ref{secTidalExamples} we specialize our post-Newtonian conservation laws to study the classic example of tidal interactions.  More precisely, we consider the spacetime of a body represented by a general multipole expansion sitting in a weak external field and derive formulas that allow one to analyze the work done by these tidal interactions as well as the transfer of angular momentum due to tidal torques.  We then use these equations to analyze two examples of tidal interactions within the familiar arena of our solar system.  Firth, by putting an RQF around Jupiter's moon Io, we compute the thermal energy transferred to Io due to the tidal forces of Jupiter.  Next, we calculate the tidal torque that the Moon exerts on the Earth.  This torque mines the Earth of angular momentum and causes the Moon to recede in its orbit.  In both case, we actually compute numerical values for these effects and in both cases we find solid agreement with observation.  It is important to keep in mind that while some of these results have been previously obtained in other contexts, the RQF approach we employ does not rely on working with pseudotensor expressions thereby providing a qualitatively more useful picture than the traditional methods.  Lastly, in {\S}\ref{secSummaryPN}, we summarize and see that the RQF approach provides a natural formalism that allows one to start from first principles (i.e., completely general quasilocal conservation laws) and, very cleanly, arrive at practical results.

\section{Post-Newtonian Expansion of the RQF Conservation Laws}\label{secPNRQFLaws}

Let us begin with the general conservation law found by integrating equation (\ref{differential_quasilocal_conservation_law}) over the worldtube of our quasilocal observers:
\begin{equation} \label{integrated_quasilocal_conservation_lawPN}
\frac{1}{c} \int\limits_{\mathcal{S}_f - \mathcal{S}_i}  d{\mathcal{S}}\,  T_{\mathcal B}^{ab} u^{\mathcal{S}}_a \psi_b = \int\limits_{\Delta\mathcal{B}}  d \mathcal{B} \, \left[ \frac{1}{\kappa} G^{ab} n_a \psi_b - T_{\mathcal B}^{ab} D_{(a} \psi_{b)} \right].
\end{equation}
As we have seen, appropriate choices of $\psi^a$ will lead to conservation laws for either energy, momentum, or angular momentum.  In particular, if we take $\psi^a = u^a$ and decompose $T_{\mathcal B}^{ab}$ according to (\ref{SurfaceSEMcomponents}) we arrive at the following energy conservation law for an RQF (equation (\ref{eq:SimpleConservationEq})):
\begin{align}\label{EnergyGCLPN}
\int\limits_{\mathcal{S}_f - \mathcal{S}_i}  d\hat{\mathcal{S}} \, & \left[ \mathcal{E} - \mathcal{P}^a v_a \right]  = - \int\limits_{\Delta{\mathcal B}}  N \, dt \, d\hat{\mathcal{S}}\, \left[  - T^{\mathtt{mat}}_{ab} u^a n^b + \alpha_a \mathcal{P}^a  \right].
\end{align}
Similarly, taking instead $\psi^a = \phi^a$, where $\phi^a$ can be either a boost or rotation CKV, yields a linear or angular momentum conservation law for an RQF (see equation (\ref{explicit_integrated_quasilocal_conservation_law})):
\begin{align}\label{MomentumGCLPN}
\int\limits_{\mathcal{S}_f - \mathcal{S}_i}  d\hat{\mathcal{S}} \, & \left[   \phi_a \left( \mathcal{P}^a + \frac{1}{c^2}\mathcal{S}^{ab} v_b\right) \right] = \qquad \nonumber \\
 & \qquad - \int\limits_{\Delta{\mathcal B}}  N \, dt \, d \hat{\mathcal{S}}\, \left[  T^{\mathtt{mat}}_{ab} \phi^a n^b  + \frac{1}{c^2} \mathcal{E} \alpha_a \phi^a + 2 \nu \varepsilon_{ab} \mathcal{P}^a \phi^b - {\rm P} \mathcal{D}_{a} \phi^{a}  \right].
\end{align}
The value of these equations, (\ref{EnergyGCLPN}) and (\ref{MomentumGCLPN}), is best seen by contrasting them with equations (2.2) in reference \cite{HT1985}
which are the basis of a similar perturbative analysis but using instead the traditional pseudotensor approach.  In doing this, one sees several advantages to the quasilocal approach: (1) In the traditional approach, one requires asymptotic flatness of the spacetime to have a well-defined mass, momentum, and angular momentum.  On the other hand, as seen in previous chapters, equations (\ref{EnergyGCLPN}) and (\ref{MomentumGCLPN}) are valid regardless without the need for any spacetime symmetries because they instead make use of the omnipresent RQF observers' four-velocity, boost CKVs, and rotation CKVs to define energy, momentum, and angular momentum respectively. (2) The quasilocal approach does not make use of pseudotensors.  All of the objects in the quasilocal conservation laws are tensors so, while they are given here in the RQF gauge, the can easily be transformed into any gauge in a straightforward manner.  (3) The quasilocal conservation laws are made up of pieces which have simple and useful physical interpretations.  For example, the fluxes which affect the energy of the system are the usual matter fluxes, $- T^{\mathtt{mat}}_{ab} u^a n^b$, plus the fluxes of gravitational energy encoded in the motions the RQF observers have to undergo to maintain rigidity, $\alpha_a \mathcal{P}^a$ - in essence, a gravitational Poynting vector.  In the pseudotensor approach, on the other hand, you have the pseudotensor ``potential'', $H^{abcd}$ - which, at best, can be described as related to a ``sort of'' effective SEM tensor.  However, given that the local SEM tensor is unable to accurately describe gravity one cannot hope for any {\it meaningful} interpretation.  Indeed, in practice, $H^{abcd}$ is simply taken to be a quadratic combination of metric density elements.  One simply casts aside any hope of physical interpretation in the early stages of a calculation.

We would now like to analyze the general conservation laws above in the post-Newtonian approximation.  To do this, we will need to find the metric for an RQF embedded in the standard post-Newtonian spacetime.  First, however, let us quickly review the set up for the post-Newtonian approximation.   Recall that, in the post-Newtonian scheme, we assume that we are dealing with non-relativistic systems that are bound by weak mutual gravitational attraction amongst constituent particles so that kinetic energies are comparable to gravitational potential energies.  This allows us to expand metric quantities in terms of a dimensionless parameter $\epsilon \sim V/c \sim \sqrt{GM/c^2R}$ where $V$, $M$, and $R$ are typical velocities, masses, and separation distances respectively of the particles comprising the system under study.  This expansion leads to the post-Newtonian metric, which can be found in many standard textbooks on general relativity (see \cite{Weinberg} for example).  In pseudo-Cartesian coordinates $X^A = (X^0=cT, X^I)$, $I=1,2,3$, this metric is given by\footnote{Note that, technically, we are not using the full first post-Newtonian approximation. We are dropping the 1PN terms in the space-space components of the metric. This tremendously simplifies the analysis, and is sufficient to obtain useful results in the case of energy and angular momentum conservation.}
\begin{align}\label{WeinbergMetric}
g_{00} &= -1 - \frac{2\Phi}{c^2}  - \left(\frac{2\Phi^2}{c^4} + \frac{2\Psi}{c^4} \right) + \mathcal{O}(\epsilon^6), \nonumber \\
g_{0J} &= \frac{\zeta_J}{c^3} + \mathcal{O}(\epsilon^5), \nonumber \\
g_{IJ} &= \delta_{IJ} \left(1 - \frac{2 \Phi}{c^2}\right) + \mathcal{O}(\epsilon^4)
\end{align}
where $\Phi/c^2 \sim \mathcal{O}(\epsilon^2)$, $\zeta_J/c^3 \sim \mathcal{O}(\epsilon^3)$, and $\Psi/c^4 \sim \mathcal{O}(\epsilon^4)$ can be functions of all of the coordinates $X^A$.

To move to the RQF frame, with adapted coordinates $x^\alpha = (t,r,x^i = (\theta, \phi))$, we apply the transformation
\begin{align} \label{PNTransformation}
cT &= ct + \on{1}{f^0} +  \on{3}{f^0} +   \mathcal{O}(\epsilon^5), \nonumber \\
X^I &= r r^I + \on{2}{f^I} +   \mathcal{O}(\epsilon^4),
\end{align}
where $r^I = (\sin \theta\, \cos \phi  , \,\sin \theta \, \sin \phi , \,\cos \theta)$ are the usual direction cosines. The number $n$ above the functions $\on{n}{f^A}$ denotes the order in $\epsilon$ of that function.  Note that the transformation of the time coordinate involves only odd powers of $\epsilon$ because it must change sign under time-reversal, whereas the spatial transformation has only even powers to keep the same sign under time-reversal. The functions $\on{n}{f^0}$ allow for an arbitrary infinitesimal perturbation in the time foliation, while the three sets of functions $\on{n}{f^I}$ introduce enough freedom in spatial perturbations of the coordinate embedding to satisfy the three RQF conditions.

Before we actually solve the RQF equations and give the full metric resulting from the transformation above, let us make an observation that will simplify the end result.  Following the transformation (\ref{PNTransformation}), one finds $\on{1}g_{tr} = -c \partial_{r}\on{1}{f^0}$ and $\on{1}g_{tj} = - c \partial_{j}\on{1}{f^0}$.  However, for a general post-Newtonian RQF, i.e., equation (\ref{WeinbergMetric}), these metric components vanish at this order.  This is a result of the fact that, at zeroth order in $\epsilon$, the standard post-Newtonian spacetime has zero acceleration and rotation.
Thus we must take
\begin{align}
\partial_r \on{1}{f^0} = 0, \quad \partial_j \on{1}{f^0} = 0.
\end{align}
In other words, $\on{1}{f^0}$ can only have time dependence. With this simplification, the three RQF conditions in equation~(\ref{eq:RigidityCondition}) can be shown to reduce to
\begin{align}\label{PNRQFEquation}
0 = 2r^2 \left( \frac{1}{r}\mathbb{B}_{I(i}\mathbb{D}_{j)}\on{2}{f^I} - \frac{\Phi}{c^2} \mathbb{S}_{ij} \right) + \mathcal{O}(\epsilon^4),
\end{align}
where we have introduced the covariant derivative operator, $\mathbb{D}_i$, associated with the unit two-sphere metric $\mathbb{S}_{ij}$, as well as the three boost CKVs $\mathbb{B}^I_i = \mathbb{D}_i r^I$.  It is straightforward to see that a particular solution to these three differential equations is given by $\on{2}{f^I_{\rm p}} = r r^I \Phi/c^2$.  We are also free to add to it the general homogeneous solution, $\on{2}{f^I_{\rm h}} = \alpha^I (t,r) + \epsilon^{I}_{\phantom{I}JK} r^J \beta^K (t,r)$, where $\epsilon_{IJK}$ is the alternating symbol.  Here, $\alpha^I (t,r)$ and $\beta^I (t,r)$ are six arbitrary functions of time (for a given $r$) that impart arbitrary $\ell=1$ acceleration and rotation to the RQF. They are the quasilocal analogues of the six degrees of freedom of arbitrarily accelerating and rotating rigid frames we are familiar with in Newtonian mechanics. However, in keeping with the fact that, at zeroth order in $\epsilon$, the standard post-Newtonian spacetime is inertial, we will suppress this freedom and take
\begin{equation}\label{RQFSolution}
\on{2}{f^I} = r r^I \frac{\Phi}{c^2},
\end{equation}
but will return to this point in {\S}\ref{secTidalExamples}. In this equation, $\Phi$ is a function of the RQF coordinates through the zeroth order version of equation~(\ref{PNTransformation}), i.e., $T=t$ and $X^I=rr^I$.

It is now straightforward to show that the metric for an RQF embedded in the post-Newtonian spacetime given in equation~(\ref{WeinbergMetric}) is
\begin{align} \label{RQFWeinberg}
g_{tt} &= -c^2 -2 \bigg[ \Phi + c \dot{\on{1}{f^0}} \bigg] \nonumber \\
& \qquad - 2 \bigg[ \frac{\Psi}{c^2} + \frac{\Phi^2}{c^2} + \frac{r}{2 c^2} (\Phi^2)^\prime + \frac{2}{c} \Phi (\dot{\on{1}{f^0}}) + \frac{1}{c} \on{1}{f^0} \dot{\Phi}  + \frac{1}{2}(\dot{\on{1}{f^0}})^2 + c(\dot{\on{3}{f^0}}) \bigg]  + \mathcal{O}(\epsilon^6), \nonumber \\
g_{tr} &= \bigg[ \frac{\zeta}{c^2} + \frac{r}{c^2} \dot{\Phi} - c \on{3}{f^{0}}^\prime \bigg] + \mathcal{O}(\epsilon^5), \nonumber \\
g_{tj} &= \bigg[ \frac{r}{c^2} \zeta_j  - c \mathbb{D}_j \on{3}{f^0} \bigg] + \mathcal{O}(\epsilon^5), \nonumber \\
g_{rr} &= 1 + \big[ \frac{2 r}{c^2} \Phi^\prime \big] + \mathcal{O}(\epsilon^4), \nonumber \\
g_{rj} &= \big[ \frac{r}{c^2} \mathbb{D}_j \Phi  \big] + \mathcal{O}(\epsilon^4), \nonumber \\
g_{ij} &= r^2 \mathbb{S}_{ij} + \mathcal{O}(\epsilon^4)
\end{align} 
where $\Phi$, $\zeta_J$, and $\Psi$ are now functions of the RQF coordinates $x^\alpha$ through the zeroth order version of equation~(\ref{PNTransformation}), and we have adopted a simplified notation of denoting radial derivatives with a prime, $\partial_r f  = f^\prime$, and time derivatives with a dot, $\partial_t f = \dot{f}$.  It is also important here to remember that time derivatives carry an order in $\epsilon$ since $\frac{1}{c}\frac{\partial}{\partial t} \sim \frac{v}{c} \frac{\partial}{\partial x^I} \sim \epsilon$.  For convenience, we have also decomposed $\zeta_J$ into a radial part, $\zeta = r^I \zeta_I$, and a part tangential to the RQF two-sphere, $\zeta_i := \mathbb{B}^I_i \zeta_I$.

It will be useful here to collect a few results that will recur throughout our analysis of the conservation laws. First, from the metric above, it is straightforward to write down the shift covector for the RQF observers,
\begin{align}\label{VelocityPN}
u_j = \frac{1}{N} g_{tj} = c \left[  r \frac{ \zeta_j}{c^3}  - \mathbb{D}_j \on{3}{f^0} \right] + \mathcal{O}(\epsilon^5).
\end{align}
Using the shift covector and equation (\ref{eq:ObserversAcceleration}) we can compute the tangential acceleration that the RQF observers must undergo to maintain rigidity,
\begin{align}\label{AccelPN}
\alpha_j = \bigg[ \mathbb{D}_j \Phi \bigg] + \bigg[ \frac{r}{2c^2} \mathbb{D}_j (\Phi^2)^\prime + \frac{r}{c^2} \dot{\zeta}_j + \frac{1}{c^2} \mathbb{D}_j \Psi + \frac{1}{c} \on{1}{f^0} \mathbb{D}_j \dot{\Phi} ) \bigg] + \mathcal{O}(\epsilon^6),
\end{align}
A lengthy calculation gives the quasilocal momentum density of the RQF,
\begin{align}\label{QuasilocalMomentum}
\mathcal{P}_j = \frac{1}{c^4 \kappa} \bigg[ \frac{1}{2} (r\zeta_j)^\prime + \frac{1}{4} \mathbb{D}_j \left( r \zeta^\prime + \mathbb{D}^k \zeta_k \right) \bigg] + \mathcal{O}(\epsilon^3).
\end{align}
Note that $c^4 \kappa \sim G$, so it is easy to see that $|\mathcal{P}|\times {\rm Area} \sim MV\epsilon$ at lowest order, where $MV$ represents a typical momentum in the system.

In {\S} \ref{secCurved}, we found that the equivalence principle can be used to relate the acceleration, twist, and quasilocal momentum to effective gravitoelectromagnetic (GEM) fields.  Let us make use of that idea now; the reasons for doing this will become apparent later on when we look at the flux of gravitational energy.  Thus we define the GEM potentials
\begin{align}
\phi^{\mathtt{GEM}} &:= \Phi +\left[\frac{r}{2c^2} (\Phi^2)^\prime +\frac{1}{c^2} \Psi + \frac{1}{c} \on{1}{f^0} \dot{\Phi} \right]  + \mathcal{O}(\epsilon^6), \nonumber \\
A_I^{\mathtt{GEM}} &:= - \frac{1}{4c} (r \zeta^\prime + \mathbb{D}^k \zeta_k ) r_I + \frac{1}{2c} \zeta_i \mathbb{B}_I^{i} + \mathcal{O}(\epsilon^5).
\end{align}
The acceleration (\ref{AccelPN}) can then be identified with a gravitoelectric field projected onto the RQF surface by the relation (see page \pageref{GEMDiscussion})
\begin{align}\label{GravitoelectricPN}
e^{\mathtt{GEM}}_j := \mathbb{B}^J_j E^{\mathtt{GEM}}_J  = - \mathbb{D}_j \phi^{\mathtt{GEM}} - \mathbb{B}^J_j \dot{A}_J^{\mathtt{GEM}}  = - \alpha_j.
\end{align}
Similarly, the quasilocal momentum density can be identified with the tangential part of the gravitomagnetic field via
\begin{align}\label{GravitomagneticPN}
b^{\mathtt{GEM}}_j = \mathbb{B}^J_j B^{\mathtt{GEM}}_J  = \mathbb{B}^J_j \epsilon_J^{\phantom{J}KL} \partial_K A_L^{\mathtt{GEM}}  = - \frac{c^3 \kappa}{r} \mathbb{E}_j^{\phantom{j}k} \mathcal{P}_k ,
\end{align}
where $\partial_I$ denotes differentiation with respect to $x^I = r r^I$, and $\mathbb{E}_{ij}$ is the volume element associated with the metric $\mathbb{S}_{ij}$.  Lastly, it is worth noting that the twist of the congruence of RQF observers, see equation~(\ref{eq:nu}), is related to the radial part of the gravitomagnetic field:
\begin{align}\label{TwistPN}
r^J B^{\mathtt{GEM}}_J = \frac{1}{2 c r} \mathbb{E}^{ij} \mathbb{D}_i \zeta_j + \mathcal{O}(\epsilon^5)=c\nu .
\end{align}
The twist starts at order $\epsilon^3$ because we have followed the usual post-Newtonian approach and assumed that, at zeroth order in $\epsilon$, the spacetime is inertial. Thus, it is only non-zero once the effects of rotational frame-dragging arise.

Finally, one can show that the quasilocal energy density is given by
\begin{align}\label{QuasilocalEnergy}
\mathcal{E} = \mathcal{E}_{\mathtt vac} + \frac{1}{c^2 \kappa} \bigg[  2 \Phi^\prime +\frac{1}{r} \mathbb{D}^2 \Phi \bigg] + \mathcal{O}(\epsilon^2)
\end{align}
with $\mathbb{D}^2 := S^{ij} \mathbb{D}_i  \mathbb{D}_j$.  Recall, $\mathcal{E}_{\mathtt vac} := - \frac{2}{\kappa r}$ is the quasilocal vacuum gravitational energy density and is of order $\epsilon^{-2}$.  This energy density, in general, will involve vacuum, matter, and gravitational contributions.  However, at these low orders, contributions due to gravitational effects do not show up.  The remaining two terms are thus regular matter contributions beginning at zeroth order in $\epsilon$ and, for later convenience, we define
\begin{align}\label{EMatter}
\mathcal{E}_{\mathtt mat} := \frac{1}{c^2 \kappa} \big[  2 \Phi^\prime +\frac{1}{r} \mathbb{D}^2 \Phi \big] + \mathcal{O}(\epsilon^2).
\end{align}
Notice that, for the simple gravitational potential $\Phi = - G M / r$ for a mass $M$, this gives $\mathcal{E}_{\mathtt mat} = M c^2 / 4 \pi r^2$.  Integrated over the spherical surface of the RQF, this quasilocal energy density gives a total matter energy $Mc^2$ just as one would expect.  

Working with the metric (\ref{RQFWeinberg}) above, we can now evaluate equations (\ref{EnergyGCLPN}) and (\ref{MomentumGCLPN}) to get conservation laws for energy, momentum, and angular momentum in the post-Newtonian limit.   

\subsubsection*{Energy}

The integral on the left-hand side of equation~(\ref{EnergyGCLPN}) gives the change in the energy inside the RQF between the surfaces of simultaneity $\mathcal{S}_i$ and $\mathcal{S}_f$, including the term $-\mathcal{P}^a v_a$ required to adjust for the motion of the RQF observers relative to $\mathcal{S}_i$ and $\mathcal{S}_f$. However, inspection of equations~(\ref{VelocityPN}) and (\ref{QuasilocalMomentum}), and the fact that $v_i=-u_i$, reveals that $\mathcal{P}^a v_a \sim \mathcal{O}(\epsilon^4)$. On the other hand, $\mathcal{E}$ is only known at vacuum and zeroth order in $\epsilon$---see  equation~(\ref{QuasilocalEnergy}). Thus, the left-hand side of the energy conservation law involves only the lowest order matter contribution:
\begin{align}
\Delta\mathrm{E}_{\mathtt{RQF}} = \int\limits_{\mathcal{S}_f - \mathcal{S}_i}  d\hat{\mathcal{S}} \, \left[ \mathcal{E} - \mathcal{P}^a v_a \right] = \int\limits_{\mathcal{S}_f - \mathcal{S}_i}  d\hat{\mathcal{S}} \, \left[ \mathcal{E}_{\mathtt mat} + \mathcal{O} (\epsilon^2) \right],
\end{align}
where $\mathcal{E}_{\mathtt mat}$ is given in equation~(\ref{EMatter}) and the vacuum contributions on $\mathcal{S}_i$ and $\mathcal{S}_f$ cancel out. Thus, to the order we are working, we cannot use the left-hand side of the energy conservation law to determine $\Delta\mathrm{E}_{\mathtt{RQF}}$ at order $\epsilon^2$, e.g., changes in the Newtonian kinetic energy of masses in motion inside the RQF. However, we can obtain such $\epsilon^2$ information from the right-hand side, as we shall now see.

On the right-hand side of equation~(\ref{EnergyGCLPN}), the first term represents the matter energy flux and the second the gravitational energy flux. The former can be used to compute $\Delta\mathrm{E}_{\mathtt{RQF}}$ when, e.g., a particle enters or leaves the RQF sphere. However, since our primary interest is the gravitational energy flux we will set the matter energy flux to zero (i.e., $ T^{ab} n_a u_b = 0$) at the RQF surface. This leaves just the gravitational energy flux term, $- \alpha_a \mathcal{P}^a $.  The presence of the lapse function accounts for time-dilation across the system (as discussed after equation (\ref{RQFPoyntingDensity})), but since the quasilocal momentum, $\mathcal{P}^a$, is known only at order $\epsilon$ in our post-Newtonian approximation, the lapse function can be ignored. Hence, the (outward) gravitational energy flux is represented by
\begin{align} \label{adotPPN}
\alpha_i \mathcal{P}^i = \frac{\mathbb{S}^{ij}}{ c^4 \kappa r^2} (\mathbb{D}_i \Phi) \left( \frac{1}{2} \left( r \zeta_j \right)^\prime + \frac{1}{4} \mathbb{D}_j \left( r \zeta^\prime + \mathbb{D}^k \zeta_k \right)   \right) + \mathcal{O}(\epsilon^5).
\end{align}
In this form it is difficult to argue that this is what one should expect for the flux of gravitational energy.  However, it becomes clear that this is a sensible result by using our GEM fields,equations~(\ref{GravitoelectricPN}) and ({\ref{GravitomagneticPN}), to calculate the GEM Poynting flux normal to the surface of the RQF sphere for comparison:
\begin{align} \label{GEMtoadotP}
r^I S^{\mathtt{GEM}}_I  = \frac{1}{c^3 \kappa} r^I \epsilon_{I}^{\phantom{I}JK} E^{\mathtt{GEM}}_J B^{\mathtt{GEM}}_K = \frac{1}{c^3 \kappa} \mathbb{E}^{ij} \, e^{\mathtt{GEM}}_i \, b^{\mathtt{GEM}}_j = \alpha_i \mathcal{P}^i + \mathcal{O}(\epsilon^5).
\end{align}
It is satisfying to see that, at leading order, our gravitational energy flux is really just the radial component of the GEM Poynting flux. It should be pointed out however that, while this is a useful tool for qualitatively understanding a cumbersome expression like equation~(\ref{adotPPN}), the GEM analogy quickly breaks down as a means of quantifying the flow of gravitational energy beyond leading order, and one should not hope to satisfy equation~(\ref{GEMtoadotP}) at higher orders  (for more details refer to the discussion on page \pageref{GEMFailureDiscussion}).  On the other hand, as we saw in Chapters \ref{chNova} and \ref{chLocalQuasilocal}, $\alpha_a \mathcal{P}^a$ {\it will} continue to capture gravitational energy flow accurately at higher orders and so must, in fact, be {\it exactly} the gravitational Poynting vector.

In summary, as $\Delta t=t_f -t_i\rightarrow 0$, the completely general RQF energy conservation law in equation~(\ref{EnergyGCLPN}) reduces, in our post-Newtonian approximation, to
\begin{align}\label{EnergyRate}
\frac{d\mathrm{E}_{\mathtt{RQF}}}{dt} =  \frac{1}{c^4 \kappa} \int \limits_{\mathcal{S}_t}  d\mathbb{S}\, \Phi \bigg[ \frac{1}{2} (r \mathbb{D}^k \zeta_k)^\prime + \frac{1}{4} \mathbb{D}^2 \left( r \zeta^\prime + \mathbb{D}^k \zeta_k \right)  \bigg] + \mathcal{O}(\epsilon^5)
\end{align}
where we have used equation~(\ref{adotPPN}) and integrated by parts, and $d\Omega = \sin \theta \, d\theta \, d\phi$ is the surface element on a unit round sphere in standard spherical coordinates.

\subsubsection*{Linear Momentum}

To obtain a linear momentum conservation law we choose $\phi^a$ in equation~(\ref{MomentumGCLPN}) to be a boost CKV, which in the RQF coordinate system means taking $\phi^i = \frac{1}{r} \mathbb{B}^{Ii}$ . Here $I=1,2,3$ corresponds to a boost in the $X^I$ direction. The left-hand side of equation~(\ref{MomentumGCLPN}) then gives the change in the corresponding $\ell=1$ spherical harmonic component of the linear momentum inside the RQF between the surfaces of simultaneity $\mathcal{S}_i$ and $\mathcal{S}_f$, including the term $\frac{1}{c^2}\mathcal{S}^{ab} \phi_a v_b $ required to adjust for the motion of the RQF observers relative to $\mathcal{S}_i$ and $\mathcal{S}_f$:
\begin{align} \label{RQFMomentumDef}
\Delta\mathrm{P}^I_{\mathtt{RQF}} = \int\limits_{\mathcal{S}_f - \mathcal{S}_i}  d\hat{\mathcal{S}} \left[   \frac{\mathbb{B}^{Ii}}{r} \left( \mathcal{P}_i + \frac{1}{c^2}\mathcal{S}_{ij} v^j \right) \right].
\end{align}
Recall from equation~(\ref{QuasilocalMomentum}) that $\mathcal{P}_i$ (which is due to frame-dragging) begins at order $\epsilon$. Based on what we saw in the energy case, one might expect the relativistic stress term $\frac{1}{c^2}\mathcal{S}_{ij} v^j$ to be higher order, and thus negligible in our post-Newtonian approximation, but this turns out not to be the case because the quasilocal stress (in particular, the pressure) has a leading order vacuum term at order $\epsilon^{-2}$,
\begin{align}\label{QuasilocalStress}
\mathcal{S}_{ij} = - \frac{r}{\kappa} \mathbb{S}_{ij} - \frac{r}{c^2 \kappa} \bigg[ \big( \mathbb{D}_{(i} \mathbb{D}_{j)} - \mathbb{S}_{ij} \mathbb{D}^2 \big) \Phi \bigg] + \mathcal{O}(\epsilon^2),
\end{align}
while $v^j$ is of order $\epsilon^3$ (recall that $v_i = - u_i$). Therefore, the relativistic stress term actually contributes at the same order as $\mathcal{P}_i$ to the left-hand side of
equation~(\ref{MomentumGCLPN})---in general, both pieces are needed to account for the linear momentum measured by the RQF observers. Evaluating the integrand in equation~(\ref{RQFMomentumDef}) and integrating over $\mathcal{S}_t$ determines the linear momentum inside the RQF at time $t$:
\begin{align}\label{PNMomentumLHS}
\mathrm{P}^I_{\mathtt{RQF}} = \frac{r}{c^4 \kappa} \int\limits_{\mathcal{S}_t}  d\Omega \left[ \frac{r}{2} ( \zeta^{I})^\prime + \mathbb{B}^{I}_i \zeta^i - \frac{2 c^3}{r} r^I \on{3}{f^0} \right] + \mathcal{O}(\epsilon^3)
\end{align}
As a quick check of this equation, we imagine a Newtonian particle of mass $M$ moving with constant velocity $V$ through an RQF sphere. In the simplest case that we choose $\on{3}{f^0}$ in equation (\ref{VelocityPN}) such that $u_j =0$, it is easy to show that (the appropriate component of) $\mathrm{P}^I_{\mathtt{RQF}}$ equals $MV$ precisely when the particle is inside the RQF, and zero when it is outside.

However, our primary interest is in the right-hand side of equation~(\ref{MomentumGCLPN}), which has four terms representing fluxes of linear momentum. Apart from the matter linear momentum flux term, which we will turn off at the boundary of the RQF, it turns out that the dominant gravitational linear momentum flux term is the one involving the quasilocal pressure, $- {\rm P} \hat{D}_a \phi^a $.  From equation~(\ref{QuasilocalStress}) we find that this pressure is
\begin{align}
\mathrm{P} = \frac{1}{2}\sigma^{ij}\mathcal{S}_{ij} = - \frac{1}{\kappa r} + \frac{1}{2c^2 \kappa r} \mathbb{D}^2 \Phi + \mathcal{O}(\epsilon^2).
\end{align}
In our post-Newtonian approximation, this pressure cannot be evaluated at order $\epsilon^2$ or higher. This in turn renders any information about the other fluxes at order $\epsilon^2$ and higher inconsequential since we cannot construct a complete picture of all of the fluxes. Unfortunately, one does not encounter non-zero net fluxes below order $\epsilon^2$.  We can see this by noting that, after integrating over time, which decreases the order in $\epsilon$ by one, the right-hand side of equation~(\ref{MomentumGCLPN}) can be evaluated at best at orders $\epsilon^{-3}$ and $\epsilon^{-1}$.  However, from equation~(\ref{PNMomentumLHS}), we already know that the left-hand side vanishes at these orders. This means that the fluxes on the right-hand side of equation~(\ref{MomentumGCLPN}) must integrate identically to zero in our post-Newtonian approximation. It is a straightforward calculation to verify this; we omit the calculation for the sake of brevity.

In summary, taking a time derivative of the linear momentum conservation law (\ref{MomentumGCLPN}) yields
\begin{align}\label{MomentumRate}
\frac{d\mathrm{P}^I_{\mathtt{RQF}}}{dt} = \mathcal{O}(\epsilon^2).
\end{align}
In other words, at this order in the post-Newtonian approximation, we cannot compute the rate of change of the RQF momentum.  This result may seem troublesome.  For example, if we consider the simple case of a particle with mass $m$ and velocity $v$ passing through our RQF and we choose $\on{3}{f^0}$ so that $u_j$ vanishes, then the momentum inside the RQF (i.e., the surface integral of $\phi^i \mathcal{P}_i$ as given by equation (\ref{RQFMomentumDef})) can be shown to be $mv + \mathcal{O}(\epsilon^3)$ when the particle is inside the RQF and zero otherwise.  Changes in the momentum should then be of order $\epsilon^2$.  What this is telling us is that, despite $\zeta_J$ (the relevant parameter in our metric) being of order $\epsilon^3$, we cannot compute this order $\epsilon^2$ effect.  This null result is unfortunate but actually not unexpected.  On page 78 of \cite{Wald1984}, Wald explains that in order to compute the acceleration of a test mass in linearized gravity, one makes use of the geodesic equation which is actually trivial in linearized gravity.  The lesson is that it is standard, when analyzing Einstein's equations by perturbing around a flat background, that to find results at a particular order you may have to do certain elements of the calculation a higher order.  The lesson is that, when analyzing Einstein's equations by perturbing around a flat background, it is standard to have to do certain elements of the calculation at higher order than desired to be able to find results at a particular lower order.

\subsubsection*{Angular Momentum}

To obtain an angular momentum conservation law we choose $\phi^a$ in equation~(\ref{MomentumGCLPN}) to be a rotation CKV, which in the RQF coordinate system means taking
$\phi^i = \mathbb{R}^{Ii} = \mathbb{E}^{i}_{\phantom{i}j}\mathbb{B}^{Ij}$. Here $I=1,2,3$ corresponds to a rotation about the $X^I$ axis. The left-hand side of equation~(\ref{MomentumGCLPN}) then gives the change in the corresponding $\ell=1$ spherical harmonic component of the angular momentum inside the RQF between the surfaces of simultaneity $\mathcal{S}_i$ and $\mathcal{S}_f$, including, as in the linear momentum case, the relativistic stress term:
\begin{align}\label{AngularMomentumPNCLLHS}
\Delta\mathrm{J}^I_{\mathtt{RQF}} = \int\limits_{\mathcal{S}_f - \mathcal{S}_i}  d\hat{\mathcal{S}}\, \left[   \mathbb{R}^{Ii} \left( \mathcal{P}_i + \frac{1}{c^2}\mathcal{S}_{ij} v^j \right) \right].
\end{align}
For the same reason as in the linear momentum case, we will, in general, need both terms in the integrand to compute the change in angular momentum inside the RQF.

On the right-hand side of equation~(\ref{MomentumGCLPN}), notice that, since the divergence of a rotation CKV is zero, the previously dominant flux, $- {\rm P} \hat{D}_a \phi^a $, is identically zero here and thus the angular momentum conservation law will contain more physics at this post-Newtonian order than the linear momentum law above. Taking the matter angular momentum flux term $-T^{ab} n_a \phi_b$ to be zero at the surface of the RQF leaves just two flux terms on the right-hand side of the conservation law. The first is the gravitational angular momentum flux $-\frac{1}{c^2} \mathcal{E} \alpha^a \phi_a$, which can be calculated at orders unity and $\epsilon^2$, while the other flux, $2 \nu \epsilon^{ab} \phi_a \mathcal{P}_b$, represents a Coriolis effect that starts at order $\epsilon^4$, and so can be neglected. Hence, the RQF angular momentum conservation law in our post-Newtonian approximation reduces to
\begin{align}\label{AngularMomentumPNCL}
\frac{d\mathrm{J}^I_{\mathtt{RQF}}}{dt} = - & \int \limits_{\mathcal{S}_t} d\hat{\mathcal{S}} \, \left[ N \frac{1}{c^2} \mathcal{E} \alpha^i \mathbb{R}^{I}_{i} + \mathcal{O}(\epsilon^4) \right].
\end{align}
The (outward) gravitational angular momentum flux is found to be
\begin{align} \label{AngularMomentumFlux}
N \frac{1}{c^2} \mathcal{E} \alpha^i \mathbb{R}^I_i & = \mathbb{D}_i \bigg[ \frac{\mathcal{E}_{\mathtt{vac}}}{c^2} \mathbb{R}^{Ii} \big( \Phi +  \frac{1}{2c^2}(r\Phi^2)^\prime + \frac{1}{c^2}\Psi + \frac{1}{c} \partial_{t} (\Phi \on{1}{f^0}) \big) \bigg] \nonumber \\
\quad & + R^{Ii} \bigg[ \frac{r}{c^4} \mathcal{E}_{\mathtt{vac}} \dot{\zeta}_i + \frac{1}{c^2} \mathcal{E}_{\mathtt{mat}} \mathbb{D}_i \Phi \bigg] + \mathcal{O}(\epsilon^4).
\end{align}
The first group of terms in square brackets involves contributions at orders unity and $\epsilon^2$;  being a divergence, they will integrate to zero in equation (\ref{AngularMomentumPNCL}). Interestingly, this means that $\Psi$ does not show up in any of our post-Newtonian conservation laws, despite being necessary to compute them. This is an example of needing to work at a higher order during the intermediate steps of a perturbative calculation than is achieved in a final answer.  Also notice that the arbitrary time re-foliation parameter, $\on{1}{f^0}$, will thus not appear in the final result, leaving the integrated flux gauge-invariant. The physically relevant fluxes are thus contained in the second set of square brackets. Substituting the flux in equation~(\ref{AngularMomentumFlux}) into equation~(\ref{AngularMomentumPNCL}) and integrating by parts then gives the rate of change of angular momentum inside the RQF:
\begin{align}\label{AngularMomentumRate}
\frac{d\mathrm{J}^I_{\mathtt{RQF}}}{dt} = \frac{r^2}{c^4 \kappa } \int  d\Omega \,\, \mathbb{R}^{Ii} \, \bigg[ 2 \dot{\zeta}_i + \Phi \, \mathbb{D}_i \left( 2 \Phi^\prime + \frac{1}{r} \mathbb{D}^2 \Phi \right) \bigg] + \mathcal{O}(\epsilon^4).
\end{align}

Equations (\ref{EnergyRate}), (\ref{MomentumRate}), and (\ref{AngularMomentumRate}) for the rates of change of energy, momentum, and angular momentum respectively are the main result of this section.  Given a metric in standard post-Newtonian form (\ref{WeinbergMetric}) one can use these relations to immediately compute the rate of change of energy and angular momentum of an RQF system (with the rate of change of linear momentum not showing up at this post-Newtonian order).  Note that all of these rates are independent of the choice of time-foliation (i.e., these equations hold for arbitrary choice of the functions $\on{n}{f^0}$).  In order to appreciate the utility of these equations let us now use them to analyze tidally interacting systems.

\section{Application to Tidal Interactions}\label{secTidalExamples}

Tidal interactions have acted as a test bed for analyzing conservation laws in general relativity by multiple authors in the past few decades,  perhaps most notably by Hartle and Thorne in 1985 \cite{HT1985}.  In this section, we will first demonstrate the validity of the RQF approach by reproducing standard results for describing the transfer of energy and angular momentum via tidal interactions.  We will then show that these equations have straightforward and practical applications by looking at two examples of tidal interactions in the solar system; in particular, the tidal heating of Jupiter's satellite Io and the mining of Earth's angular momentum by the Moon.  It will be clear from the calculation that the RQF approach does not rely on working with pseudotensor expressions, thereby providing a qualitatively more useful picture than  traditional methods \cite{Purdue,Booth}.
  
To begin, let us consider the spacetime describing a body at rest with centre of mass at the origin immersed in the field of some arbitrary external body.  We can characterize the field of the internal body with a typical multipole expansion (see \cite{Weinberg} for example) where we denote its mass $M$, quadrupole moment $Q_{IJ}$, angular momentum $J_I$, and angular momentum current $K_{IJ}$.  Meanwhile, the external field is described by the electric and magnetic parts of the Weyl tensor: $E_{IJ} := C_{0 I 0 J} $ and $B_{IJ} := \frac{1}{2} \epsilon_{I}^{\phantom{I}KL} C_{0JKL}$ respectively.  In the de Donder gauge, the metric then takes the form of equation (\ref{WeinbergMetric}) with parameters \cite{Zhang}
\begin{align}\label{PhiZeta}
\Phi &= - \frac{GM}{r} -\frac{3 G}{2 r^3} Q_{KL} r^K r^L +\frac{c^2 }{2} r^2  E_{KL} r^K r^L, \nonumber \\
\zeta_J &= -\frac{2G}{r^2} \epsilon_{JKL} J^K r^L - \frac{4G}{r^3} \epsilon_{JKL} K^K_{\phantom{K}M} r^L r^M - \frac{2 c^3 }{3} r^2 \epsilon_{JKL} B^K_{\phantom{K}M} r^L r^M \nonumber \\ & \quad\quad -\frac{2G}{r^2} \dot{Q}_{JK} r^K - \frac{10}{21} c^2 r^3 \dot{E}_{KL} r_J r^K r^L + \frac{4}{21} c^2 r^3 \dot{E}_{JK} r^K.
\end{align}
Note that all of the rank two tensors in (\ref{PhiZeta}) are symmetric and trace-free.  Furthermore, the internal quadrupole moment is defined with the convention $Q_{IJ} := \int d^3 x \, \rho \left( x^I x^J - \frac{1}{3} r^2 \delta^{IJ} \right)$.

In this spacetime, we now embed an RQF enclosing and centred on the internal body, but not enclosing the external body (i.e., $ \mathcal{L} \ll r \ll \mathcal{R}$ where $\mathcal{L} \sim GM/c^2$ is the size of the internal body and $\mathcal{R}$ is the radius of curvature of the external field which is related to the Ricci scalar by $R \sim \mathcal{R}^{-2}$).   As promised above, we can now simply substitute the metric functions (\ref{PhiZeta}) into the conservation laws (\ref{EnergyRate}) and (\ref{AngularMomentumRate}).  Let us first look at the rate of change of energy.  After carrying out the integration we obtain
\begin{align}
\frac{d\mathrm{E}_{\mathtt{RQF}}}{dt} &= -\frac{c^2}{2} E_{IJ} \dot{Q}^{IJ} - \frac{1}{10} \frac{d}{dt}\big[ c^2 E_{IJ} Q^{IJ} -\frac{9G}{r^5} Q_{IJ} Q^{IJ} -\frac{c^4}{6G} r^5 E_{IJ} E^{IJ} \big] + \mathcal{O}(\epsilon^5). \label{TidalEnergyRate}
\end{align}
This equation can now be used to calculation the power transferred from the external field to the internal body via tidal interactions.  To do this though, we need to separate the secular variations in the energy inside the RQF (those that continuously accumulate over time) from the periodic ones which reset after a complete orbit.  Looking at the terms in square brackets in equation (\ref{TidalEnergyRate}) we see that it is made up of three contributions: first, the term proportional to $\frac{d}{dt}(E_{IJ} Q^{IJ})$ represents the rate of change of the interaction energy, the second term is due to the change of the internal body's own field at the RQF surface, and the last term characterizes the rate of change of the external field at the surface.  All three of these terms are periodic in nature \cite{Booth}.  This leaves just the first term on the right hand side of equation (\ref{TidalEnergyRate}) to describe the power due to tidal heating,
\begin{align}\label{TidalPower}
P_{\mathtt{tidal}} = -\frac{c^2}{2} E_{IJ} \dot{Q}^{IJ}.
\end{align}
It is this term that will be of interest in practical discussions regarding the transfer of thermal energy via tidal interactions as we will see in {\S}\ref{secsubSolarExamples} below.

It is worthwhile to take a moment to compare this analysis to that from the traditional pseudotensor approach (see Purdue, for example, \cite{Purdue}).  This involves analyzing an equation qualitatively identical to (\ref{TidalEnergyRate}) above, but the implications are very different.  In particular, in the pseudotensor approach one isolates the tidal heating, a gauge independent quantity, from the gauge dependent periodic fluctuations (as we essentially did above).  However, the coefficients of the terms in square brackets in equation (\ref{TidalEnergyRate}) will depend on how you choose to localize gravitational energy when you define your stress-energy-momentum pseudotensor.  In our approach, the final result does not necessitate localizing gravity a particular way.  For comparison though, it is interesting to observe that, in the language of Purdue \cite{Purdue}, a gauge choice which corresponds to the Landau-Lifshitz way of localizing gravitational energy (i.e, $\alpha = -3$ in her notation) matches the results of the RQF approach.

Consider next how the angular momentum inside the RQF changes.  For the tidal metric above, equation (\ref{AngularMomentumRate}) becomes, after integration,
\begin{align}\label{TidalAngularMomentumRate}
\frac{d\mathrm{J}^I_{\mathtt{RQF}}}{dt} &=   c^2 \epsilon^{IJK} E_{JL} Q_K^{\phantom{K}L} - \frac{4}{3} \dot{J}^I + \mathcal{O}(\epsilon^4).
\end{align}
Similar to the energy equation above, this equation gives the change in angular momentum inside the RQF system. The first term on the right hand side describes the physical torque associate with to tidal heating
\begin{align}\label{TidalTorque}
\tau^I_{\mathtt{tidal}} = c^2 \epsilon^{IJK} E_{JL} Q_K^{\phantom{K}L}.
\end{align}
This is the important piece for physical applications.  From equation (\ref{AngularMomentumFlux}) we can see the mechanism which leads to this torque - specifically, this flux comes from the term $ \frac{1}{c^2} \mathcal{E}_{\mathtt{mat}} \mathbb{R}^{Ii} \mathbb{D}_i \Phi$ where we recognize that $R^{Ii} \mathbb{D}_i \Phi$ is the component of the RQF observers' acceleration, $\alpha_{j}$, about the $x^I$-axis.  Therefore, the angular momentum of the system is increasing the same way it would in the linear case - the RQF observers are accelerating relative to a mass so, from their point of view, they see its momentum increase.  Only now the acceleration is in the angular direction so after integrating over the surface they will be taking a measure of the {\it angular} momentum.  Of course, both $\mathcal{E}_{\mathtt{mat}}$ and $\alpha_{j}$ depend on the quadrupole moment and gravitoelectric field so we could attempt to explain the quasilocal flux in terms of these quantities but such an explanation would be messy and rely on a Newtonian way of interpreting gravity as a force.  The general relativistic explanation as given above is fundamentally different and requires us to think non-locally but, at the end of the day, it is simpler.  Furthermore, it is very economical in the sense that it explains momentum and angular momentum transfer with a single mechanism.

The second flux on the right hand side of (\ref{TidalAngularMomentumRate}), $-\frac{4}{3} \dot{J}^I$, actually has an interesting interpretation as well.  As the rotation rate of the internal body changes, it drags the surrounding spacetime around with it.  In order to remain fixed relative to the distant stars, the RQF observers must then rotate in the opposite direction.  In particular, they must undergo an acceleration $\alpha_j = \frac{r}{c^2} \dot{\zeta}_j = -\frac{2G}{c^2r} \mathbb{R}^K_j \dot{J}_K$ (see equation (\ref{AccelPN})). Recalling that we have seen evidence that the gravitational vacuum is a {\it real} source of mass,  this implies that they are effectively accelerating along a surface with negative mass density $\rho_{\mathtt{vac}} = \frac{1}{c^2} \mathcal{E}_{\mathtt{vac}} = - \frac{c^2}{4\pi G r}$.  This leads to a perceived change in the angular momentum of the system much in the same way that accelerating linearly relative to an object at rest changes the momentum that the accelerating observer ascribes to that object.  If we look back at equation (\ref{AngularMomentumFlux}), we can see that it is this combination that is the origin of the flux $ R^{Ii} \frac{r}{c^4} \mathcal{E}_{\mathtt{vac}} \dot{\zeta}_i$.  As a verification of this argument, one can move to the frame that freely rotates with respect to the distant stars (i.e., the frame-dragged frame) and show that this frame is locally inertial - that is, there is no angular acceleration.  As a result, the rigid observers in this frame rotate {\it with} the gravitational vacuum and do not perceive any change in the angular momentum due to this effect; in their frame, the flux on the right hand side of (\ref{TidalAngularMomentumRate}) will simply be the tidal torque (\ref{TidalTorque}).

In order to compare our result to previous work, it is important to make the distinction here between the angular momentum inside the RQF, $\mathrm{J}^I_{\mathtt{RQF}}$, and the angular momentum of the internal body, $J^I$.  In \cite{HT1985}, Hartle and Thorne use a pseudotensor approach to derive the rate of change of the angular momentum of the internal body as solely arising from the tidal torque (\ref{TidalTorque}), $\dot{J}^I = c^2 \epsilon^{IJK} E_{JL} Q_K^{\phantom{K}L}$.  Since our conservation law yields the rate of change of the angular momentum inside the RQF it is not surprising, based on the argument above, that we find an additional flux.  However, it turns out that we can reproduce Hartle and Thorne's result by evaluating the left hand side of the conservation law (\ref{AngularMomentumPNCLLHS}).  Making use of (\ref{QuasilocalMomentum}) and (\ref{QuasilocalStress}) we find
\begin{align}\label{AngularMomentumLHS}
\oint\limits_{\mathcal{S}_t}  d\mathcal{S} \, \bigg[   \mathbb{R}^I_i \frac{d}{dt} \left( \mathcal{P}^i + \frac{1}{c^2}\mathcal{S}^{ij} v_j \right) \bigg]  = \frac{1}{2 c^4 \kappa} \oint d\mathbb{S} \, \mathbb{R}^{Ii} \,\big( r^3  \dot{\zeta}_i \big)^\prime = - \frac{1}{3} \dot{J}^I + \mathcal{O}(\epsilon^4).
\end{align}
The factor of $-\frac{1}{3}$ can be understood by noting that in the RQF frame the observers consider the angular momentum to be made up of two contributions.  First, there is the angular momentum of the internal body, $\mathrm{J}^I_{\mathtt{body}} = J^I$.  In addition to this, as we argued above, they also perceive the rotating gravitational vacuum to have an angular momentum $\mathrm{J}^I_{\mathtt{vac}} = -\frac{4}{3} J^I$.  Therefore, the total angular momentum that the RQF observers see is the sum of the two: $\mathrm{J}^I_{\mathtt{RQF}} = \mathrm{J}^I_{\mathtt{body}} + \mathrm{J}^I_{\mathtt{vac}} = -\frac{1}{3} J^I$.  Conveniently, we can now equate (\ref{TidalAngularMomentumRate}) and (\ref{AngularMomentumLHS}) to obtain
\begin{align}
\frac{d J^I}{dt} &=   c^2 \epsilon^{IJK} E_{JL} Q_K^{\phantom{K}L} + \mathcal{O}(\epsilon^4) \label{TidalAngularMomentumHT}.
\end{align}
in agreement with equation (3.17b) in \cite{HT1985}.\footnote{Note that in reference \cite{HT1985}, Hartle and Thorne perform a slightly different expansion than we do.  In particular, they treat the gravitoelectric and gravitomagnetic fields to be formally of the same order.  The same is done for the mass quadrupole momentum and angular momentum current.  However, in the post-Newtonian approximation, the gravitomagnetic and angular momentum current are each {\it smaller} by an order in $\epsilon$ than their counterparts.  As a result, a term of the form $\epsilon^{IJK} B_{JL} K_K^{\phantom{K}L}$ which appears in their equation (6.23) would be of order $\epsilon^6$ in our expansion.  This is why it does not appear in equation (\ref{TidalAngularMomentumHT}).}

Despite being able to accurately interpret equation (\ref{TidalAngularMomentumRate}) and isolate the physical tidal torque, one may find the appearance of the $-\frac{4}{3} \dot{J}^I$ term unappealing.  In fact, we are inclined to agree with this statement, but we would like to stress that this is not inherently the fault of taking a quasilocal approach or even making use of RQFs.  This extra term arises because we embedded our RQF in the standard post-Newtonian spacetime which, by convention, is non-rotating with respect to the distant stars.  However, at the surface of the RQF we have a non-zero twist (see equation (\ref{TwistPN})).  This means that our RQF is undergoing rotational frame-dragging in the static post-Newtonian background spacetime so this is, perhaps, not the most natural frame to work in.  If we instead moved to a locally non-rotating frame ($\nu = 0$), then the frame-dragging vanishes and one finds $\dot{\mathrm{J}}^I_{\mathtt{RQF}} = \dot{J}^I = c^2 \epsilon^{IJK} E_{JL} Q_K^{\phantom{K}L}$ - precisely as one would expect.  This is all easily accomplished using an RQF approach by either of the strategies explained after equation (\ref{TwistPN}) in {\S} \ref{secPNRQFLaws}.

\subsection{Examples: Solar System Dynamics} \label{secsubSolarExamples}

We will now look at two examples  within our solar system to test the utility of the results above.  Specifically, we want to test the formulas for tidal power (\ref{TidalPower}) and tidal torque (\ref{TidalTorque}).  It is well-known \cite{Lopes} that Jupiter's satellite Io is volcanically active and that this activity cannot be explained without taking into account the enormous tidal forces exerted on Io in its eccentric orbit around Jupiter.  This scenario is an ideal one to apply the equation for tidal work to quantify the amount of energy transferred and compare with observation.  Another well documented phenomenon is the recession of the Moon in its orbit around the Earth  \cite{StaceyDavis}.  This is because the tidal field of the Moon creates bulges on the Earth, but the rotation of the Earth causes these bulges to rotate ahead of the common axis joining the two bodies.  In turn, the Moon then pulls on the forward (closer) bulge more than it does on the trailing (farther) bulge.  This net torque, which we will calculate using our results above, converts rotational angular momentum of the Earth to the orbital angular momentum of the Earth-Moon system resulting in an increase in the orbital radius.
	
In both of these examples  it will be useful to employ a result of Love's \cite{Love} which will allow us to relate the quadrupole potential from the internal body, $\Phi_{\mathtt{quad}}$, to the the tidal potential of the external body, $\Phi_{\mathtt{tidal}}$.  The basic idea is that the quadrupole moment exists only because the squashed shape of the internal body is due to the external body's tidal potential.  As such, one should be able to relate the two potentials by a numerical factor, $k_2$, called the Love number, which characterizes how easily the internal body is deformed.  Specifically, the potentials should satisfy the relation \cite{StaceyDavis}
\begin{align}\label{LoveRelationBasic}
\Phi_{\mathtt{quad}}(t,\psi) = k_2 \frac{R_{\mathtt{int}}^5}{r^5} \Phi_{\mathtt{tidal}}(t-\tau,\psi-\delta)
\end{align}
where $R_{\mathtt{int}}$ is the radius of the internal body, $\psi$ is the angle between the common axis joining the two bodies and a point in space, $\delta$ is the angle that the quadrupole is carried ahead of the common axis due to the rotation of the internal body, and $\tau$ is the lag in the tide due to the finite time that it takes the internal body to deform.  For a tidally locked satellite like Io, the absence of rotation means that $\delta = 0$.  On the other hand, for orbits with negligible eccentricity like the Earth-Moon system, the shape of the body is constant in time and thus, not only is $\tau =0$, but the time-dependence in (\ref{LoveRelationBasic}) can be ignored altogether.

It will be useful to recast equation (\ref{LoveRelationBasic}) in terms of our notation - that is, in terms of the quadrupole moment tensor, $Q_{IJ}$, and external tidal field tensor, $E_{IJ}$.  If we momentarily define new spatial coordinates $\bar{X}^{\bar{I}}$ that are related to the standard $X^I$ by a rotation through an angle $\delta$ ahead of the common axis between the two bodies then we should be able to write $\bar{Q}_{\bar{I}\bar{J}}(t) = - \lambda k_2 E_{IJ}(t-\tau)$ where $\lambda$ is a dimensionful constant.  If we substitute this relation into the metric function for the full potential $\Phi$ in equation (\ref{PhiZeta}) it is straightforward to show that, in order to satisfy Love's original relation (\ref{LoveRelationBasic}), we must take $\lambda = \frac{1}{3} \frac{c^2}{G} R_{\mathtt{int}}^5$ which leads to the relation\footnote{Note that this relation differs from recent work on tidal effects in neutron stars (see references \cite{Damour} and \cite{Poisson} for example).  In these references, the term `Love number' is used to refer to the apsidal constant which actually differs from the standard Love number by a factor of two, $k_{2,\mathtt{Love}} = 2k_{2,\mathtt{apsidal}}$.  Here we choose to maintain Love's original definition of the Love number and that used by the geophysics community where one typically has to turn to find values of the Love number for bodies in our solar system.}
\begin{align}\label{LoveRelation}
\bar{Q}_{\bar{I}\bar{J}}(t) = - \frac{1}{3} \frac{c^2}{G} R_{\mathtt{int}}^5 k_2 E_{IJ}(t-\tau).
\end{align}

\subsubsection*{Tidal Power in the Jupiter-Io System}

Let us now put an RQF around Jupiter's moon Io and use equation (\ref{TidalPower}) to compute the work done on Io during one orbit.  As mentioned above, Io is tidally locked to Jupiter ($\delta = 0$) so the coordinates $\bar{X}^I$ are just the $X^I$ coordinates.  However, we {\it do} need to take into account that the tidal bulge from the external field at a given time induced on Io does not occur instantaneously - there is a lag time $\tau$ (see figure \ref{JupiterIo}).  The tidal work per unit time is then
\begin{align}\label{TidalPowerIo}
P_{\mathtt{tidal}} & = -\frac{c^2}{2} E_{IJ}(t) \dot{Q}^{IJ}(t) \nonumber \\
& = \frac{c^4}{4G} k_{2,\mathtt{Io}} R_{\mathtt{Io}}^5 E_{RR}(t) \dot{E}_{RR}(t-\tau)
\end{align}
where we have made use of equation (\ref{LoveRelation}).  Here, $E_{RR}(t) = -\frac{2G M_J}{c^2 R(t)^3}$ is the radial-radial component of the tidal field of Jupiter.  The lag time, $\tau$, is unfortunately not well known but, according to \cite{Yoder}, should go like the period of the orbit divided by the dissipation factor, $Q$, which is typically assumed to have the value $Q \simeq 100$ for Io \cite{MurrayDermott}.  This means $\tau$ should be approximately $\tau \simeq 25$ minutes for Io.  The Love number for Io is also not well-known because its calculation relies on knowing the rigidity, $\mu$, of Io.  The usual way to get around this is to assume that Io has the rigidity of a typical rocky body, $\mu \simeq 5 \times 10^{10} \, Pa$.  This yields an approximate Love number for Io of $k_{2,\mathtt{Io}} = 0.03$ \cite{MurrayDermott}.  Using these parameters, we then numerically integrate the power over one full orbit and divide by the orbital period to find the average power transferred to Io.  This yields an average power of $\langle \mathtt{P} \rangle \simeq 1.6 \times 10^{14} \, W$.  The currently accepted value based on various models and observations for the heat generated through tidal interactions in Io is $0.6-1.6 \times 10^{14} \, W$ \cite{Lopes}.   Our value agrees well with observation and we expect that with better knowledge of the Love number, tidal lag time, and a more accurate description of the time dependence of the quadrupole moment our set up could be used to compute the even {\it more} accurately the actual amount of tidal heating in Io.  The main lesson from this exercise, however, is that equation (\ref{TidalPower}) {\it is} the correct expression for tidal power in general relativistic notation and has utility in real-life problems like the one above.

\begin{figure}
\begin{center}
\includegraphics[scale=0.45]{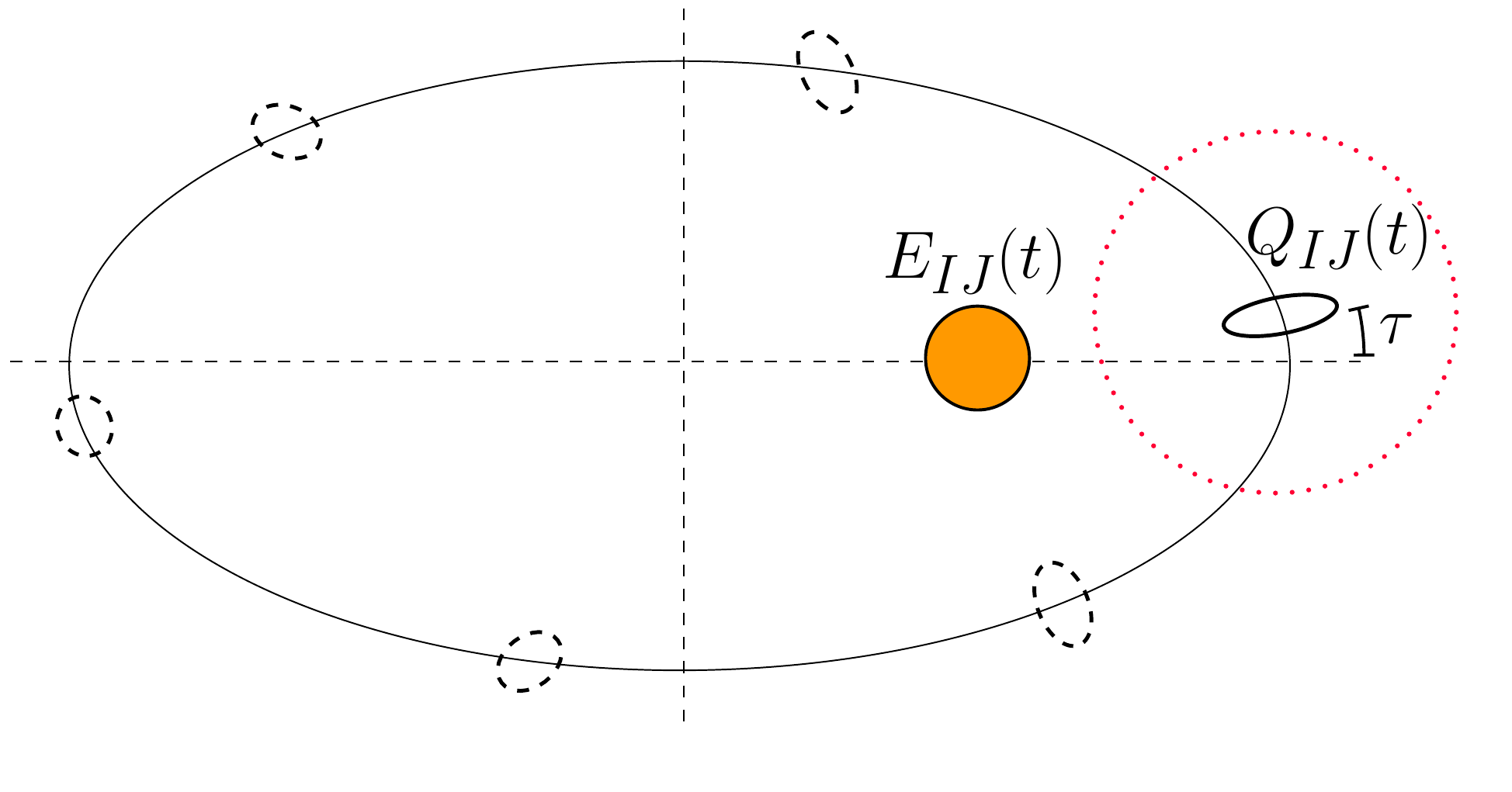}
\caption{The tidal field of Jupiter, $E_{IJ}$, induces a quadrupole moment, $Q_{IJ}$, in Io which is delayed by a lag time $\tau$ due to the finite time it takes for deformation to fully set in.  Furthermore, the elliptical nature of the orbit means that the strength of Jupiter's tidal field varies at Io and, in turn, the degree to which Io gets squashed varies too.  This results in a continuous transfer of gravitational energy from Jupiter to Io which is dissipated thermally.  By placing an RQF around Io and averaging equation (\ref{TidalPower}) over one full orbit, we determine the average power of this tidal heating.}\label{JupiterIo}
\end{center}
\end{figure}

\subsubsection*{Recession of the Moon}

In our next example, we consider the Earth-Moon system and centre our RQF around the Earth (ignoring the motion of the Earth relative to the distant stars).  Since the Earth-Moon orbit has negligible eccentricity, the quadrupole moment and tidal field do not vary appreciably during the orbit.  However, the Earth {\it does} rotate ahead of the Moon and, as a result, it carries its bulge ahead of their common axis by an angle $\delta$ (see figure \ref{EarthMoon}).  For concreteness, let us take the Earth-Moon system to be connected by the $X$-axis, with the $Z$-axis running along the rotational axis of the Earth.  We will then be interested in computing the rate at which the $Z$-component of the angular momentum of the Earth, $J^Z$, changes.  This is just given by the equation for the tidal torque (\ref{TidalTorque}).  We find that
\begin{align}\label{TidalTorqueEarthStep1}
\tau^Z_{\mathtt{tidal}} = \frac{3}{2} c^2 E_{XX} Q_{XY}
\end{align}
where we have used the fact that $E_{IJ}$ is diagonal and $E_{YY} = -\frac{1}{2} E_{XX}$.

To relate the quadrupole moment to the tidal field tensor we first   need to work out how $Q_{XY}$ is related to the components of $\bar{Q}_{\bar{I}\bar{J}}$.  The transformation between $X^I$ and $\bar{X}^{\bar{I}}$ coordinates is simply a rotation about the $Z$-axis by an angle $\delta$.  Specifically, $X = \bar{X} \cos \delta - \bar{Y} \sin \delta$ and $Y = \bar{X} \sin \delta + \bar{Y} \cos \delta$.  Therefore, a quick calculation leads to $Q_{XY} = \frac{3}{4} \bar{Q}_{\bar{X}\bar{X}} \sin (2 \delta)$.  Now we can use equation (\ref{LoveRelation}) which relates $\bar{Q}_{\bar{X}\bar{X}}$ to $E_{XX}$, the tidal field from the Moon at Earth, to show
\begin{align}\label{TidalTorqueEarth}
\tau^Z_{\mathtt{tidal}} = -\frac{3}{2} k_{2,\mathtt{E}} G M_M^2 \frac{R_E^5}{R_{EM}^6} \sin (2\delta).
\end{align}
This reproduces the Newtonian result exactly (see equation 8.20 in reference \cite{StaceyDavis}) and, using the measured values of $\delta = 2.89^\circ$ and $k_{2,\mathtt{E}} = 0.245$ along with standard values for all other parameters, gives a net torque of $4.4 \times 10^{16} \, kg \, m^2 \, s^{-2}$.  As discussed above, this torque transfers angular momentum to the orbital motion of the Earth-Moon system and results in a recession rate for the Moon of $37$ millimetres per year which agrees precisely with the observed value \cite{StaceyDavis}.

\begin{figure}
\begin{center}
\includegraphics[scale=0.65]{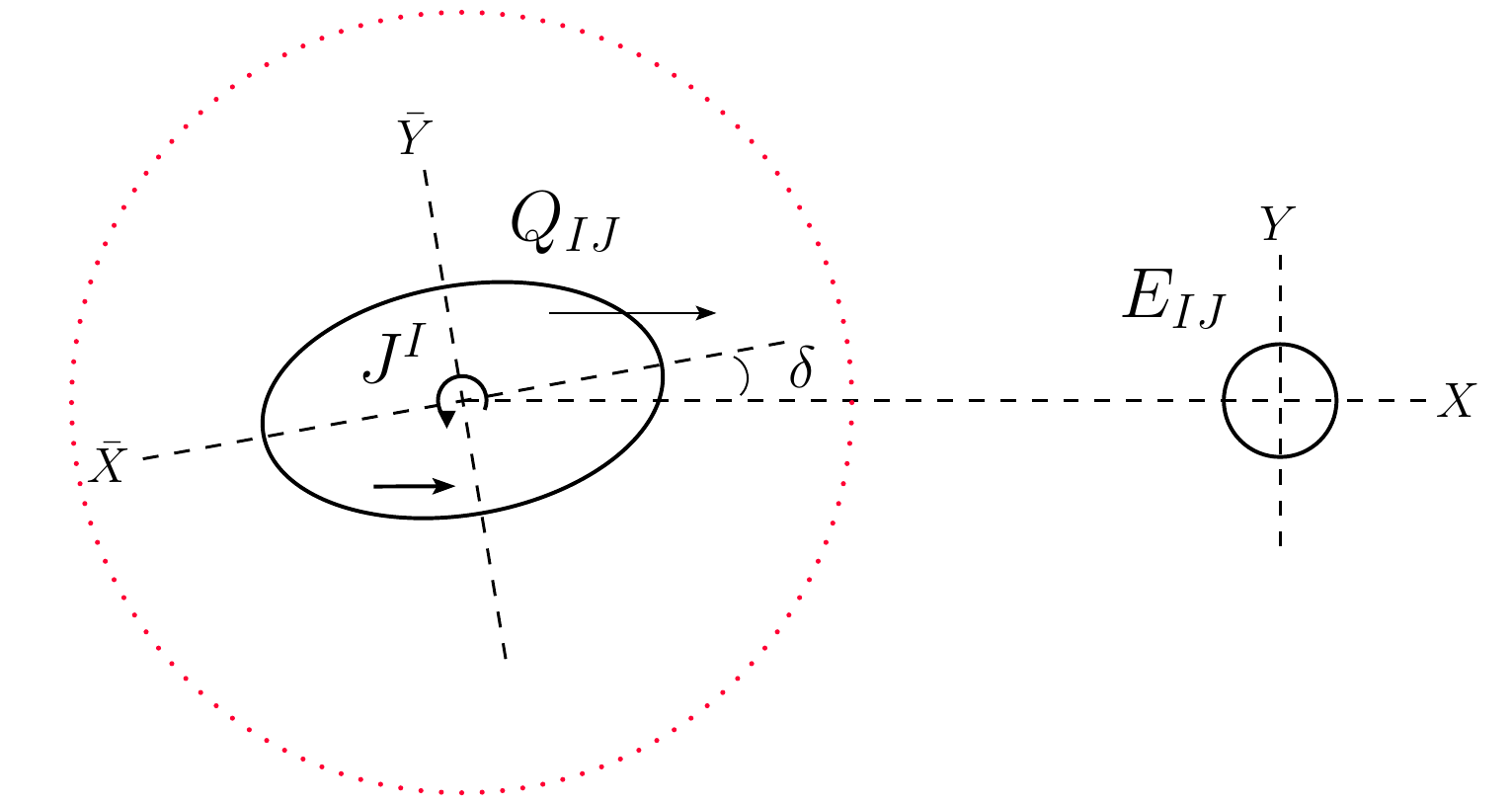}
\caption{The Moon's gravitational field induces a tidal bulge on the Earth, $Q_{IJ}$, which is carried ahead of the Earth-Moon axis by an angle $\delta$ due to the rotation of the Earth, $J^I$.  The tidal field of the Moon, $E_{IJ}$, then pulls on the near-side bulge more strongly than the far-side bulge which results in a net torque slowing down the rotation of the Earth.  Angular momentum is thus transferred to the Earth-Moon orbit resulting in the recession of the Moon.  We analyze this angular momentum transfer by centering an RQF around the Earth and computing the instanteneous torque using equation (\ref{TidalTorque}).}\label{EarthMoon}
\end{center}
\end{figure}

\section{Summary}\label{secSummaryPN}

In this chapter, we used our quasilocal conservation laws to derive a simplified set of post-Newtonian conservation laws.  We achieved this by embedding a rigid quasilocal frame in the general post-Newtonian spacetime and expanding the conservation laws in the slow-motion, weak-field limit.  Immediately, we found that this perturbative quasilocal approach allowed for an elegant description of various quantities in terms of gravitoelectromagnetic fields.  In particular, the gravitational energy flux that the RQF observers measure is simply given by the radial component of the GEM Poynting vector.  More importantly, however, we were able to derive this simple set of conservation laws without the need to introduce pseudotensors.

To this order in the post-Newtonian approximation, we found that one could not compute the change in linear momentum inside the RQF.  It should be reiterated that this was a consequence of the need to go to higher orders in ones calculation than expected when analyzing gravity by perturbing around flat spacetime and not a drawback of the quasilocal approach.  We were, however, able to derive expressions for the rate of change of the energy and angular momentum inside the RQF.  In both cases we found that the change could be expressed in terms of one simple gravitational flux (when matter fluxes were set to zero at the RQF boundary).  To demonstrate the utility of these results, we next looked at them in the context of tidal interactions.

For the energy case, our post-Newtonian conservation law led to the well-known expression for the power of tidal heating, $P_{\mathtt{tidal}} = -\frac{c^2}{2} E_{IJ} \dot{Q}^{IJ}$.  While this is not a new result, it was instructive to see it derived without needing to resort to the pseudotensor approach which includes the unphysical working assumption that gravitational energy can be localized.  In the case of angular momentum, we computed the expression for the torque due to the transfer of gravitational angular momentum via tidal interactions, $\tau^I_{\mathtt{tidal}} = c^2 \epsilon^{IJK} E_{JL} Q_K^{\phantom{K}L}$.  More importantly, however, we found that the quasilocal approach provided us with a deeper understanding of why this torque exists without the need to interpret gravity as a force.  In particular, it is a result of the {\it universal} mechanism that a mass appears to gain momentum when you accelerate towards it.  Furthermore, in order to isolate $\tau^I_{\mathtt{tidal}} $ as the physically relevant torque, we needed to treat the quasilocal vacuum energy density as a {\it real} source of mass which the locally non-inertial RQF observers found added an additional contribution to the change in the angular momentum of the system.  This lends further credibility to the notion encountered in previous chapters that gravitational vacuum contributions, while seemingly unnatural at first sight, may in fact be entirely physical.

Finally, we looked at two examples of tidal interactions within our solar system to further demonstrate that the RQF formalism is not only conceptually interesting but, at the end of the day, can be used for {\it practical} applications too.  Specifically, we first estimated the average rate at which gravitational energy is transferred from Jupiter to Io and dissipated via friction; the predicted rate of tidal heating matched closely with observation.  Next, we looked at the Earth-Moon system to compute the torque that the Moon exerts on the Earth.  Our resulting expression reproduced the Newtonian result exactly and thus agreed precisely with observation.  While many of these results have been calculated by other means already, we present them here to demonstrate how one would arrive at them from a general relativistic starting point.  In doing so, we set the stage for easily computing relativistic corrections to these Newtonian effects.  In the bigger picture, it also shows how one can begin with completely general quasilocal conservation laws and obtain results of everyday importance.

\chapter{Conclusions}\label{chConclusions}

In this thesis, we have introduced the concept of a rigid quasilocal frame (RQF).  We have provided convincing evidence for the generic existence of RQFs, not just in flat spacetime, but in arbitrary curved spacetimes.  Moreover, we have seen that this frame exhibits the full six motional time-dependent degrees of freedom we are familiar with from Newtonian mechanics.  This result is intimately related to the fact that the RQF construct always admits precisely three boost and three rotation conformal Killing vector fields associated with the action of the Lorentz group on a two-sphere.  Replacing the local matter stress-energy-momentum tensor with the Brown-York matter {\it plus} gravity boundary SEM tensor, we further made use of these CKVs, along with the RQF observers' four-velocity, to uniquely define the energy, momentum, and angular momentum inside an RQF without relying on the pre-general relativistic practice of appealing to spacetime symmetries.  This makes the quasilocal approach significantly more useful than a traditional local approach for studying spatially extended, dynamical general relativistic systems.  Notably, since gravitational contributions to energy, momentum, and angular momentum are inherently non-local, the RQF structure naturally includes both matter {\it and} gravitational effects.

With physically sensible definitions of energy, momentum, and angular momentum in hand, we have also shown that the RQF construction gives rise to a set of completely general conservation laws which describe the changes in these quantities in a system in terms of fluxes across the boundary.  Since an RQF is a congruence with zero expansion and shear this allows for a clean identification of only the most relevant fluxes crossing the boundary - that is, fluxes due merely to changes in the size or shape of the boundary are eliminated.  The resulting fluxes are simple, exact, and quantified in terms of operationally-defined geometrical quantities on the boundary.  Furthermore, they explain at a deeper level the mechanisms behind gravitational energy and momentum transfer across a boundary by way of the equivalence principle.  In particular, when we accelerate relative to a mass, the {\it energy} changes at a rate proportional to our acceleration times the {\it momentum} inside - this idea is captured in the gravitational Poynting vector, $\alpha_a \mathcal{P}^a$.  Similarly, the {\it momentum} inside the RQF changes at a rate proportional to our acceleration times the {\it energy} inside, which we see in the gravitational momentum flux $\frac{1}{c^2} \mathcal{E} \alpha_a \phi^a$.  Remarkably, this new insight has consequences for how we should understand everyday occurrences like a falling apple - that is, the change in energy of the apple involves frame dragging, which is proportional to $\mathcal{P}^a$, while the change in momentum involves the extrinsic curvature near the apple, which is proportional to $\mathcal{E}$.  Our naive general relativistic intuition tells us that these effects should be so tiny that they should not matter but they are multiplied by {\it huge} numbers (like $c^4/G$) to give rise to macroscopic effects.  This is how general relativity {\it universally} explains the transfer of energy and momentum but we needed rigid quasilocal frames to uncover this beautiful property of nature.

By investigating RQFs perturbatively or under certain simplifying conditions we have also demonstrated the practical utility of this approach to general relativity.  Highlights include:

\vspace{-2pt}\noindent (a) In Chapter \ref{chRigidRevisited}, while demonstrating the existence of RQFs in flat spacetime, we analyzed time-dependent rotations and discovered that the reason these are so difficult to treat relativistically and, relatedly, the reason that Ehrenfest's rigid rotating disk paradox has gone unsolved for so long is that rotation introduces a non-locality in time.  In other words, in order to maintain rigidity, one needs to know, not only the instantaneous rotation rate, but {\it infinitely} many of its derivatives - that is, the entire history of the system.  This is doable for the boundary of a system (i.e., an RQF) but not for the three-dimensional volume inside.

\vspace{-2pt}\noindent (b) We also considered RQFs in the small-sphere limit to derive many of our results.  One particular example that we analyzed in Chapters \ref{chNova} and \ref{chLocalQuasilocal}, that of Bell's spaceship accelerating through an electromagnetic field, led to a very interesting conclusion.  We found that the change in electromagnetic energy inside the spaceship was made up of two pieces: the usual electromagnetic Poynting flux (combined with an interesting time-dilation effect) accounted for half the change while the {\it gravitational} Poynting vector equally contributed to make up the other half.  General relativistic effects were {\it not} just a small correction.  The upshot of this observation is that, in general, electromagnetism in flat spacetime does {\it not} tell you what is actually going on.  This is perhaps not so surprising when you think about it in terms of Einstein's equations, $G_{ab} = 8\pi T_{ab}$;  when an electromagnetic field is present $T_{ab} \neq 0$, so we cannot work as though $G_{ab} = 0$ and expect to correctly explain what is going on.

\vspace{-2pt}\noindent (c) In Chapter \ref{chArch}, we considered the RQF linear momentum conservation law in the context of stationary observers and fields and, in doing so, derived for the first time an {\it exact} fully general relativistic analogue of Archimedes' law.  In essence, this law demonstrates that the weight of the matter and gravitational fields contained in a finite region of space is supported by the stresses (buoyant forces) acting on the boundary of that region.

\vspace{-2pt}\noindent (d) Expanding our conservation laws next in a post-Newtonian approximation in Chapter \ref{chPN}, we derived a simple set of laws that describe non-relativistic systems bound by mutual gravitational attraction amongst its constituent particles.  In turn, we used these laws in the context of tidal interaction to obtain expressions for the rates of gravitational energy and angular momentum transfer between two bodies - that is, the tidal heating and tidal torque - without the need to define unphysical pseudotensors.  Testing these formulas out in the solar system we were able to accurately predict both the tidal heating of Io by Jupiter as well as the rate at which the Earth's spin angular momentum is converted into Earth-Moon orbital angular momentum.  More importantly, however, the RQF approach explained the transfers of energy and momentum not as the difference of forces acting on a tidal bulge, but rather more fundamentally in the language of the equivalence principle in terms of ``accelerations relative to mass''.

There was a recurring theme in trying to study relativistic systems by moving from the RQF formalism to a perturbative approximation.  We found that we were able to explain things in terms of whatever scenario we were perturbing around (e.g. post-Newtonian being Newtonian physics, small-sphere being general relativity with a {\it local} viewpoint) but that this was not a natural way to interpret general relativity.  The terms that you encounter do not individually have any geometrical significance.  On the other hand, the quasilocal approach {\it always} gave very simple, exact, geometrical descriptions of the mechanisms at work.  This is precisely what we want if we are going to properly attack the problem of motion in general relativity.

To summarize, before we can advance on to things like quantum gravity, we first need to properly understand what general relativity is telling us about the nature of gravity and the universe.  And what we are seeing here is strong evidence that the universe is holographic.  The most basic entities in physics - energy, momentum, and angular momentum - are {\it not} localizable quantities.  If spacetime is not dynamical then, sure, we can get away with doing things locally but, at a fundamental level, forgetting about spacetime dynamics is a huge step away from reality.  We are really deeply ingrained with a local way of thinking.  It is time to start thinking about things from a quasilocal perspective and try to understand what comes with this completely different way of thinking.  One potential new insight is that the gravitational vacuum may actually be physical.  If this is true, then it would be crucially important to accurately understand; perhaps it has effects similar to that of the cosmological constant or maybe it plays an important role as a Poincar\'{e} stress for describing stable fundamental particles.  A quasilocal approach could potentially be used to understand the weird non-local nature of quantum particles too.  These are things that should be built into a theory of quantum gravity from the outset.  Whatever the case may be, if there can be only one lesson to carry away from this thesis, it is that locality is limited and, once we accept this, we can employ rigid quasilocal frames as a powerful tool for {\it properly} understanding the universe.


\appendix

\chapter*{Appendices}
\addcontentsline{toc}{chapter}{APPENDICES}

\chapter{Useful Properties of the Boost and Rotation Conformal Killing Vectors}\label{AppendixCKV}

As we have seen throughout this work, the three boost and three rotation conformal Killing vectors (CKVs) that are admitted on the RQFs boundary play an important roles in both demonstrating that an RQF exhibits the full six Newtonian motional degrees of freedom as well as in defining physically sensible definitions of energy, momentum, and angular momentum inside the RQF.  As such, these CKVs come up frequently enough that it will be useful to clearly define these objects and summarize some of their useful properties.  For simplicity we will limit specific examples in our discussion to the case of a round-sphere RQF but all of these results can be generalized to an arbitrary closed two-surface. 

We begin by taking the three functions $r^I = ( \sin \theta \cos \phi, \sin \theta \sin \phi , \cos \theta)$ as a basis for the $\ell=1$ spherical harmonics.  As usual, we raise and lower indices in the Cartesian $x^I$ coordinates with the Kronecker delta function (e.g., $r_I = \delta_{IJ} r^J$).  It is straightforward to show that these functions satisfy the identity, $\delta_{IJ} r^{I} r^{J} = 1$.  On a close spatial two-surface with metric unit round-sphere metric $\mathbb{S}_{ij} = \mathrm{diag} (1,\, \sin^2 \theta)$ in coordinates $x^i = (\theta, \phi)$ we then construct, using the $r^I$ functions, two sets of $\ell=1$ spherical harmonic covector fields.  The first set are the three boost generators,
\begin{align}\label{boost}
\mathbb{B}^I_i (\theta, \phi) := \mathbb{D}_i r^I = \left( \begin{array}{ccc}
\cos \theta \cos \phi & \cos \theta \sin \phi & - \sin \theta \\
- \sin \theta \sin \phi & \sin \theta \cos \phi & 0
\end{array} \right),
\end{align}
where $\mathbb{D}_i$ is the covariant derivative operator associated with the unit round sphere metric, $\mathbb{S}_{ij}$.  The second set are the three rotation generators,
\begin{align}\label{rotation}
\mathbb{R}^I_i (\theta, \phi) := \mathbb{E}_i^{\phantom{i}j} \mathbb{B}^I_j = \left( \begin{array}{ccc}
-\sin \phi & \cos \phi & 0 \\
- \sin \theta \cos \theta \cos \phi & - \sin \theta \cos \theta \sin \phi & \sin^2 \theta
\end{array} \right),
\end{align}
where $\mathbb{E}_{ij}$ denotes the volume element associated with $\mathbb{S}_{ij}$ (and $\mathbb{E}_{\theta \phi} = \sin \theta$).  The contravariant form of these generators is given by $\mathbb{B}_I^i:=\delta_{IJ}\mathbb{S}^{ij}\mathbb{B}^J_j$ and $\mathbb{R}_I^i:=\delta_{IJ}\mathbb{S}^{ij}\mathbb{R}^J_j$ where $\mathbb{S}^{ij}$ is the matrix inverse of $\mathbb{S}_{ij}$.  Notice that the sign of the rotation generator has been chosen to agree with the ``right-hand rule'' of physics.  For example, a rotation about the $X^3$-axis (or $z$-axis) should be in the positive $\phi$ direction (with no $\theta$ component).  The trade off is that the commutators of our generators satisfy a Lorentz algebra that differ by a minus sign from the the usual form \footnote{See equation (11.99) of \cite{Jackson3rdEdition} for the usual form of the Lorentz algebra.}.  In particular,
\begin{align}
\left[ \mathbb{R}^I , \mathbb{R}^J \right] = - \epsilon^{IJ}_{\phantom{IJ}K} \mathbb{R}^K, \qquad \left[ \mathbb{R}^I , \mathbb{B}^J \right] = - \epsilon^{IJ}_{\phantom{IJ}K} \mathbb{B}^K, \qquad \left[ \mathbb{B}^I , \mathbb{B}^J \right] = \epsilon^{IJ}_{\phantom{IJ}K} \mathbb{R}^K
\end{align}
where $B^I := B^{Ii} \frac{\partial}{\partial x^i}$, $R^I := R^{Ii} \frac{\partial}{\partial x^i}$, and $\epsilon^{IJK}$ is the alternating symbol.

The definition of the rotation generator, equation (\ref{rotation}), can be shown to be equivalent to the (often more useful) relation
\begin{align}\label{defnRotationCKV}
\mathbb{R}^I_i = \epsilon^I_{\phantom{I}JK} r^J \mathbb{B}^K_i.
\end{align}

It is straightforward to show that these vectors satisfy the following additional properties:
\begin{align} \label{RandBmiscprops}
\begin{array}{lll}
r_I B^I_i = 0, & \qquad \mathbb{D}_i \mathbb{B}^I_j = - \mathbb{S}_{ij} r^I, & \qquad \mathbb{D} \cdot \mathbb{B}^J = -2 r^J, \\
r_I R^I_i = 0, & \qquad \mathbb{D}_i \mathbb{R}^I_j = \mathbb{E}_{ij} r^I, & \qquad \mathbb{D} \cdot \mathbb{R}^J = 0.
\end{array}
\end{align}

Furthermore, using these relations, it is simple verify that these are conformal Killing vectors as expected\footnote{In fact, $\mathbb{R}^I_i$ is actually a pure Killing vector since  $\mathbb{D}_{(i}\mathbb{R}_{j)}^J = 0$}:
\begin{align}
\mathbb{D}_{<i}\mathbb{B}_{j>}^J=0=\mathbb{D}_{<i}\mathbb{R}_{j>}^J.
\end{align}
where angle-brackets are used to denote the symmetric trace-free part. 

Finally, we give relations for the various contractions of these conformal Killing vectors
\begin{align}\label{completeness}
\begin{array}{ll}
\mathbb{S}^{ij} \mathbb{B}_{i}^{I}\mathbb{B}_{j}^{J} = \mathbb{S}^{ij} \mathbb{R}_{i}^{I}\mathbb{R}_{j}^{J} = P^{IJ}, & \qquad \mathbb{S}^{ij} \mathbb{B}^I_i \mathbb{R}^J_j = \epsilon^{IJK} r_K, \\
\mathbb{E}^{ij} \mathbb{B}_{i}^{I}\mathbb{B}_{j}^{J} = \mathbb{E}^{ij} \mathbb{R}_{i}^{I}\mathbb{R}_{j}^{J} = \epsilon^{IJK} r_K,  & \qquad \mathbb{E}^{ij} \mathbb{B}^I_i \mathbb{R}^J_j = - P^{IJ}, \\
\delta_{IJ} \mathbb{B}_{i}^{I} \mathbb{B}_{j}^{J} = \delta_{IJ} \mathbb{R}_{i}^{I}\mathbb{R}_{j}^{J} = \mathbb{S}_{ij}, & \qquad \delta_{IJ} \mathbb{B}_{i}^{I} \mathbb{R}_{j}^{J} = - \mathbb{E}_{ij}, \\
\epsilon_{IJK} \mathbb{B}_{i}^{I} \mathbb{B}_{j}^{J} = \epsilon_{IJK} \mathbb{R}_{i}^{I}\mathbb{R}_{j}^{J} = \mathbb{E}_{ij} r_K, & \qquad \epsilon_{IJK} \mathbb{B}_{i}^{I} \mathbb{R}_{j}^{J} = \mathbb{S}_{ij}.
\end{array}
\end{align}
where $P^{IJ} := \delta^{IJ} - r^{I} r^{J}$ projects vectors perpendicular to the radial direction.


\bibliographystyle{abbrv}
\cleardoublepage 
\phantomsection  
\renewcommand*{\bibname}{References}

\addcontentsline{toc}{chapter}{\textbf{References}}

\bibliography{ThesisBib}


\end{document}

%% file: FrontPages.tex
\pagestyle{empty}
\pagenumbering{roman}

\begin{titlepage}
        \begin{center}
        \vspace*{1.0cm}

        \Huge
        {\bf Rigid Quasilocal Frames}

        \vspace*{1.0cm}

        \normalsize
        by \\

        \vspace*{1.0cm}

        \Large
        Paul McGrath \\

        \vspace*{3.0cm}

        \normalsize
        A thesis \\
        presented to the University of Waterloo \\ 
        in fulfillment of the \\
        thesis requirement for the degree of \\
        Doctor of Philosophy \\
        in \\
        Physics \\

        \vspace*{2.0cm}

        Waterloo, Ontario, Canada, 2014 \\

        \vspace*{1.0cm}

        \copyright\ Paul McGrath 2014 \\
        \end{center}
\end{titlepage}

\pagestyle{plain}
\setcounter{page}{2}

\cleardoublepage
 
  \noindent
I hereby declare that I am the sole author of this thesis. This is a true copy of the thesis, including any required final revisions, as accepted by my examiners.

  \bigskip
  
  \noindent
I understand that my thesis may be made electronically available to the public.

\cleardoublepage


\begin{center}\textbf{Abstract}\end{center}

This thesis begins by introducing the concept of a rigid quasilocal frame (RQF) as a geometrically natural way to define an extended system in the context of the dynamical spacetime of general relativity.  An RQF is defined as a two-parameter family of timelike worldlines comprising the worldtube boundary (topologically $\mathbb{R} \times S^2$) of the history of a finite spatial volume with the rigidity conditions that the congruence of worldlines is expansion-free (the ``size'' of the system is not changing) and shear-free (the ``shape'' of the system is not changing).  We demonstrate that this frame exists in flat {\it and} arbitrary curved spacetimes and, moreover, exhibits the full six motional time-dependent degrees of freedom we are familiar with from Newtonian mechanics.  The latter result is intimately connected with the fact that a spatial slice through the RQF - having a two-sphere topology - always admits precisely six conformal Killing vector (CKV) fields (three boosts and three rotations) associated with the action of the Lorentz group on a two-sphere.  These CKVs, along with the four-velocity of observers on the RQF, are then used to quasilocally define the energy, momentum, and angular momentum inside an RQF without relying on the pre-general relativistic practice of appealing to spacetime symmetries.  These quasilocal definitions for energy, momentum, and angular momentum also involve replacing the local matter-only stress-energy-momentum (SEM) tensor with the Brown-York matter {\it plus} gravity boundary SEM tensor.  This allows for the construction of completely general conservation laws which describe the changes in a system in terms of fluxes across the boundary.  Furthermore, since an RQF is a congruence with zero expansion and shear only relevant fluxes appear in these conservation laws - that is, fluxes due merely to changes in the size or shape of the boundary are eliminated.

These resulting fluxes are simple, exact, and quantified in terms of operationally-defined geometrical quantities on the boundary and we show that they explain at a deeper level the mechanisms behind gravitational energy and momentum transfer by way of the equivalence principle.  In particular, when we accelerate relative to a mass, the {\it energy} changes at a rate proportional to our acceleration times the {\it momentum} (and we propose an {\it exact} gravitational analogue of the electromagnetic Poynting vector to capture this idea).  Similarly, the {\it momentum} of that object changes at a rate proportional to our acceleration times the {\it energy}.  This new insight has fascinating consequences for how we should understand everyday occurrences like a falling apple - that is, the change in energy of the apple involves frame dragging while the change in momentum involves extrinsic curvature effects near the apple.  Our naive general relativistic intuition tells us that these quantities should be so tiny that they should be negligible and, indeed, they {\it are} tiny but they are multiplied by {\it huge} numbers (like $c^4/G$) to give rise to macroscopic effects.  This is how general relativity {\it universally} explains the transfer of energy and momentum but we needed rigid quasilocal frames to uncover this beautiful property of nature.

Using the RQF formalism we also investigate a variety of specific problems.  In particular, while looking at time-dependent rotations we discover that the reason Ehrenfest's rigid rotating disk paradox has gone unsolved for so long is that rotation introduces a subtle non-locality in time.  By this we mean that, in order to maintain rigidity while undergoing time-dependent rotation, one needs to know, not only the instantaneous rotation rate, but the {\it entire} history of the motion.  This makes it impossible to keep a volume of observers rigid but is doable with an RQF.  We also consider RQFs in the small-sphere limit to derive many of our results and one example with particularly interesting consequences involves Bell's spaceship accelerating through an electromagnetic field.  Here, we show that the change in electromagnetic energy inside the spaceship is made up of two pieces: the usual electromagnetic Poynting flux accounts for half the change while the {\it gravitational} Poynting vector equally contributes to make up the other half.  This means that electromagnetism in flat spacetime generically does {\it not} tell you what is actually going on.  Rather, the curvature due to the electromagnetic field necessitates a fully general relativistic treatment to get the whole story.  We also use the RQF linear momentum conservation law in the context of stationary observers and fields to derive, for the first time, an exact fully general relativistic analogue of Archimedes' law.  In essence, this law demonstrates that the weight of the matter and gravitational fields contained in a finite region of space is supported by the stresses (buoyant forces) acting on the boundary of that region.  Furthermore, in a post-Newtonian approximation, we derive a simple set of quasilocal conservation laws which describe non-relativistic systems bound by mutual gravitational attraction.  In turn, we use these laws to obtain expressions for the rates of gravitational energy and angular momentum transfer between two tidally interacting bodies - that is, the tidal heating and tidal torque - without the need to define unphysical pseudotensors.  Moreover, the RQF approach explains these transfers of energy and momentum again, not as the difference of forces acting on a tidal bulge, but instead more fundamentally in the language of the equivalence principle in terms of ``accelerations relative to mass''.

Throughout this work we demonstrate that the RQF approach always gives very simple, geometrical descriptions of the physical mechanisms at work in general relativity.  Given that this approach also includes both matter {\it and} gravitational energy, momentum, and angular momentum and does not rely on spacetime symmetries to define these quantities, we argue that we are seeing here strong evidence that the universe is actually quasilocal in nature.  We are really deeply ingrained with a local way of thinking, so shifting to a quasilocal mindset will require great effort, but we contend that it ultimately leads to a deeper understanding of the universe.

\cleardoublepage

\begin{center}\textbf{Acknowledgements}\end{center}

This thesis is a reflection not only of my own hard work but also of the extraordinary academic and emotional support I have been so fortunate to receive on my journey.  I am eternally grateful to my supervisor, Robert Mann.  Robb welcomed me into the world of gravitational physics research as an undergraduate nearly eight years ago.  At the time, all I knew about general relativity was that it sounded cool.  That was no problem for Robb.  He invited me into his office, described a bunch of advanced projects in a way that even {\it I} could understand, told me to think about which one I was most interested in, and then offered me a choice between two books: Weinberg or Wald?\footnote{I chose Wald.}  Over the next few weeks I got a crash course in general relativity, began doing {\it real} theoretical research for the first time, and discovered that Robb must have figured out a way to slow down the passage of time.  How else could he do all that he does and still be able to dedicate so much time to his students?  In all seriousness though, the instruction, guidance, and friendship I have received from Robb over the years have played a vital role in crafting me into the researcher and person I am today.

I am also forever indebted to my co-supervisor, Richard Epp, who I can most certainly say this thesis would not be possible without - after all, rigid quasilocal frames were his idea!  It has been a privilege and a pleasure to learn from and work with Richard.  He has a way of explaining even the most intimidating concepts in a simple and natural way.  Much of my physics intuition I owe to the countless hours I have spent in his office discussing research and physics in general.  I truly feel that I lucked out in having the opportunity to study his idea of rigid quasilocal frames; not only do I enjoy it immensely, but I think it will, at the very least, play an important role in taking our understanding of the universe to the next level.

On a more personal level, I certainly would not be where I am today without the love and support of my fianc\'{e}e, Eleanor.  Whenever I doubted my abilities as a researcher, she was there to encourage me and get me back on track.  Knowing that she is proud of me and my accomplishments has been a huge driving force in writing my thesis and finishing my degree.  Eleanor also helped me find the right balance between work and play and made sure I never became too disconnected from the ``real'' world.  I still can't believe how lucky I am to have found her.

Of course, I would never have made it this far without the unwavering support of my parents, Linda and William.  Even though it looks like another language to them, they still get flush with pride every time I try to show them something I'm working on.  They often ask where I get my ``smarts'' from; I know it was from them.  It is only because I have my mum's patience and perseverance that I'm able to keep pushing forward when a calculation starts running into the tens (or even hundreds) of pages.  And I have my dad to thank for teaching me not to take any new idea for granted - to always scrutinize it first and be sure that it really {\it does} make sense.

I also have to thank the rest of my family for their continual support and encouragement.  My brothers, David and Adam, have always been there for me and, being much more extroverted than myself, have kept my social skills more or less up to par over the years.  They are not with us any more, but my grandparents also played a very important role in making me the person I am today.  In fact, I may have never pursued my passion for physics if not for the last advice my dad's mum gave me, ``You don't need to be rich, you just need to be happy."  I would also like to thank {\it all} of the Kohlers for welcoming me into their family with open arms.

In the academic world, I am also grateful to my committee members, Achim Kempf and Niayesh Afshordi, for their valuable feedback throughout my degree.  I would also like to thank Stefan Idziak both for getting me hooked on research after my first undergraduate year and his friendship over the years since then.  My gratitude also goes out to a particularly unique individual I met during my years at the University of Alberta, Don Page.  From my interactions with Don I saw first hand the importance of approaching problems unconventionally.  My high school physics teacher, Daniel Muttiah, also deserves praise for being the first to show me just how beautiful physics could be.  Last, but certainly not least, I would also like to thank Judy McDonnell for her tireless efforts assisting in administrative matters. 

One last group who undoubtedly have helped me along the way are my friends and peers and a few deserve special mention.  Andrew Louca is just an amazing friend in all respects; our shared love of science, music, comedy, and much more has made him someone I can turn to for anything.  Sean Stotyn is another person I have clicked with since we first met; our unusual shared sense of humour has made him someone I can always be myself around.  Moreover, being a few steps ahead of me in the academic world meant that I could always turn to him for much needed advice.   I would also like to thank Alex Venditti for his friendship and many interesting physics discussions and Chris Saayman for providing much needed distractions from research.  Thanks also go out to my research group - in particular, Miok Park, Danielle Leonard, Melanie Chanona (whose hard work was crucial for obtaining many of the results in chapter 6), Wilson Brenna, Eric Brown, Aida Ahmadzadegan, and, last but not least, Alex Smith.





\cleardoublepage

\renewcommand\contentsname{Table of Contents}
\tableofcontents
\cleardoublepage
\phantomsection


\addcontentsline{toc}{chapter}{List of Figures}
\listoffigures
\cleardoublepage
\phantomsection		


\pagenumbering{arabic}